\documentclass[a4paper,11pt]{article}
\pdfoutput=1 % if your are submitting a pdflatex (i.e. if you have
\usepackage{jheppub}                                         
\usepackage{color}
\usepackage[svgnames,table, usenames,dvipsnames]{xcolor}
\usepackage{tensor}
\usepackage{epstopdf}
\usepackage[section]{placeins}
\usepackage{longtable}
\usepackage{multirow}
\usepackage{pdflscape}
\usepackage{booktabs}
\usepackage{tabularx}

\newcommand{\be}{\begin{equation}}
\newcommand{\ee}{\end{equation}}
\newcommand{\ba}{\begin{eqnarray}}
\newcommand{\ea}{\end{eqnarray}}

\newcommand{\beq}{\begin{equation}}
\newcommand{\eeq}{\end{equation}}
\newcommand{\beqa}{\begin{eqnarray}}
\newcommand{\eeqa}{\end{eqnarray}}
\newcommand{\nn}{\nonumber}

\title{\boldmath Thermodynamics of hairy black holes in Lovelock gravity}

\author[a]{Robie A. Hennigar,}
\author[b,a]{Erickson Tjoa}
\author[a]{and Robert B. Mann}

\affiliation[a]{Department of Physics and Astronomy, University of Waterloo, \\
Waterloo, Ontario, Canada, N2L 3G1}
\affiliation[b]{Division of Physics and Applied Physics, School of Physical and Mathematical Sciences, Nanyang Technological University, 
\\ Singapore, Singapore, 637371}
\emailAdd{rhennigar@uwaterloo.ca}
\emailAdd{etjoa002@e.ntu.edu.sg}
\emailAdd{rbmann@uwaterloo.ca}

\abstract{We perform a thorough study of the thermodynamic properties of a class of Lovelock black holes with conformal scalar hair arising from coupling of a real scalar field to the dimensionally extended Euler densities.  We study the linearized equations of motion of the theory and describe constraints under which the theory is free from ghosts/tachyons.  We then consider, within the context of black hole chemistry, the thermodynamics of the hairy black holes in the Gauss-Bonnet and cubic Lovelock theories.  We clarify the connection between isolated critical points and thermodynamic singularities, finding a one parameter family of these critical points which occur for well-defined thermodynamic parameters.  We also report on a number of novel results, including `virtual triple points' and the first example of a `$\lambda$-line'---a line of second order phase transitions---in black hole thermodynamics.   }
 
\begin{document} 
\maketitle
\flushbottom

\section{Introduction}

The study of higher curvature corrections to the Einstein-Hilbert action is an active area of study motivated primarily by attempts to quantize the gravitational field.  While general relativity is non-renormalizable as a quantum field theory, the addition of higher derivative terms to the action can lead to a power-counting renormalizable theory while having negligible influence in the low energy domain \cite{Stelle:1976gc}.  Higher curvature corrections are also present in string theory where the Gauss-Bonnet term appears in the low energy effective action~\cite{Zwiebach:1985uq}. Within the context of the AdS/CFT correspondence, higher curvature corrections appear as $1/N_c$ corrections within the dual CFT or, alternatively, as new couplings between operators in the dual CFT yielding a broader universality class of dual CFTs~\cite{Buchel:2008vz, Hofman:2009ug, Myers:2010jv}.

While well-motivated, higher curvature gravities bring a number of difficulties and potential pathologies along with them, making their investigation a non-trivial undertaking.  For example, the resulting equations of motion will generically feature derivatives of fourth order or higher, and the linearized equations of motion for a graviton perturbation often reveal that the graviton is a ghost.  A number of these issues are alleviated in \textit{Lovelock gravity}~\cite{Lovelock:1971yv}.  Lovelock gravity is the most general torsionless theory of gravity for which the field equations are second order and is the natural generalization of Einstein gravity to higher dimensions.  The essential idea is to augment the Einstein-Hilbert action with the dimensionally continued Euler densities.  The $k^{th}$ term is then either topological or vanishes identically below the critical dimension $d = 2k+1$ where it becomes gravitationally non-trivial.  The theory is ghost-free for a  Minkowski vacuum~\cite{Zumino:1985dp} and also in other maximally symmetric backgrounds provided the coupling constants are constrained. 
Lovelock gravity thus provides a natural testbed for exploring the effects of higher curvature terms on  gravitational physics.   

Recently, Oliva and Ray have shown that it is possible to conformally couple a scalar field to the Lovelock terms while maintaining second order field equations for both the metric and the scalar field~\cite{Oliva:2011np}.  In subsequent work, these authors, along with collaborators, have demonstrated that this theory admits black hole solutions where the scalar field is regular everywhere outside of the horizon and the back-reaction of the scalar field onto the metric is captured analytically~\cite{Giribet:2014bva, Giribet:2014fla, Galante:2015voa}.  This work provided the first example of black holes with conformal scalar hair in $d > 4$ where no-go results had been reported previously~\cite{nogo_hairy}. The obtained solutions are valid for positive, negative and vanishing cosmological constant; however, the AdS case is of especial interest due to the role scalar hair plays in holographic superconductors~\cite{Hartnoll:2008vx,Gubser:2008px}.  

In the present work, our primary concern is the thermodynamics of the AdS hairy black hole solutions of this theory within the context of \textit{black hole chemistry}.  In this framework the cosmological constant is promoted to a thermodynamic parameter~\cite{Creighton:1995au, Caldarelli:1999xj, Dolan:2010ha, Dolan:2011xt} in the first law of black hole mechanics, a result supported by geometric arguments~\cite{Kastor:2009wy, Kastor:2010gq}.  One of the major results to follow was the discovery of critical behaviour for AdS black holes similar to the Hawking-Page transition~\cite{Hawking:1982dh}, but analogous to that seen in everyday thermodynamic systems.  For example, in \cite{Kubiznak:2012wp} it was shown that the charged Schwarzschild-AdS black hole undergoes a small/large black hole phase transition  with the phase diagram and critical exponents identical to those of the van der Waals liquid/gas system.  A cascade of subsequent work established further examples of van der Waals behaviour, triple points, and reentrant phase transitions for AdS black holes \cite{Gunasekaran:2012dq, Hendi:2012um, Altamirano:2013ane, Cai:2013qga, Altamirano:2014tva, Wei:2014hba, Frassino:2014pha}, developed entropy inequalities for AdS black holes \cite{Cvetic:2010jb, Hennigar:2014cfa}, and discussed the notion of a holographic heat engine \cite{Johnson:2014yja}.  For a broad class of metrics these phenomena can be understood in a more general framework \cite{Mandal:2016anc,Majhi:2016txt}. However in the particular case of higher curvature gravity, this framework is not adequate: not only are many of these phenomena observed \cite{Wei:2012ui, Cai:2013qga, Xu:2013zea, Mo:2014qsa, Wei:2014hba, Mo:2014mba, Zou:2014mha, Belhaj:2014tga, Xu:2014tja, Frassino:2014pha, Dolan:2014vba, Belhaj:2014eha, Sherkatghanad:2014hda, Hendi:2015cka, Hendi:2015oqa, Hennigar:2015esa, Hendi:2015psa, Nie:2015zia, Hendi:2015pda, Hendi:2015soe, Johnson:2015ekr, Hendi:2016njy, Zeng:2016aly}, but new behaviour such as multiple reentrant phase transitions \cite{Frassino:2014pha} and isolated critical points \cite{Frassino:2014pha, Dolan:2014vba, Hennigar:2015esa} are seen, the latter corresponding to critical exponents which differ from the mean field theory values.  Isolated critical points have so far only been observed in cubic (and higher) theories of gravity with finely tuned coupling constants~\cite{Dolan:2014vba}.   

Despite the growing literature on the critical behaviour of AdS black holes, relatively little investigation has been carried out for theories coupled to scalar fields.  Those cases which have been studied report van der Waals behaviour \cite{Belhaj:2013ioa, Toldo:2016nia, Miao:2016aol, Miao:2016ieh}.  Previous studies of black hole thermodynamics within the conformal coupling model of Oliva and Ray have focused primarily on the case where the gravitational sector consists of only the Einstein-Hilbert term \cite{Giribet:2014bva, Giribet:2014fla, Hennigar:2015wxa}.  The resulting black holes have been shown to exhibit Hawking-Page type transitions~\cite{Giribet:2014bva, Giribet:2014fla} along with reentrant phase transitions and van der Waals behaviour~\cite{Hennigar:2015wxa}.  More recent work has focused on the determination of boundary terms and the evaluation of the Euclidean action~\cite{Chernicoff:2016jsu} and enforcing causality constraints on the scalar field coupling inherited from the AdS/CFT~\cite{Chernicoff:2016uvq}.  In this work we aim to fill gaps in the existing literature by considering the thermodynamics of these hairy black holes in both Gauss-Bonnet and cubic Lovelock gravity and addressing the issue of vacuum stability.

Our paper is organized as follows.  In Section~\ref{sec:soln} we review the conformal coupling model of Oliva and Ray, derive the resulting charged black hole solutions for general Lovelock gravity, and obtain the thermodynamic quantities which satisfy the first law of thermodynamics.   In Section~\ref{sec:ghosts} we consider the linearized field equations about a maximally symmetric background and derive the constraints on the coupling constants which ensure the graviton is neither a ghost nor a tachyon.  In Section~\ref{sec:gb_case} we specialize to the case of Gauss-Bonnet gravity where we study the criticality of hairy black holes in $5$ and $6$ dimensions with and without electric charge.  In Section~\ref{sec:cl_case} we consider the criticality of hairy black holes in cubic Lovelock gravity in $7$ and $8$ dimensions with and without electric charge.  In each case we observe a rich structure of critical phenomena, including novel examples of ``virtual triple points" and pockets of local thermodynamic stability.  In Section~\ref{sec:cl_ICPs} we study the case of cubic Lovelock gravity.   We find that the hair gives rise to a one-parameter family of isolated critical points which occur under much more general conditions than in previous work, clarifying the relationship between isolated critical points and thermodynamically singular points.  Finally, we show that under certain circumstances the hairy black holes in cubic Lovelock gravity exhibit $\lambda$-lines~\cite{Hennigar:2016xwd}.  These lines of second order phase transitions are the first such examples in black hole thermodynamics, and due to an interesting connection with superfluid $^4$He, we have termed the black holes with this property `superfluid black holes'.

\section{Exact Solution \& Thermodynamics}\label{sec:soln}

We consider a theory containing a Maxwell field and a real scalar field conformally coupled to gravity through a non-minimal coupling between the scalar field and the dimensionally extended Euler densities.  The theory is conveniently written in terms of the rank four tensor~\cite{Oliva:2011np},
\ba 
\tensor{S}{_\mu_\nu^\gamma^\delta} &=& \phi^2 \tensor{R}{_\mu_\nu^\gamma^\delta} - 2 \delta^{[\gamma}_{[\mu} \delta^{\delta]}_{\nu]}\nabla_\rho\phi\nabla^\rho\phi 
- 4\phi \delta^{[\gamma}_{[\mu} \nabla_{\nu]} \nabla^{\delta]} \phi + 8 \delta^{[\gamma}_{[\mu}\nabla_{\nu]}\phi\nabla^{\delta]}\phi \,,
\ea
which transforms homogeneously under the conformal transformation, $g_{\mu\nu} \to \Omega^2 g_{\mu\nu}$ and $\phi \to \Omega^{-1}\phi$ as $\tensor{S}{_\mu_\nu^\gamma^\delta} \to \Omega^{-4} \tensor{S}{_\mu_\nu^\gamma^\delta}$. 
The action for the theory in $d$ spacetime dimensions is given by
\be\label{action} 
{\cal I} =  \frac{1}{16 \pi G}\int d^dx \sqrt{-g} \, \left(\sum_{k=0}^{k_{\rm max}} {\cal L}^{(k)} - 4 \pi G F_{\mu\nu}F^{\mu\nu} \right)
\ee
with $F_{\mu\nu} = \partial_\mu A_\nu - \partial_\nu A_\mu$, the Lagrangian densities given by
\ba
{\cal  L}^{(k)} =   \frac{1}{2^k} \delta^{(k)} \left(a_k R^{(k)} + b_k \phi^{d-4k} S^{(k)} \right)
\ea
where 
\be 
k_{\rm max} = \left[ \frac{d-1}{2}
\right]
\ee
(the brackets denoting the integer part of $(d-1)/2$), with  
\be 
R^{(k)} = \prod_r^k \tensor{R}{^{\alpha_{r}} ^{\beta_{r}} _{\mu_r} _{\nu_r}}\,,  \quad S^{(k)} = \prod_r^k \tensor{S}{^{\alpha_{r}} ^{\beta_{r}} _{\mu_r} _{\nu_r}}\, ,
\ee
and $\delta^{(k)}$ is the generalized Kronecker tensor,
\be 
\delta^{(k)} = (2k)! \, \delta_{[\alpha_1}^{\mu_1} \delta_{\beta_1}^{\nu_1}\cdots \delta_{\alpha_k}^{\mu_k} \delta_{\beta_k ]}^{\nu_k} \, .
\ee
One can obtain the equations of motion for the gravitational field in the standard way, and they can be conveniently written in terms of the generalized Einstein tensor, 
\be 
G_\mu^\nu = - \sum_{k=0}^{k_{\rm max}} \frac{a_k}{2^{k+1}} \delta_{\mu \rho_1 \cdots\rho_{2k}}^{\nu \lambda_1 \cdots \lambda_{2k}} \tensor{R}{^{\rho_1} ^{\rho_2} _{\lambda_1} _ {\lambda_2}} \cdots \tensor{R}{^{\rho_{2k-1}} ^{\rho_{2k}} _{\lambda_{2k-1}} _ {\lambda_{2k}}} \, .
\ee

The theory has stress-energy associated with both the scalar and Maxwell fields, with the former given by
\be 
(T_1){_\mu ^\nu} = \sum_{k=0}^{k_{\rm max}} \frac{b_k}{2^{k+1}} \phi^{d-4k} \delta_{\mu \rho_1 \cdots\rho_{2k}}^{\nu \lambda_1 \cdots \lambda_{2k}}\tensor{S}{^{\rho_1} ^{\rho_2} _{\lambda_1} _ {\lambda_2}} \cdots \tensor{S}{^{\rho_{2k-1}} ^{\rho_{2k}} _{\lambda_{2k-1}} _ {\lambda_{2k}}}
\ee
and the latter,
\be 
(T_2){_\mu ^\nu} = \frac{1}{2}\left( F_{\mu\rho}F^{\nu \rho} - \frac{1}{4} F_{\lambda \rho} F^{\lambda\rho} \delta_\mu^\nu \right) \, .
\ee
The gravitational field equations then read,
\be 
G_{\mu\nu} = (T_1){_{\mu \nu}} + 16 \pi G (T_2)_{\mu \nu} \, .
\ee

By varying the action with respect to the scalar field, one can show that the scalar field must obey the following equation of motion:
\be 
\sum_{k=0}^{k_{\rm max}} \frac{(d-2k)b_k}{2^k} \phi^{d-4k-1} \delta^{(k)} S^{(k)} = 0 \, .
\ee
Note that the above equation of motion ensures that the trace of the stress-energy tensor of the scalar field vanishes on shell, as expected for this conformally invariant theory.  Similarly, varying the action with respect to the Maxwell gauge field $A_\mu$, we obtain the Maxwell equations, 
\be 
\nabla_\mu F^{\mu\nu} = 0 \, .
\ee

Here we are interested in spherically symmetric topological black hole solutions to this theory with a metric of the form  
\be 
ds^2 = -f dt^2 + f^{-1}dr^2 + r^2d\Sigma_{(\sigma)d-2}^2  
\ee
where  $d\Sigma_{(\sigma)d-2}^2$ represents the line element on a  hypersurface of constant scalar curvature equal to $(d-2)(d-3)\sigma$ corresponding to flat, spherical and hyperbolic horizon topologies for $\sigma = 0, +1, -1$, respectively.  We denote the volume of this submanifold as $\Sigma^{(\sigma)}_{d-2}$, which for the case of $\sigma=+1$ reduces to the volume of a sphere,
\be 
\Sigma^{(+1)}_{d-2}  = \frac{2 \pi^{(d-1)/2}}{\Gamma\left(\frac{d-1}{2}\right)} \, .
\ee
 We obtain a solution to the field equations provided $f$ solves the polynomial equation,
\ba\label{mastereqn1}
\sum_{k=0}^{k_{\rm max}} \frac{(d-1)!}{(d-2k-1)!} a_k \left(\frac{\sigma-f}{r^2} \right)^k &=& \frac{16 \pi G (d-1)  M}{\Sigma^{(\sigma)}_{d-2} r^{d-1}} + \frac{(d-1)(d-2)H}{r^d} 
\nn\\
&&\qquad - \frac{8 \pi G (d-1)}{(d-3)}\frac{ Q^2}{r^{2d-4}}
\ea
while the scalar field is given by
\be 
\phi = \frac{N}{r}
\ee
where, in order to satisfy the equations of motion, $N$ must satisfy the constraints
\ba 
\sum_{k=1}^{k_{\rm max}} k b_k \frac{(d-1)!}{(d-2k-1)!} \sigma^{k-1}N^{2-2k} &=& 0 \, ,
\nn\\
\sum_{k=0}^{k_{\rm max}} b_k \frac{(d-1)! \left(d(d-1)+4k^2 \right)}{(d-2k-1)!} \sigma^k N^{-2k} &=& 0 \, .
\label{N-constr}
\ea
Note that in the preceding equations there is only a single unknown, $N$.  Thus, one of these equations serves as  a constraint on the allowed coupling constants.

The electromagnetic field strength is given by
\be 
F = \frac{Q}{r^{d-2}} dt \wedge dr
\ee
where $Q$ is the electric charge and is related to the quantity $e$ used in \cite{Hennigar:2015wxa, Galante:2015voa, Giribet:2014fla} via
\be 
e = -\frac{Q}{(d-3)\sqrt{d-2}} \, .
\ee
The scalar hair enters into eq.~\eqref{mastereqn1} through the term $H$, which is given by,
\be\label{defn_of_H} 
H = \sum_{k=0}^{k_{\rm max}} \frac{(d-3)!}{(d-2(k+1))!}b_k \sigma^k N^{d-2k} \, .
\ee
It is easy to see for the planar case ($\sigma = 0$) that $H=0$, i.e. that planar solutions have no hair.  
 
%In general, $H$ will be a parameter which takes its values in the real numbers.  

The polynomial \eqref{mastereqn1} can be put into a more convenient form with the following rescaling
\ba 
\alpha_0 = \frac{a_0}{(d-1)(d-2)}\, , \quad \alpha_1 = a_1 \, , \quad \alpha_k = a_k \prod_{n=3}^{2k} (d-n) \, \textrm{ for } k \ge 2\, ,
\ea
yielding
\be\label{mastereqn} 
\sum_{k=0}^{k_{\rm max}} \alpha_k \left(\frac{\sigma-f}{r^2} \right)^k = \frac{16 \pi G M}{(d-2)\Sigma^{\sigma}_{d-2} r^{d-1}} + \frac{H}{r^d} - \frac{8 \pi G}{(d-2)(d-3)}\frac{Q^2}{ r^{2d-4}} \, .
\ee

Having the solution in hand, we now turn to a discussion of the thermodynamics.  We construct the first law in extended thermodynamic phase space in analogy with \cite{Kastor:2010gq} where we treat all dimensionful couplings as thermodynamic quantities.  For the mass, temperature and electric potential, straightforward calculation yields
\ba\label{temp_etc} 
M &=& \frac{(d-2) \Sigma_{d-2}^{\sigma}}{16 \pi G} \sum_{k=0}^{k_{\rm max}} \alpha_k \sigma^k r_+^{d - 2k -1} - \frac{(d-2)\Sigma^{\sigma}_{d-2} H}{16 \pi G r_+} + \frac{\Sigma^{\sigma}_{d-2} Q^2}{2(d-3) r_+^{d-3}}
\nn\\
T &=& \frac{f'(r_+)}{4 \pi} = \frac{1}{4 \pi r_+ D(r_+)} \left[\sum_k \sigma \alpha_k(d-2k-1) \left(\frac{\sigma}{r_+^2}\right)^{k-1} + \frac{H}{r_+^{d-2}} - \frac{8 \pi G Q^2}{(d-2) r_+^{2(d-3)}} \right]
\nn\\
\Phi &=& \frac{\Sigma^{\sigma}_{d-2} Q}{(d-3) r_+^{d-3}}
\ea
where
\be 
D(r_+) = \sum_{k=1}^{k_{\rm max}} k \alpha_k (\sigma r_+^{-2})^{k-1} \, .
\ee

Employing Wald's method for determining the entropy \cite{Wald:1993nt, Iyer:1994ys}, we compute
\be 
S = -2 \pi \oint d^{d-2} x \sqrt{\gamma_{\rm h}} \, Y^{abcd} \hat{\varepsilon}_{ab} \hat{\varepsilon}_{cd} \quad \text{where} \quad Y^{abcd} = \frac{\partial {\cal L}}{\partial R_{abcd}}
\ee
where ${\cal L}$ is the Lagrangian density, $\gamma_{\rm h}$ is the determinant of the induced metric on the horizon and $\hat{\varepsilon}_{ab}$ is the binormal to the horizon.  For the solution \eqref{mastereqn} we find
\ba 
\tensor{Y}{_a _b ^c ^d} = \frac{1}{16 \pi G} \sum_{k=1}^{k_{\rm max}} \frac{k (2k!)}{2^k}  \delta_{[a}^{c} \delta_{b}^{d} \delta_{\alpha_2}^{\mu_2} \delta_{\beta_2 }^{\nu_2}\cdots \delta_{\alpha_k}^{\mu_k} \delta_{\beta_k ]}^{\nu_k}  \left[a_k \prod_{r=2}^{k_{\rm max}} \tensor{R}{^{\alpha_r} ^{\beta_r} _{\mu_r} _{\nu_r}} \right. 
\nn\\
\left. + b_k \phi^{d-4k+2} \prod_{r=2}^{k_{\rm max}} \tensor{S}{^{\alpha_r} ^{\beta_r} _{\mu_r} _{\nu_r}}\right] \, .
\ea
The part of this expression explicitly containing products of the Riemann tensor leads to the standard Lovelock black hole entropy, while the remaining part represents a contribution to the black hole entropy due to the scalar hair.  Explicitly, the full black hole entropy is given by
\be \label{entropy-B} 
S = \frac{(d-2)\Sigma^{(\sigma)}_{d-2}}{4 G} \sum_{k=1}^{k_{\rm max}} k \sigma^{k-1} \left[ \frac{ \alpha_k }{d-2k} r_+^{d-2k} +  \frac{ b_k (d-3)! }{\left(d-2k \right)!}  N^{d-2k}\right] \, .
\ee
In the case where $b_k = 0 $ for $k > 2$, the hairy contribution to the entropy takes a particularly nice form
\be\label{nice_entropy} 
S = \frac{\Sigma^{(\sigma)}_{d-2}}{4 G}   \left[ \sum_{k=1}^{k_{\rm max}} \frac{(d-2) k \sigma^{k-1}  \alpha_k  }{d-2k} r_+^{d-2k}  - \frac{d}{2\sigma (d-4)} H\right] \, \quad \textrm{if } b_k = 0 \,\,\,  \forall k > 2.
\ee
Since the fall off is the same to all orders in the hairy terms, and the contribution to the entropy is always just an additive constant, in this paper we will employ this latter form of the entropy.  Since the hairy contribution to the entropy is independent of $r_+$, we shall employ \eqref{nice_entropy} henceforth.
 Doing so allows for calculational convenience and one does not miss out on any of the new physics that the extra $b_k$-dependent terms  in \eqref{entropy-B}  contribute.

Employing the extended first law
\be 
\delta M = T\delta S + \Phi \delta Q + \sum_k\Psi^{(k)} \delta \alpha_k + \sum_k {\cal K}^{(k)} \delta b_k 
\ee
we find it is satisfied provided
\begin{align}
\Psi^{(k)} &= \frac{\Sigma^{(\sigma)}_{d-2} (d-2) }{16 \pi G} \sigma^{k-1} r_+^{d-2k} \left[\frac{\sigma}{r_+} - \frac{4 \pi k T}{d-2k} \right] \, ,
\nn\\
{\cal K}^{(k)} &= -\frac{\Sigma^{(\sigma)}_{d-2} (d-2)!   }{ 16 \pi G } \sigma^{k-1} N^{d-2k} \left[\frac{\sigma}{\left(d-2(k+1)\right)!r_+} + \frac{4\pi k T}{(d-2k)!} \right]
\end{align}
and the Smarr relation which follows from scaling
\be 
(d-3)M = (d-2)TS + (d-3)\Phi Q +  \sum_k 2(k-1) \Psi^{(k)} \alpha_k + (d-2)\sum_k {\cal K}^{(k)} b_k \, ,
\ee
also holds.  We point out that in the situation when black hole solutions are considered, the couplings $b_k$ are not all independent, but are constrained by Eq.~\eqref{N-constr}.  As a result, in these cases, one must keep in mind that the variations of $b_k$ in the first law above are not all independent. Henceforth we shall set $\alpha_1=1$ so that we recover general relativity in the limit $\alpha_k \to 0$ for $k > 1$ and we will also set $G=1$.

\section{Equations of motion and ghosts}\label{sec:ghosts}

In this paper we will report on hairy black hole solutions in Lovelock gravity sourced by non-minimal conformal scalar hair.  The black hole solutions produced by this theory are remarkably simple, providing an excellent arena for investigating the effects of scalar hair on higher curvature black holes.   As mentioned in the introduction, higher curvature theories often suffer from having unstable vacua, i.e. the graviton is a ghost in maximally symmetric backgrounds.  The absence of ghosts is a necessary condition for a stable ground state and thus for a sensible theory.  Here we address this problem for the hairy black holes studied in this work, where the non-minimal coupling leads to modifications of the standard Lovelock result. Our approach is to consider the linearized theory for a constant scalar field $\Phi$ in a maximally symmetric AdS background.  Naively, one might expect trivial results for the following reason: For a constant scalar field, the action reduces to Lovelock gravity with redefined couplings.  Thus, one might expect that the conditions for a ghost free vacuum would reduce to the standard Lovelock conditions but for these redefined couplings.  However, this is not the case since the equations of motion for the scalar field must also be respected, leading to a non-trivial result.  

The main results of this section are the following: The theory is ghost free when expanded about a background in which the scalar field vanishes.  However, we point out that when the scalar field takes on a constant value, there will generically be ghosts for the couplings corresponding to spherical black holes. We report first on the case of Einstein gravity non-minimally coupled to a scalar field, before moving on to a detailed analysis of cubic Lovelock gravity.  

\subsection{Einstein case}

As a warm up exercise, let us consider the theory that contains only the Ricci scalar and cosmological constant in the gravitational sector, and scalar hair coupled up to the Gauss-Bonnet term in the matter sector.  We wish to expand the equations of motion about a ground state of the theory.  To this end, we explore the situation where the scalar field takes on a constant value (denoted by $\Phi$) and the spacetime is asymptotically AdS with curvature radius $L$.  That is,
\be
R_{abcd} = -\frac{1}{L^2} \left( g_{ac}g_{bd} - g_{ad}g_{bc} \right) \, .
\ee
  In this case, the equations of motion for the scalar field and the gravitational field are given (respectively) by
\begin{align}\label{lefe:scalar_field}
0 =&  (d-4)(d-3)(d-2)(d-1)d b_2\Phi^{d-5} -  L^2 \Phi^{d-3} (d-2)(d-1)d b_1  + d b_0 L^4 \Phi^{d-1} \, ,
\\
0 =&  -\frac{16 \pi }{2} b_0 \Phi^{d} + \frac{16 \pi (d-1)(d-2) b_1 \Phi^{d-2}  }{2 L^2} - \frac{16 \pi (d-1)(d-2)(d-3)(d-4) b_2 \Phi^{d-4}}{2 L^4}
\nn\\ &+ \Lambda + \frac{(d-1)(d-2)}{2 L^2}        \, .        
\end{align}
The second equation, given the first, is nothing more than the statement
\be\label{lefe:cosmological_constant} 
\Lambda = -\frac{(d-1)(d-2)}{2 L^2}
\ee
indicating that, for this configuration, the cosmological constant sets the curvature for the AdS space, as we might well expect.  Taken together, Eqs.~\eqref{lefe:scalar_field} and \eqref{lefe:cosmological_constant} comprise the background about which we shall expand.  

We consider perturbations to our background metric of the form $g_{ab} = g_{ab}^{[0]} + h_{ab}$ and work to first order in this perturbation (here $g_{ab}^{[0]}$ represents the maximally symmetric background metric which solves the field equations).  Without fixing to any particular gauge, the linearized equations of motion take the form,
\begin{align}
 -&\frac{1}{2}\left(1 + 16 \pi b_1 \Phi^{d-2} - \frac{32 \pi (d-4)(d-3) b_2 \Phi^{d-4}}{L^2} \right) \left[\square h_{ab} + \nabla_{b}\nabla_{a}h^{c}{}_{c} -  \nabla_{c}\nabla_{a}h_{b}{}^{c} -  \nabla_{c}\nabla_{b}h_{a}{}^{c}  \right. 
 \nn\\
 &\left. + g^{[0]}_{ab} \left(  \nabla_{d}\nabla_{c}h^{cd} -   \square h^{c}{}_{c} \right)  + \frac{(d-1)}{ L^2} g_{ab}^{[0]}h^{c}{}_{c} \right]
 \nn\\
&+  \left[ \frac{(d-1)}{L^2} \left(1 + 8 \pi d b_1 \Phi^{d-2} - \frac{8 \pi (d-4)(d-3)(d+2) b_2  \Phi^{d-4}}{ L^2} \right)- 8 \pi b_0 \Phi^{d} \right]h_{ab} \, =8\pi T_{ab}.
\end{align}
Making use of the scalar field equations of motion, this equation can be further simplified to
\begin{align}
 -&\frac{1}{2}\left(1 + 16 \pi b_1 \Phi^{d-2} - \frac{32 \pi (d-4)(d-3) b_2 \Phi^{d-4}}{L^2} \right) \left[\square h_{ab} + \nabla_{b}\nabla_{a}h^{c}{}_{c} -  \nabla_{c}\nabla_{a}h_{b}{}^{c} -  \nabla_{c}\nabla_{b}h_{a}{}^{c}  \right. 
 \nn\\
 &\left. + g^{[0]}_{ab} \left(  \nabla_{d}\nabla_{c}h^{cd} -   \square h^{c}{}_{c} \right)  + \frac{(d-1)}{ L^2} g_{ab}^{[0]}h^{c}{}_{c} - \frac{2(d-1)}{ L^2}h_{ab} \right]
 =8\pi T_{ab}.
\end{align}
Thus, the linearized field equations are of the same form as Einstein gravity, and we need only check the sign of the pre-factor to ensure the absence of ghosts.
In the limit $\Phi=0$, these linearized equations reduce to the expected form for perturbations of the Einstein equation about an AdS background, that is, the theory is ghost-free in this limit.  

We wish to use these linearized equations to place constraints on the coupling constants of the scalar field and we would like these constraints to be relevant, not only in the trivial case of constant scalar field, but also in the case considered in this work where the coupling constants of the scalar field obey certain fixed relationships set by Eq.~\eqref{N-constr}.  That is, we consider the linearized equations subject to Eqs.~\eqref{lefe:scalar_field} and also
\begin{align}\label{b2-constraint}
b_2 =& \frac{(d-2)(d^2-d-8)}{4d(d-3)(d-4)}\frac{b_1^2}{b_0} \, ,
\end{align}
which is a consequence of Eqs.~\eqref{N-constr} for this setup.  To ensure that the graviton is not a ghost we require
\begin{align}\label{eqn:Ein_ghost_constr}
1 + 16 \pi b_1 \Phi^{d-2} - \frac{32 \pi (d-4)(d-3) b_2 \Phi^{d-4}}{L^2} &> 0 \, .
\end{align}
In this case we can make significant progress analytically.  We find that condition~\eqref{lefe:scalar_field} is satisfied if

\begin{equation}\label{phiGB}
\Phi = \epsilon \frac{\sqrt{2b_0 b_1 d(d-2) (d^2-d+ 2\lambda\sqrt{2d(d-1)})}}{2 d |b_0| L}
\end{equation}
where $\epsilon,\lambda = \pm 1$. Since the dimension dependent part of the square root is always positive for $d \ge 4$, we can see that $b_0$ and $b_1$ are required to have the same sign for $\Phi$ to be real. Since here we assume $\Phi\neq 0$, Eq.~\eqref{eqn:Ein_ghost_constr}  can be written as
\begin{equation}\label{Einstein_ghost}
\Phi^{d-4}\left(\Phi^{4-d}+\frac{16\pi(d-2)\left(\lambda\sqrt{2d(d-1)}+4\right)}{dL^2}\frac{b_1^2}{b_0}\right)>0
\end{equation}

For even $d$, we have $\Phi^{d-4}>0$ so we can ignore the multiplicative prefactor $\Phi^{d-4}$. We see that if $b_0>0$, then the second term is positive whenever we choose $\lambda=1$ and if $b_0<0$, then we may choose $\lambda=-1$ to ensure the second term is still positive. Eq.~\eqref{Einstein_ghost} can be satisfied independent of our choice for $\epsilon$. For odd $d$, note that if we choose $\epsilon=1$ (so that the first term is positive), then $b_0<0$ is allowed for $\lambda=-1$ and $b_0>0$ is allowed for $\lambda=1$. The key point here is that there is always an appropriate choice of $(\epsilon,\lambda)$ specifying $\Phi$ that allows the ghost constraint to be satisfied for any $d\geq 3$. Therefore, we conclude that for any given choice of $b_0, \, b_1$ such that $b_0 b_1>0$ we can always make an appropriate choice of $(\epsilon,\lambda)$ for $\Phi$ in order to make the gravitational theory free from ghosts. Thus in the Einstein case, the theory is free from ghosts if expanded about an AdS background with the scalar field vanishing.  If instead a constant scalar field is used, the only constraint for vacuum stability is that $b_0$ and $b_1$ have the same sign.  Unfortunately, this restriction is in contradiction to what is required for the existence of spherical black holes, where $b_0 b_1< 0$ is required.

\subsection{Gauss-Bonnet and Cubic Lovelock cases}
Having completed our study of this relatively simple case, we are now poised to consider higher order Lovelock terms in the gravitational sector.  For concreteness (and for applicability in this work) we shall present the linearized equations for third order Lovelock gravity. We can then obtain the Gauss-Bonnet case by setting $\alpha_3 = 0$.

The first noteworthy change when considering Lovelock gravity is that the curvature scale of the AdS background is no longer simply the the length scale associated with the cosmological constant, but contains contributions from the higher curvature terms.  Explicitly, the Riemann curvature of the AdS background is written as
\be 
R_{abcd} = -\frac{F_\infty}{L^2} \left( g_{ac}g_{bd} - g_{ad}g_{bc} \right)
\ee
where $L$ is the length scale associated with the cosmological constant and $F_\infty$ represents the leading order behaviour of the metric function at large r, i.e. $F_{\infty}$ is the leading order term of $ f(r)/(r^2 \alpha_0)$ as $r \to \infty$ and is required to be positive for AdS asymptotics.  The particular value of $F_\infty$ can be obtained using the fact that it solves the equation
\be \label{metricF}
1 - F_\infty + \frac{\alpha_2}{L^2} F_\infty^2 - \frac{\alpha_3}{L^4} F_\infty^3 = 0 \, 
\ee
which is what one obtains from evaluating the field equations on this maximally symmetric background.  We now repeat the calculations from earlier but for this more complicated theory.  We find that the equations of motion for the (constant) scalar field reduces to the constraint,
\be\label{scalar_field_eom_3rd} 
0 = d b_0 \Phi^{d-1} - \frac{d(d-1)(d-2) F_\infty b_1 \Phi^{d-3}}{L^2} + \frac{d(d-1)(d-2)(d-3)(d-4) F_\infty ^2 b_2 \Phi^{d-5}}{L^4}
\ee
while the linearized gravitational field equations take the form,
\begin{align}
&-\frac{1}{2}\left(1-2\frac{\alpha_2}{L^2} F_\infty + 3 \frac{\alpha_3}{L^4} F_\infty^2 + 16 \pi b_1 \Phi^{d-2} - \frac{32 \pi (d-3)(d-4) b_2 F_\infty \Phi^{d-4}}{L^2}  \right) \left[\square h_{ab} \right. \nn\\
&\left. + \nabla_{b}\nabla_{a}h^{c}{}_{c} -  \nabla_{c}\nabla_{a}h_{b}{}^{c} -  \nabla_{c}\nabla_{b}h_{a}{}^{c} 
  + g^{[0]}_{ab} \left(  \nabla_{d}\nabla_{c}h^{cd} -   \square h^{c}{}_{c} \right) + \frac{(d-1)F_\infty}{L^2} g_{ab}^{[0]}h^{c}{}_{c} 
 \right.
 \nn\\
 &- \left.\frac{F_\infty (d-1)}{L^2} h_{ab}\right] = 8 \pi T_{ab}\,,
%\nn\\
%%&-\left[\frac{F_\infty (d-1)}{ L^2} \left(1 -2\alpha_2 F_\infty + 3 \alpha_3 F_\infty^2 -  16 \pi b_1 F_\infty \Phi^{d-2} \right) + \frac{(d^3 -4d^2 + 7d+4)  b_2 F_\infty^2 \Phi^{d-4}}{L^4} \right] g_{ab}^{[0]}h^{c}{}_{c}
%%\nn\\
%&+  \left[ \frac{F_\infty (d-1)}{L^2} \left(1  -2\frac{\alpha_2}{L^2} F_\infty + 3 \frac{\alpha_3}{L^4} F_\infty^2  + 8 \pi d b_1 \Phi^{d-2} - \frac{8 \pi (d-4)(d-3)(d+2) b_2 F_{\infty} \Phi^{d-4}}{ L^2} \right) \right. 
%\nn\\& \left. - 8 \pi b_0 \Phi^{d} 
%  \right]h_{ab}  = 8 \pi T_{ab} \, ,
\end{align} 
where we have included on the right-hand side a stress energy tensor which may arise from minimal coupling to other matter fields and have once again used the equations of motion of the scalar field.    We then have the constraint,
\begin{align}\label{cubic_ghost_constraint}
&1-2\frac{\alpha_2}{L^2} F_\infty + 3 \frac{\alpha_3}{L^4} F_\infty^2 + 16 \pi b_1 \Phi^{d-2} - \frac{32 \pi (d-3)(d-4) b_2 F_\infty \Phi^{d-4}}{L^2}  >0
\end{align}
where $\Phi$ is a solution of Eq.~\eqref{scalar_field_eom_3rd}  and we again enforce Eq.~\eqref{b2-constraint}. After some algebra, we see that Eq.~\eqref{scalar_field_eom_3rd} is satisfied if
\be\label{phiLovelock}
\Phi=\epsilon\frac{\sqrt{2b_0 b_1 d(d-2)\left(d^2-d + 2\lambda\sqrt{2d(d-1)}\right)F_\infty}}{2d|b_0| L}
\ee
where $\epsilon,\lambda=\pm 1$. This is identical to Eq.~\eqref{phiGB} for Einstein case except for the extra factor of $F_\infty$ in the square root. Therefore in both the Gauss-Bonnet and cubic Lovelock cases we see that $b_0,b_1,b_2$ also must have the same signs.\footnote{This is in fact a general requirement if one fixes $b_k = 0$ for $k \ge 3$.}  In these more complicated cases, it is difficult to make additional meaningful progress analytically.  We have investigated these conditions under a number of circumstances and report the results here.  

In the case of Gauss-Bonnet gravity, the situation is quite simple and depends only on the ratio $\alpha_2/L^2$.  In particular, if $\alpha_2/L^2 < 0$, there will be one asymptotically AdS branch and one asymptotically dS branch; the AdS branch is free from ghosts provided $b_0$ and $b_1$ have the same sign.  If $0 \le \alpha_2/L^2 \le 1/4$, both branches are asymptotically AdS and in each case the vacuum will be free of ghosts instabilities provided only that $b_0$ and $b_1$ are of the same sign.  For $\alpha_2/L^2 > 1/4$, there exist no solutions  to the field equations with asymptotic regions;  in thermodynamic language, the maximal pressure constraint is violated: in Gauss-Bonnet gravity this entire parameter range is unphysical.  These results are true independent of the spacetime dimension $d \ge 5$.

\begin{figure}[htp]
\centering
\includegraphics[width=0.6\textwidth]{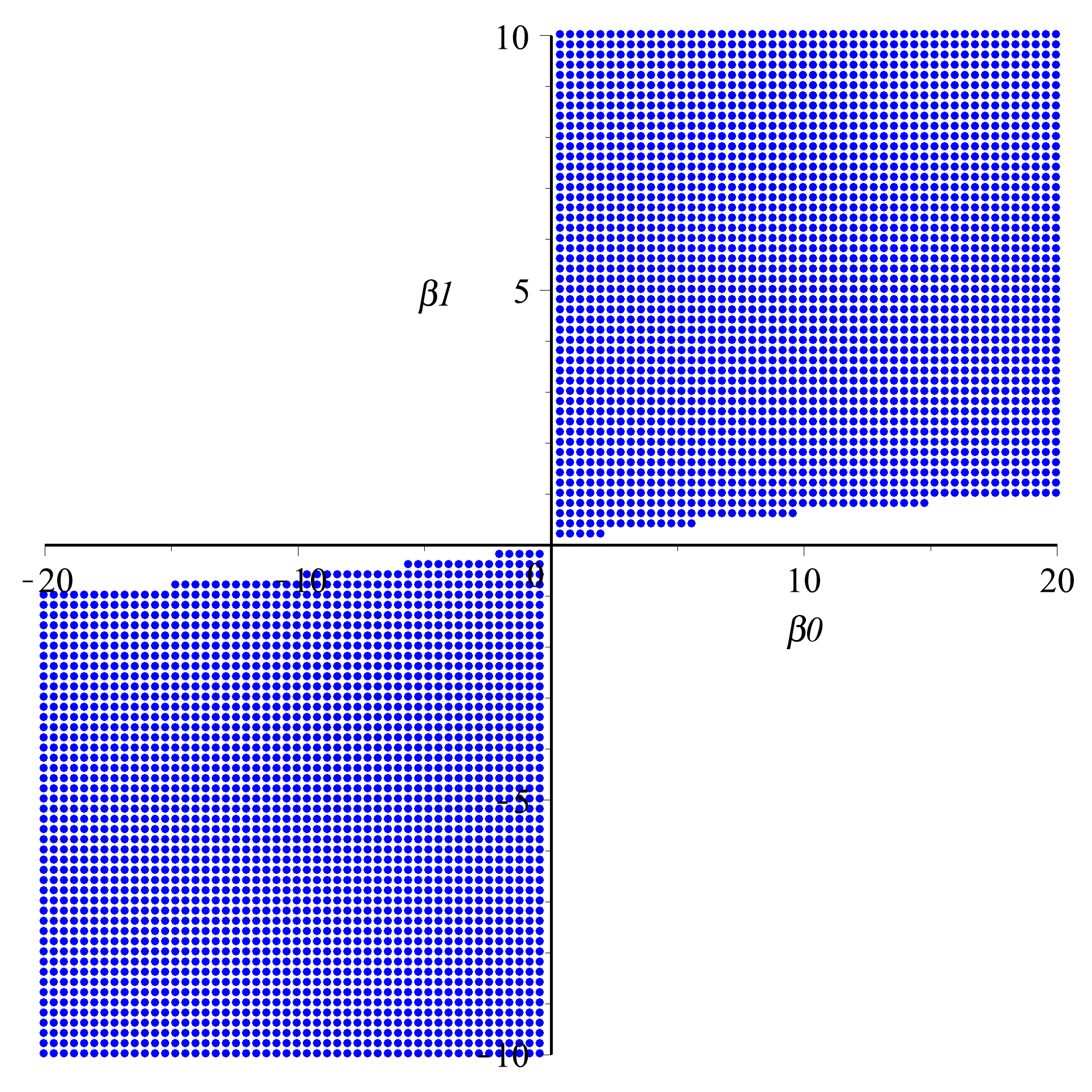}
\caption{{\bf Ghost free parameters for Einstein branch}: $(\beta_0, \beta_1)$ parameter space with $p=0.01$ and $\alpha = 4$ for the Einstein branch of cubic Lovelock gravity with blue dots corresponding to viable values of $\beta_0$ and $\beta_1$ (defined by Eq.~\eqref{dimless_hair}).  This is a tighter constraint than for the other branches where a sufficient condition to satisfy ghost-free condition is $b_0b_1 > 0$.  While this plot was constructed for $d=7$, higher dimensions are qualitatively similar.}
\label{fig:ein_branch_ghost}
\end{figure}

For cubic Lovelock gravity there is the additional parameter $\alpha_3$.  To simplify matters, we work with the dimensionless variables defined in Eq.~\eqref{dimless_pres_coupling_cl} and additionally define, 
\be\label{dimless_hair} 
b_k = \alpha_3^{\frac{d-2}{4}} \beta_k \, , \quad \hat{\Phi} = L \Phi \, , \quad \alpha_0 = \frac{1}{L^2}
\ee
so that all of the expressions are dimensionless.  In subsequent sections we will see that the existence of  AdS asymptotics for all three Lovelock branches translates to a pressure constraint,  $p\in(p_-,p_+)$.  For $(p, \alpha)$ combinations that yield AdS asymptotics, we find that the Lovelock and Gauss-Bonnet branches are ghost free provided $b_0$ and $b_1$ have the same sign, though there are additional constraints on for the Einstein branch, as highlighted in Figure~\ref{fig:ein_branch_ghost}.  For $p \not \in (p_-, p_+)$ only the Lovelock branch has well-defined asymptotics.  As it turns out, if $b_0$ and $b_1$ have the same sign, then this branch will be free from ghosts.

Thus we have seen that in both Gauss-Bonnet and cubic Lovelock gravity  with a scalar field conformally coupled to the first three Euler densities (i.e. up to the Gauss-Bonnet term), a necessary condition for the absence of ghosts is that $b_0b_1 > 0$.  Furthermore, this is also a sufficient condition in all cases except for the Einstein branch in the cubic Lovelock case.  From this we can conclude that the hyperbolic black holes studied previously in \cite{Hennigar:2015wxa} and those considered in this paper are free from ghost instabilities regardless of whether the scalar field vanishes or is taken to be a constant.  This follows because, for $\sigma = -1$ it is always possible to solve the equations of motion for the scalar field with $b_0 b_1 > 0$.   However, if one limits the coupling to the first three Euler densities, then the restrictions forced upon the $b_k$ couplings by a solution of the scalar field equations of motion (see Eq.~\eqref{N-constr}) require $b_0b_1 <0$ in the case of spherical symmetry.  This contradicts the conditions required for the theory to be ghost free in a constant scalar field background.  Therefore, for this setup, for the couplings required in the spherical case, the theory suffers from ghost instabilities if the scalar field is taken to be a non-vanishing constant in the background.

This conclusion does not mean that the spherical black hole solutions are pathological.  For example, if the constant scalar field were taken to be zero, then all of the above constraints could be simultaneously satisfied.  Furthermore, the above analysis was for the very specific case where the scalar field couples only to the first three Euler densities, such as that considered in~\cite{Giribet:2014bva, Giribet:2014fla, Hennigar:2015wxa}.  More generally, one could couple the scalar field to additional terms, e.g. the cubic Lovelock term (by keeping $b_3$ non-zero) or even, for example, to the quasi-topological term~\cite{Myers:2010jv, Oliva:2010eb}.  We now demonstrate that by coupling to the cubic Lovelock term the ghost conditions can be satisfied in the case of spherical symmetry.

\begin{figure}[htp]
\centering
\includegraphics[width=0.4\textwidth]{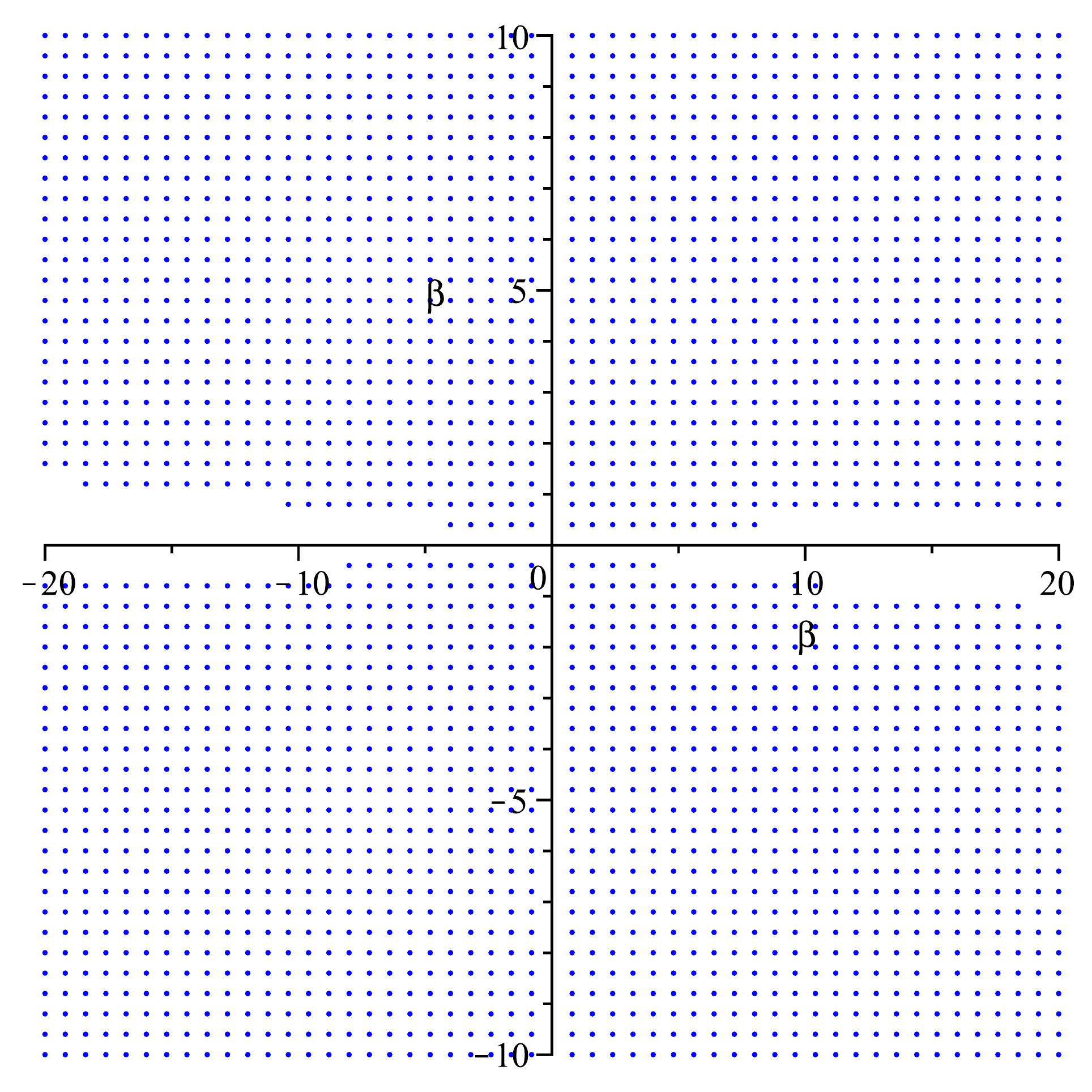}
\caption{{\bf Ghost conditions}: Plots of allowed coupling constants when the scalar field is coupled to the cubic Lovelock term for the Einstein branch (the Lovelock and Gauss-Bonnet branches satisfy the constraint for all couplings).  The colored region corresponds to allowed values of the coupling. Thus, $b_0 b_1 < 0$ is permitted, which is required for spherical black holes.  This plot has been constructed for $d=7$ cubic Lovelock gravity with $p=0.01$ and $\alpha=4$ in terms of the dimensionless parameters given in Eqs.~\eqref{dimless_pres_coupling_cl} and ~\eqref{dimless_hair}.  Here $b_2 = 0$ for simplicity in this representative example.  }
\label{fig:cl_cubic_hair_ghosts}
\end{figure}

Consider the case where $b_0,b_1, b_3 \neq 0$ with $b_2 = 0$.  While clearly not the most general case, this instance will allow us to illustrate the point most clearly without having to perform a full analysis of a four-dimensional parameter space.  For these couplings, Eqs.~\eqref{N-constr}~and~\eqref{defn_of_H} reduce to,
\begin{align}
N &=  \frac{\varepsilon}{3 b_0}\sqrt{\frac{-\sigma b_0 b_1 6(d-2)(d^2 - d - 12)}{d}} \, , \\
b_3 &= -\frac{4}{27} \frac{(d-2)^2( d^3 + 2 d^2 - 15 d - 36)}{d^2 ( d-3)(d-5)(d-6)} \frac{b_1^3}{b_0^2} \, , \\
H &= N^d(d-3)!\left[ \frac{b_0}{(d-2)!} + \frac{ \sigma b_1 N^{-2}}{(d-4)!} + \frac{\sigma^3 b_3 N^{-6}}{(d-8)!} \right] \, ,  
\end{align}
with $\varepsilon \in \{-1, 0, +1 \}$.  The constant scalar field and ghost constraints are (respectively)
\begin{align}
& b_0 \Phi^{d-1} - \frac{(d-1)(d-2) b_1 F_\infty \Phi^{d-3}}{L^2} - \frac{(d-1)! b_3 F_\infty ^3 \Phi^{d-7}}{(d-7)! L^6} = 0\, , \nn\\
&1 - 2 \frac{ \alpha_2 F_\infty}{L^2} + 3\frac{\alpha_3 F_\infty^2}{L^4} + 16 \pi \left[b_1 \Phi^{d-2} + 3\frac{(d-6)(d-5)(d-4)(d-3) b_3 F_\infty^2 \Phi^{d-6}}{L^4} \right] > 0 \, .
\end{align}
Considering the expression for $N$ above, it is clear that for $\sigma = +1$ a sensible solution for the scalar field would require $b_0 b_1 < 0$.  Exploring the constraints above we find that solutions of this type are certainly possible, though not for the entire $(b_0, b_1)$ parameter space. Figure~\ref{fig:cl_cubic_hair_ghosts} illustrates the situation for cubic Lovelock gravity in seven dimensions. 

For the case of Gauss-Bonnet gravity in $d \ge 7$, non-zero $b_0, b_1, b_2, b_3$ will cure the ghost problems in the spherical case for a general, constant scalar field background. Of course, since non-zero $b_3$ indicates coupling to the cubic Lovelock term, this approach is only valid in $d \ge 7$.   However, one could also couple to the cubic quasi-topological term~\cite{Myers:2010jv, Oliva:2010eb} to achieve this in $d=5$ Gauss-Bonnet gravity.  It is not clear how one could alleviate the problem in $d=6$ Gauss-Bonnet gravity for general constant scalar field backgrounds. 

  In summary, we have shown that if $b_k = 0 $ for $ k \ge 3$, the hyperbolic solutions (with $\sigma=-1$) are free from ghost instabilities for all constant scalar field configurations. The spherical black hole solutions require coupling constants  that give rise to an unstable vacuum for a generic constant scalar field background, a result which is true in any dimension $ d \ge 5$.   However, for a vanishing scalar field background or if one allows  additional $b_k$ to be non-zero, this will not be true in general, and sensible solutions exist.  One may naturally wonder what effect this has on the thermodynamics of the spherical black holes.  The overall effect of the scalar hair is to add a term to the equation of state which falls off as $1/v^d$ and to add a constant shift to the entropy.  This is true irrespective of the number of non-zero $b_k$.  Thus the qualitative results will be the same regardless of the number of $b_k's$ included, and only shifts in the precise values of $h$ will occur. It would be interesting to consider more general perturbations about the black hole background to determine if these black holes are unstable in general.  

\section{Gauss-Bonnet gravity}\label{sec:gb_case}

In this section we discuss the effects of the conformal hair on the thermodynamics of black holes in Gauss-Bonnet gravity. 

\subsection{Thermodynamic equations and constraints}

The first issue we address is the existence of asymptotically AdS regions.  Due to the presence of the non-linear curvature, it is not a generic feature that asymptotically AdS regions exist.  We consider the polynomial 
\be 
\alpha_0 + \frac{\sigma-f}{r^2} + \alpha_2 \left(\frac{\sigma-f}{r^2}\right)^2- \frac{h}{r^d} - \frac{16\pi M}{(d-2)\Sigma^{(\sigma)}_{d-2} r^{d-1}} + \frac{8\pi}{(d-2)(d-3)} \frac{Q^2}{r^{2d-4}}  = 0
\ee
for the metric function.
Assuming an asymptotic region exists for both solutions of the above equation it must be the case that
\be\label{gb_discrim} 
1-4\alpha_0\alpha_2 \ge 0 \, .
\ee
We therefore restrict our focus to parameter values that satisfy this inequality.  We shall refer to the solution that reduces to 
that of Einstein gravity in the limit $\alpha_2\to 0$ to be the `Einstein branch' (denoted $f_E$) and the other solution to be the `Gauss-Bonnet branch' (denoted $f_{GB}$).
From Figure~\ref{gaussmetric} we see that the presence of conformal hair is compatible with existence of AdS asymptotics and event horizons if the above inequality is satisfied.

We shall employ the  dimensionless quantities $(v,t,q,p,h)$ in our thermodynamic analysis, where
\begin{align} 
r_+ &= v \sqrt{\alpha_2} \, , \quad T = \frac{t}{\sqrt{\alpha_2} (d-2)} \, , \quad Q = \frac{q}{\sqrt{2}} \alpha_2^{\frac{d-3}{2}} \, , \quad \frac{(d-1)(d-2) \alpha_0}{16\pi} = \frac{p}{4 \alpha_2} \, , 
\nn\\
H &= \frac{4 \pi h}{d-2} \alpha_2^{\frac{d-2}{2}}  \, .
\end{align}
Condition  \eqref{gb_discrim} thus becomes a constraint on the pressure 
\be 
p \le \frac{(d-1)(d-2)}{16 \pi}
\ee
which must be satisfied for well defined asymptotics\footnote{Note that unlike cubic Lovelock gravity where there is always at least one well-defined asymptotic, Gauss-Bonnet gravity has no well-defined asymptotics if $p>p_{max}$.}.  Furthermore, positivity of the entropy requires from Eq.~\eqref{nice_entropy}  that 
\be\label{gbentropy}
2 \sigma^2 (d-2)^2 v^{d-4} + \sigma(d-2)(d-4) v^{d-2} - 2 \pi d h > 0  
\ee 
assuming $\alpha_2>0$.
While for $\sigma=1$ this is satisfied trivially provided $ h < 0$, care is required otherwise.  With the addition of the hair it is possible for the entropy to be negative for $\sigma = 1$ and we shall investigate if this results in any new and interesting thermodynamic behaviour. 

Solving the expression for the temperature in Eq.~\eqref{temp_etc} for $\alpha_0$  we obtain 
\be\label{eqn:GB_eos} 
p = \frac{t}{v} -  \frac{ (d-2)(d-3) \sigma }{4 \pi v^2} + \frac{2\sigma t}{v^3} - \frac{(d-2)(d-5) \sigma^2}{4\pi v^4} - \frac{h}{v^d}  + \frac{q^2}{v^{2(d-2)}}
\ee 
for the equation of state.\footnote{Note that the $\alpha_2 \to 0$ limit is subtle in this parameterization (since we have assumed non-zero $\alpha_2$ in formulating this equation of state).  One must take $\alpha_2 \to 0$ while simultaneously holding $p / \alpha_2$ constant.  This limit results in the third and fourth terms of Eq.~\eqref{eqn:GB_eos} vanishing.}  We also find
\begin{align}
g &= \frac{\alpha_2^{\frac{3-d}{2}}}{\Sigma^{(\sigma)}_{d-2}} \left(M-TS \right)
\nn\\
&= -\frac{1}{16 \pi \left(v^2 + 2\sigma\right)} \bigg[\frac{4\pi p v^{d+1}}{(d-1)(d-2)}  - \sigma v^{d-1} + \frac{24 \pi \sigma p v^{d-1}}{(d-1)(d-4)}  - \frac{(d-8)\sigma^2v^{d-3} }{d-4} 
\nn\\
&- \frac{2(d-2) \sigma^3}{d-4} v^{d-5}  \bigg] + \frac{q^2 \left[(2d-5)(d-4)v^2 + 2\sigma(d-2)(2d-7) \right]}{4(d-2)(d-3)(d-4)(v^2 + 2\sigma) v^{d-3}} 
\nn\\
& + \frac{h \left[4\pi d\left(pv^4 - q^2 v^{2(4-d)} + h v^{4-d} \right)  - 3(d-2)(d^2-5d + 8) \sigma^2  - (d-2)(d^2 -7d + 8) \sigma  v^2  \right]}{8(d-2)^2(d-4) (v^2 + 2\sigma) \sigma v}  
\end{align}
for the Gibbs free energy. The state of the system will be that which minimizes $g$ at fixed $t, p, q$ and $h$.  

 In what follows we shall specialize to five and six dimensions and perform a detailed study of the thermodynamics of these black holes.  We have found that all of the interesting thermodynamic behaviour produced by the scalar hair is already present in five dimensions.  That is, going to higher dimensions does not produce any novel phenomena.  We consider both spherical and hyperbolic black holes in five dimensions, and include a short discussion of  hyperbolic black holes in six dimensions.

\subsection{$P-v$ criticality in $d=5$}

\begin{figure}[tp]
\includegraphics[scale=0.25]{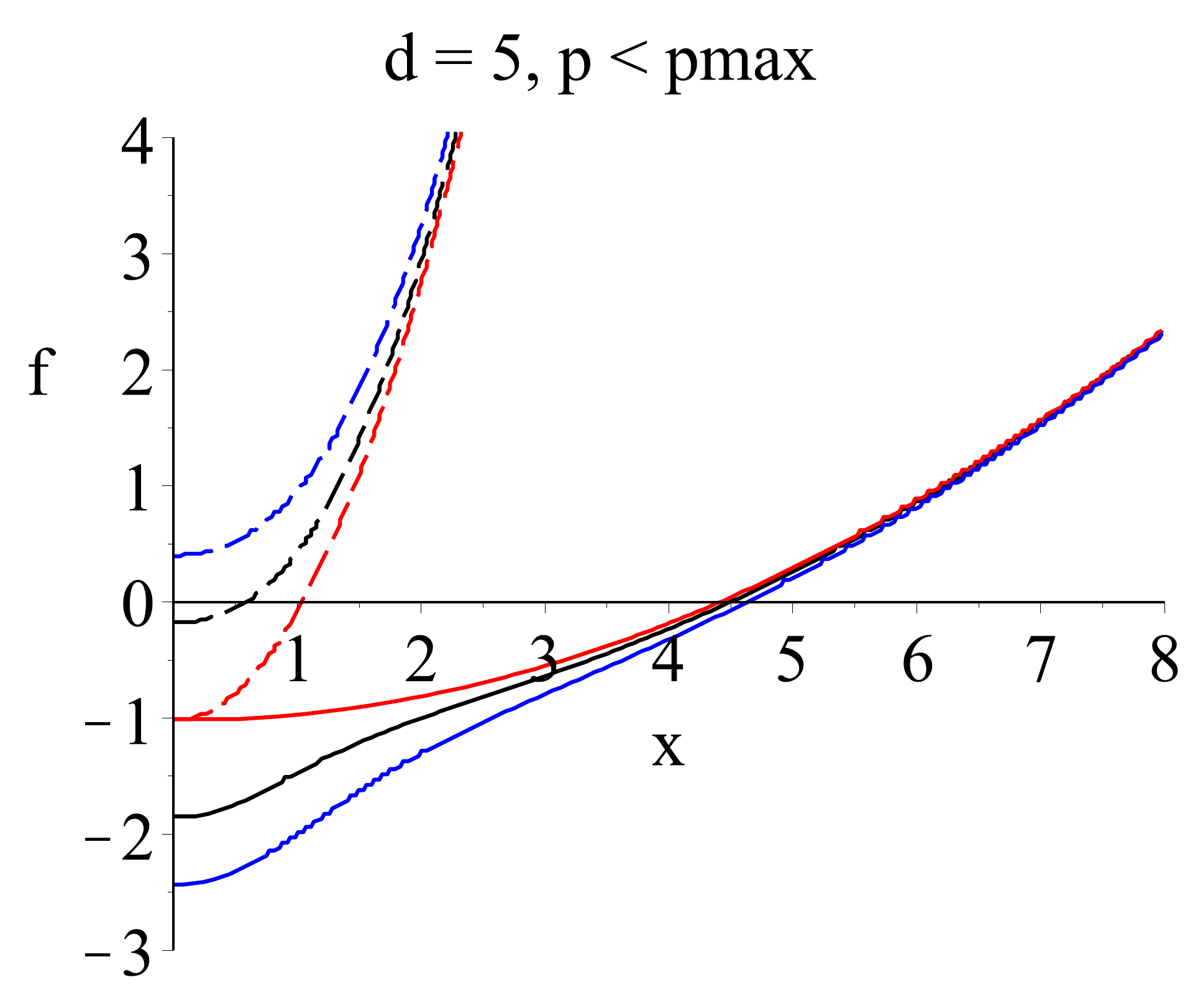}
\includegraphics[scale=0.25]{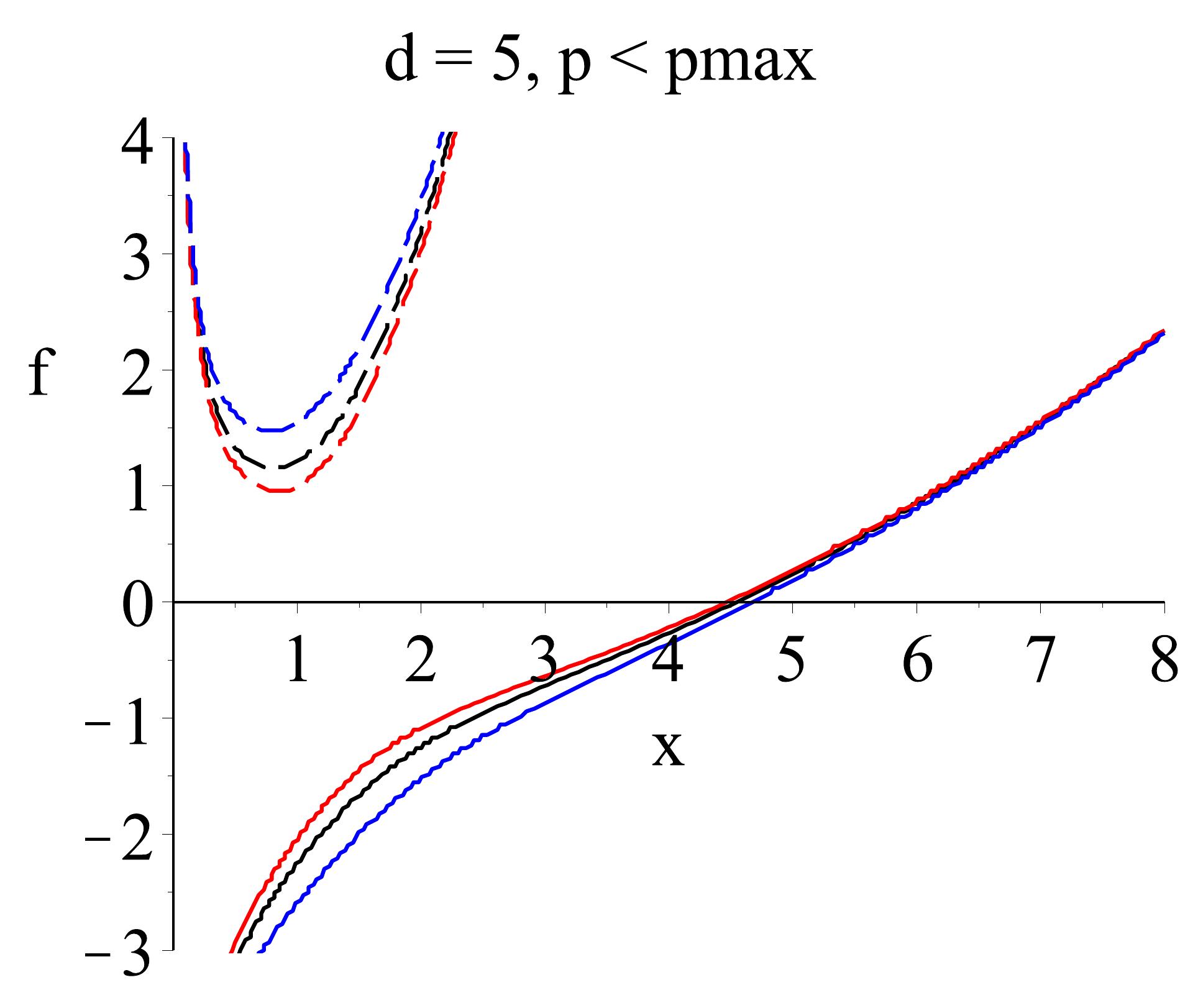}
\includegraphics[scale=0.25]{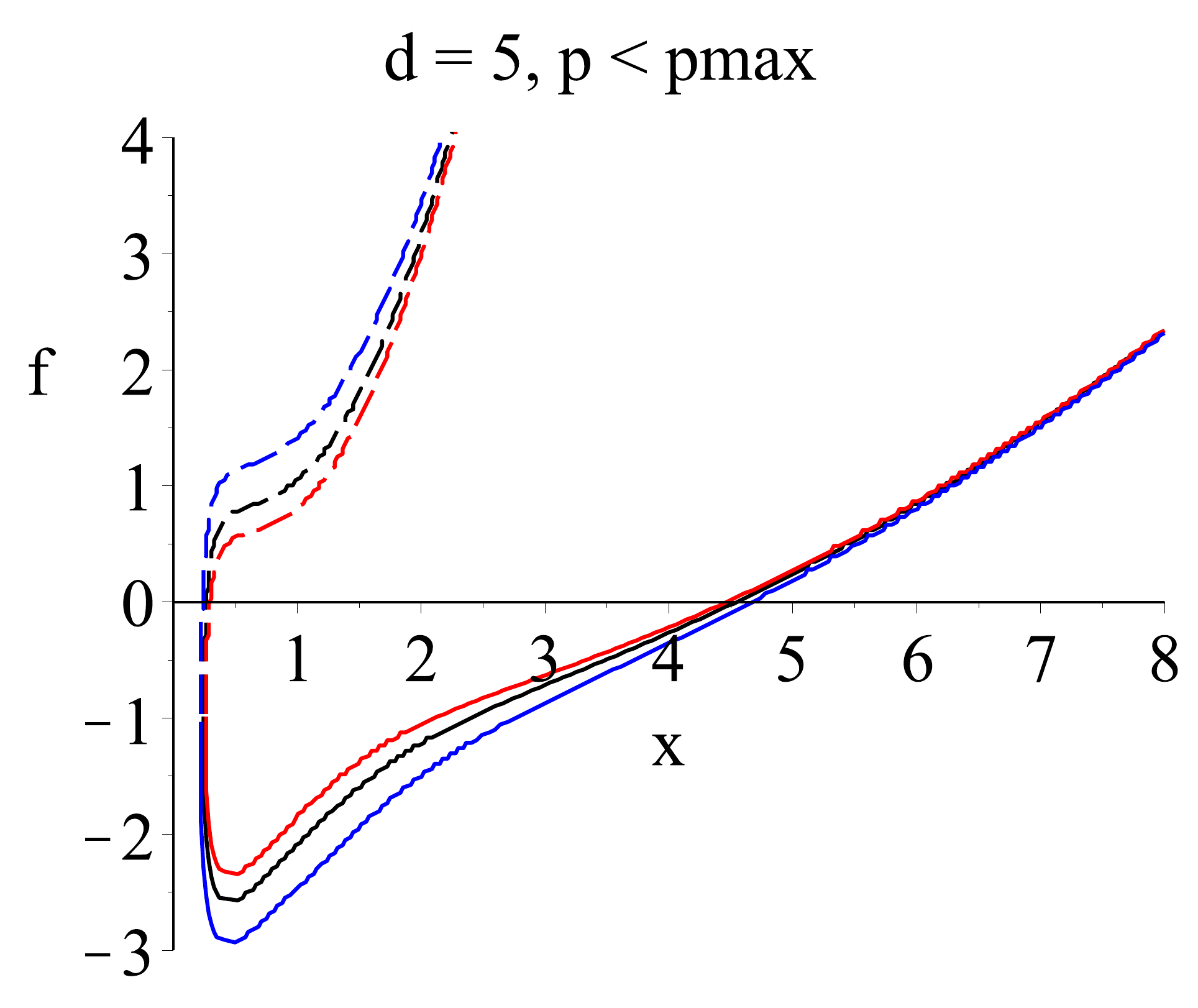}
\includegraphics[scale=0.25]{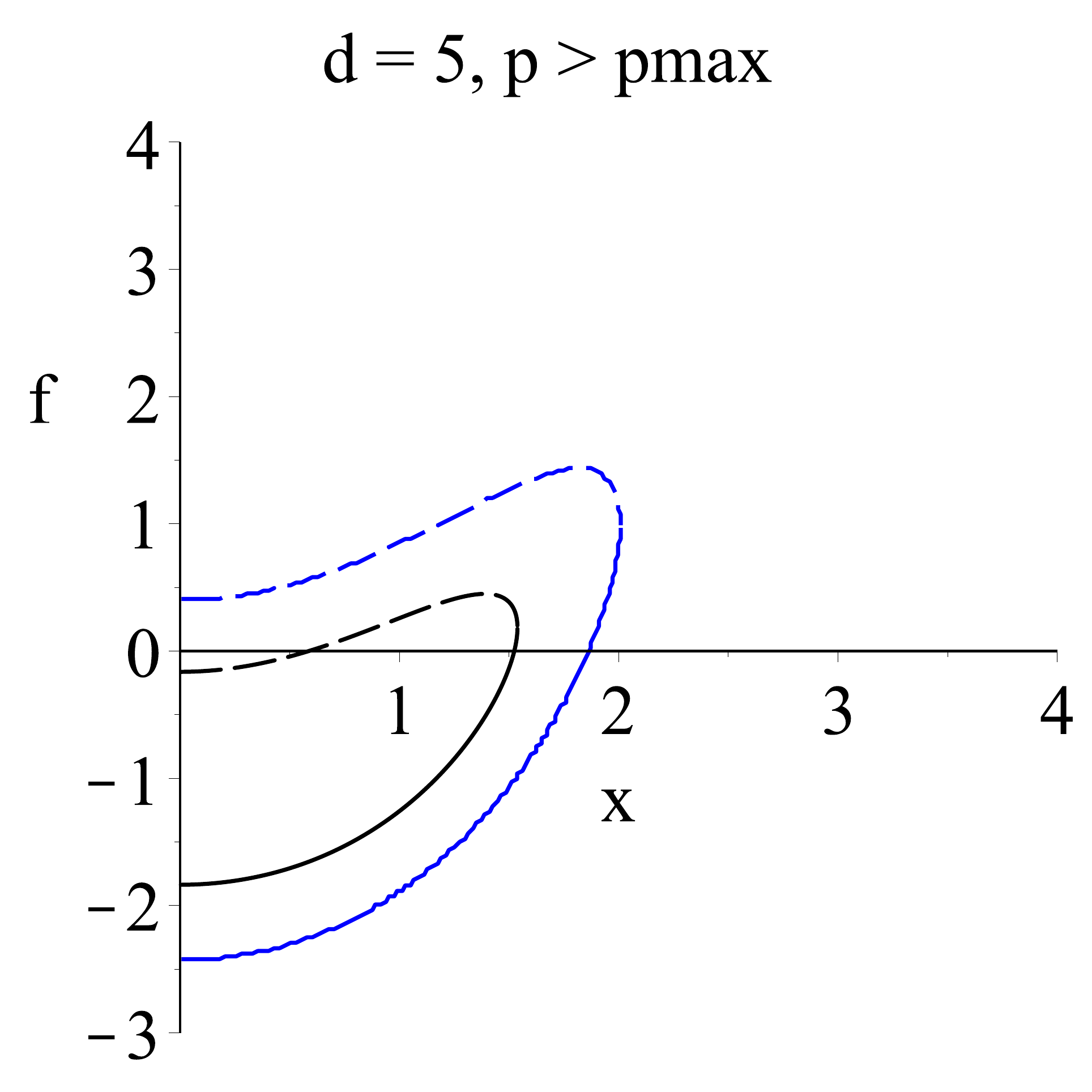}
\includegraphics[scale=0.25]{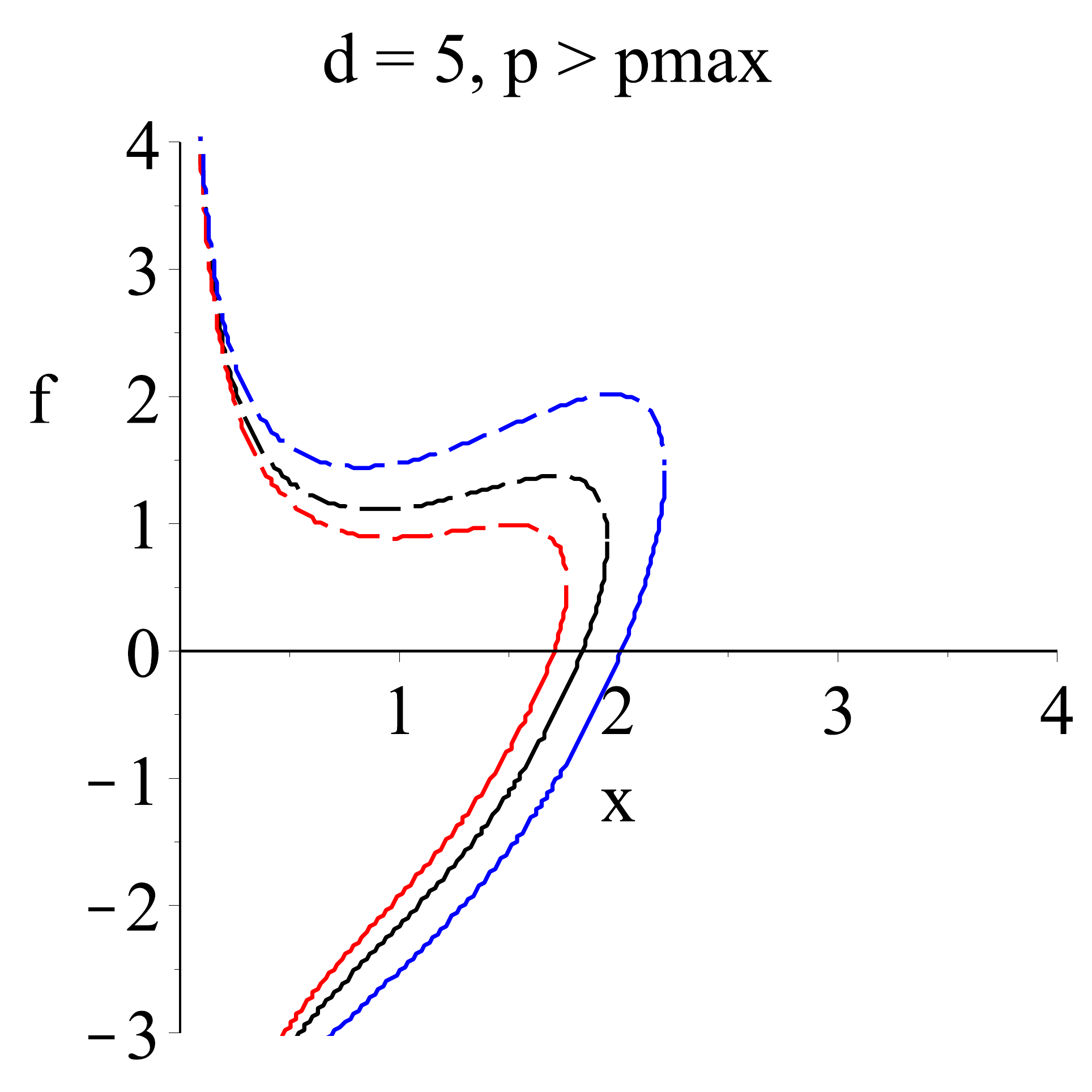}
\includegraphics[scale=0.25]{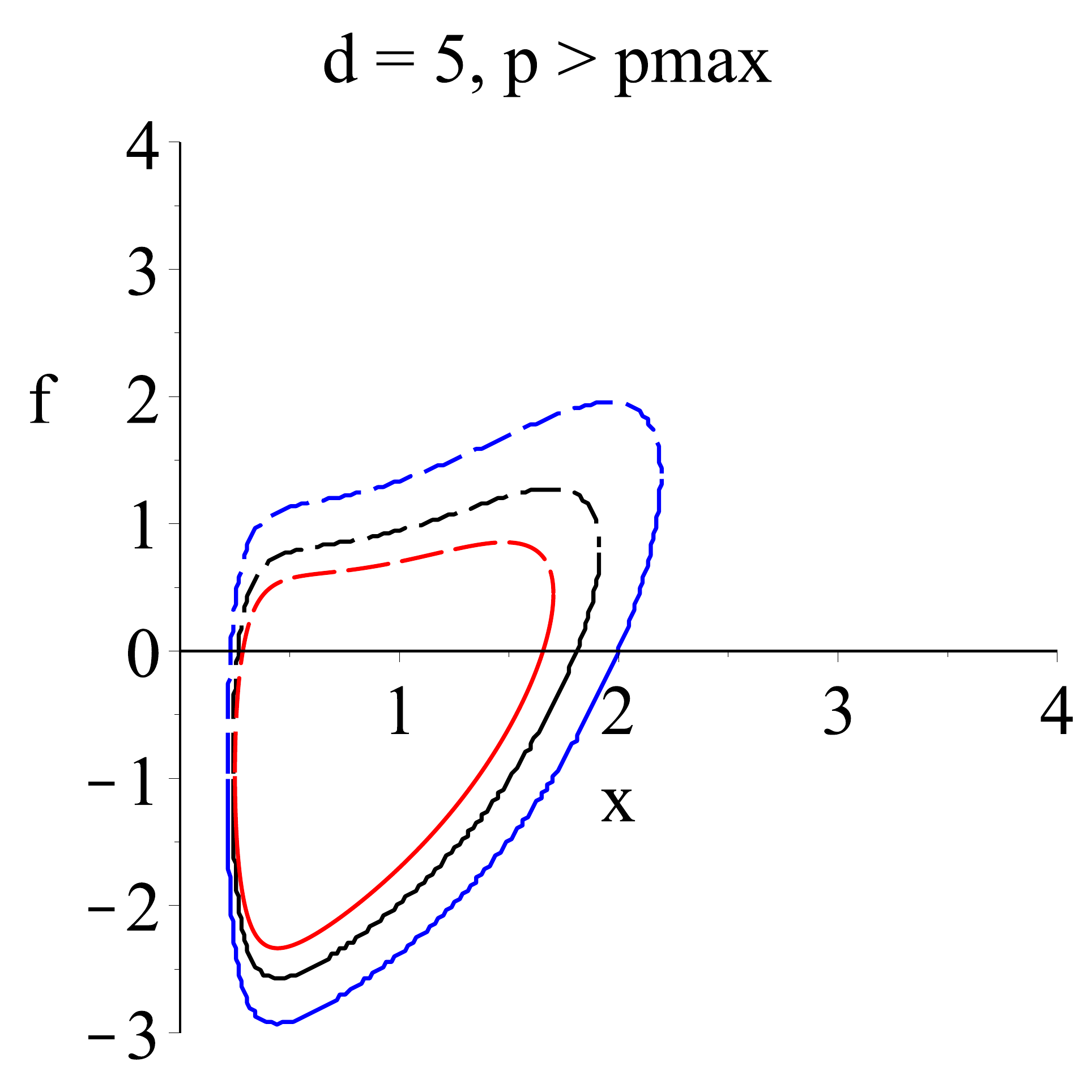}
\caption{{\bf Metric function for $d=5$ for hyperbolic ($\sigma=-1$) black holes}: plots of $f$ as a function of rescaled radial coordinate $x = r/\sqrt{\alpha_2}$. In each plot $q,h$ are fixed and different curves represent different masses of the black hole where the red, black and blue curves refer to $m = 0, \, 0.7 , \, 2.0$ respectively. The dashed curves represent the Gauss-Bonnet branch $f_{GB}$, while solid curves represent the Einstein branch $f_E$. \textit{Top left}: $h=0, \, q=0, \, p=p_{max}/5$. \textit{Top center}: $h=0.5, \, q=0, \, p=p_{max}/5$. \textit{Top right}: $h=0.5, \, q=0.5, \, p=p_{max}/5$. \textit{Bottom left}: $h=0, \, q=0, \, p=1.5p_{max}$. Note that in the uncharged case with no hair, the compact radial region approaches zero as the mass tends to zero, hence there is no red curve in this plot. \textit{Bottom centre}: $h=0.5, \, q=0, \, p=1.5p_{max}$. \textit{Bottom right}: $h=0.5, \, q=0.5, \, p=1.5p_{max}$. For uncharged solutions, the scalar hair generally results in $f_{\pm}\to\pm\infty$ as $r\to 0$ while the behaviour of the metric function at infinity is unaffected. Since charge term has the highest fall-off rate, the presence of charge will result in $f$ terminating at finite, nonzero $x$ regardless of $h$ and $m$. Note that if the maximum pressure constraint is not satisfied, then the radial coordinate  becomes compact and we do not have AdS asymptotics.}
\label{gaussmetric}
\end{figure}

In the case of five dimensions, we require that 
\be 
p < p_{max} = \frac{3}{4\pi}
\ee
to ensure the existence of AdS asymptotics (c.f. Eq.~\eqref{gb_discrim}).  If this inequality is violated the nonlinear curvature becomes too strong and the spacetime becomes compact in the radial coordinate, as illustrated in Figure~\ref{gaussmetric}.

\begin{figure}[tp]
\centering
\includegraphics[width=0.4\textwidth]{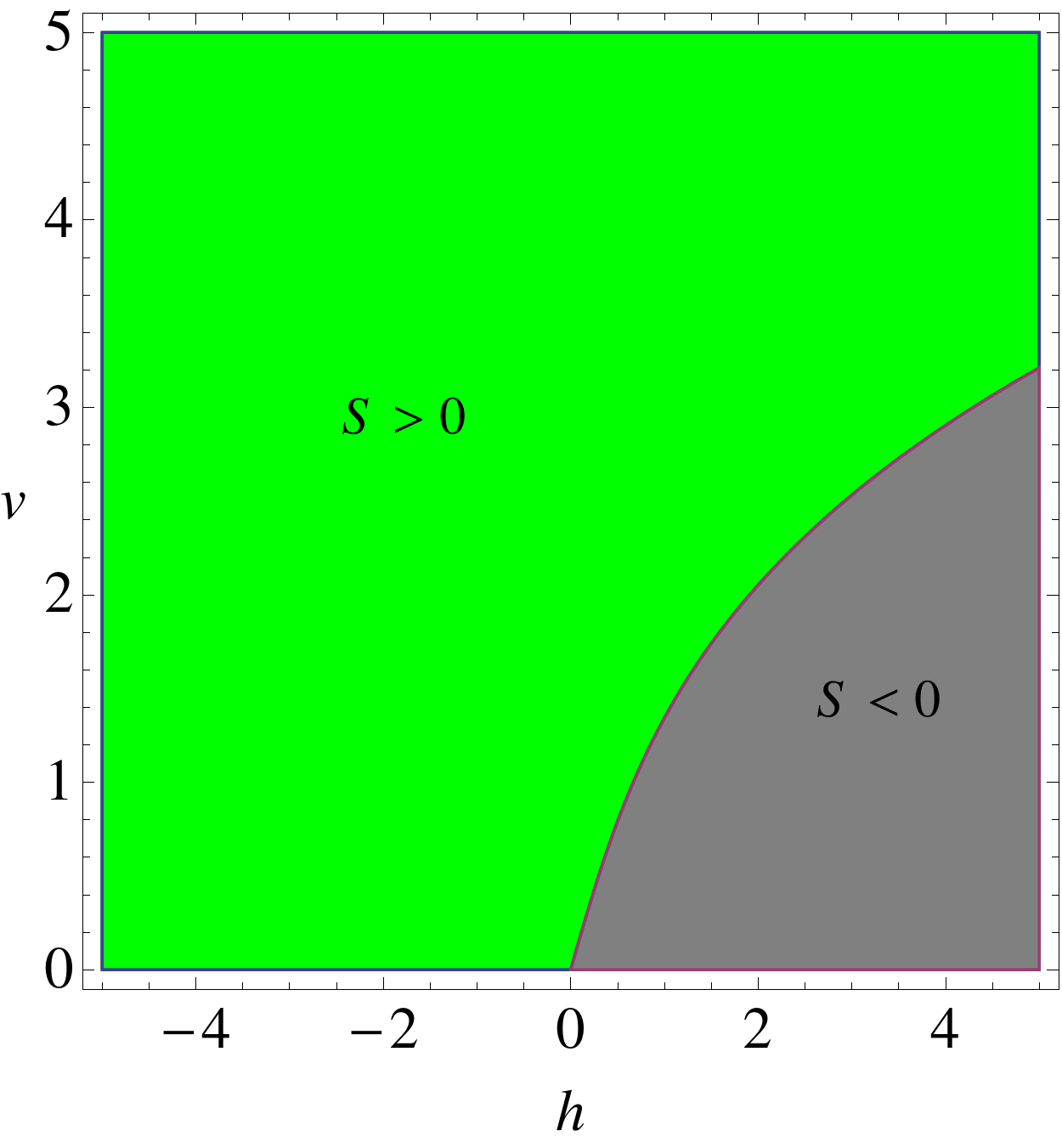}
\quad
\includegraphics[width=0.4\textwidth]{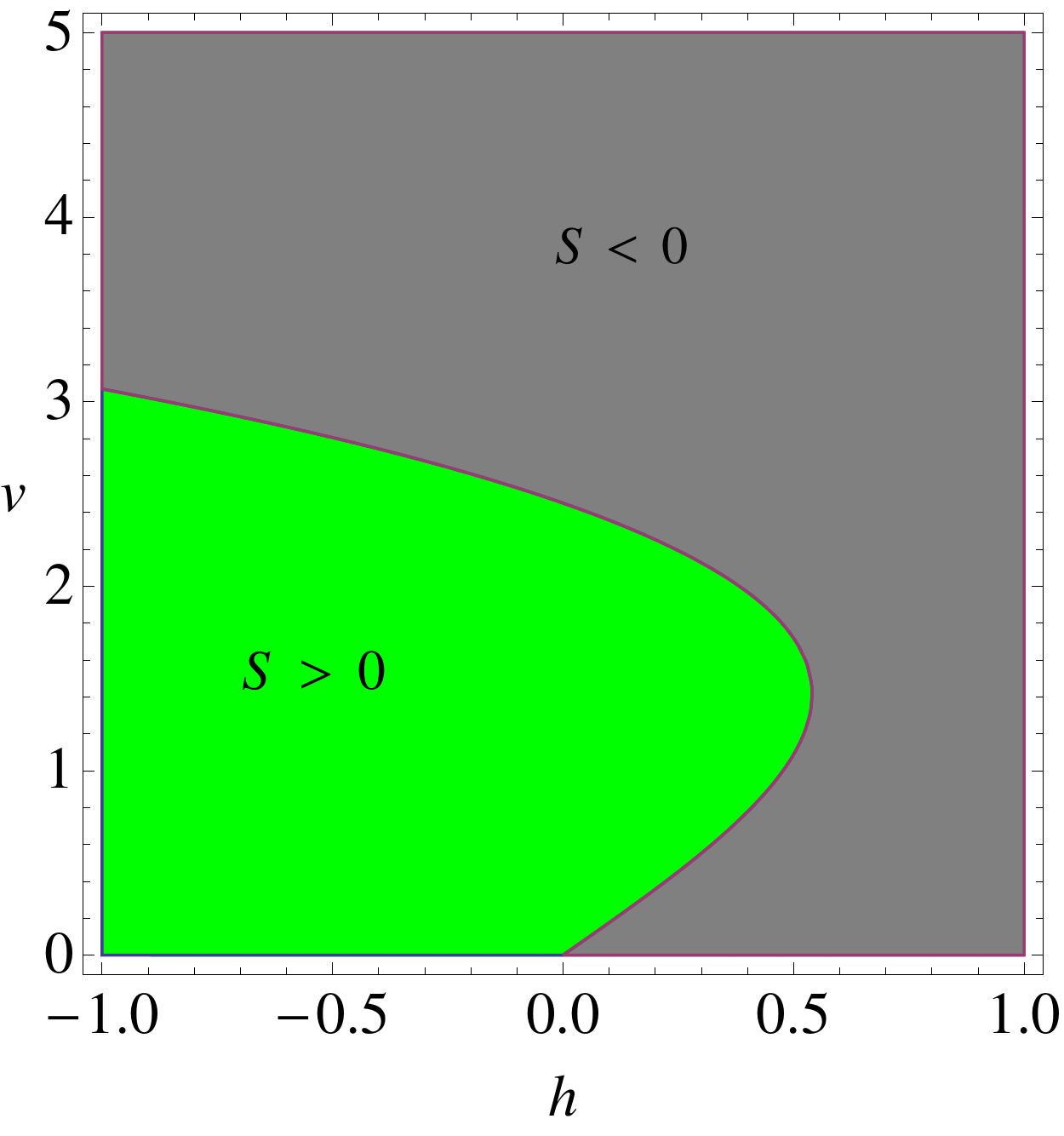}
\caption{{\bf The allowed volume in presence of hair}: \textit{Left}: For spherical horizons ($\sigma=1$).  \textit{Right}: For hyperbolic horizons ($\sigma=-1$). In each case, the green shaded region corresponds to positive entropy, while in the gray shaded region the entropy is negative. It is interesting to note that, in the hyperbolic case, positivity of entropy results in a maximum black hole volume for any given hair parameter, $h$. Note that, since Eq.~\eqref{d5entropy} in independent of the electric charge, these plots are valid for any value of $q$. }
\label{entropy1}
\end{figure}

From Eq.~\eqref{gbentropy}, the positive entropy condition for $d=5$ is given by
\be\label{d5entropy}
3\sigma v^3+18v-10\pi h\geq 0 \, .
\ee
For spherical horizons ($\sigma=1$) the inequality is trivially satisfied for $h\leq 0$. For $h>0$, the entropy is positive if $v$ is not too small; that is, there is a lower bound on $v$ depending on $h>0$ for spherical black holes. Figure~\ref{entropy1} shows the region of positive entropy as function of $h$. For hyperbolic horizons ($\sigma=-1$), the situation is slightly more complicated, and is shown graphically in Figure~\ref{entropy1}.  The figure makes it possible to see the general result: large black holes with lots of hair
have negative entropy and so are not physically allowed.  In specific terms,
for $h<0$, there is an upper bound on the allowed volumes which depends on the specific value of $h$.  For $0<h < 6 \sqrt{2}/(5 \pi) \approx 0.540$ the hyperbolic black hole volume is bounded from above and below. For $h\gtrsim 0.540$, the entropy is always negative and hyperbolic black holes of positive entropy cannot exist. 

The equation of state is given by
\be
p=\frac{t}{v}-\frac{3\sigma}{2\pi v^2}+\frac{2\sigma t}{v^3}+\frac{q^2}{v^6}-\frac{h}{v^5}
\ee
and a critical point occurs when $p=p(v)$ has an inflection point, i.e.
\be 
\frac{\partial p}{\partial v}=\frac{\partial^2 p}{\partial v^2}=0 \, .
\ee
The critical points satisfy
\begin{align}\label{d5criticalpoints}
t_c&=\frac{3\sigma v_c^4-6\pi q^2+5\pi hv_c}{\pi v_c^3(v_c^2+6\sigma)} \, , 
\\
0 &=v_c^6-6\sigma v_c^4-10\pi\sigma q^2v_c^2-36\pi q^2+\left(20\pi +\frac{20}{3}\pi\sigma v_c^2\right)h v_c
\end{align}
and we see that the
qualitative effect of the hair is to introduce new terms of odd powers in the volume $v$.

As usual, planar black holes do not have any critical points and this is true for both Gauss-Bonnet and higher-order Lovelock gravity in arbitrary dimensions.  Furthermore, as noted earlier, in the planar case the black holes cannot support hair. For these reasons we disregard the planar solutions in our analysis  and henceforth assume $\sigma\neq 0$ for both
the equation of state and Gibbs free energy\footnote{In particular, we set $\sigma^{2n}=1$ since these even powers of $\sigma$ do not feature in these equations.}. In the following sections we will consider the critical behaviour for spherical and hyperbolic topologies separately.

\subsubsection{Spherical case}

\begin{figure}[tp]
\includegraphics[scale=0.25]{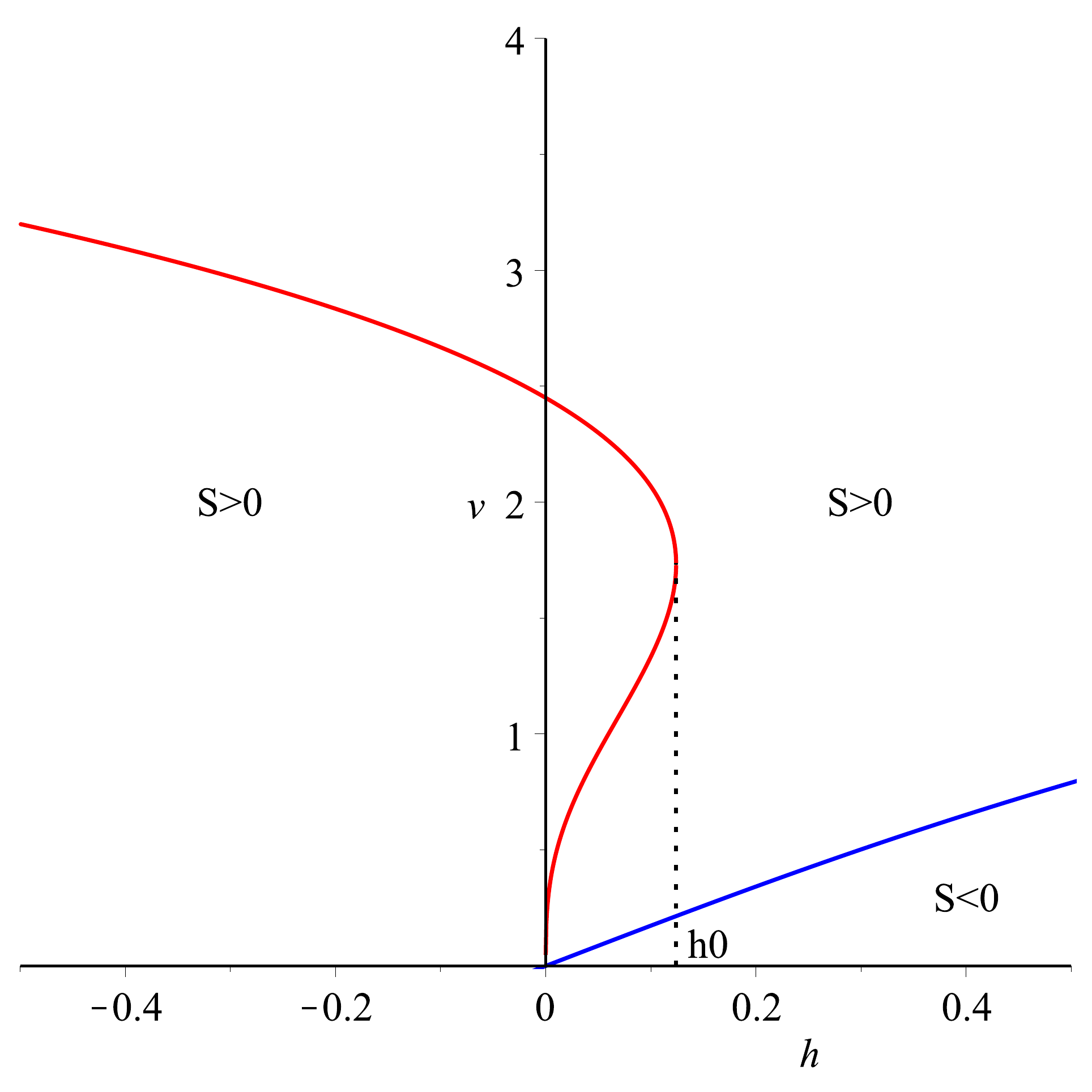}
\includegraphics[scale=0.25]{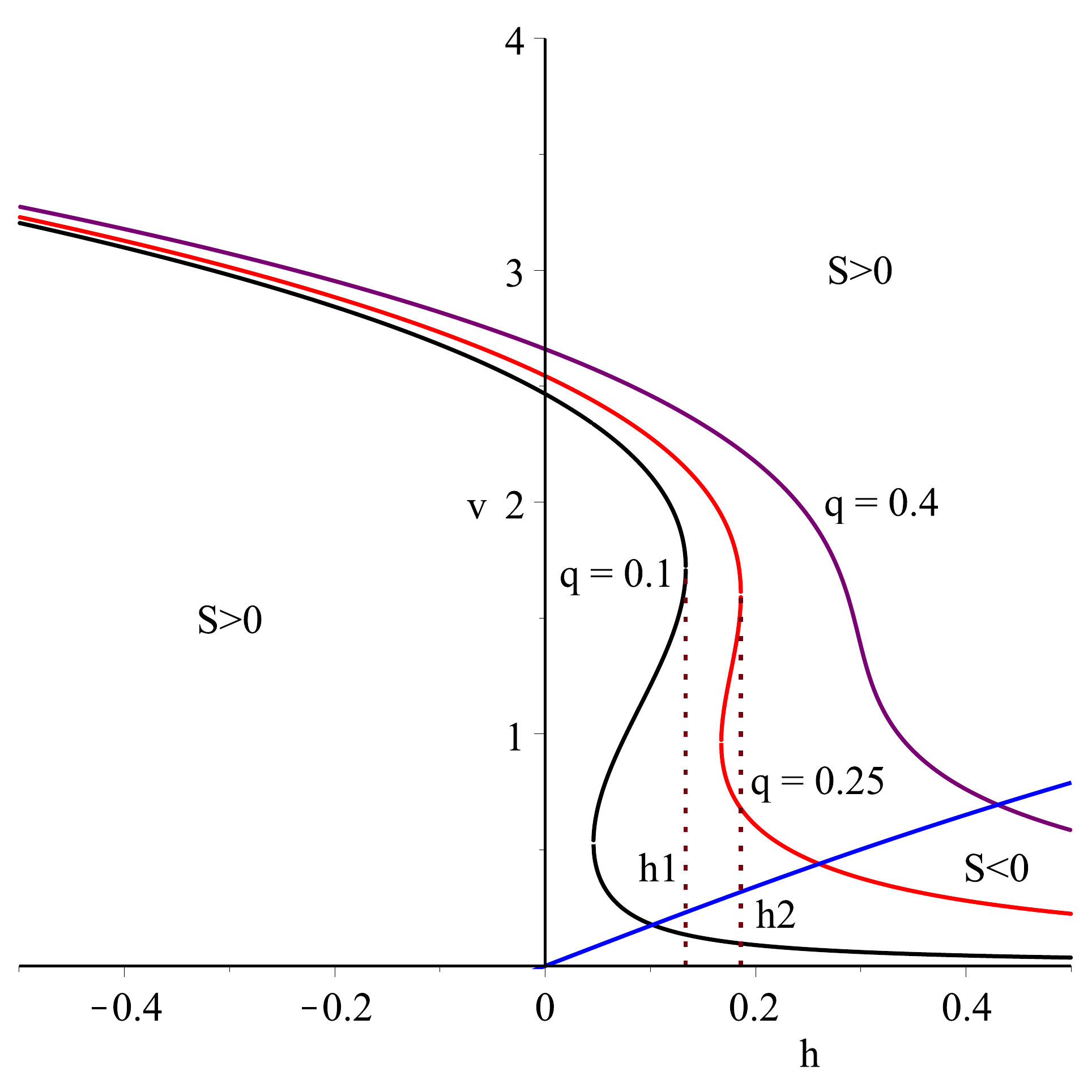}
\includegraphics[scale=0.25]{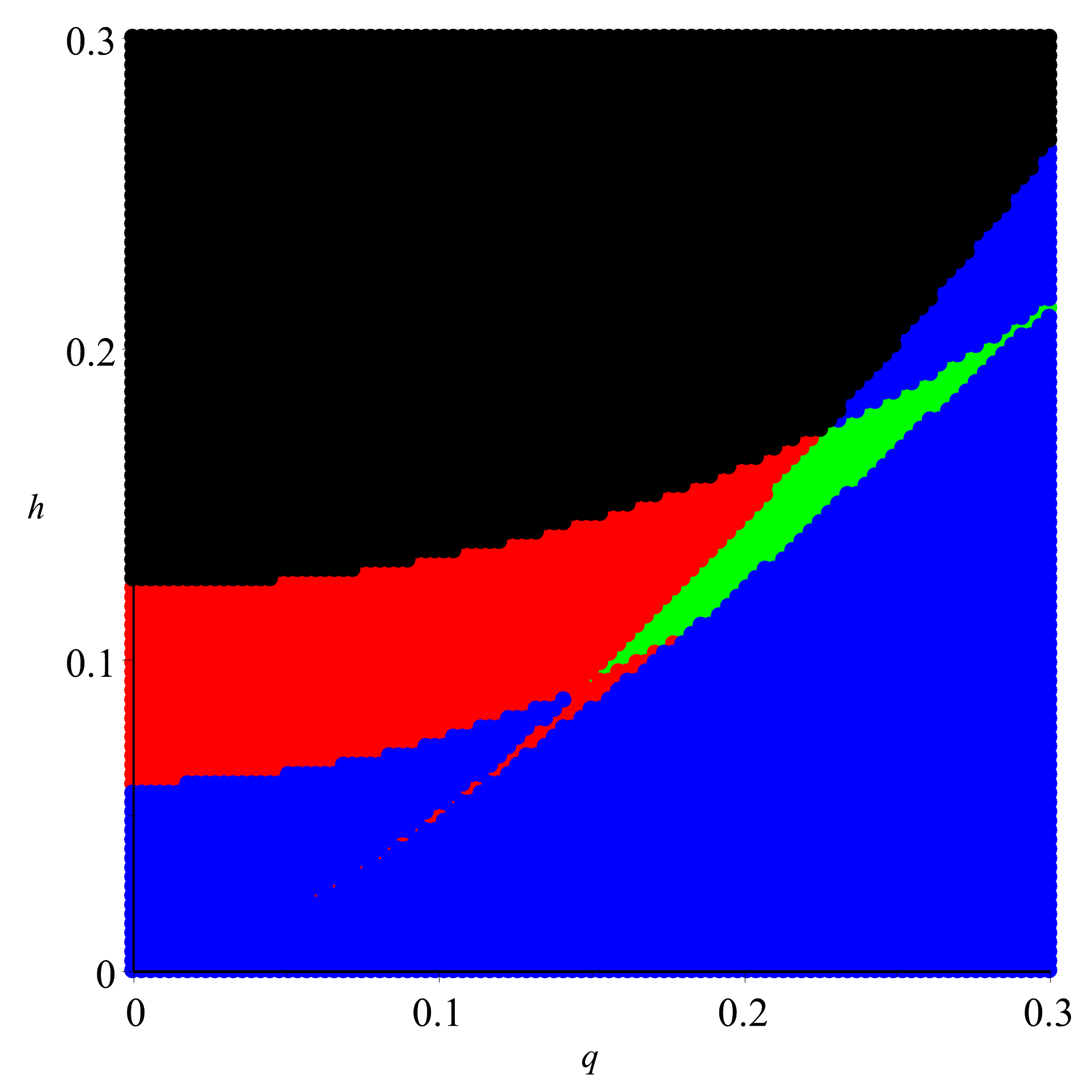}
\caption{{\bf Number of critical points for $\sigma=+1$ in $d=5$}: \textit{Left}: Uncharged ($q=0$) case. The red curve is the locus of all critical values $v_c$ as a function of $h$. For $h\gtrsim h_0\approx 0.1240$, there can be no critical points. For $h\in(0,0.1240)$, there can be up to two positive critical volumes (not necessarily all physical) while for negative $h$, there is exactly one physical critical point which exhibits standard VdW behaviour. \textit{Center}: Charged ($q\neq 0$) case. Depending on the value of $q$, as many as three critical points are possible.  For the weakly charged case $q\in(0.1414,0.3073)$, there is a small interval  $(h_2,h_{*})$ with standard VdW behaviour, with $h_{*}$ given by the intersection of the zero-entropy curve (blue) and critical volume curves.
In both of these plots, the region below the blue, diagonal line corresponds to negative entropy. \textit{Right}: a plot of $(q,h)$ parameter space showing the number of possible physical critical points, taking into account positive entropy and maximum pressure constraints. We ignore $h<0$ in this plot since it always has just one critical point with VdW character. Here, black, blue, red and green regions represent zero, one, two and three physical critical points, respectively. Note that the plot does not exclude critical points associated with, for example, a `cusp' in the Gibbs free energy, for which no phase transitions occur.}
\label{gb_crit1}
\end{figure}

\begin{figure}[htp]
\includegraphics[scale=0.25]{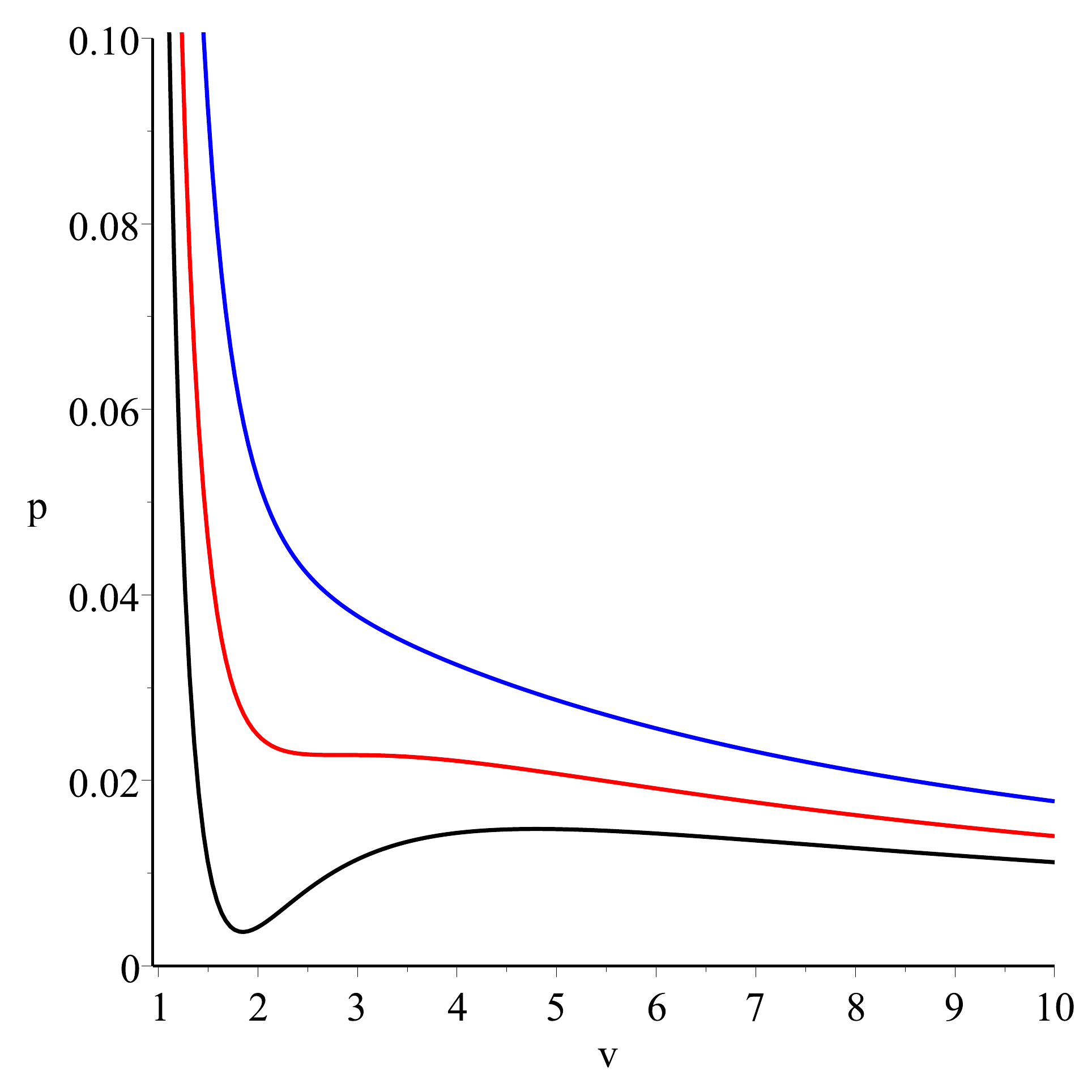}
\includegraphics[scale=0.25]{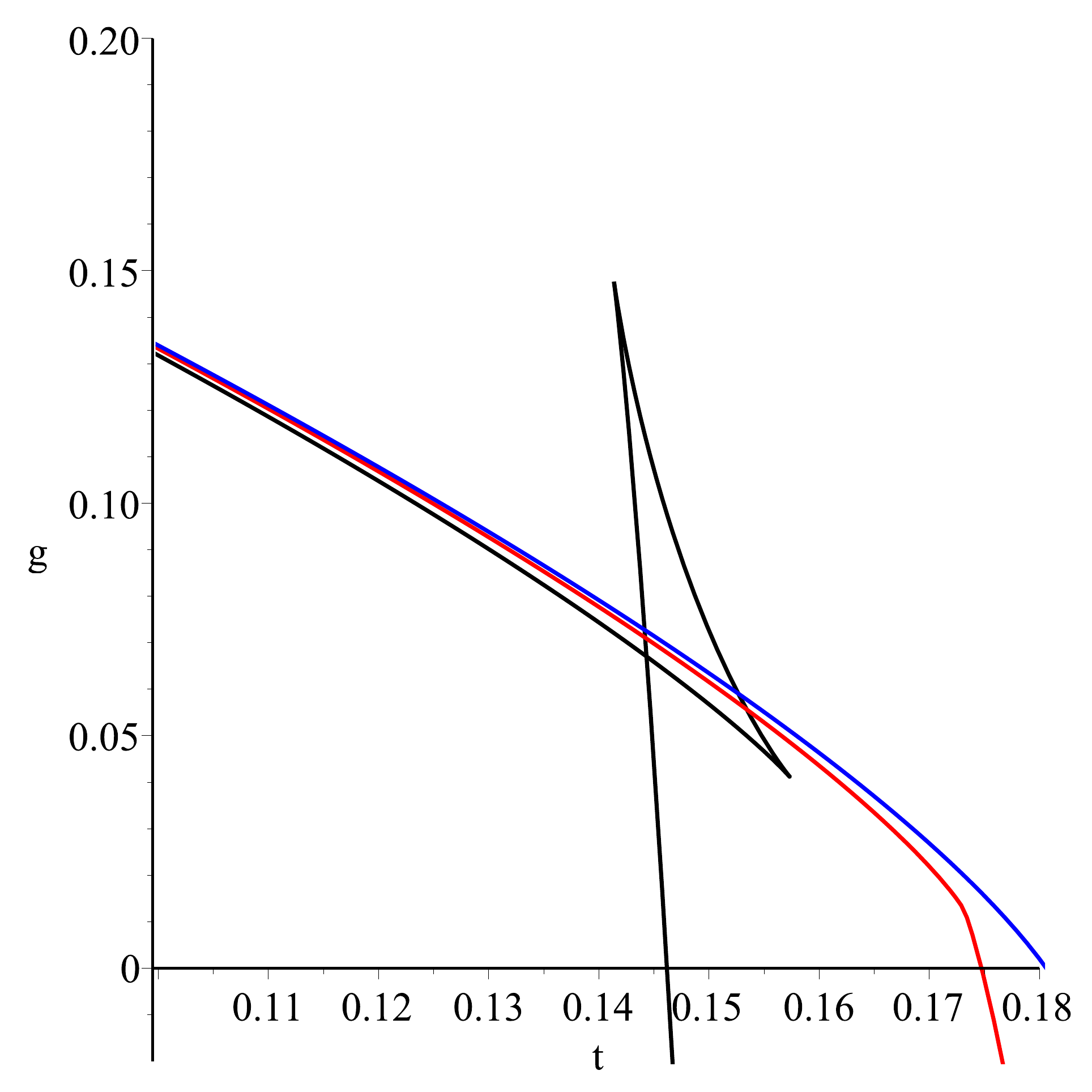}
\includegraphics[scale=0.25]{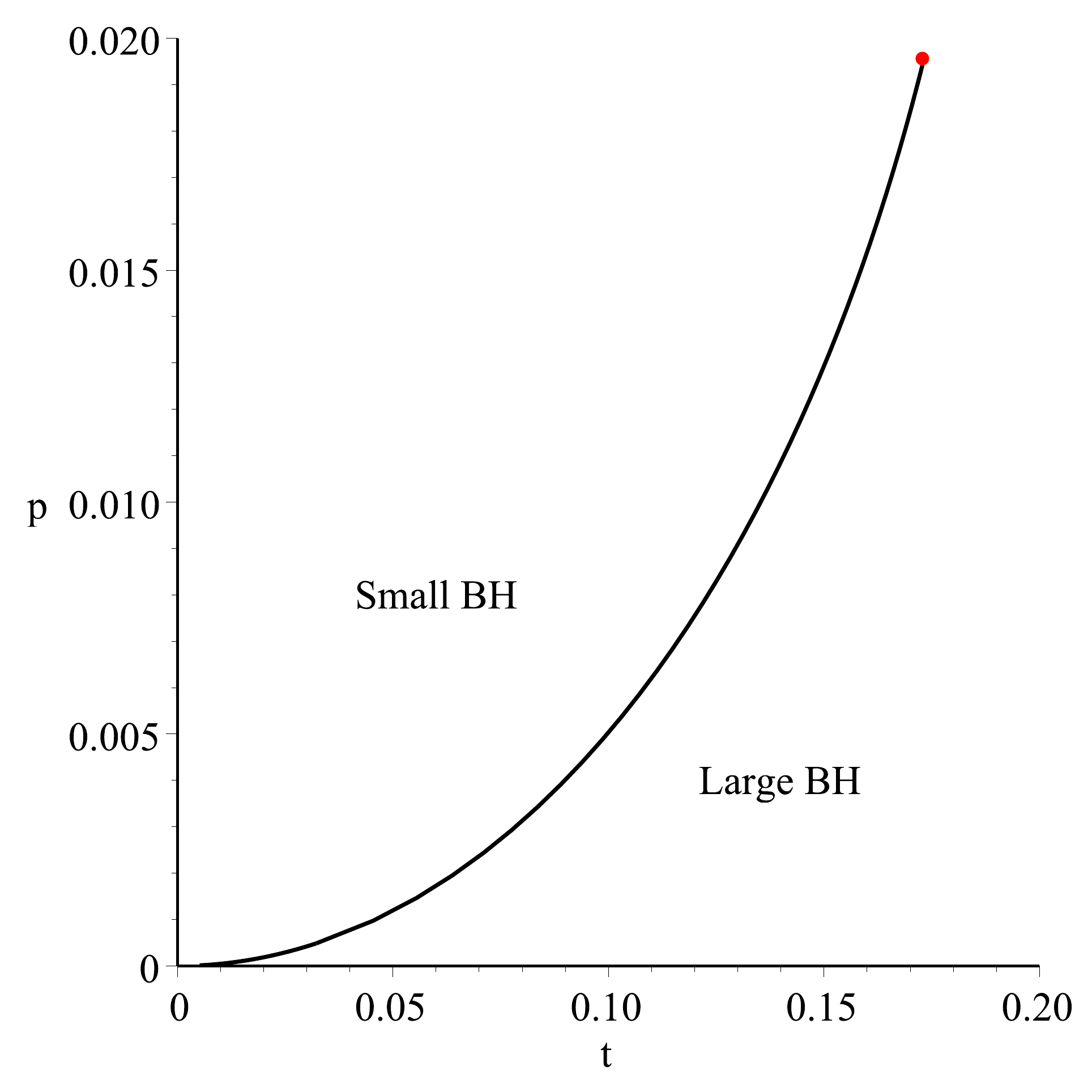}
\caption{{\bf Uncharged hairy black hole}: the case for $h=-0.5, \, q=0, \, \sigma = +1$. \textit{Left:} the $p-v$ diagram, showing the usual VdW oscillation. The red curve represents critical isotherm at $t=t_c$. The blue and black curves correspond to $t>t_c$ and $t<t_c$, respectively. \textit{Centre}: the $g-t$ diagram. The black curve represents $p<p_c$, the blue curve correspond to $p>p_c$ and the red curve is for $p=p_c$. We observe standard swallowtail behaviour. \textit{Right:} The $p-t$ diagram, showing the coexistence line of the first-order phase transition terminating at a critical point (shown here as a red dot). These plots are analogous to typical behaviour of the liquid-gas phase transition of the Van der Waals' fluid.}
\label{gb_VdW}
\end{figure}

Since analytic methods are not possible in our analysis due to the complexity of the Gibbs free energy and equation of state, we resort to graphical and numerical methods. In Figure~\ref{gb_crit1}, we plot the set of critical volumes $v_c$ as function of $h$ and superimpose the condition for positive entropy. We see that in the uncharged ($q=0$) case, there are no critical points for $h\gtrsim 0.1240$ while $h<0$ admits at most one physical critical point. For $h\in(0,0.1240)$ we can have up to a maximum of two critical points which are not necessarily physical (for example $p_c$ or $t_c$ may be negative, or $p_c>p_{max}$). It is not difficult to show that for $q=0,h<0$ we have exactly one physical critical point which displays standard Van der Waals' behaviour, as shown in Figure~\ref{gb_VdW}.

\begin{figure}[htp]
\includegraphics[scale=0.38]{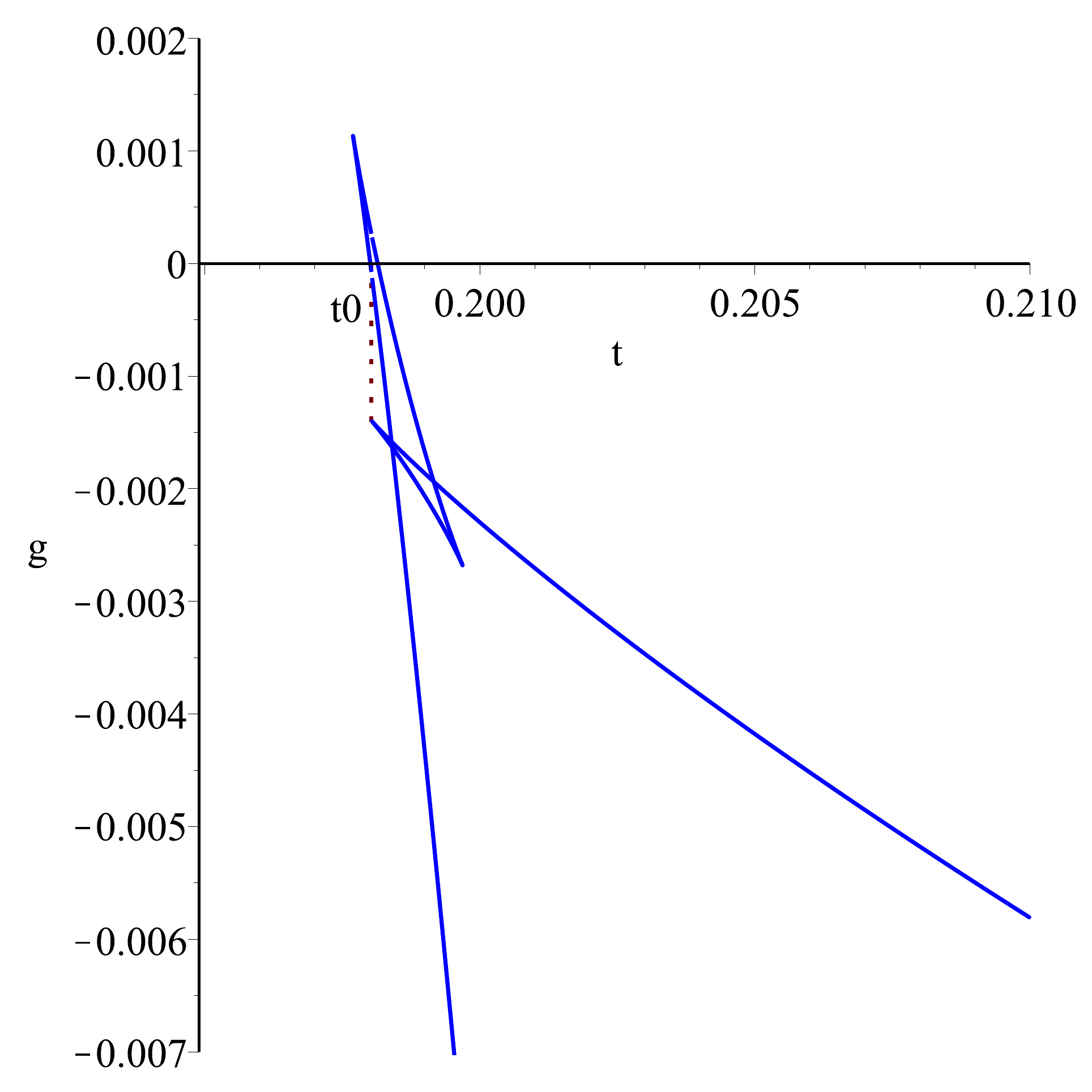}
\includegraphics[scale=0.38]{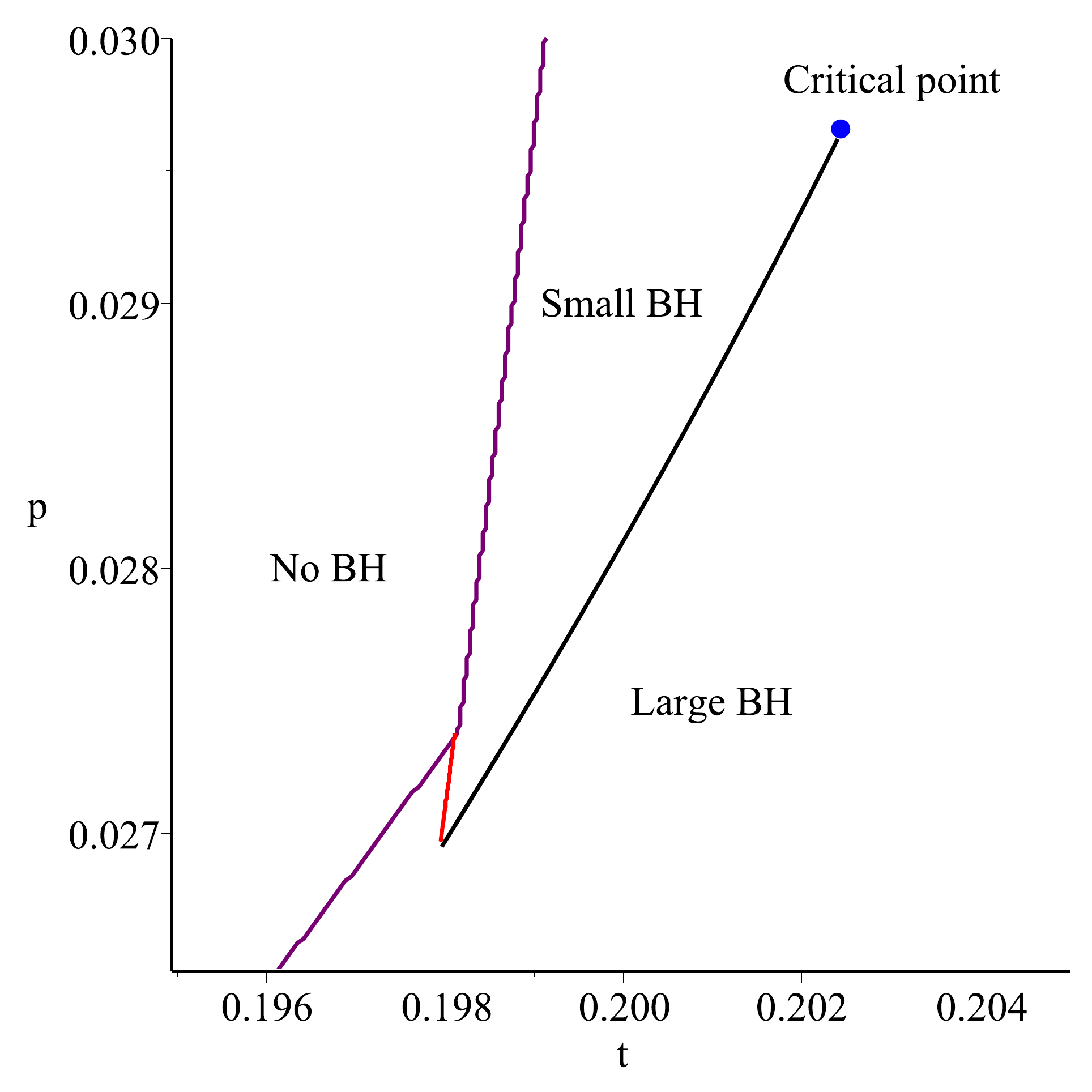}
\caption{{\bf Uncharged hairy black holes}: the case for $h>0 , \, \sigma = +1$. \textit{Left}: Gibbs free energy for $h=0.09, \, p\approx 0.0272$. There is a reentrant phase transition corresponding to the zeroth-order phase transition at $t=t_0$  followed by a first-order phase transition at the intersection with the swallowtail structure. \textit{Right}: the corresponding $p-t$ phase diagram. The red curve represents the zeroth-order phase transition, while the purple curve marks the boundary where no black hole solution exists. The black curve is the first-order coexistence curve, which terminates at the critical point, marked here by a blue circle.}
\label{gb_cusp}
\end{figure}

For $0<h\lesssim 0.1240$ there can be up to two physical critical points. Already in the uncharged case we can see a new thermodynamic phenomenon in $d=5$ spherical black holes, namely a \textit{reentrant phase transition}\footnote{A reentrant phase transition is said to occur if, through a monotonic variation of a thermodynamic parameter,  the system is observed to change phase two or more times, with the final phase and initial phase being macroscopically identical. These transitions were first observed in the miscibility of nicotine/water mixtures as temperature is varied \cite{hudson1904mutual}.} (RPT), as shown in Figure~\ref{gb_cusp}. This type of phase transition, though now common to observe in black hole systems, has not been found previously in $5$ dimensional, spherical black holes in Gauss-Bonnet gravity. Since critical points exist only up to $h\approx 0.1240$, this exhausts all possibilities for uncharged spherical hairy black holes. A more complete and organized classification of the possible thermodynamic behaviour for Gauss-Bonnet hairy black holes is given in Table~\ref{tab:gb_5d_spherical}.

For the charged case, we observe (c.f. Figure~\ref{gb_crit1}) that critical points can exist, in principle, for any $h$;  however, for any given $q$, there will always be an  $h_{*}$ such that for $h > h_{*}$ all of the critical volumes correspond to negative entropy black holes. For $h<0$, the critical behaviour is analogous to that of the standard Van der Waals' fluid. If the charge is sufficiently small i.e. $q\lesssim 0.3073$ then there is a small range of $h$ where up to three critical points are possible. For $q\in (0.1414,0.3073)$,  there is a small range of positive $h_2<h<h_{*}$ (c.f. Figure~\ref{gb_crit1}) with standard VdW character, noting that $h_{*}$ is determined by both the positive entropy condition and charge, given by the intersection of the zero-entropy curve and the critical volume curves. Since there can be up to three physical critical points for some range of $q$ and $h$, we can expect some new thermodynamic behaviour which cannot be found for $d=5$ Gauss-Bonnet spherical black holes in the absence of hair.  It turns out, this expectation is justified; the complete list of possible thermodynamic behaviour is presented in Table~\ref{tab:gb_5d_spherical}, and we now proceed to illustrate some of these thermodynamic phenomena.

\begin{figure}[tp]
\centering
\includegraphics[scale=0.38]{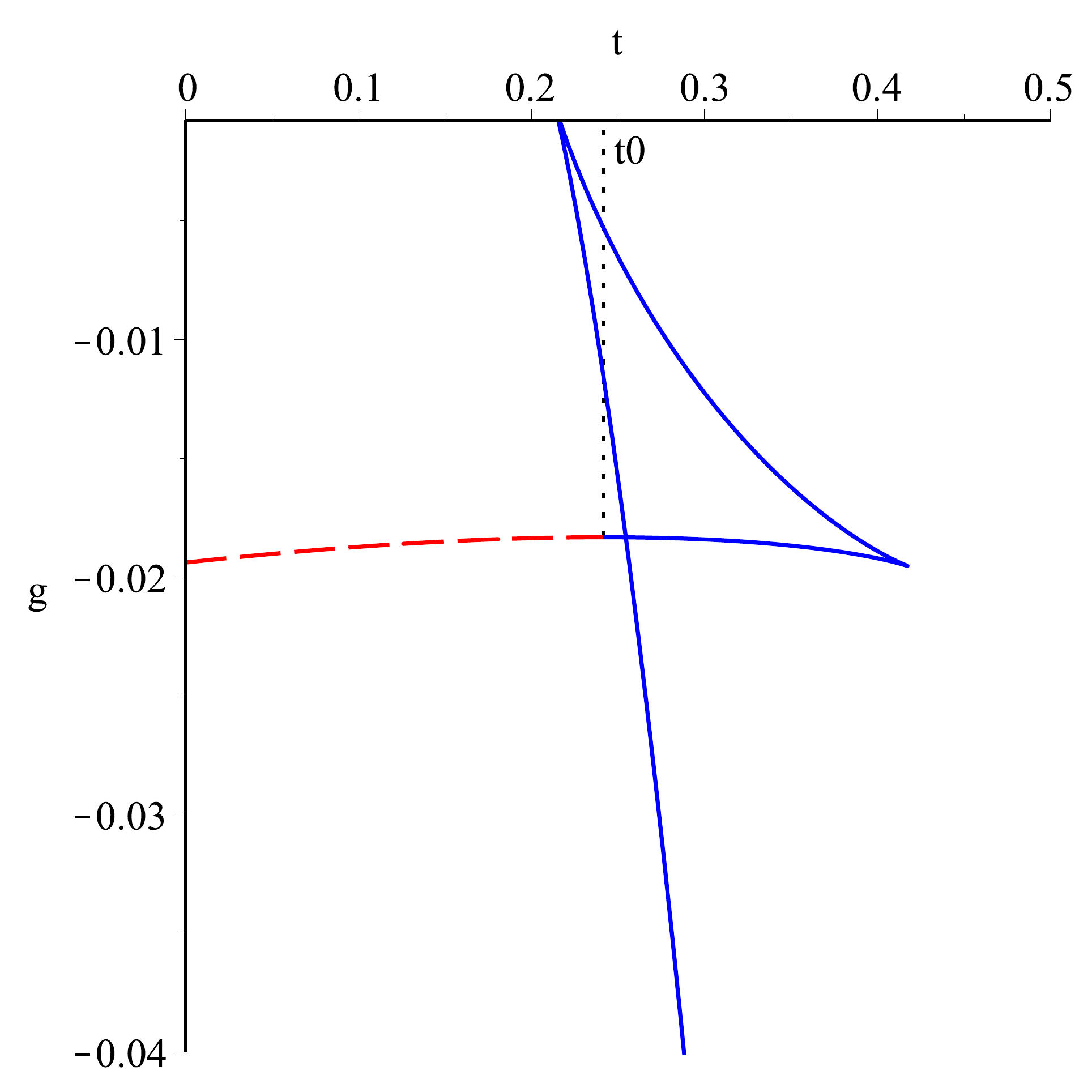}
\includegraphics[scale=0.38]{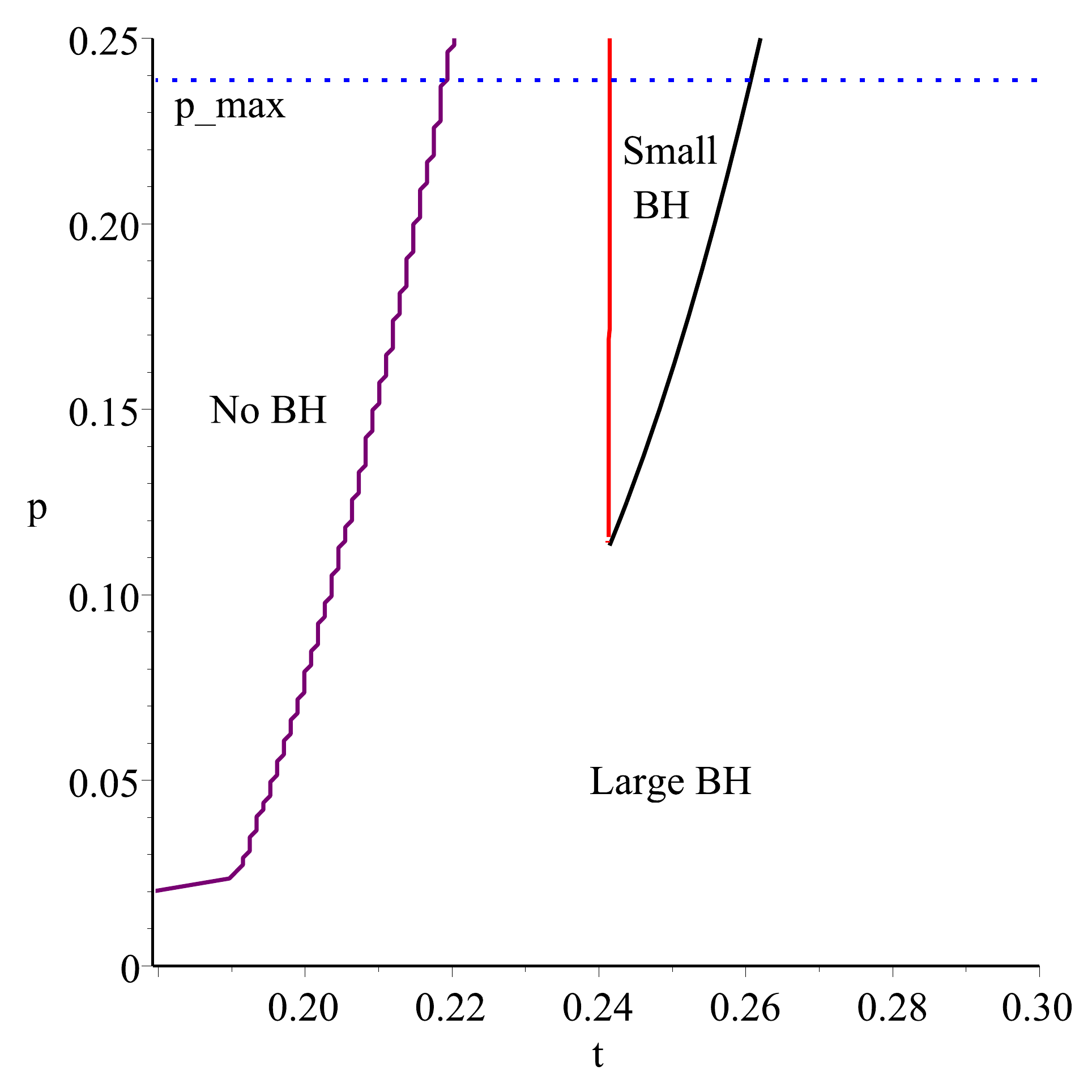}
\caption{{\bf Reentrant phase transition in charged spherical black holes}: \textit{Left}: Gibbs free energy for $h=0.08, \, q=0.1, \, p=3/(5\pi)=0.8p_{max}$. The dashed red curve highlights portions the Gibbs free energy  corresponding to negative entropy black holes.  Regarding the negative entropy black holes as unphysical gives rise to a zeroth-order phase transition which occurs at $t=t_0$. \textit{Right}: The $p-t$ phase diagram. The red curve denotes $(p,t)$ at which zeroth-order phase transition occurs, while the black line corresponds to a first order phase transition. The purple curve marks the border of the region admitting no physical (i.e. positive entropy) black hole solutions. It is clear that there is a large BH/small BH/large BH reentrant phase transition for $0.11\lesssim p\leq p_{max}$. Note that in this case, none of the critical points are physical---while the first order phase transition does terminate at a critical point, this occurs for $p_c>p_{max}$.}
\label{gb_RPT}
\end{figure}

\begin{figure}[htp]
\includegraphics[scale=0.25]{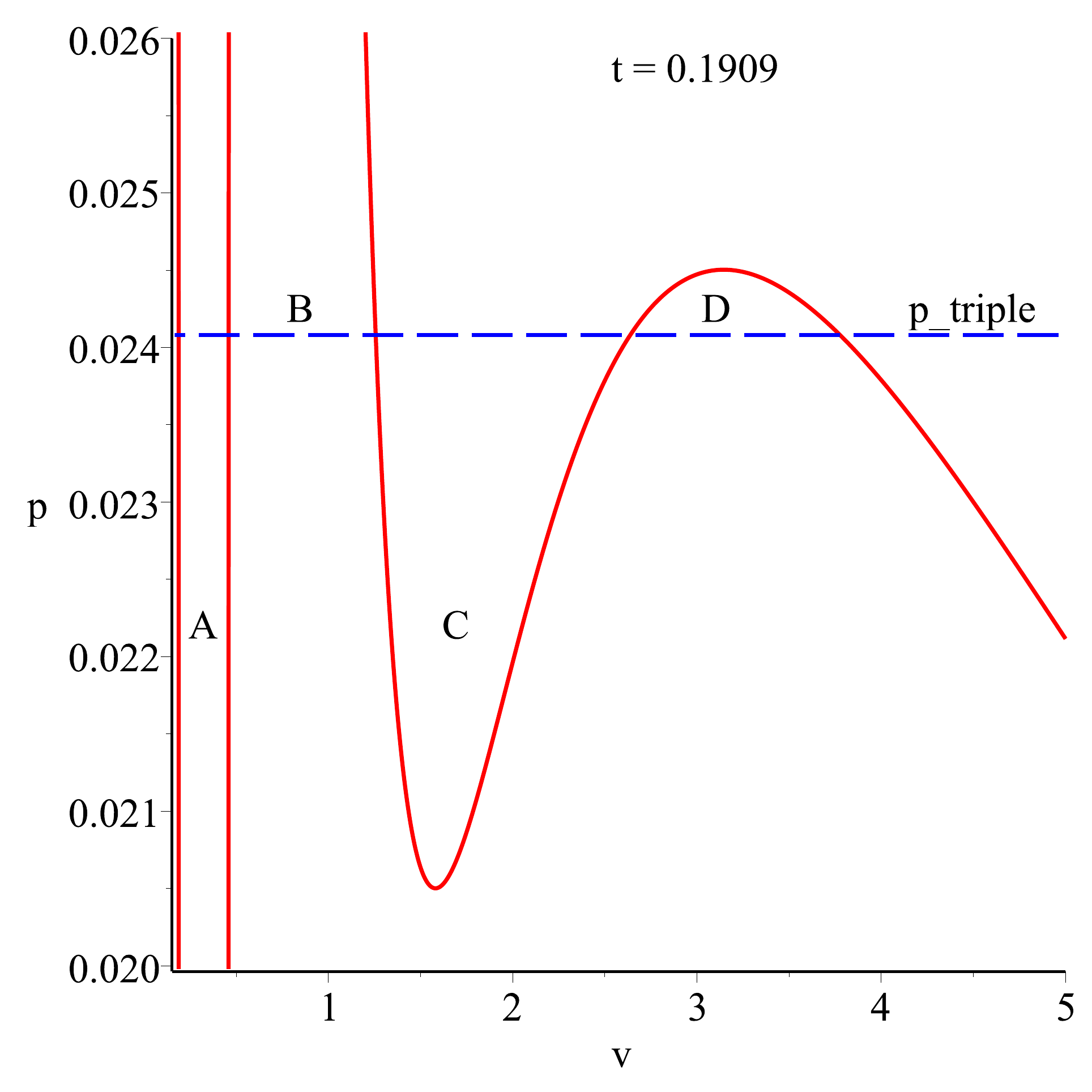}
\includegraphics[scale=0.25]{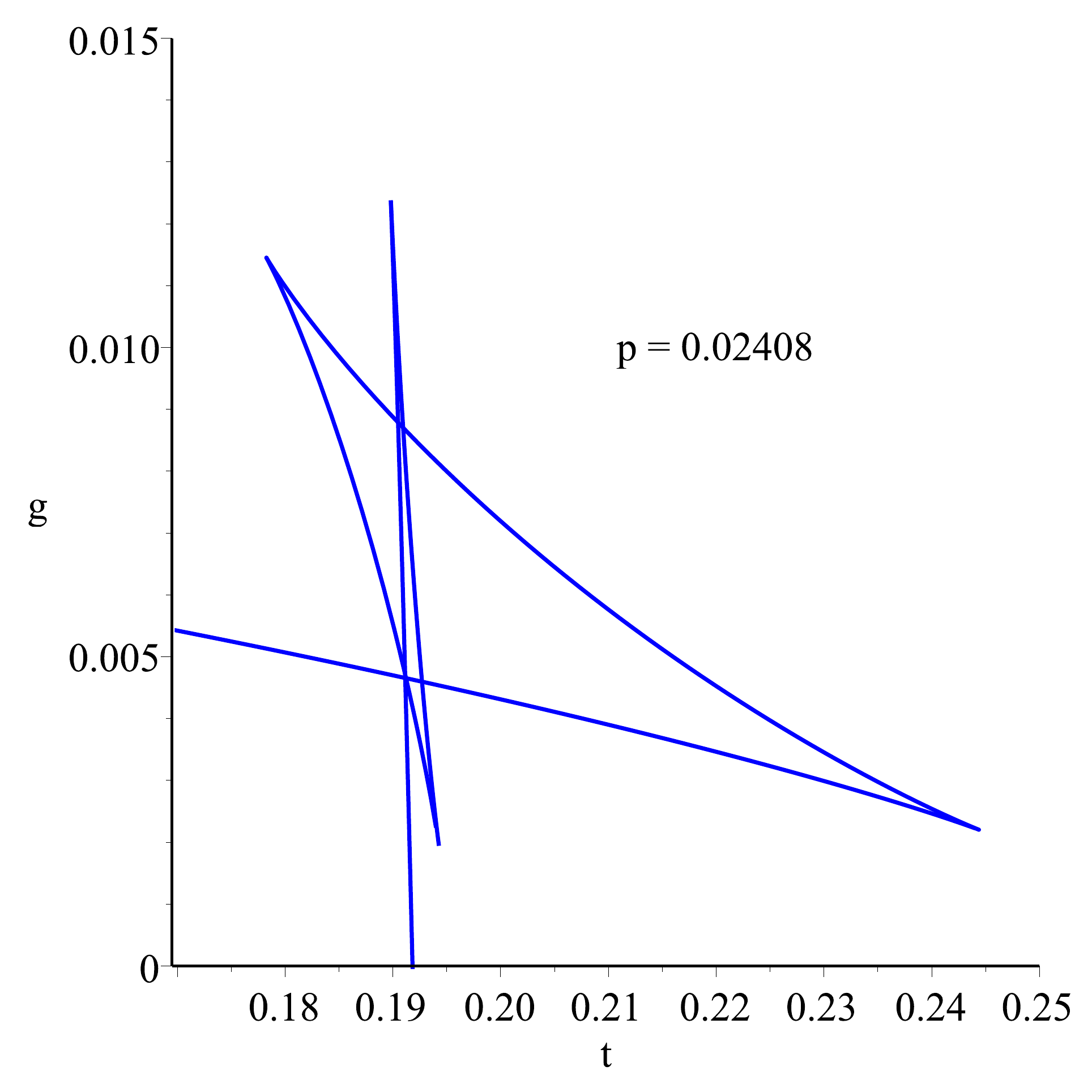}
\includegraphics[scale=0.25]{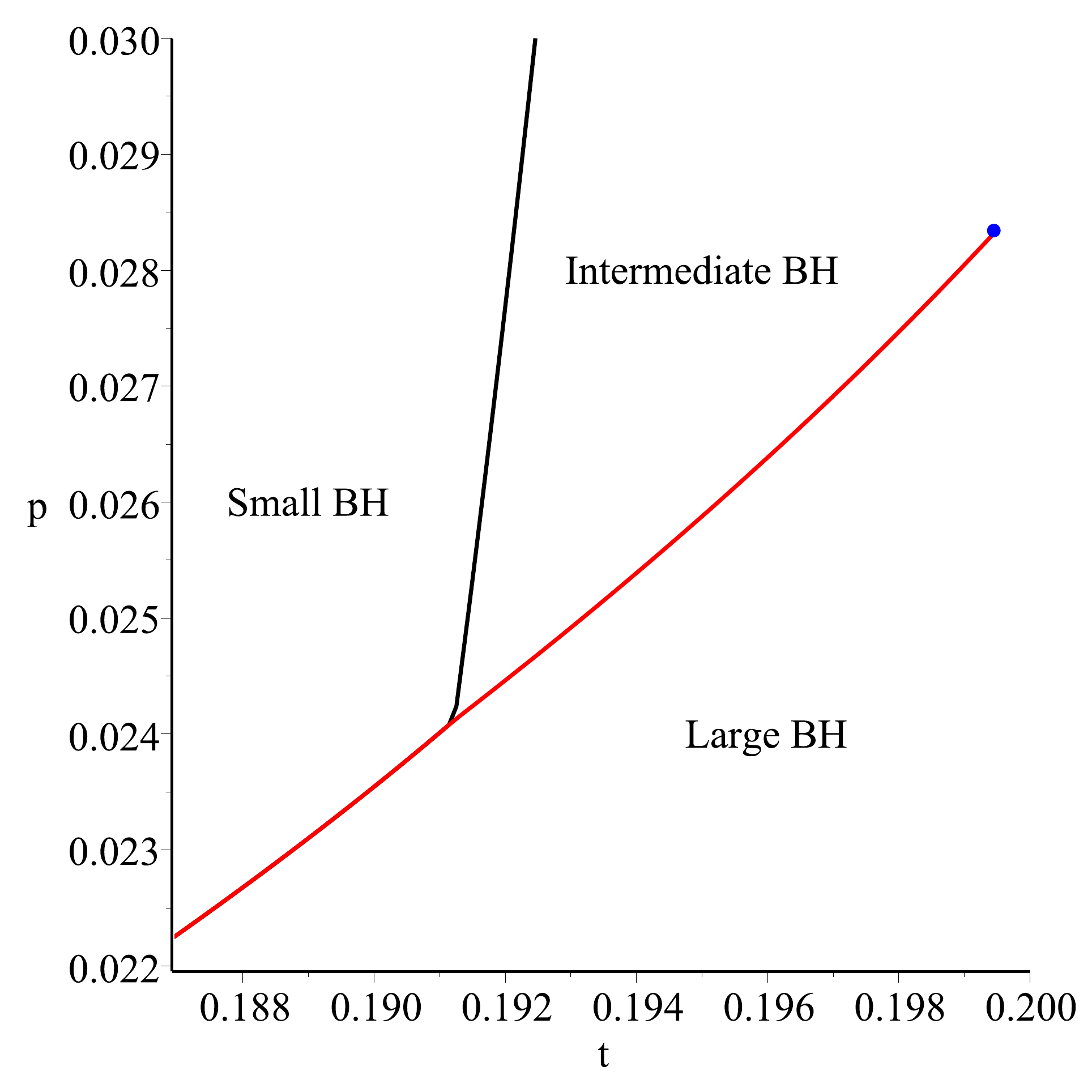}
\caption{{\bf Triple point for $h=0.064, \,  q=0.1 , \, \sigma = +1 $}: \textit{Left}: $p-v$ diagram showing a triple point situation, where there is a `double' Maxwell equal-area region for an isotherm where the triple point occurs, marked by region $A,B,C,D$. \textit{Centre}: The Gibbs free energy, showing three Gibbs branches intersecting at a point where a first-order phase transition occurs (i.e. triple point). \textit{Right}: The $p-t$ phase diagram showing a triple point at the intersection of the two coexistence lines. There is only one physical critical point (shown here as a small circle) as the other critical point (at which the second coexistence line terminates) has $p_c>p_{max}$ and hence is unphysical. The triple point occurs at pressure $p_{triple}\approx 0.02408$.}
\label{gb_triple}
\end{figure}

 Overall, we find that the presence of conformal hair has allowed a plethora of new thermodynamic behaviour for $d=5$ spherical Gauss-Bonnet charged black holes including: \textit{reentrant phase transitions} (RPT), \textit{triple points} (TP), and a new behaviour that we refer to as a \textit{virtual triple point} (VTP). Figure~\ref{gb_RPT} shows a particular case of RPT, where for some range of pressures there is a large/small/large black hole reentrant phase transition. Figure~\ref{gb_triple} shows a particular example illustrating a triple point, where the triple point occurs at $p_{triple}\approx 0.02408$ for $h=0.064, \,  q=0.1$. Instances of reentrant phase transitions in higher curvature gravity have been reported previously for cubic Lovelock gravity, while a triple point was previously found for electrically charged Gauss-Bonnet black holes in $d=6$~\cite{Frassino:2014pha}. Therefore the presence of conformal hair has given rise to genuinely new thermodynamic phenomena in five-dimensional Gauss-Bonnet gravity.

\begin{figure}[htp]
\includegraphics[scale=0.25]{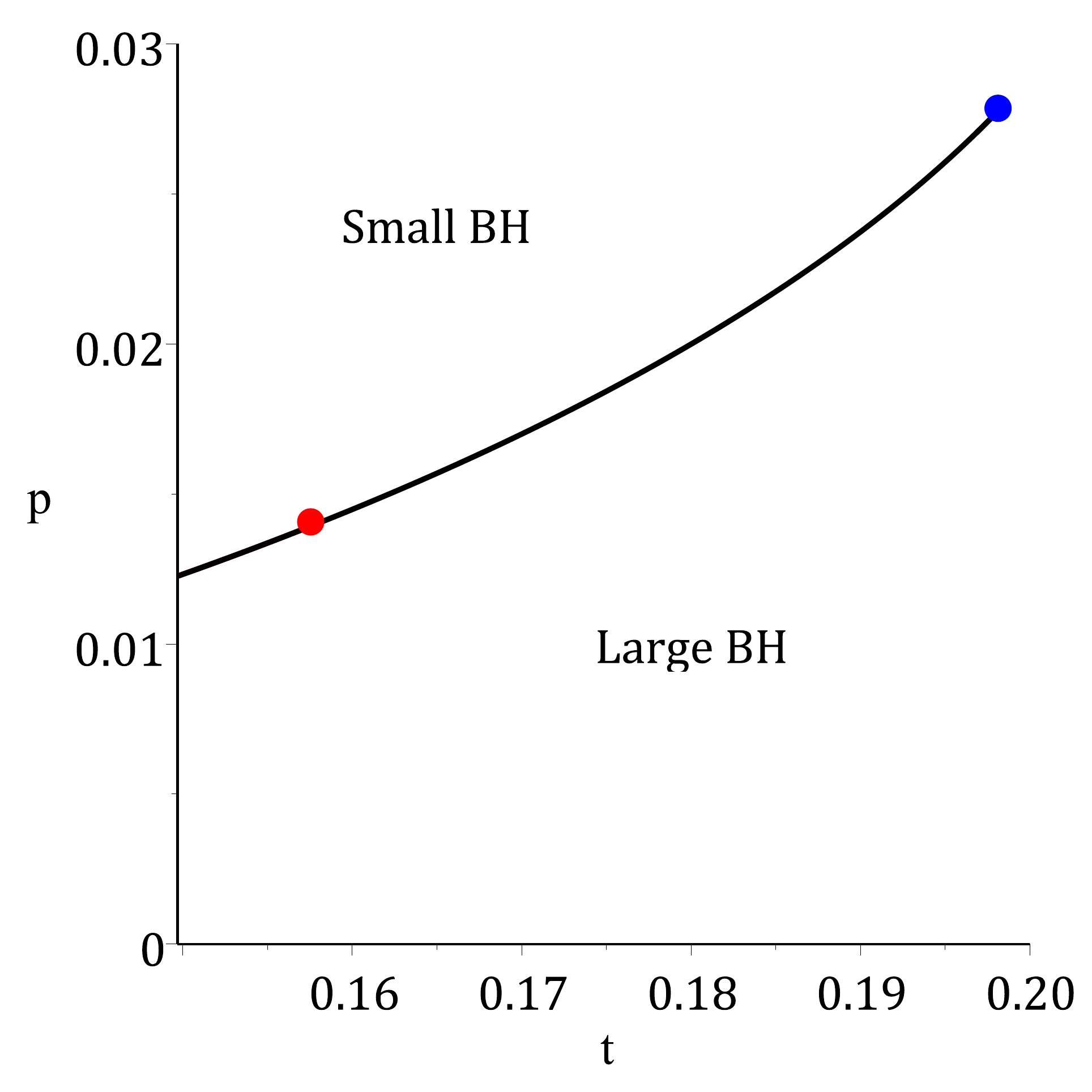}
\includegraphics[scale=0.25]{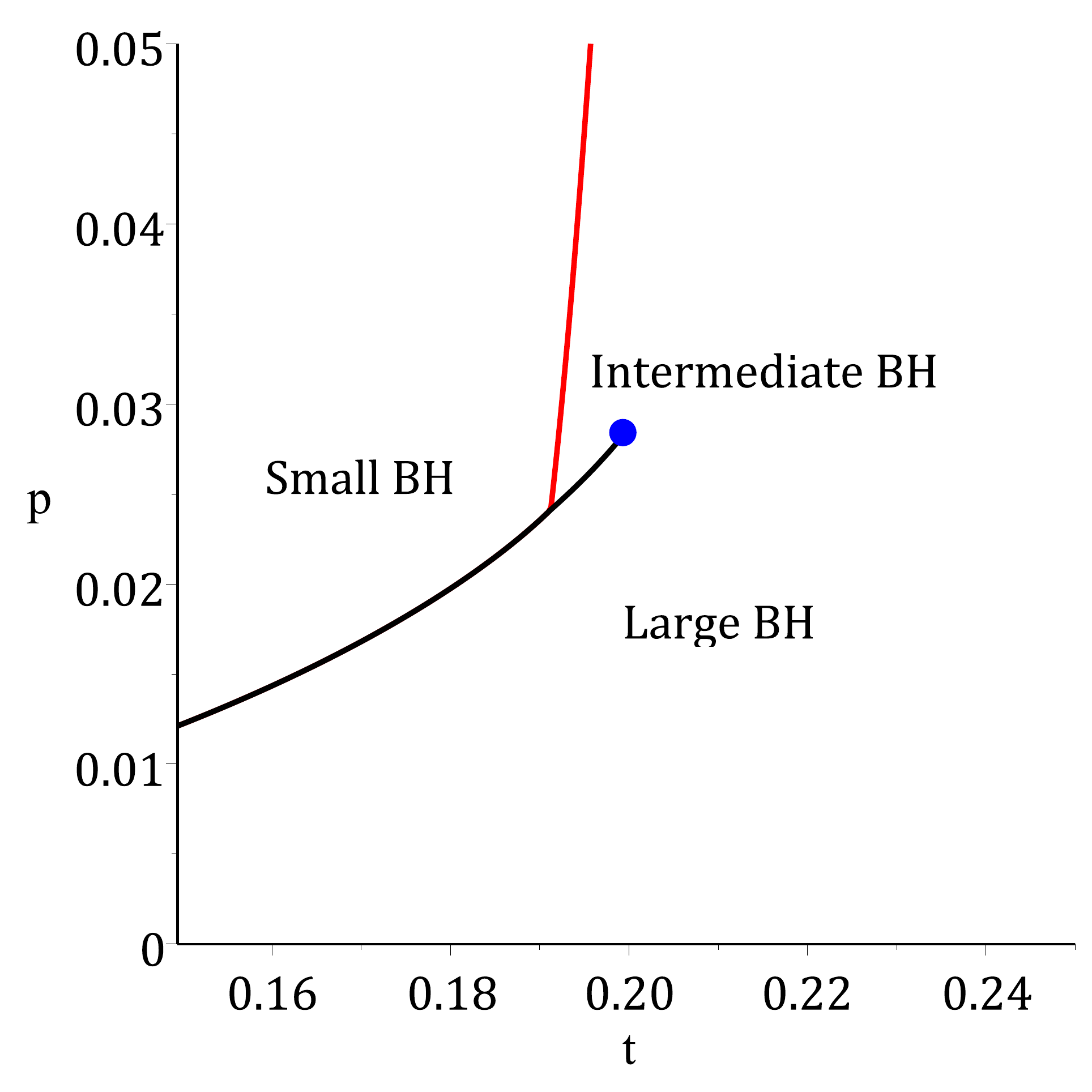}
\includegraphics[scale=0.25]{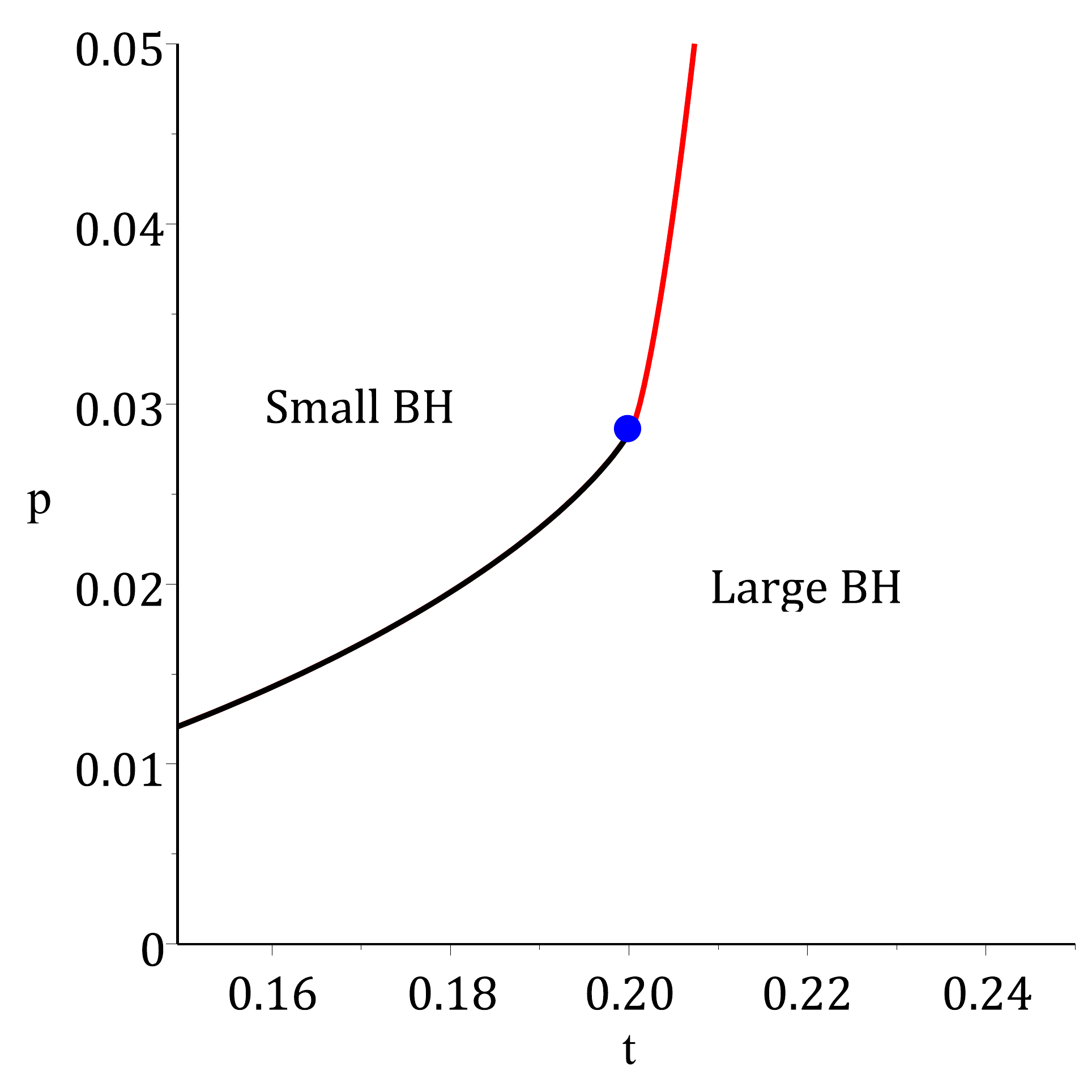}
\caption{{\bf Virtual triple point for $q=0.1 , \, \sigma = +1$}: \textit{Left}: The virtual triple point at $h\approx 0.0487682$. \textit{Centre}: triple point at $h=0.064$. \textit{Right}: another virtual triple point at $h\approx 0.06998$. Loosely speaking, $h$ acts as a tuning parameter which `moves' the position of the critical points in the phase diagram, thus altering the overall phase behaviour of the black hole spacetime. We see that as $h$ increases, starting from a virtual triple point, one of the coexistence curves extends upwards while the other shrinks giving rise to a triple point, and eventually producing another virtual triple point. }
\label{gb_vtp}
\end{figure}

\begin{figure}[tp]
\includegraphics[scale=0.25]{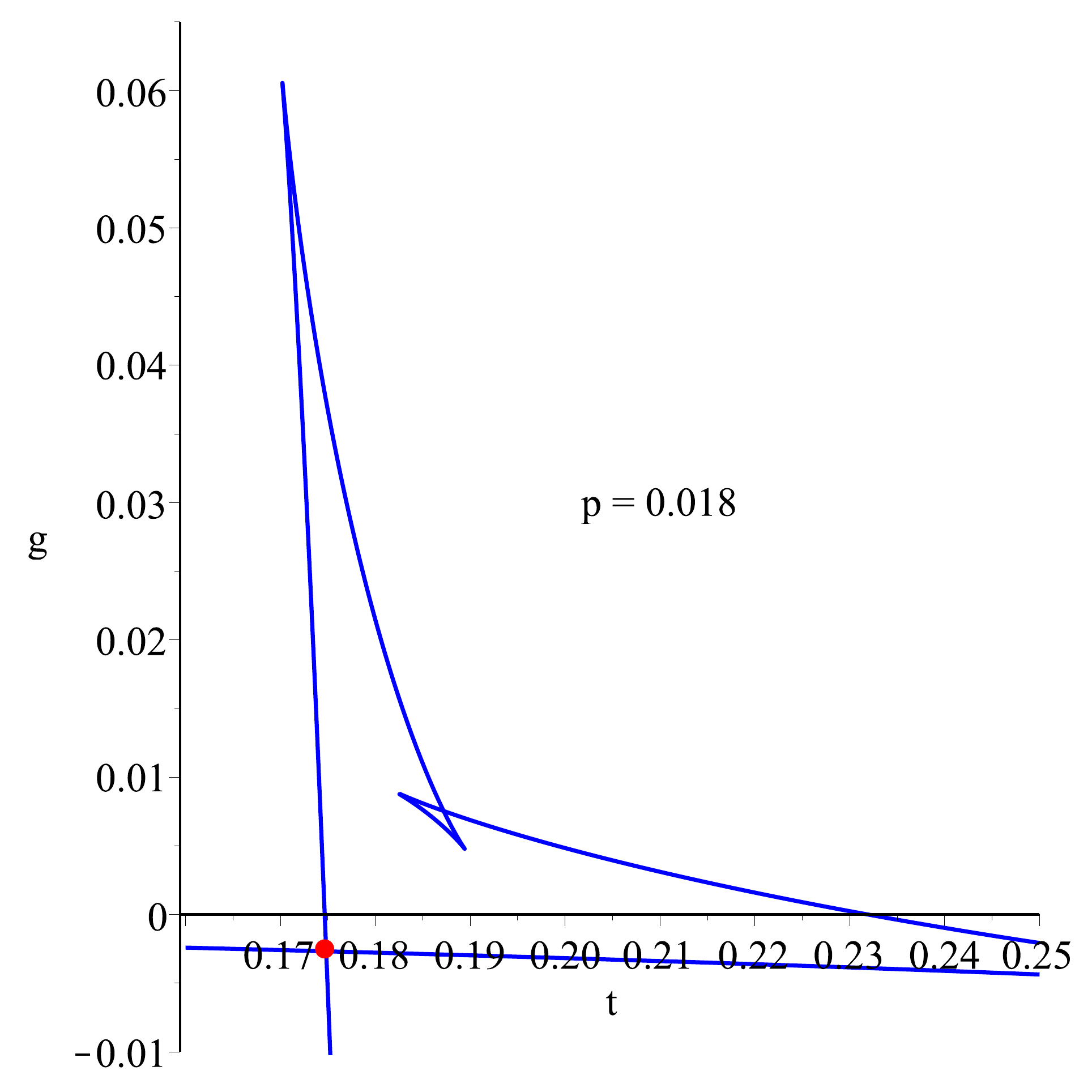}
\includegraphics[scale=0.25]{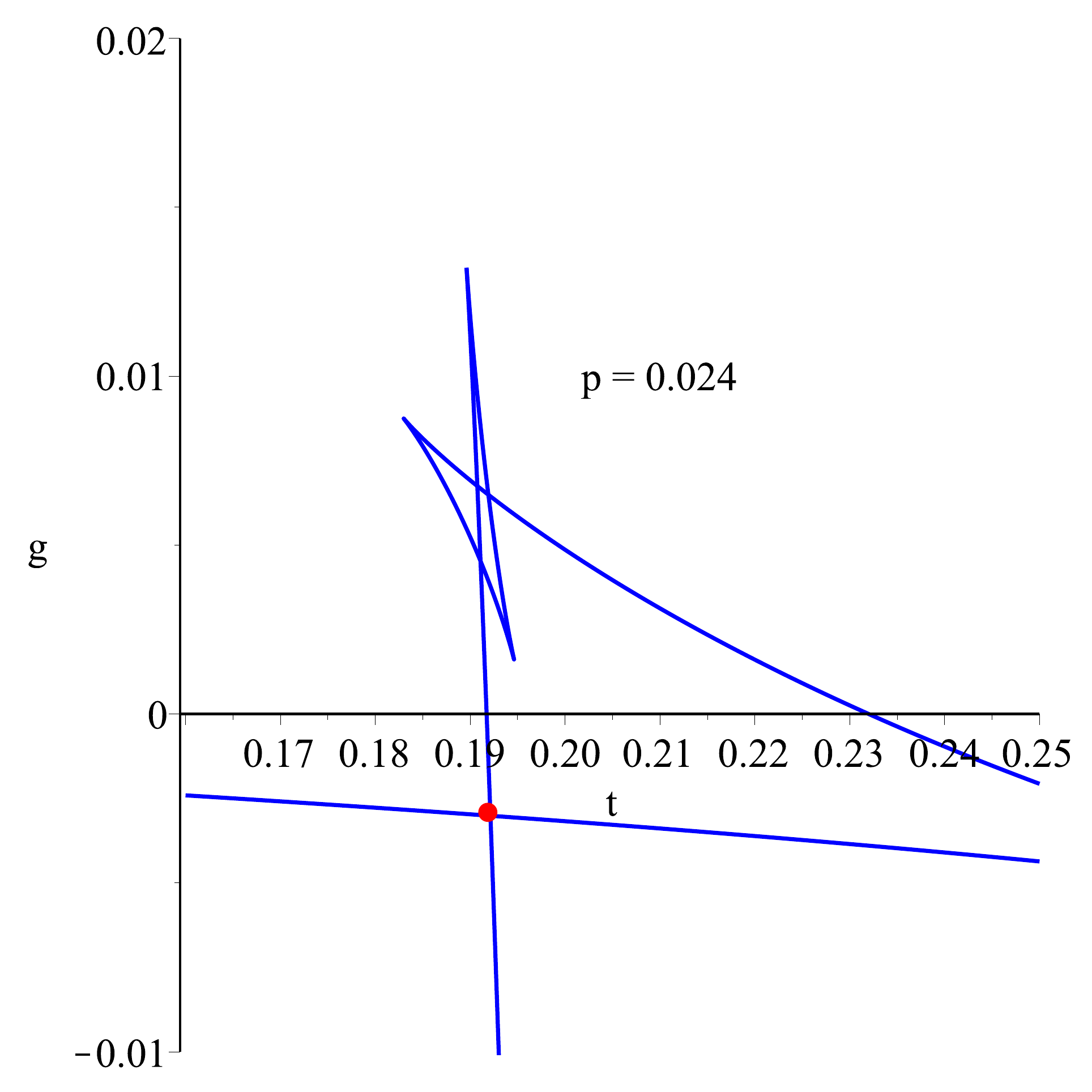}
\includegraphics[scale=0.25]{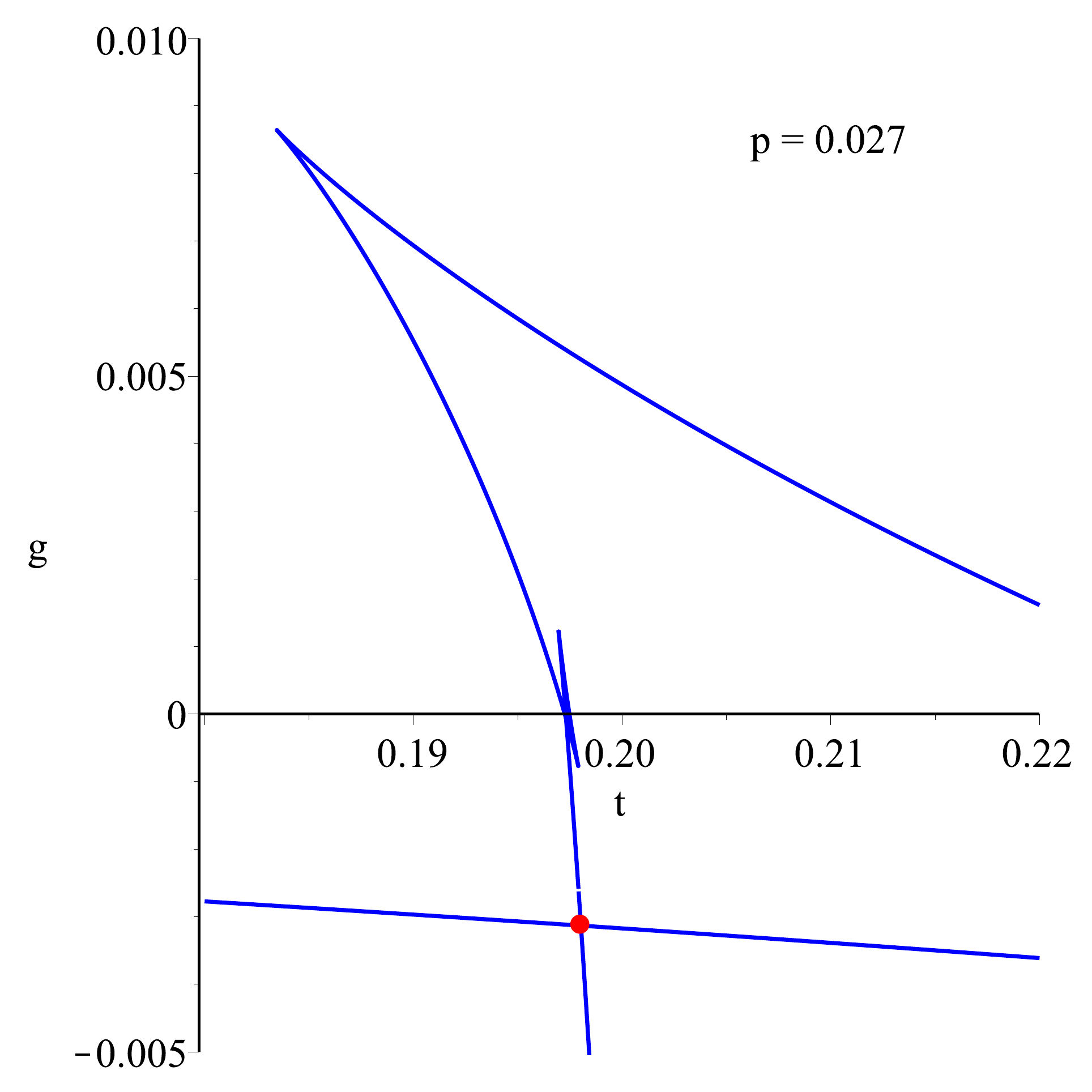}
\caption{{\bf Double-swallowtail behaviour for spherical case in $d=5$}: \textit{Left}: $g$-$t$ plot for $\sigma=1, \, h=0.06998, \, p=0.018$. There is a locally thermodynamically stable black hole branch at the smaller swallowtail structure, as the curvature of the locally minimizing Gibbs free energy branch indicates positive specific heat capacity. \textit{Centre}:  $g$-$t$ plot for $p=0.024$. As the pressure is increased, the smaller swallowtail `moves counterclockwise' towards the red point where the global first-order phase transition occurs. \textit{Right}: $g$-$t$ plot for $p=0.027$. Observe that the swallowtail has become very small.  In general, as $p$ is further increased, the smaller swallowtail will completely vanish at its critical point before merging with the first order phase transition (red dot). However, there are two precise values of $h$ ($h\approx0.0487682, 0.06998$) at which the critical point of the smaller swallowtail coincides exactly with the red point where global first-order phase transition occurs.  We call this a `virtual triple point'.}
\label{twinswallow1}
\end{figure}

In Figure~\ref{gb_vtp} we show an example of a \textit{virtual triple point} (VTP). This is a situation where one has a triple-point type phase diagram but one of the coexistence lines terminates exactly at the other coexistence line. In this scenario the only point in the phase diagram where three phases coexist is exactly at the triple point which is also a critical point. This is different from the usual triple point scenario where the coexistence line `branches out' of another coexistence line before terminating, producing a region where `intermediate-size black holes' can exist (cf. Figure~\ref{gb_triple}). The Gibbs free energy plot in Figure~\ref{twinswallow1} shows a locally thermodynamically stable branch (a `small' swallowtail) which moves `counter-clockwise' as pressure increases until it eventually disappears exactly at where the Gibbs-minimizing first-order phase transition occurs (red dot). In contrast, the usual triple point will have three Gibbs branches intersecting at the Gibbs-minimizing first-order phase transition as the small swallowtail shape does not vanish as it crosses that point (cf. centre plot of Figure~\ref{gb_triple}).

 These thermodynamic phenomena follow a general trend:  for small values of the charge $q$, a triple point occurs for some $h\in(h_1,h_2)$ (where the particular values of $h_1, h_2$ depend on $q$) and  VTPs occur at precisely $h=h_1$ and $h=h_2$. For $h\lesssim h_1, h\gtrsim h_2$ the system exhibits VdW type behaviour where there is a locally stable swallowtail structure of the type shown in Figure~\ref{twinswallow1} that never globally minimizes the Gibbs free energy. Therefore in some sense, a VTP marks the point where a transition from VdW to TP behaviour (vice versa) takes place. Intuitively speaking, we can view a VTP in two ways:
\begin{itemize}
\item As the limit where a second-order phase transition (of the non-minimal Gibbs swallowtail structure) coincides with first-order phase transition (of the physical swallowtail structure).
\item As the limit where a triple point is also a critical point. In the usual triple point phenomenon, the triple point is a bifurcation point of two coexistence curves; in the VTP case, this bifurcation point is also the endpoint of one of the coexistence curves.
\end{itemize}
Figure~\ref{twinswallow1}, which plots the Gibbs free energy, is most naturally interpreted from the first viewpoint, while for Figure~\ref{gb_vtp}, which shows the phase diagrams, the second interpretation is more natural.

Here we pause to remark on the nomenclature used in this paper.  The term \textit{virtual triple point} has appeared in a previous instance in the literature on this subject \cite{Altamirano:2014tva} where it was used to describe the situation where the coexistence lines of a zeroth and first order phase transition meet.  In this article we refer to this latter phenomenon as a \textit{pseudo-triple point} (PTP).  We believe that the term virtual triple point is more appropriate for the phenomena highlighted in Figure~\ref{gb_vtp} because, except for a single finely tuned value of $h$, the system exhibits a bonafide triple point.  Furthermore, we distinguish this result from the isolated critical points discussed in \cite{Frassino:2014pha, Dolan:2014vba} since in those cases two critical points merge to create the isolated critical point, which has non-standard critical exponents.  Here we only ever have individual critical points which have standard, mean field theory critical exponents.

\subsubsection{Hyperbolic case}
\begin{figure}[htp]
\includegraphics[scale=0.38]{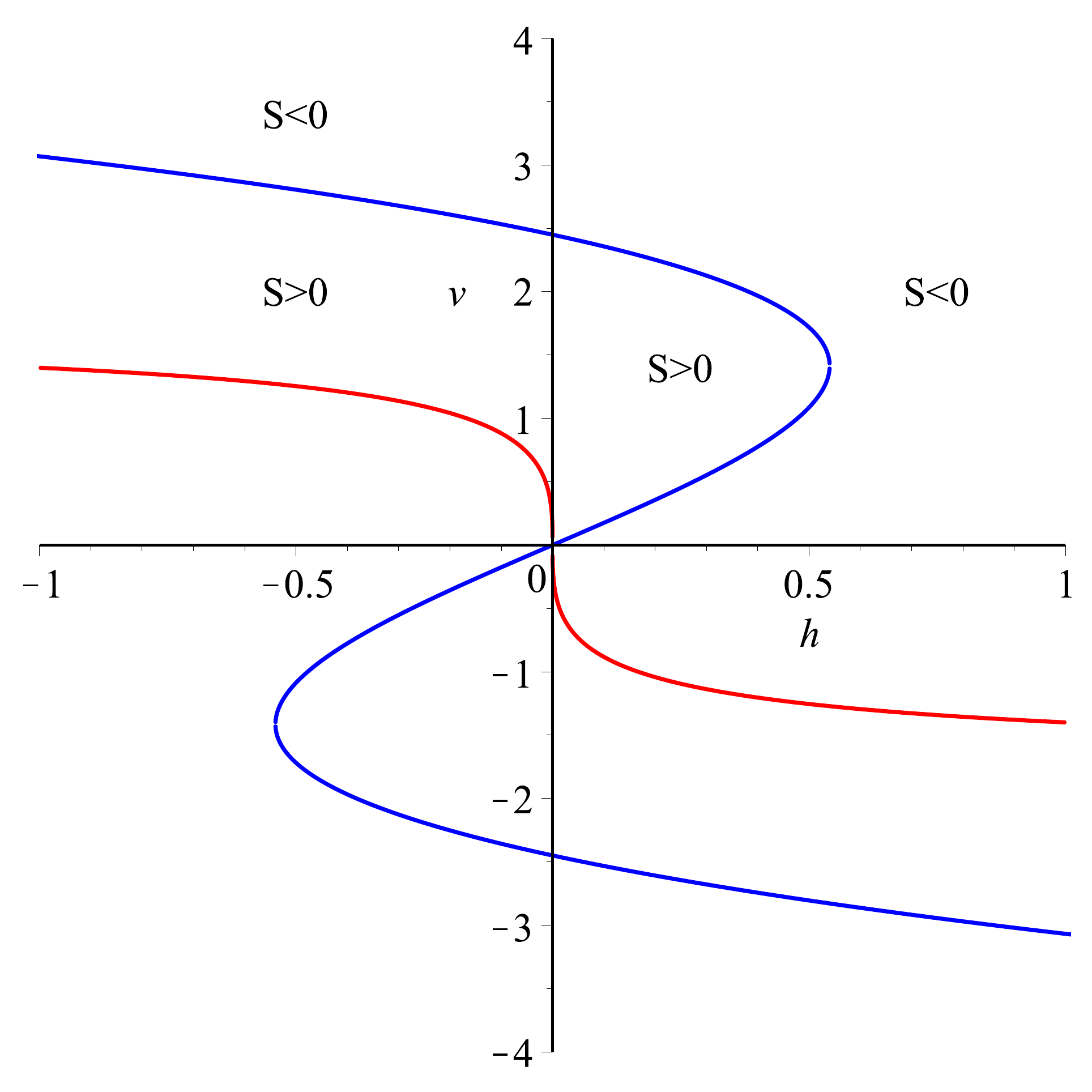}
\includegraphics[scale=0.38]{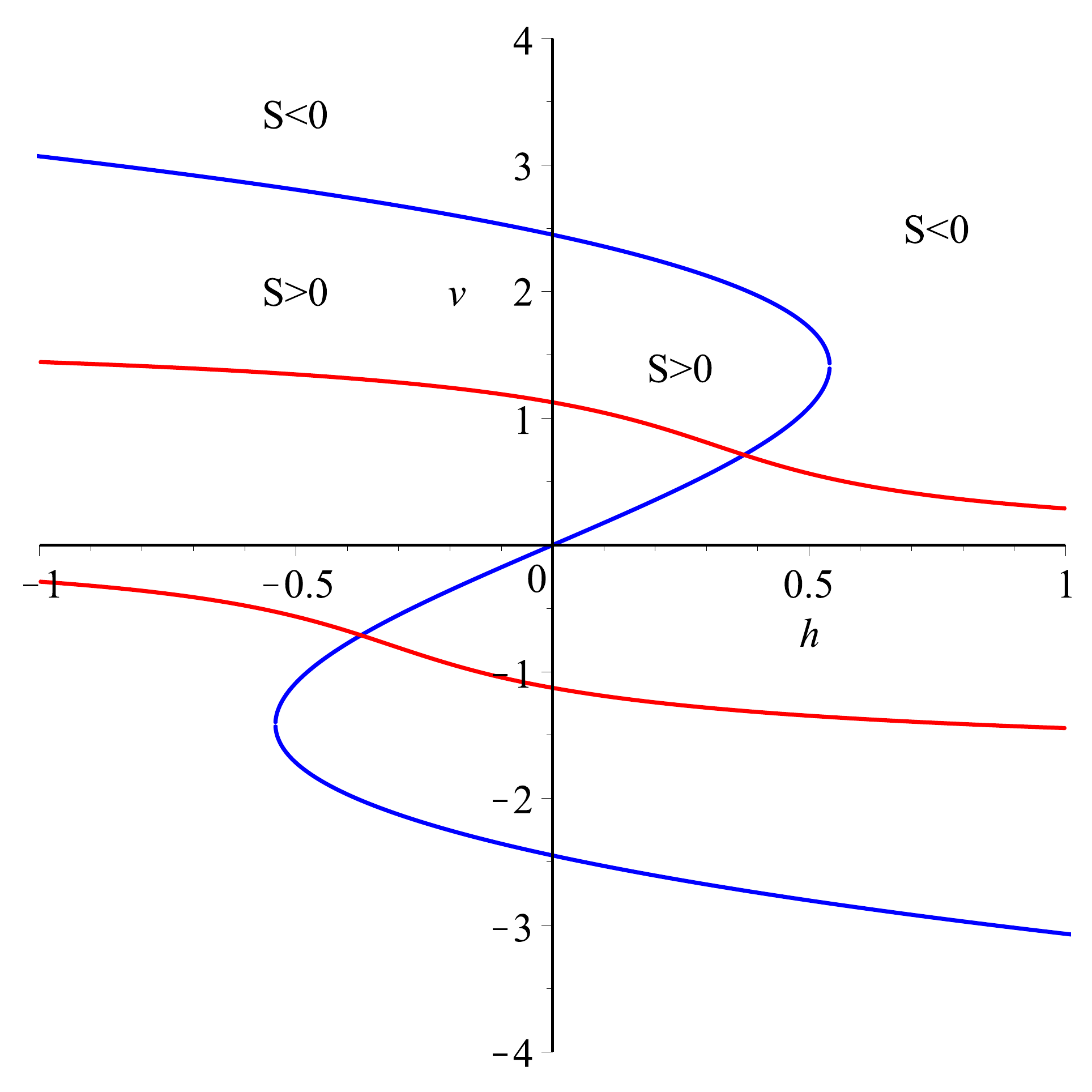}
\caption{{\bf Critical points and positive entropy for $\sigma = -1$}: \textit{Left}: The plot of critical volumes as a function of $h$ for $q=0$. The blue curve is the entropy bound, where for $h<0$ positive entropy is achieved when $v\leq v_{blue}$. This shows that positive critical volumes only occur for $h<0$. \textit{Right}: for $q=0.4$, highlighting the effect of electric charge. Enforcing all physicality constraints ($S>0, \, p<p_{max}$), none of these apparent critical volumes will be physical critical points.}
\label{gb_crit2}
\end{figure}

For the hyperbolic case we only need to consider $h<6\sqrt{2}/(5\pi)\approx 0.540$, otherwise the entropy is negative for any $v$. We also note from Figure~\ref{entropy1} that for $\sigma=-1$ the thermodynamic volume always has an upper bound (and for $h\in(0,0.540)$ also a lower bound). The possible critical volumes are simpler than in the spherical case: for uncharged hyperbolic black holes no criticality is observed for $h> 0$ while at most one critical point is possible for $h<0$, as illustrated in Figure~\ref{gb_crit2}. However, by enforcing the pressure and entropy constraints ($S>0, \, p<p_{max}$),  one can show that \textit{there are no physical critical points for any $q$ and $h$}. Nonetheless, there is still interesting thermodynamic  behaviour,  which we discuss below.

\begin{figure}[tp]
\includegraphics[scale=0.25]{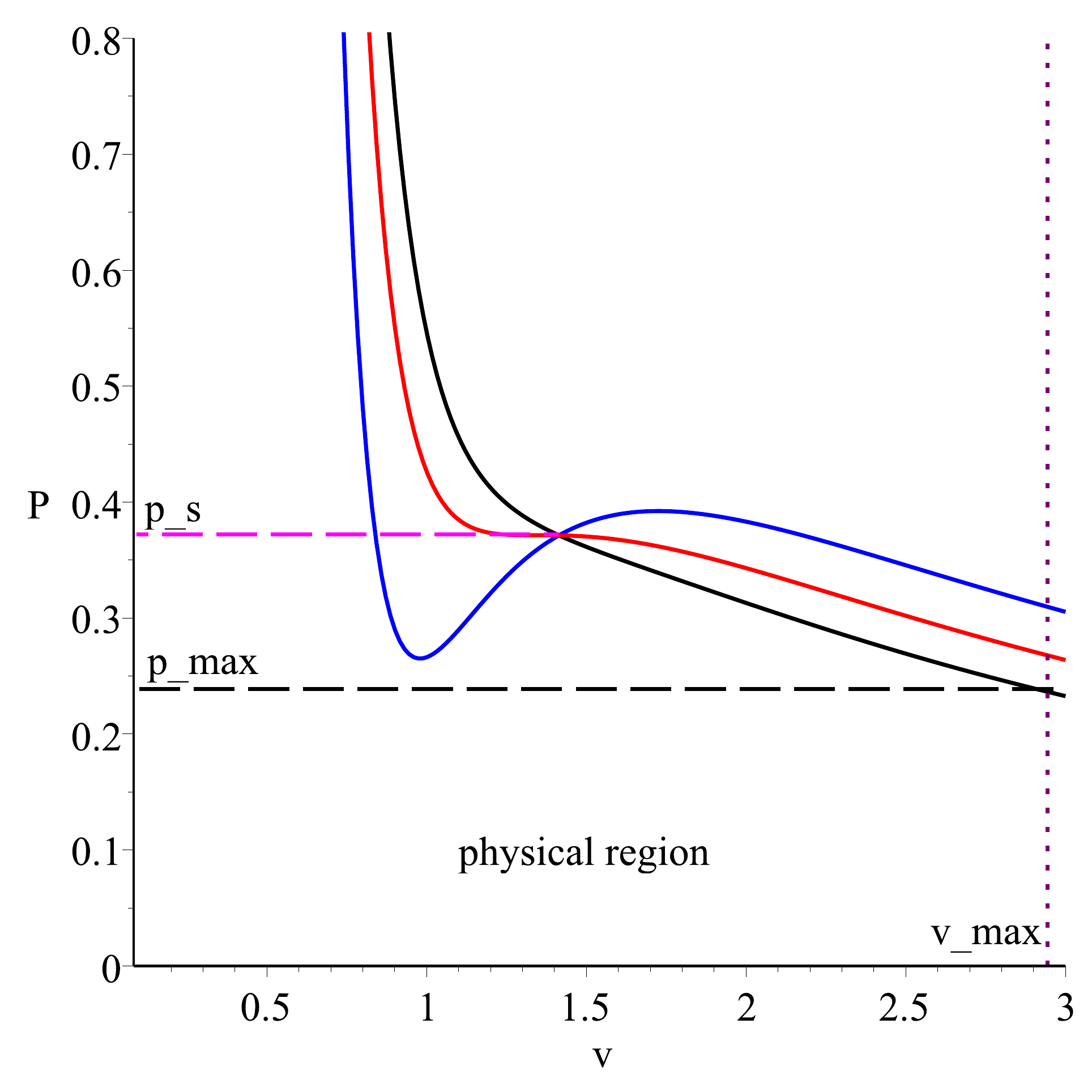}
\includegraphics[scale=0.25]{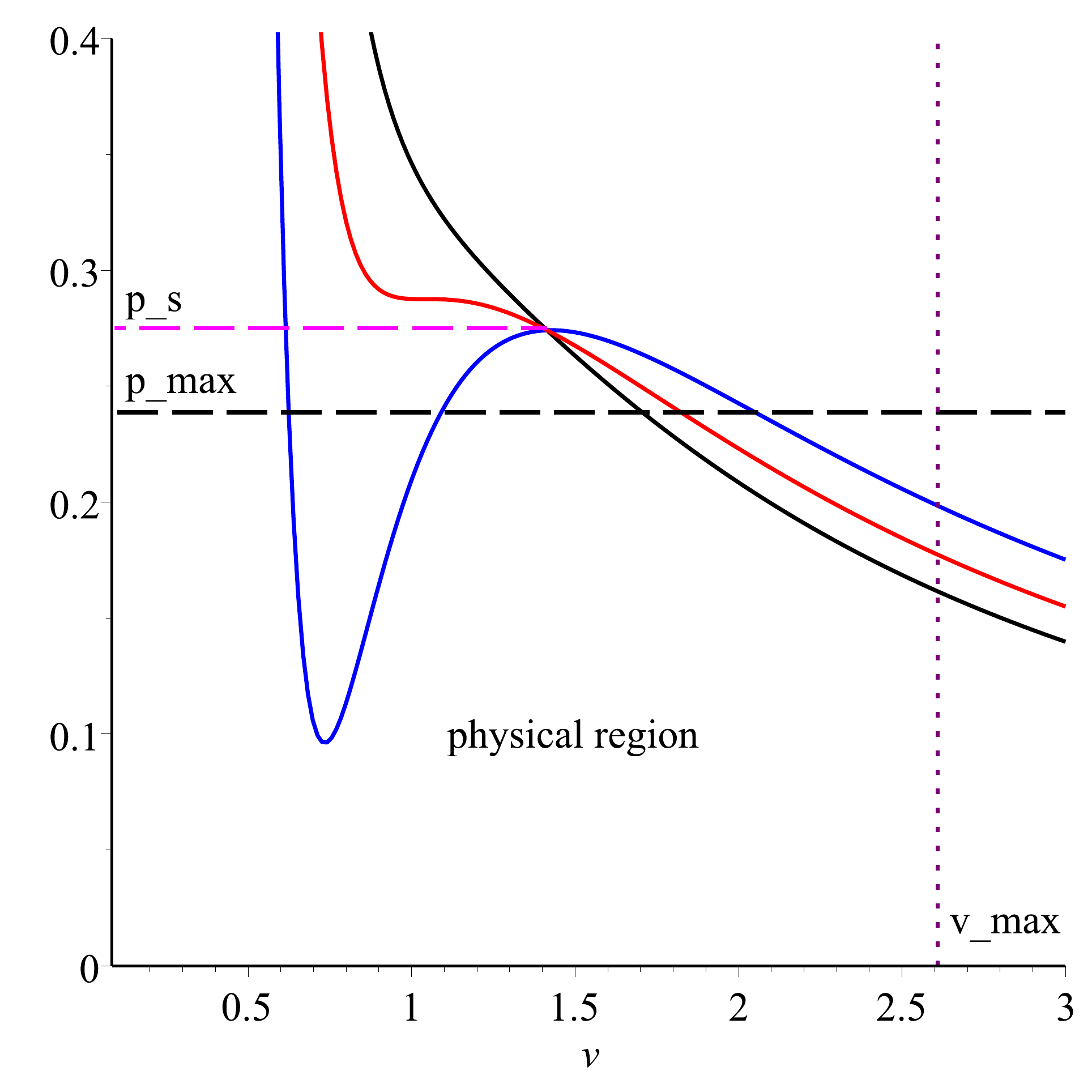}
\includegraphics[scale=0.25]{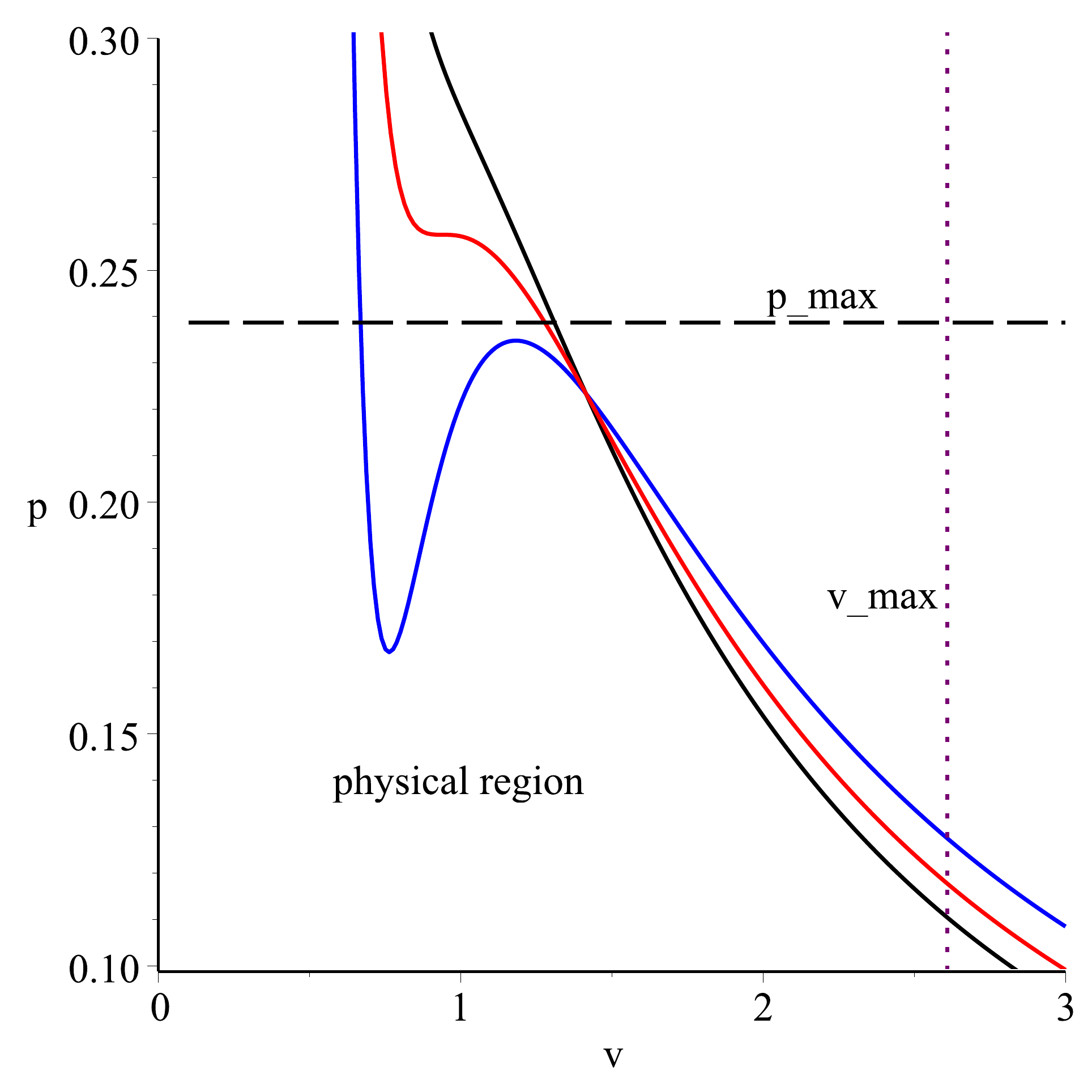}
\caption{{\bf Thermodynamic singularity}: \textit{Left}: the case for $h=-0.75, \, q=0, \, \sigma=-1$. The $p-v$ diagram is `shifted' up such that below some threshold $h_0\approx -0.750$, there is minimum temperature below which no black hole solution exists. \textit{Centre}: $p-v$ diagram for $h=-0.2,q=0,\sigma=-1$. Note the VdW-type oscillations, which allows a first-order phase transition to occur even if there is no physical critical point (1PT). \textit{Right}: $p-v$ diagram for $h=0.2,q=0.4,\sigma=-1$. Note that the VdW-type oscillation is now completely within the physical region, and so the Gibbs free energy can display swallowtail structure even if the critical point itself is unphysical. We also classify this as 1PT.  In each plot, the point where the isotherms cross marks the thermodynamic singularity.  This occurs in the physical region in the right-most plot, a result which was previously unobserved for $d=5$ Gauss-Bonnet gravity.}
\label{thermosing1}
\end{figure}

As in the no-hair case, there is a \textit{thermodynamic singularity} in hyperbolic black holes which can be identified by the presence of `crossing isotherms' in the $p-v$ diagram, as seen in Figure~\ref{thermosing1}. Formally, a thermodynamic singularity occurs at the point where
\be
\frac{\partial p}{\partial t}\Big |_{v=v_s}=0, \ \ \frac{\partial p}{\partial v}\Big |_{t=t_s}=0
\ee
This gives $v_s=\sqrt{2}$ and we can compute the singular pressure $p_s$ which is given by
\be\label{singularpressure}
p_s=p_{max}+\frac{q^2}{8}-\frac{\sqrt{2}h}{8}=\frac{3}{4\pi}+\frac{q^2}{8}-\frac{\sqrt{2}h}{8}
\ee
Eq.~\eqref{singularpressure} shows that the \textit{thermodynamic singularity can be physical} for  positive $h$, provided that ${q^2} <  {\sqrt{2}h}$ so that $p_s<p_{max}$.  There is also the further restriction $0<h<6\sqrt{2}/(5\pi)$ so that entropy is positive. 
 This is contrary to the no-hair case where $p_s\geq p_{max}$ with $p_s=p_{max}$ only when $q=0$ and the thermodynamic singularity occurs at negative entropy \cite{Frassino:2014pha}. Therefore, in the no-hair case, the thermodynamic singularity   is always unphysical (unless $p_s= p_{max}$) even if we allow for negative entropy. 

Since here the thermodynamic singularity can be physical for some combinations of $(q,h)$, it is interesting to check the case when thermodynamic singularity coincides with a critical point.  In the context of third (and higher) order Lovelock gravity, this situation gives rise to critical exponents that differ from the mean field theory values, and are (so far) the only such examples in black hole physics \cite{Frassino:2014pha, Dolan:2014vba}.   However, as we shall see below, it turns out to be impossible to tune $q$ and $h$ such that a critical point occurs at the thermodynamic singularity while still respecting the various physicality constraints.   Nonetheless, the thermodynamic singularity itself can occur within physical constraints so we still have the peculiar situation where at the `singular' volume $v_s$ the pressure is constant at all temperatures or at the `singular' temperature $t_s$, the volume is constant at all pressures. This peculiarity is very different from conventional fluids. 

Unlike the spherical case, here physicality constraints put an upper bound on \textit{both} volume and pressure. Consequently  the $p-v$ diagram shows that physical black hole solutions are confined to a compact region in $(p,v)$ space. As a result, the $p-v$ diagram will be discontinuous, as one can see from Figure~\ref{thermosing1}. In particular, as $h\to 0^{-}$, $p_s\to p_{max}+ q^2/8$ and the oscillatory portion of the $p-v$ diagram moves to lower pressures,  while as $h\to-\infty$, the oscillatory portion of the $p-v$ diagram is `shifted' upwards. Therefore, we have a situation where for a given fixed $h<0$, there is a \textit{maximum temperature} $t_{max}$ above which the isotherms are not physical, i.e. no physical black hole solution exists for $t>t_{max}$. For example, $q=0, \, h=-0.750$ we have $t_{max}\approx 0.69$ and above $t_{max}$ the physical region contains no isotherms at all --  hence no black holes can exist, as shown in Figure~\ref{thermosing1}. This does not happen for spherical black holes because in that case the black hole volumes are bounded only from below and hence the physical $(p,v)$ region is not a compact set. Therefore for hyperbolic black holes with any fixed $(q,h)$ with $h<0$, the entire thermodynamic phase space $(p,v,t)$ is compact since $p\in[0,p_{max}], v\in[0,v_{max}], t\in[0,t_{max}]$. We note that this is generic feature of hairy hyperbolic black holes in Gauss-Bonnet gravity in any dimension. Here in five dimensions, for $0<h\lesssim 0.540$, there is always a black hole solution but still with discontinuous $p-v$ diagram.

\begin{figure}[tp]
\includegraphics[scale=0.25]{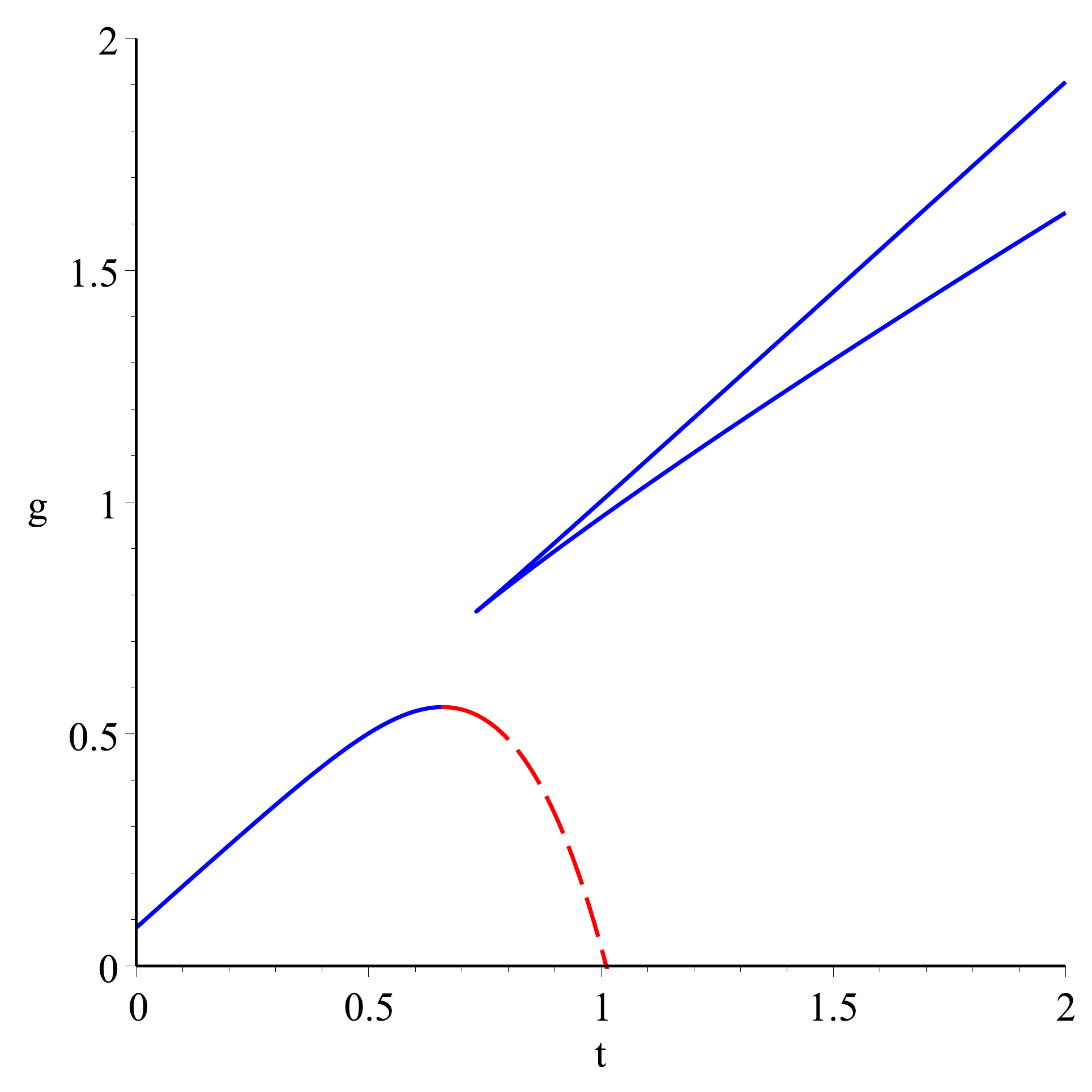}
\includegraphics[scale=0.25]{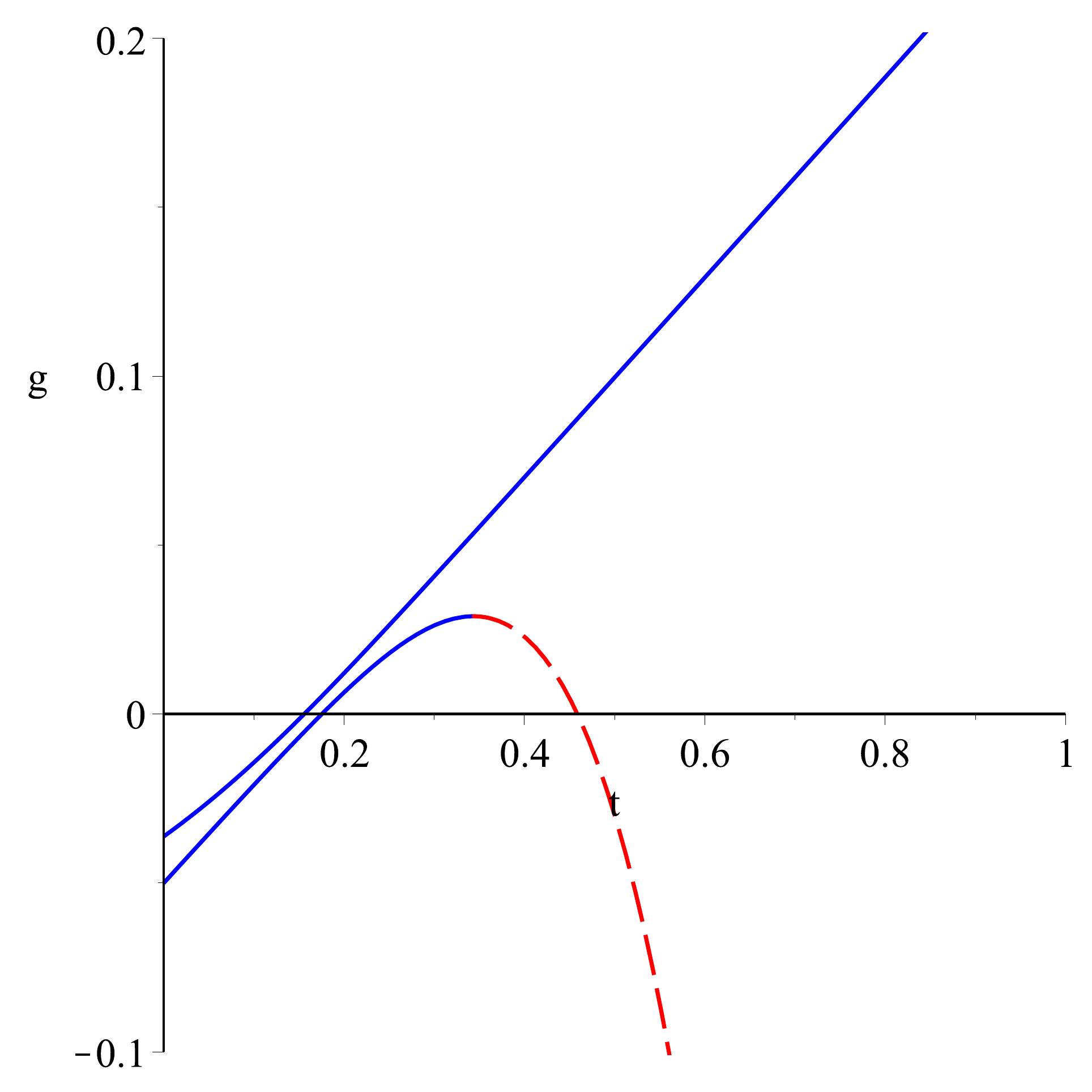}
\includegraphics[scale=0.25]{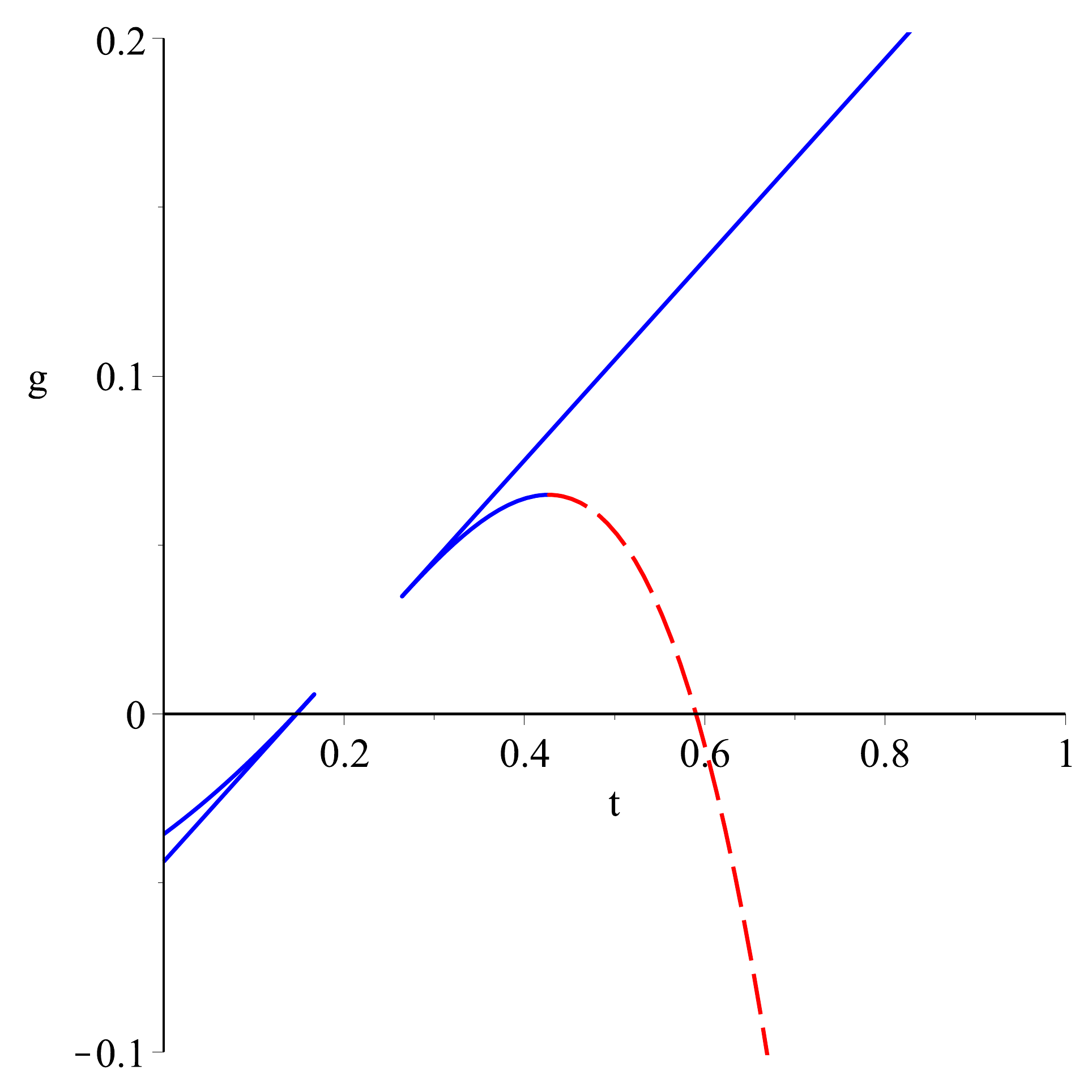}
\caption{{\bf Gibbs reconnection for hyperbolic uncharged black holes}: The behaviour near and at the thermodynamic singularity. \textit{Left}: $g-t$ diagram for $h=-0.5, \, p=p_{max}<p_s|_{h=-0.5}$. \textit{Center}: $g-t$ diagram for $h=0.2, \, p=0.9 p_s|_{h=0.2}$. \textit{Right}: $g-t$ diagram for $h=0.2, \, p=1.01 p_s|_{h=0.2}$. At $p=p_s$ the disjoint branches of the Gibbs free energy `reconnect' (center plot) and then split once more as pressure is increased, forming new asymptotic behaviour. Note that $p_s$ depends on $q$ and $h$ hence the value of $p_s$ for the left $g-t$ diagram is different from the centre and the right $g-t$ diagrams. In all cases, a dashed line indicates a branch of black holes with negative entropy.}
\label{gbreconnect1}
\end{figure}

\begin{figure}[tp]
\includegraphics[scale=0.25]{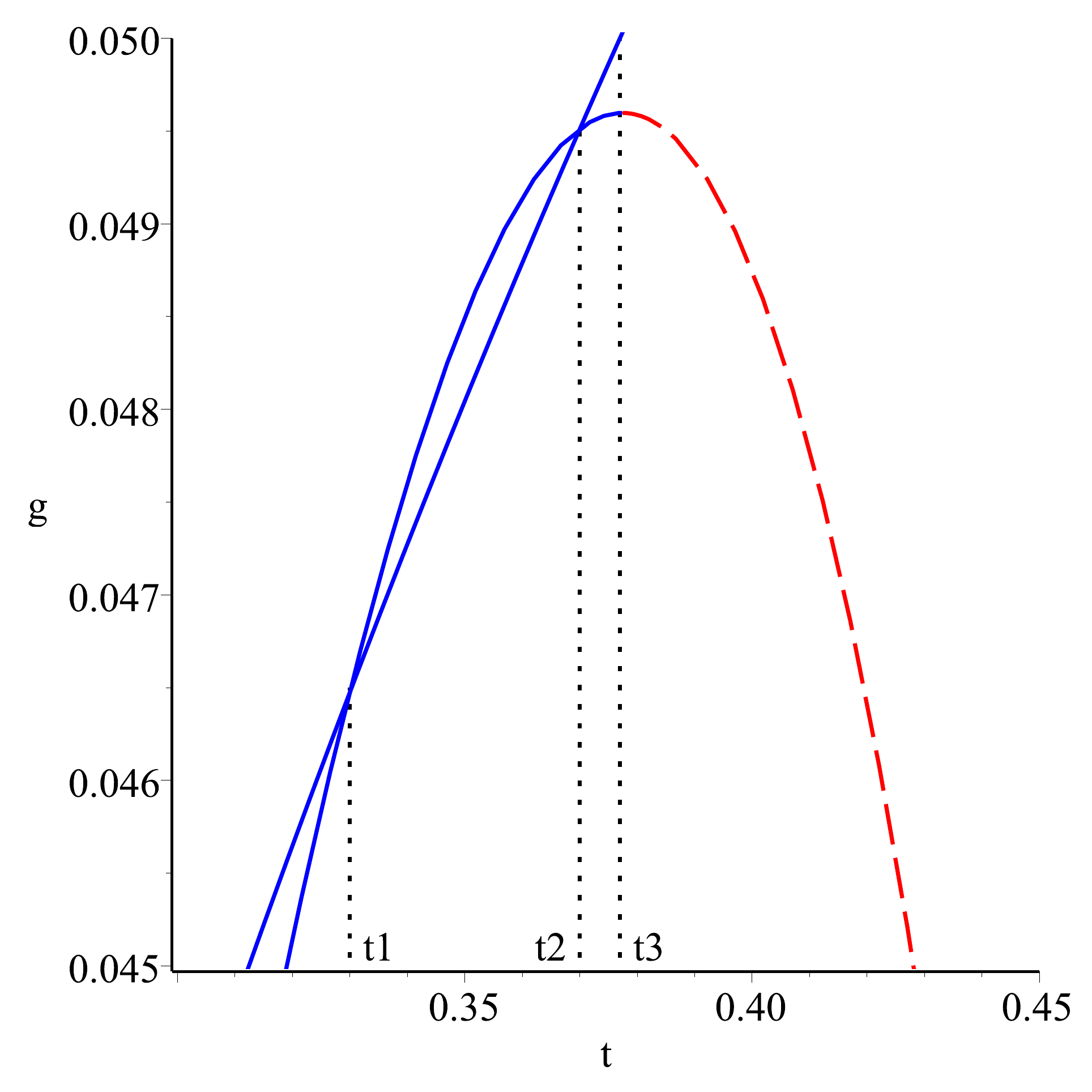}
\includegraphics[scale=0.25]{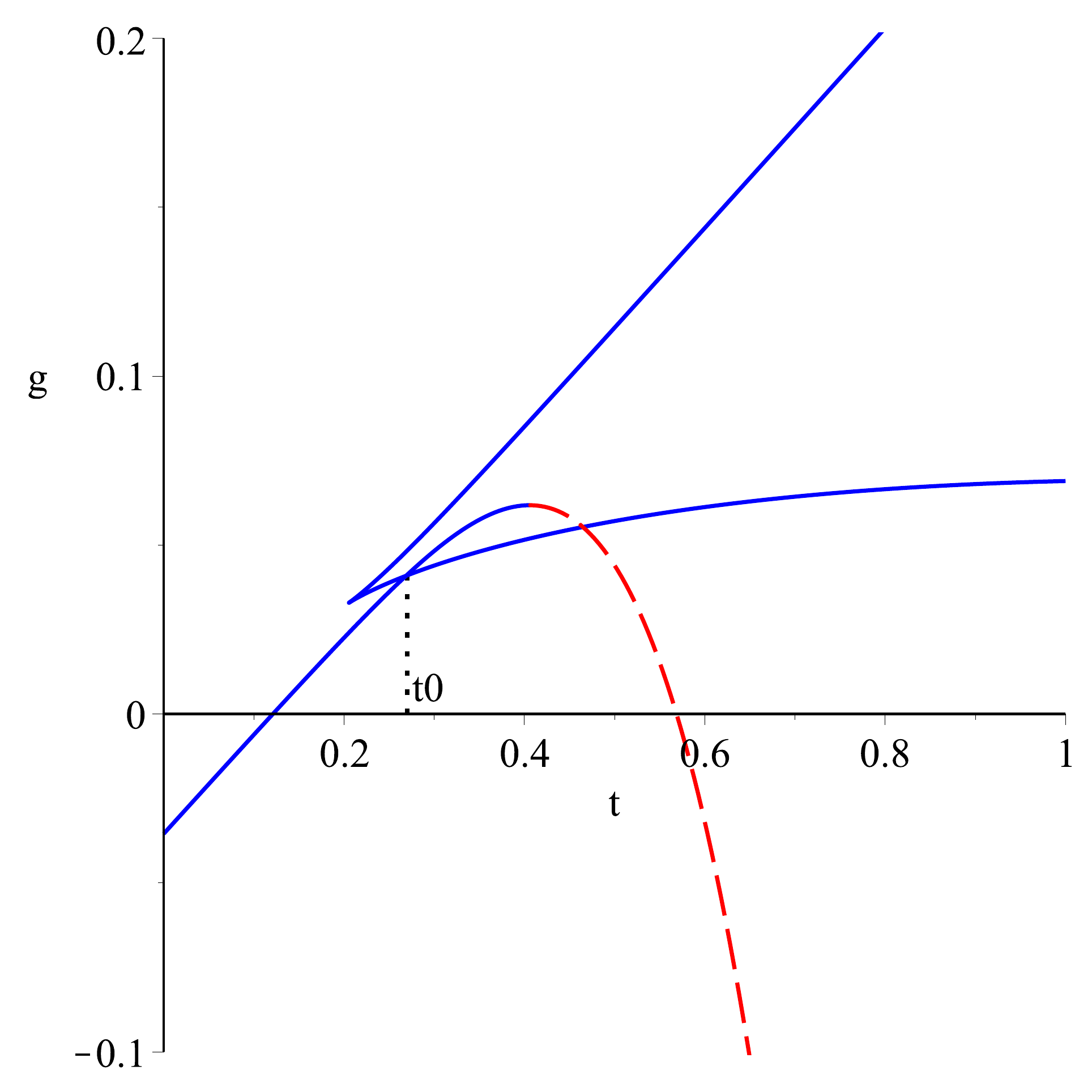}
\includegraphics[scale=0.25]{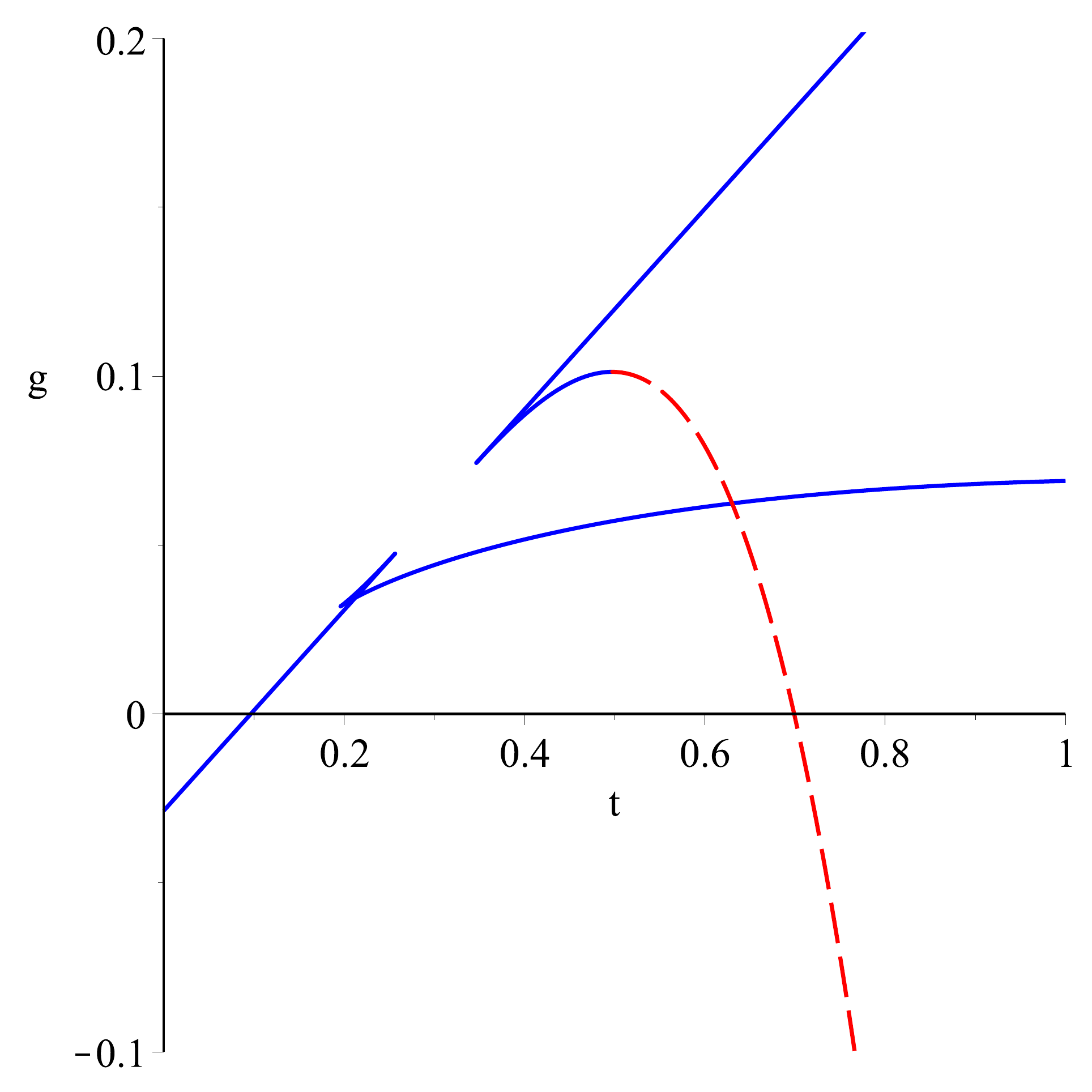}
\caption{{\bf Various phase transitions for hyperbolic charged black holes}: \textit{Left}: $g-t$ diagram for $h=0.2, \, q=0.4, \, p\approx 0.1935<p_s$, showing a \textit{double reentrant phase transition} (RPT2). At $t=t_1$, there is first order phase transition followed by another first order phase transition at $t=t_2$ back to original branch. At $t=t_3$, there is zeroth order phase transition due to negative entropy. Thus we have two reentrant phase transitions. \textit{Centre}: $g-t$ diagram for $h=0.2, \, q=0.4, \, p=0.9p_s\approx 0.201<p_{max}$, showing a first-order phase transition that is not of the VdW type.  \textit{Right}: $g-t$ diagram for $h=0.2, \, q=0.4, \, p=1.01 p_s\approx 0.226<p_{max}$. Note the swallowtail structure at the lower left. Since the critical point is unphysical for these hyperbolic black holes, this first order phase transition differs from the VdW type.}
\label{gbreconnect2}
\end{figure}

Recall that for small, positive $h$ the thermodynamic singularity occurs in the physical region of $p-v$ space and the $g-t$ diagram will show the characteristic `reconnection' near the thermodynamic singularity, illustrated in Figure~\ref{gbreconnect1}. We found that for $h<0$, one can observe the following thermodynamic behaviour: \textit{zeroth-order phase transitions} (0PT), first-order phase transitions without critical points (which we denote 1PT), and a \textit{double reentrant phase transition} (which we denote as RPT2). Figure~\ref{gbreconnect1} shows how positive entropy and `Gibbs reconnection' allows a zeroth-order phase transition to occur, since the Gibbs free energy is discontinuous. For $h>0$, only a zeroth-order phase transition occurs. Representative plots of the Gibbs free energy for the 1PT and RPT2 cases are shown in Figure~\ref{gbreconnect2} in the context of charged hyperbolic black holes.  It is interesting that such a variety of thermodynamic behaviour is present even in the absence of critical points.

Earlier it was mentioned that it is of interest to find out if the thermodynamic singularity can coincide with critical points, since this may lead to non-standard critical exponents. We will now formally show that if all maximum pressure and positive entropy constraints are enforced, there is no physical critical point that coincides with the thermodynamic singularity for any $h$ and $q$. A situation such as this would occur when thermodynamically singular points also satisfy Eq.~\eqref{d5criticalpoints}, giving
\be\label{singularcrit}
v_{s,c}=\sqrt{2},\ \ h_{s,c}=\frac{6\sqrt{2}(\pi q^2-2)}{5\pi},\ \ t_{s,c}=\frac{-3\sqrt{2}(\pi q^2-6)}{8\pi}, \ \ p_{s,c}=-\frac{7\pi q^2-54}{40\pi} \, .
\ee
To maintain positive entropy, Eq.~\eqref{gbentropy} requires that 
\be 
S=12\sqrt{2}(3-\pi q^2)\geq 0\ \  \Rightarrow \ \ |q|<\sqrt{\frac{3}{\pi}} \, .
\ee
Demanding that $h_{s,c}, t_{s,c}, p_{s,c}, S>0$ and $p<p_{max}=3/(4\pi)$, we see that the physical thermodynamic singularity can also be a critical point if it satisfies Eq.~\eqref{singularcrit} and 
\be 
\sqrt{\frac{24}{7\pi}}<|q|<\sqrt{\frac{6}{\pi}} \text{ and } |q|<\sqrt{\frac{3}{\pi}}
\ee
which \textit{cannot be simultaneously satisfied}. Therefore, we are forced to conclude that a critical point at the thermodynamic singularity is \textit{unphysical} in $d=5$. 

However, suppose for argument's sake that we relax the entropy condition. Then we only need to ensure that
\be 
\sqrt{\frac{24}{7\pi}}<|q|<\sqrt{\frac{6}{\pi}}
\ee
We shall consider the most general case in $d=5$. If we do a transformation
\be 
\omega=\frac{v}{v_c}-1, \, \, \tau=\frac{t}{t_c}-1
\ee
and do a Taylor expansion of the equation of state about the critical point, we obtain
\be\label{gb_critexponent0}
\frac{p}{p_c}=1-\left(\frac{30\pi q^2-180}{7\pi q^2-54}\right)\tau\omega-\left(\frac{5\pi q^2-150}{7\pi q^2-54}\right)\omega^3+ O(\tau\omega^2,\omega^4) \, .
\ee
From this expansion, it is easy to calculate the critical exponents using standard techniques (see, for example, \cite{Kubiznak:2012wp}.) Doing so we find that, 
\be \label{gb_critexponent3}
\alpha=0,\, \, \beta=\frac{1}{2},\, \, \gamma=1,\, \, \delta=3
\ee
which are the usual Van der Waals mean-field critical exponents. Therefore, when the critical point occurs at the thermodynamic singularity (relaxing the entropy constraint), it is characterized by standard mean-field theory critical exponents.  This demonstrates that a critical point occurring at a thermodynamic singularity is not a sufficient condition to observe the non-standard critical exponents found in \cite{Frassino:2014pha}.  This argument is true for all values of the electric charge that satisfy the above constraints; however, when $q = \sqrt{6/ \pi}$ (corresponding to $p = p_{max}$), the critical point is indeed an isolated critical point of the type reported in \cite{Frassino:2014pha}.  We shall return to this topic later in our discussion of third order Lovelock gravity.
 
Finally we note that since the thermodynamic singularity is in the interior of the physical region, the crossing isotherms imply that for $v<v_s$ we have usual VdW property while for $v_s<v<v_{max}$ we have \textit{reverse} VdW behaviour (RVdW), because for volumes within this range, lower temperature implies \textit{higher} pressure (c.f. Figure~\ref{thermosing1}). We emphasize that both VdW and RVdW can occur within the same physical sector because conformal hair allows a physical thermodynamic singularity where isotherms cross at that point.

Overall, we observe that charged hyperbolic hairy black holes can exhibit other interesting thermodynamic behaviour in $d=5$ even without physical critical points. For $h<0$, we observe 0PT, 1PT, RPT2 for both the charged and uncharged cases, while for $h>0$ only 0PT is observed for the uncharged case while all three phenomena can be seen in charged case. Furthermore, we can have a first-order phase transition with a VdW-type swallowtail structure for the charged case. These are   all highlighted in Figure~\ref{gbreconnect2}. We also have a physical region which can accommodate both VdW and reverse VdW-type behaviour. These are all new features of five-dimensional hyperbolic Gauss-Bonnet black holes made possible by the conformal hair. A full classification can be found in Table~\ref{tab:gb_5d_spherical}.

\subsection{$P-v$ criticality in $d=6$}

 We will discuss the six-dimensional case very briefly for hyperbolic black holes, in particular clarifying the role of the thermodynamic singularity. We will see that all the  behaviour in $d=6$ is similar to that in $d=5$,  as seen by comparing the list of thermodynamic phenomena in Table~\ref{tab:gb_5d_spherical} and Table~\ref{tab:gb_5d_hyperbolic} in Section~\ref{sec:gb_summary}. 

In six dimensions, the equation of state is given by
\be
p=\frac{t}{v}-\frac{3\sigma}{\pi v^2}+\frac{2\sigma t}{v^3}-\frac{1}{\pi v^4}+\frac{q^2}{v^8}-\frac{h}{v^6}
\ee
and the critical points satisfy the following:
\be 
t_c=\frac{2(3\sigma v_c^6+2v_c^4-4\pi q^2+3\pi v_c^2 h)}{v_c^5\pi(v_c^2+6\sigma)}
\ee
\be 
3v_c^8-12\sigma v_c^6+12v_c^4-4\pi q^2(7\sigma v_c^2+30)+3\pi v_c^2 h(5\sigma v_c^2+18)=0
\ee
The maximum pressure is given by $p_{max}=5/4\pi\approx 0.3979$. Positivity of entropy implies the constraint:
\be 
8\sigma v^4-12\pi h+32v^2\geq 0 \, .
\ee
As before, the entropy enforces a lower bound for the volume in the spherical case and an upper bound for the hyperbolic case.  As mentioned earlier we will not discuss at length the six-dimensional spherical black holes but we emphasize that \textit{nothing thermodynamically interesting is lost} as the five-dimensional case captures all the interesting thermodynamic behaviour for Gauss-Bonnet hairy black holes in higher dimensions. We have verified that the thermodynamics for the six-dimensional spherical case adds  nothing new compared to the five-dimensional case. We will discuss very briefly the six-dimensional hyperbolic case in particular to explore the \textit{thermodynamic singularity} in these black holes. The full classification of possible thermodynamic phenomena is presented in Table~\ref{tab:gb_5d_hyperbolic} in Section~\ref{sec:gb_summary}.

\subsubsection{Hyperbolic case}
\begin{figure}[tp]
\includegraphics[scale=0.38]{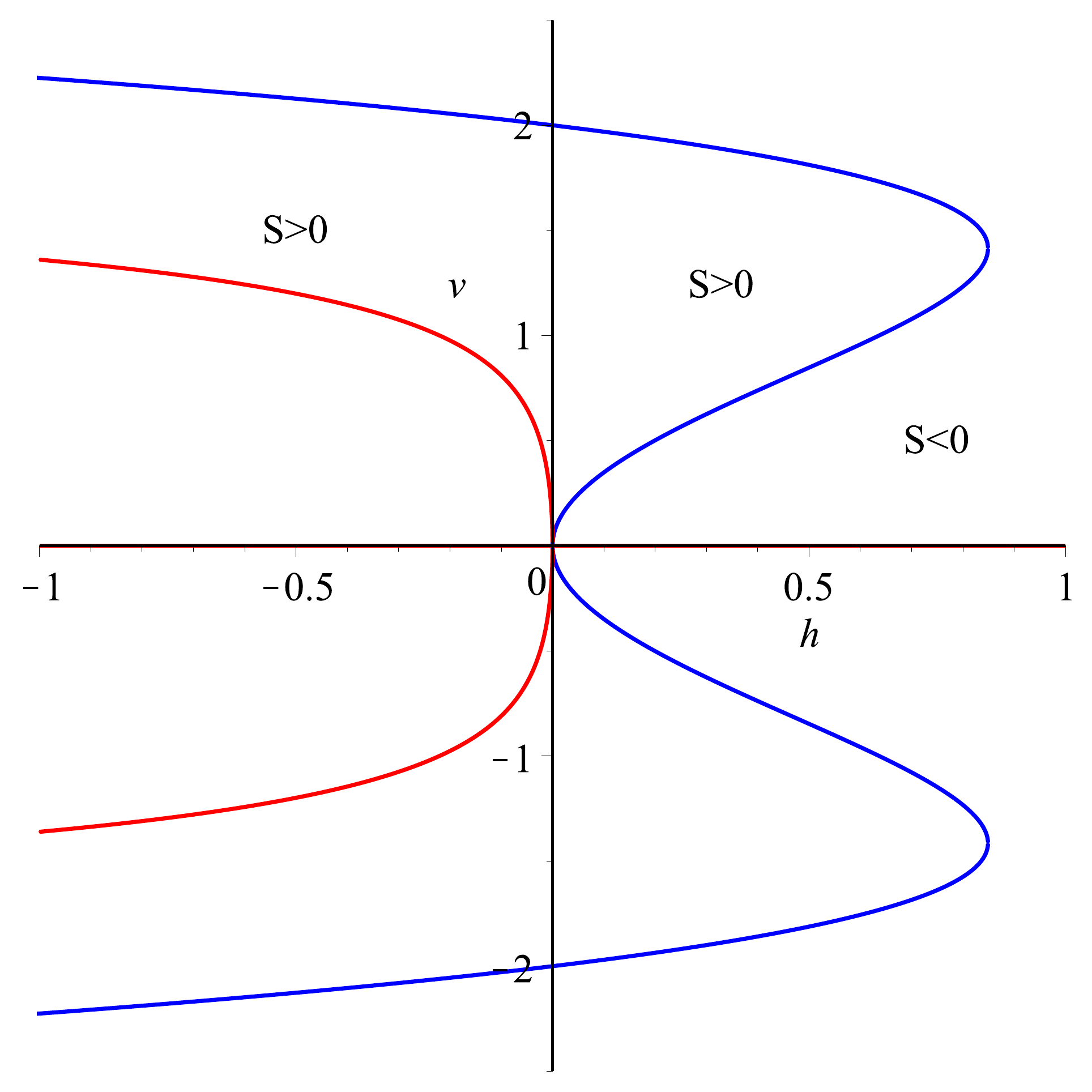}
\includegraphics[scale=0.38]{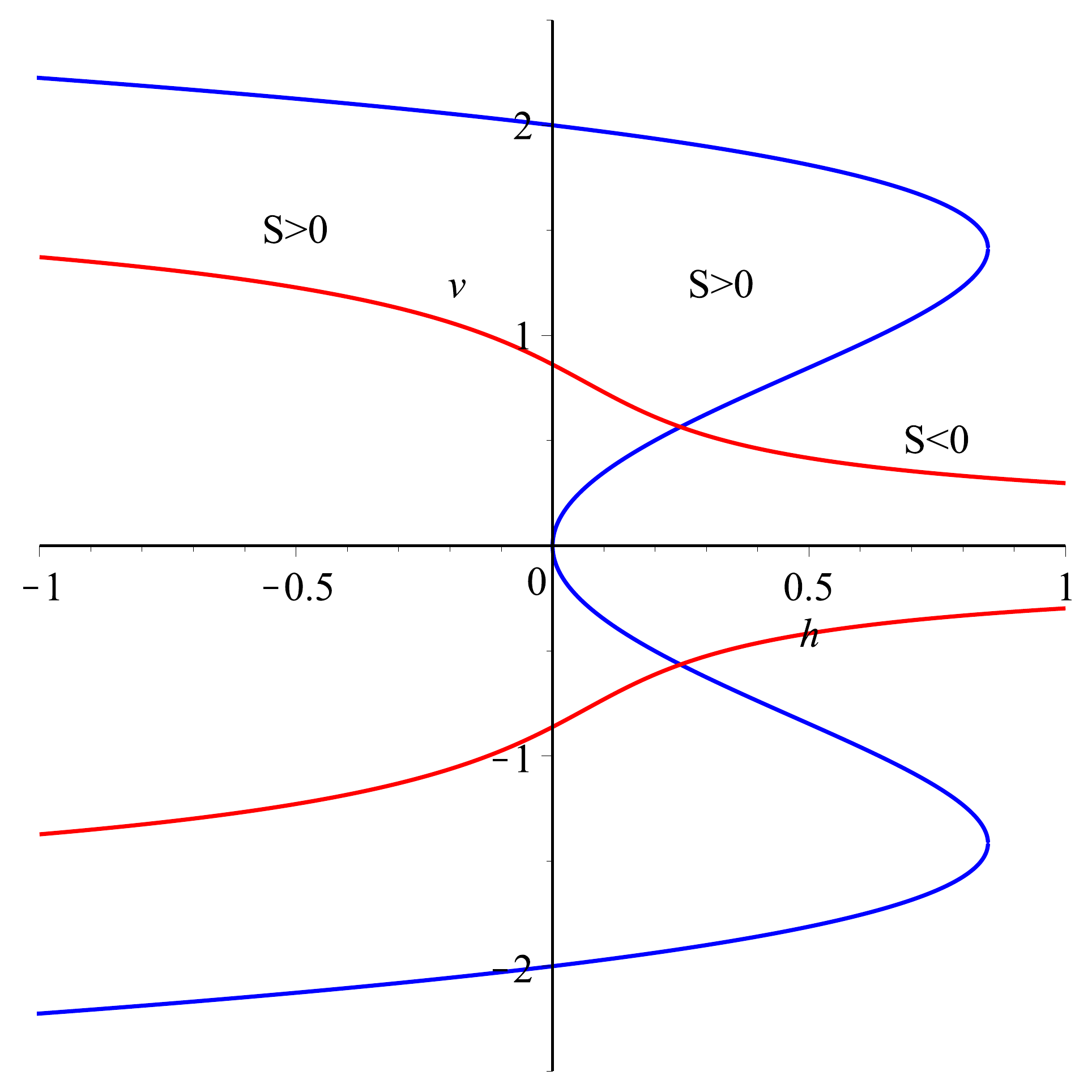}
\caption{{\bf Critical points and entropy for hyperbolic case in $d=6$}: \textit{Left}: $v$ vs. $h$ for uncharged hyperbolic hairy black holes. \textit{Right}: $v$ vs. $h$ for charged hyperbolic hairy black holes. The red curve is the locus of critical points and the blue curve is the locus of zero-entropy points. Here we observe that there is no critical point for $h>0$ for uncharged case, but presence of charge allows critical points to exist for all $h$ (not necessarily physical, e.g. we have not imposed maximum pressure constraints).}
\label{gb_crit4}
\end{figure}

For the hyperbolic case, Figure~\ref{gb_crit4} shows that there is at most one physical critical point for the  hyperbolic case in six dimensions. In fact, one can go further and show that by enforcing all physicality constraints, there are no physical critical points for any $q$ and $h\neq0$. (The $(q,h)$ space plot similar to Figure~\ref{gb_crit1} would be entirely black).  However there are still interesting thermodynamic results, albeit none of which are new.  A full classification of these is provided in Section~\ref{sec:gb_summary}. Similar to the $d=5$ case, we can have reverse VdW, zeroth-order phase transition (0PT), first-order phase transition without criticality (1PT), as well as reentrant phase transition (RPT) for $d=6$, and qualitatively similar to the $d=5$ case. Overall we observe that the conformal scalar hair enables various thermodynamic behaviours  that would otherwise not be possible without conformal hair in their respective dimensions.

These black holes also exhibit a thermodynamic singularity, which occurs at a pressure given by
\be 
p_s=\frac{5}{4\pi}+\frac{q^2}{16}-\frac{h}{8}=p_{max}+\frac{q^2}{16}-\frac{h}{8}
\ee
and we can once again show formally that a physical thermodynamic singularity will never occur at any physical critical point. By performing the same calculation as in $d=5$ we obtain
\be\label{singularcrit2}
v_{s,c}=\sqrt{2},\ \ h_{s,c}=\frac{4(\pi q^2-3)}{3\pi},\ \ t_{s,c}=-\frac{\sqrt{2}(\pi q^2-10)}{4\pi}, \ \ p_{s,c}=-\frac{5\pi q^2-84}{48\pi} \, .
\ee
To maintain positive entropy, Eq.~\eqref{gbentropy} implies
\be 
S=-16\pi q^2+80 > 0\ \  \Rightarrow \ \ |q|<\sqrt{\frac{5}{\pi}} \, .
\ee
Demanding that $h_{s,c}, t_{s,c}, p_{s,c}, S>0$ and $p<p_{max}=5/(4\pi)$, we can show that these constraints \textit{cannot be simultaneously satisfied}. In particular, the first four can be satisfied but $p<p_{max}$ is incompatible with the rest.  

Therefore we are again forced to conclude that a critical point at the thermodynamic singularity is \textit{unphysical} in $d=6$. We conjecture that this situation holds for this theory in arbitrary dimensions. However, the underlying reason is slightly different for $d=5$ and for $d\geq 6$. For $d=5$,  $S<0$ is incompatible with other constraints while for $d\geq 6$ we expect that $p<p_{max}$  is incompatible with the rest and this incompatibility gets worse as $d\to\infty$.  In some sense, the incompability in $d>6$ is worse because there one does not have the correct asymptotics to speak of a proper AdS black hole  whereas the physical status  of $S\leq 0$ is less clear.

\subsection{Summary: thermodynamics of Gauss-Bonnet hairy black holes \label{sec:gb_summary}}

We summarize here in the form of tables the thermodynamic behaviour of Gauss-Bonnet hairy black holes for $d=5$ and $d=6$, 
along with our nomenclature (Table \ref{tab:nomen}) for the various kinds of transitions we observe.  While the values of $q_0$, $q_1$, and $q_2$ are
universal, the other  specific values  of electric charge $q$ are for exemplary purposes; any value of charge chosen within the allowed range would yield the same qualitative phase behaviour.  For given value of $q$ in the table  adjusting $h$ leads to a continuous deformation of the phase diagram with distinct values  $h_a, \, h_b, \, h_c, \, h_d, \, h_e, \, h_f$ marking the boundaries between 
different kinds of criticality. 
 
Furthermore, note that the summary only suggests that the said criticality occurs in \textit{some} parameter range between the threshold values. For example, for $d=5, \, \sigma=1, \, q=0.1$ (Table \ref{tab:gb_5d_spherical}), RPT does not occur for the entire interval $(h_c,h_d)$ but only in a subset within the interval. However it is the only distinct thermodynamic behaviour within this region.

\begin{table}[!htp]
\centering
\begin{tabularx}{\textwidth}{c X }
\toprule
Abbreviation & Meaning\\
\midrule
TP & Triple Point \\
PTP & Pseudo Triple Point (due to intersection of the zeroth and first order coexistence lines)\\
VTP & Virtual Triple Point \\
VdW &Van der Waals with exactly one physical critical point\\
VdW* & VdW with more than one positive critical point \\
& (but only one which occurs on the minimizing branch of the Gibbs free energy)\\
RVdW& Reverse VdW behaviour\\
RPT& Reentrant Phase Transition\\
1PT& First-order Phase Transition without critical point\\
& (either because it is VdW-type but $p_c>p_{max}$, or because two Gibbs branches  happen to intersect without swallowtail behaviour)\\
 RPT2& Double Reentrant Phase Transition
\\ \bottomrule
\end{tabularx}
\caption{Nomenclature used in the tables.  A double reentrant phase transition is typically induced by one zeroth-order phase transition and two first-order phase transitions. Note that 
if  a system exhibits RPT2 at some pressure, it implies the system can exhibit RPT as well (at lower pressure).}
\label{tab:nomen}
\end{table}

\rowcolors{2}{White}{LightBlue!30}
\begin{table}[!htp]
\centering
\begin{tabular}{c  c c c}
\toprule
 \cellcolor{white} Charge ($q$) & Hair ($h$) \cellcolor{white} & \# Physical CP \cellcolor{white} & Behaviour  \cellcolor{white}\\
 \midrule
  \cellcolor{white} & $h\leq 0$   & 1 & VdW \\
 
\cellcolor{white} & $0<h\leq h_a$ & 0 & none (cusp) \\
 
 \multirow{-3}{0.35 \hsize}{\cellcolor{white}
 ${q=0}$}& $h>h_a\approx 0.1240$  & 0 & RPT \\
 \midrule
 \cellcolor{white}  &  $h\leq h_a\approx 0.0445$ & 1 & VdW \\
 
\cellcolor{white} & $h_a<h<h_b $              & 1 & VdW*\\
 
\cellcolor{white} & $h_b\approx 0.0487682$    & up to 3 & VTP\\
 
\cellcolor{white} & $h_b<h<h_c$               & up to 3 & TP \\
 
\cellcolor{white} & $h_c\approx 0.06998$       & up to 3 & VTP\\
 
\cellcolor{white} & $h_c<h< h_d$              & 1 & VdW*, RPT\\
 
 \multirow{-7}{0.35 \hsize}{\cellcolor{white} ${q=0.1 < q_1\approx 0.1414}$}& $h>h_d\approx 0.1336$     & 0 & none (cusp) \\
 \midrule
\cellcolor{white}   & $h\leq h_a\approx 0.166$     & 1 & VdW \\
 
\cellcolor{white}  & $h_a<h<h_b\approx 0.170$     & 1 & VdW*\\
 
\cellcolor{white}  & $h_b\approx 0.171$           & up to 3 & VTP \\
 
\cellcolor{white}  & $h_b<h<h_c\approx 0.175$     & up to 3 & TP \\
 
\cellcolor{white}  & $h_c\approx 0.176$           & up to 3 & VTP\\
 
\cellcolor{white}  & $h_c<h\leq h_d\approx 0.184$ & 1 & VdW*\\
 
\cellcolor{white}  & $h_d<h<h_e$                  & 1 & VdW, RPT \\
 
 \multirow{-8}{0.35 \hsize}{ \cellcolor{white}${q_1<q=0.25<q_2, q_2\approx 0.3073}$} & $h>h_e\approx 0.263$         & 0 & none (cusp)
 \\ \midrule
 \cellcolor{white} &   $0<h<h_a$              & 1 & VdW \\
 
\multirow{-2}{0.35 \hsize }{ \cellcolor{white}$q=0.4>q_2 \approx 0.3073$} & $h>h_a\approx 0.430$   & 0 & $S<0$ \\
\bottomrule
\end{tabular}
\caption{{\bf Critical behaviour for spherical black holes in $d=5$}: This table summarizes the various phase transitions that can take place for $d=5$ charged, hairy black holes with $\sigma = +1$.  }
\label{tab:gb_5d_spherical}
\end{table}

%%%%%%%%%%%%%%%%%%%%%%%%%%%%%%%%%%%%%%%%%%%%% Hyperbolic Horizon %%%%%%%%%%%%%%%%%%%%%%%%%%%%%%%%%%%%%%

\begin{table}[!htp]
\centering
\begin{tabular}{c c c c}
\toprule
Charge ($q$) & Hair ($h$) & \# Physical CP & Behaviour
\\
\midrule
\cellcolor{white} & $h\leq 0$                & 0 & 0PT, 1PT, RPT2 \\

\cellcolor{white} & $0<h\leq h_b$            & 0 & 0PT, RVdW \\
 \multirow{-3}{*}{\cellcolor{white}$q=0$}& $h>h_b=6\sqrt{2}/(5\pi)$ & 0 & $S<0$ \\
 \midrule
\cellcolor{white} & $h< 0$                    & 0 & 0PT, 1PT, RPT2 \\
 
\cellcolor{white} & $0<h\leq h_b$             & 0 & 0PT, 1PT, RPT2, RVdW \\ 
\multirow{-3}{*}{\cellcolor{white}$q=0.2$} &  $h>h_b=6\sqrt{2}/(5\pi)$  & 0 & $S<0$ 
\\ \bottomrule
\end{tabular}
\caption{{\bf Critical behaviour for hyperbolic black holes in $d=5$}: This table summarizes the various phase transitions that can take place for $d=5$ charged, hairy black holes with $\sigma = -1$.  Note that all non-zero charges within a given row
yield the same phase behaviour;  the value $q=0.2$ describes a representative case. }
\label{tab:gb_5d_hyperbolic}
\end{table}

%%%%%%%%%%%%%%%%%%%%%%%%%%%%%%%%%%%%%%%%%%%%%%%%%%%%%%%%%%%%%%%%%%%%%%%%%%%%%%%%%%%%%%%%%%%%%%%%%%%%%%%
%%%%%%%%%%%%%%%%%%%%%%%%%%%%%%%%%%%%%%%%%%%%%%%%%%%%%%%%%%%%%%%%%%%%%%%%%%%%%%%%%%%%%%%%%%%%%%%%%%%%%%%

\begin{table}[!htp]
\centering
\begin{tabular}{c c c c}
 \toprule
 \cellcolor{white} Charge ($q$) & \cellcolor{white} Hair ($h$) & \# Physical CP \cellcolor{white} & \cellcolor{white} Behaviour \\ \midrule
\cellcolor{white}& $h\leq 0$         & 0 & 0PT, 1PT, RPT2 \\

\cellcolor{white}& $0<h\leq h_a$     & 0 & 0PT, RVdW \\
\multirow{-3}{*}{\cellcolor{white} $q = 0$} & $h>h_a=16/(6\pi)$ & 0 & $S<0$ \\ \midrule

\cellcolor{white}& $h\leq 0$         & 0 & 0PT, 1PT, RPT2 \\

\cellcolor{white}& $0<h\leq h_a$     & 0 & 0PT, 1PT, RPT2, RVdW \\
 \multirow{-3}{*}{\cellcolor{white} $q = 0.2 > 0$} & $h>h_a=16/(6\pi)$ & 0 & $S<0$ \\ \bottomrule
\end{tabular}
\caption{{\bf Critical behaviour for hyperbolic black holes in $d=6$}: This table summarizes the various phase transitions that can take place for $d=6$ charged, hairy black holes with $\sigma = -1$.  Note that all non-zero charges within a given row
yield the same phase behaviour;  the value $q=0.2$ describes a representative case.}
\label{tab:gb_6d_hyperbolic}
\end{table}

\section{$3^{rd}$ order Lovelock gravity}\label{sec:cl_case}

In this section we consider the thermodynamic behaviour of $U(1)$ charged hairy black holes in cubic Lovelock gravity.  We begin by  making some general considerations and then proceed to focus on $d=7$ and $d=8$, which are the lowest dimensions for which cubic Lovelock terms are gravitationally active.

\subsection{Thermodynamic equations and  constraints}

In Gauss-Bonnet gravity, requiring the existence of AdS asymptotics implies a maximum pressure induced by the cosmological constant. Similar constraints apply to cubic Lovelock theory. To check for the existence of an asymptotic AdS region, we consider the polynomial 
\begin{align}
\alpha_3\left(\frac{\sigma-f}{r^2}\right)^3+\alpha_2\left(\frac{\sigma-f}{r^2}\right)^2+\left(\frac{\sigma-f}{r^2}\right)+\alpha_0=&\frac{16\pi GM}{(d-2)\Sigma_{d-2}^{(\sigma)}r^{d-1}}+\frac{H}{r^d}\nn\\&-\frac{8\pi G}{(d-2)(d-3)}\frac{Q^2}{r^{2d-4}}
\end{align}
for the metric function in 3rd order Lovelock gravity.  Being a cubic polynomial in $f$, this equation has three distinct solutions.  One of these solutions will not have a smooth $\alpha_3 \to 0$ limit; we refer to this as the \textit{Lovelock branch} $(f_{L}$).  Of the remaining two branches, we denote the \textit{Gauss-Bonnet branch} $(f_{GB})$ as the one  that does not permit a smooth $\alpha_2 \to 0$ limit and the \textit{Einstein branch} ($f_{E}$) as the one that does.  Each will have
AdS asymptotics  provided  the Lovelock coupling parameters obey the following inequality
\be\label{3-disc}
4\alpha_3-\alpha_2^2+27\alpha_0^2\alpha_3^2-18\alpha_0\alpha_2\alpha_3+4\alpha_0\alpha_2^2\le 0 \, .
\ee
If this inequality is violated, then only the Lovelock branch has valid asymptotic AdS structure, whilst the Gauss-Bonnet and Einstein branches terminate at finite $r$. 
Constructing the   dimensionless quantities $(v,t,h,a,m)$ via
\begin{align}
r_+ &= v\alpha_3^{1/4} \, , \quad  T = \frac{t\alpha_3^{-1/4}}{d-2} \, , \quad  H =\frac{4\pi h}{d-2}\alpha_3^{\frac{d-2}{4}}\\
Q &= \frac{q}{\sqrt{2}}\alpha_3^{\frac{d-3}{4}} \, , \quad  m = \frac{16\pi M}{(d-2)\Sigma_{d-2}^{(\kappa)}\alpha_3^{\frac{d-3}{4}}}
\end{align}
the condition \eqref{3-disc}, a quadratic equation in $\alpha_0$, yields the maximal and minimal pressure conditions
$p_- \leq p \leq p_+$ (c.f. Figure (13) from \cite{Frassino:2014pha}), where
\be\label{pressurecond}
p_\pm=\frac{(d-1)(d-2)}{108\pi}\left[9\alpha-2\alpha^3\pm 2(\alpha^2-3)^\frac{3}{2}\right] 
\ee
and where
\be\label{dimless_pres_coupling_cl}
p=\frac{\alpha_0(d-1)(d-2)\sqrt{\alpha_3}}{4\pi} \, , \quad \alpha=\frac{\alpha_2}{\sqrt{\alpha_3}}\, .
\ee
Positivity of entropy implies that
\be\label{entropy2}
v^4(d-4)(d-6)+2\sigma\alpha v^2(d-2)(d-6)+3(d-2)(d-4)-\frac{2\pi hd(d-6)}{\sigma v^{d-6}(d-2)}\geq 0 
\ee
from  Eq.~\eqref{nice_entropy}. 
The equation of state\footnote{Recall that we have assumed $\sigma\neq 0$ because planar black hole solutions cannot admit nonzero hair. Therefore in cubic Lovelock gravity we also express the equation of state and Gibbs free energy in a simplified form where $\sigma^{2k}=1$. This is an invalid simplification if the planar case is included.} can be shown to be
\begin{align}\label{lovelockeos}
p = &\frac{t}{v}-\frac{\sigma(d-3)(d-2)}{4\pi v^2}+\frac{2\alpha\sigma t}{v^3}-\frac{\alpha(d-2)(d-5)}{4\pi v^4}+ \frac{3t}{v^5}\nn\\&-\frac{\sigma(d-7)(d-2)}{4\pi v^6}+\frac{q^2}{v^{2(d-2)}}-\frac{h}{v^d}
\end{align}
which reduces to the usual no-hair case \cite{Frassino:2014pha} by setting $h=0$. The Gibbs free energy is given by
\begin{align}
g=
&-\frac{1}{16\pi(3+3\alpha\sigma v^2+v^4)}\bigg[\frac{4\pi pv^{d+3}}{(d-1)(d-2)}-\sigma v^{d+1}+\frac{24\pi\sigma p\alpha v^{d+1}}{(d-1)(d-4)}\nn\\
&-\frac{\alpha v^{d-1}(d-8)}{d-4}+\frac{60\pi v^{d-1}}{(d-1)(d-6)}-\frac{2\sigma\alpha^2 v^{d-3}(d-2)}{d-4}+\frac{4\sigma v^{d-3}(d+3)}{d-6}\nn\\
&-\frac{3\alpha v^{d-5}(d-2)(d-8)}{(d-4)(d-6)}-\frac{3\sigma v^{d-7}(d-2)}{d-6}\bigg]\nn\\
&+\frac{q^2}{4(3+2\sigma\alpha v^2+v^4)(d-3)v^{d-3}}\bigg[\frac{v^4(2d-5)}{d-2}+\frac{2\alpha\sigma(2d-7)v^2}{d-4}+\frac{3(2d-9)}{d-6}\bigg]\nn\\
&+\frac{h}{8\sigma v(3+3\alpha\sigma v^2+v^4)(d-2)(d-4)}\bigg[\frac{4\pi dpv^6}{(d-1)(d-2)}-\sigma(d^2-7d+8)v^4\nn\\
&-3\alpha(d^2-5d+8)v^2-\frac{\sigma(5d^3-53d^2+186d-240)}{d-6}\nn\\
&+\frac{4d\pi v^{6-d}h}{d-2}-\frac{4\pi dq^2}{(d-2)v^{2(d-5)}}\bigg]
\end{align}
The physical state of the system will be those values of $t, p, q$ and $h$ that globally minimize $g$  for a given fixed $\alpha$.

Here in cubic Lovelock gravity we are confronted with an additional difficultly compared to the Gauss-Bonnet case considered earlier: there are now three parameters $(\alpha, q, h)$ that can be adjusted.  This makes it cumbersome to completely characterize the thermodynamics in terms of the effects of $q$ and $h$.   We shall instead describe the salient effects of $q$ and $h$ in different regions of $\alpha$ that bear particular significance.  
%\begin{figure}[tp]
%\includegraphics[scale=0.38]{plots/cl/cl_allowBH1.eps}
%\includegraphics[scale=0.38]{plots/cl/cl_allowBH2.eps}
%\caption{\textbf{AdS asymptotics}: the regions where black holes are allowed for various values of $\alpha$ depending on the asymptotic behaviour. \textit{Left}: $\sigma=-1$. \textit{Right}: $\sigma=+1$. \tcbv{These plots have been derived in \cite{Frassino:2014pha} as well.}} 
%\label{ADS}
%\end{figure}

\subsection{$P-v$ criticality in $d=7$}

For the time being we assume $\alpha>0$ for simplicity. In seven dimensions, the entropy condition \eqref{entropy2} becomes
\be 
3v^4+10\sigma\alpha v^2+45-\frac{14\pi h}{5\sigma v}\geq 0
\ee
and it is clear that for negative $h$ this is trivially satisfied for spherical black holes. The equation of state is given by
\be
p=\frac{t}{v}-\frac{5\sigma}{\pi v^2}+\frac{2\alpha\sigma t}{v^3}-\frac{5\alpha}{2\pi v^4}+\frac{3t}{v^5}+\frac{q^2}{v^{10}}-\frac{h}{v^7}
\ee
and the critical points satisfy the following equations
\be\label{tempcritcub}
t_c=\frac{10\left(\sigma v_c^8 + \alpha v_c^6 -\pi q^2\right) +7\pi h v_c^3}{\pi v_c^5(v_c^4+6\alpha\sigma v_c^2 +15)}
\ee
\be
\begin{aligned}
v_c^{12}-3\alpha\sigma v_c^{10}+3(2\alpha^2-15)v_c^8-15\alpha\sigma v_c^6&-3\pi\sigma q^2(3v_c^4+14\sigma\alpha v_c^2+25)\\
&+\left(\frac{21}{5}\pi\sigma v_c^7+\frac{84}{5}\alpha\pi v_c^5+21\pi\sigma v_c^3\right)h=0
\end{aligned}
\ee
where these expressions all reduce to the cubic Lovelock case studied in \cite{Frassino:2014pha} when $h=0$.  As before, the
qualitative effect of the hair is to introduce new terms of odd powers in the volume $v$.

\subsubsection{Spherical case}

For $h=0$ it was shown in \cite{Frassino:2014pha} that for $\alpha\in(0,10)$ there is exactly one physical VdW-type critical point for both charged and uncharged spherical black holes. Furthermore, for fixed charge $q$ this critical point may become unphysical for sufficiently large $\alpha$. From our study of Gauss-Bonnet hairy black holes, we have seen that $h$ acts as a `tuning parameter' which gives a continuous one-parameter family of phase diagrams allowing for more thermodynamic possibilities. We will show that this also occurs in cubic Lovelock gravity: for hairy black holes in $d=7$, we will recover all results previously found for $d=7$ and $d=8$ from \cite{Frassino:2014pha}, along with all of the thermodynamic behaviour we described for hairy Gauss-Bonnet black holes in the previous section.

\begin{figure}[tp]
\begin{center}
\includegraphics[scale=0.25]{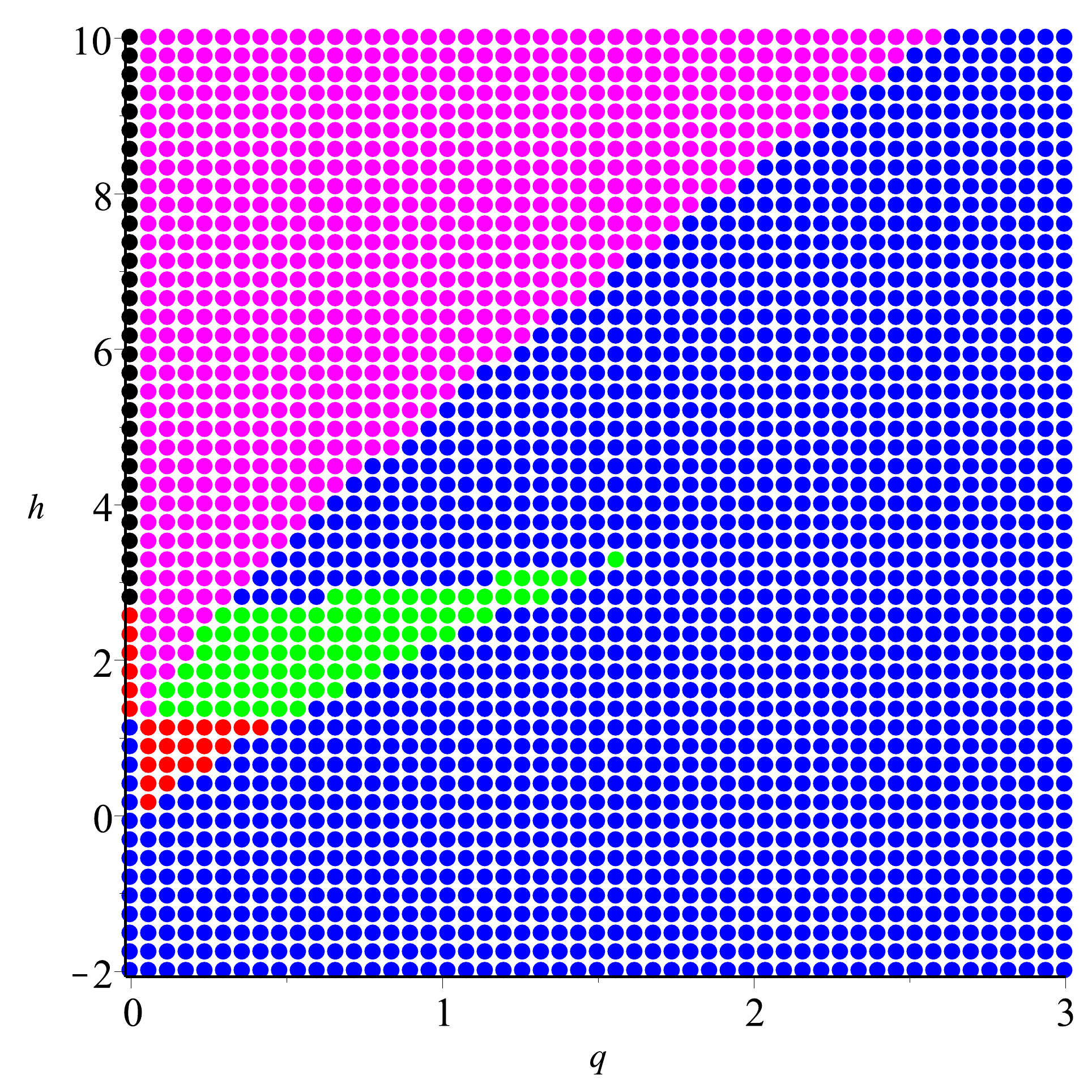}
\includegraphics[scale=0.25]{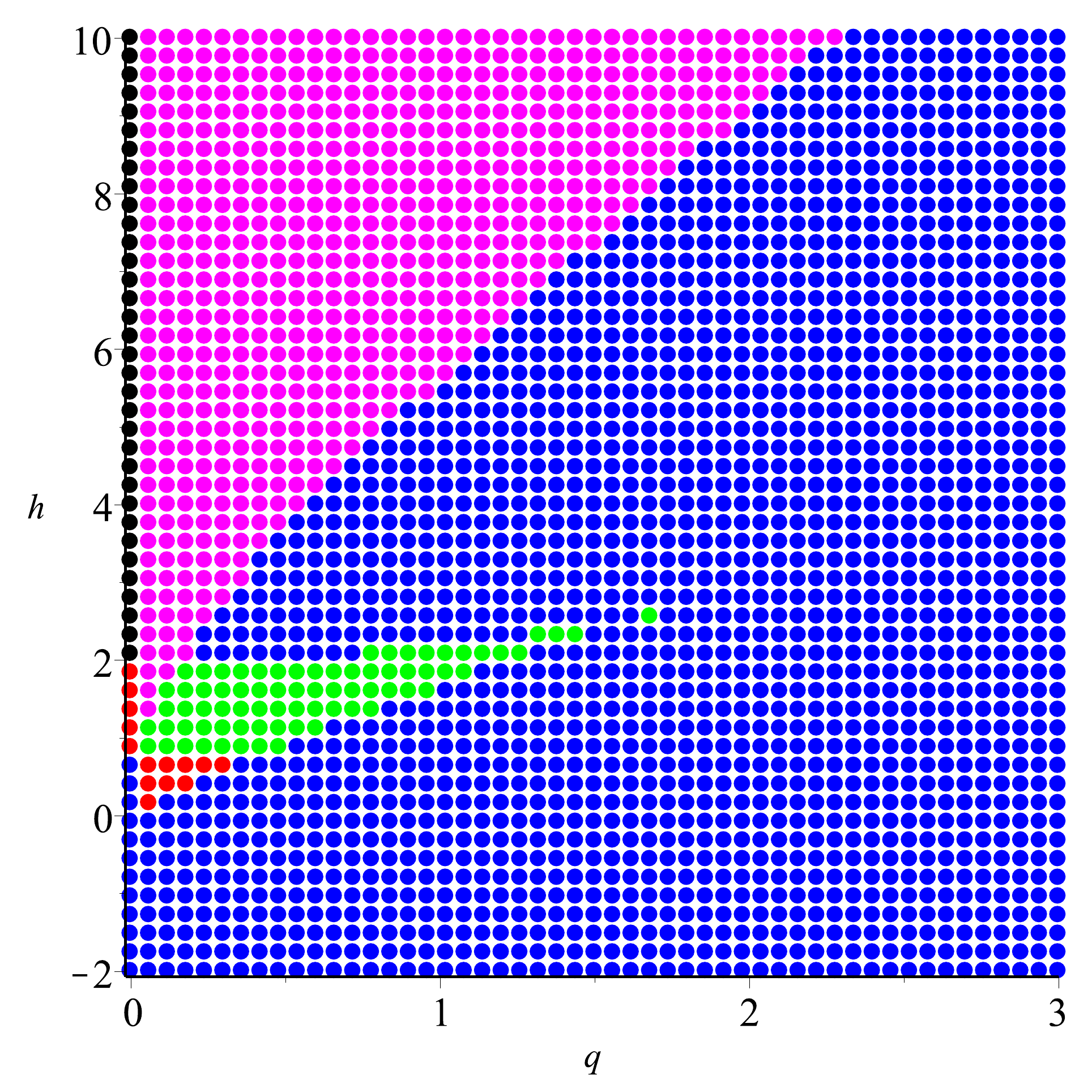}
\includegraphics[scale=0.25]{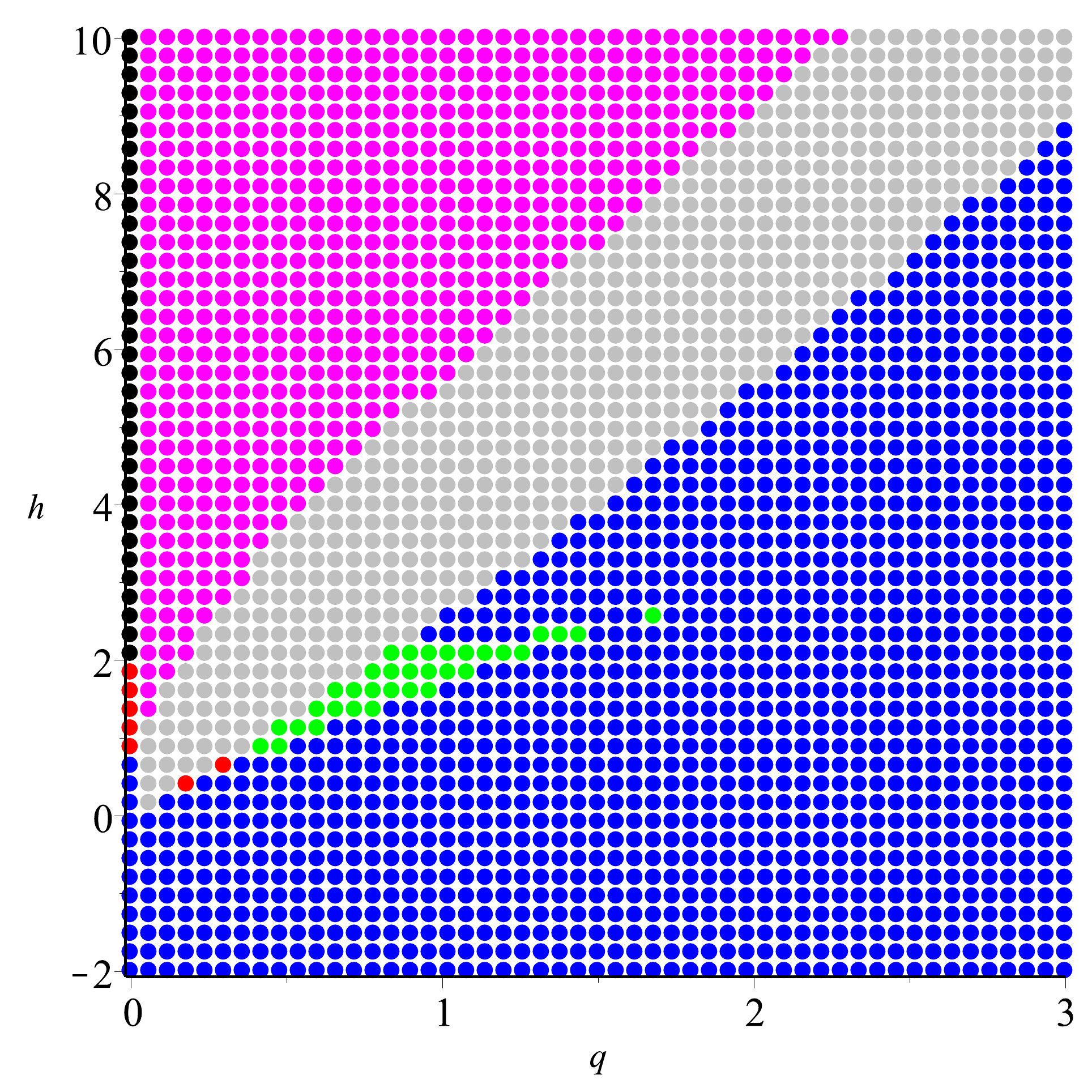}
\includegraphics[scale=0.25]{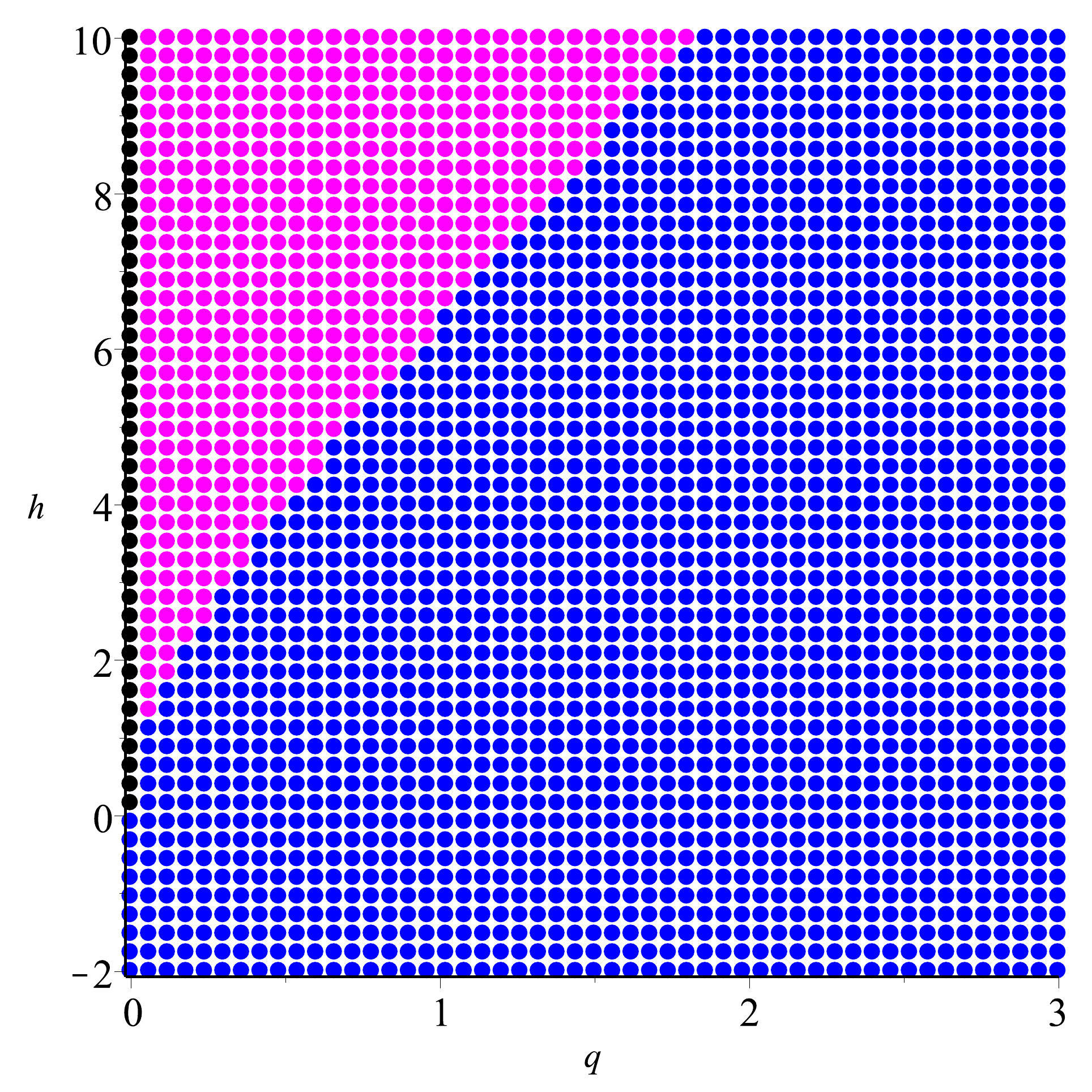}
\includegraphics[scale=0.25]{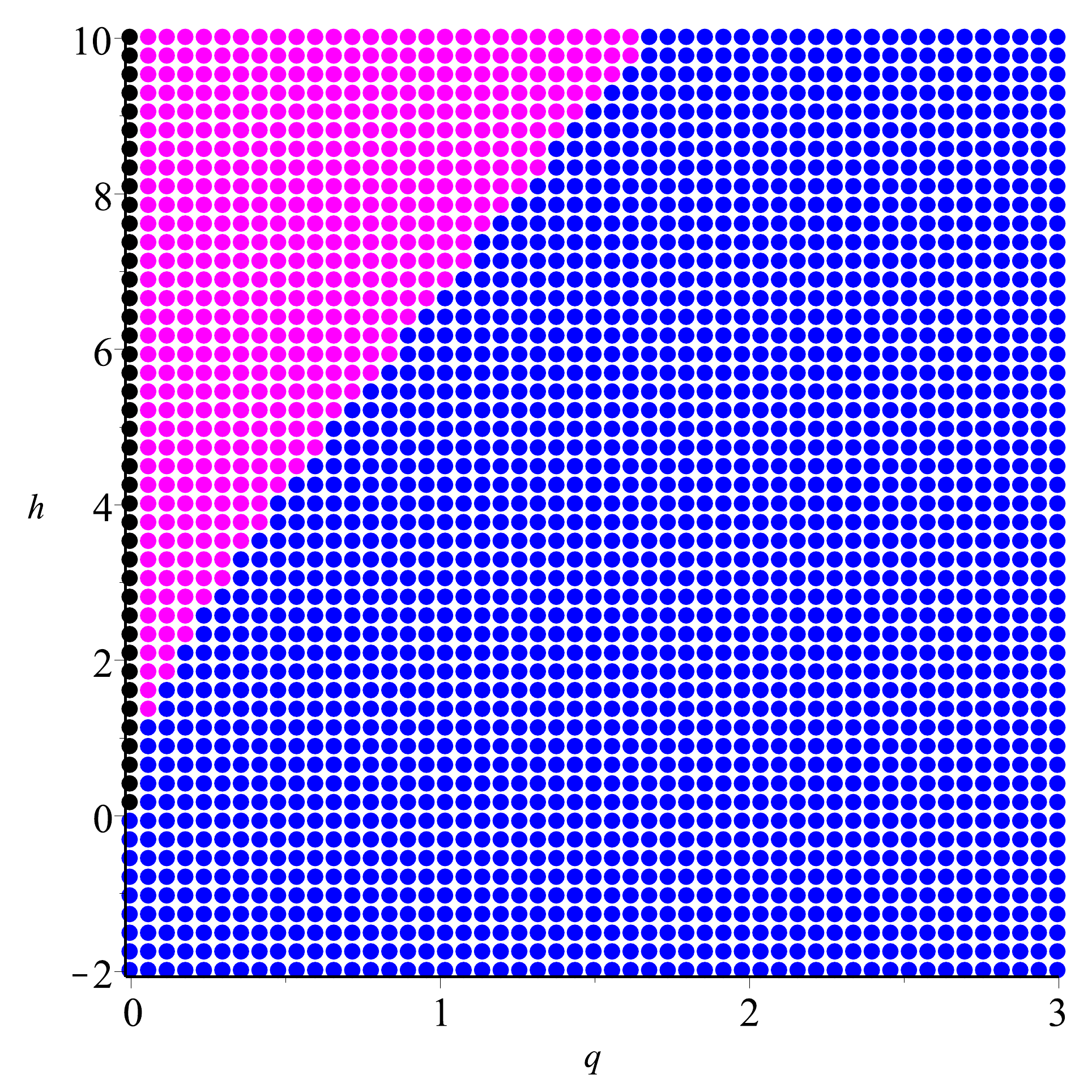}
\includegraphics[scale=0.25]{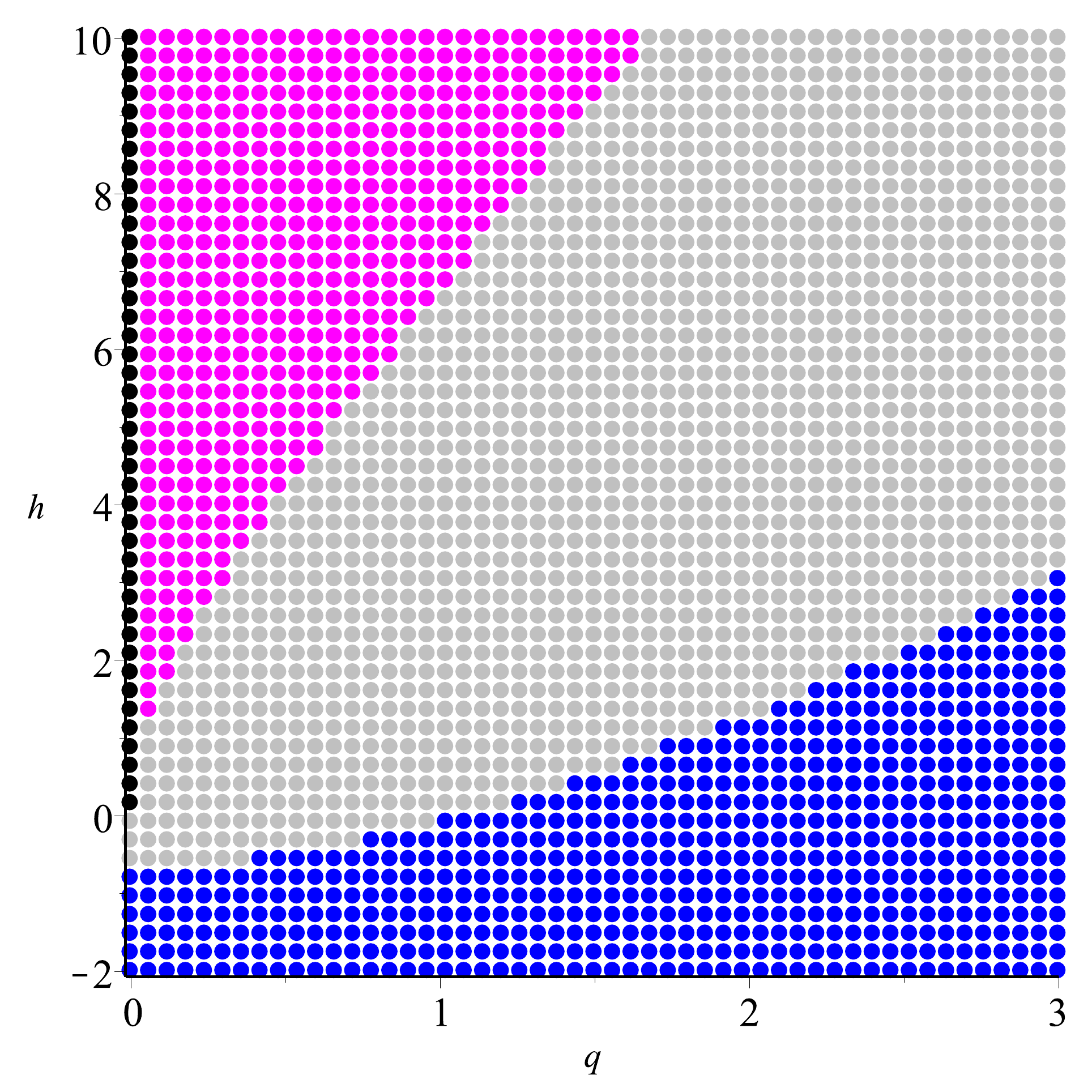}
\end{center}
\caption{{\bf The $(q,h)$ parameter space for various $\alpha$}: $d=7$ and $\sigma = +1$: 
A plot of the possible number of physical critical points for various $\alpha$.  Black corresponds to no critical point, blue to one critical point, red to two, green to three and magenta to  positive pressure but negative entropy.  In these figures, there is a sliver of black region on the vertical axis, since for $q=0$ the parameter space as a  function of $h$ is qualitatively different from   when $q\neq 0$.
Grey corresponds to the region of parameter space where the physical critical point is outside of the maximum pressure bound for the Einstein and Gauss-Bonnet branches; it is only physical for the Lovelock branch $f_L$.
\textit{Top left}: $\alpha=1<\sqrt{3}$. Note that since there is only one branch with valid AdS asymptotics for $\alpha<\sqrt{3}$, there is no maximum pressure constraint and hence there is no grey region. 
\textit{Top centre}:  $\alpha=2$ without the maximum pressure constraint. This is the parameter space for the Lovelock branch $f_L$.
\textit{Top right}: $\alpha=2$ with the pressure constraints imposed.
\textit{Bottom left}: $\alpha=4.55$.  This is approximately the value of $\alpha$ at which the blue region ends at $q=h=0$ in $d=7$. In particular, for $h=0$ there is minimum nonzero charge $q_{min}$ given by the boundary between the blue and grey regions such that if $q<q_{min}$, then the critical pressure will exceed the maximum allowed pressure \cite{Frassino:2014pha}. This special value of $\alpha$ depends on the dimension $d$. For $\alpha\gtrsim 4.55$, the grey region extends to negative $h$.
\textit{Bottom center}: $\alpha=6\gtrsim 4.55$, without the pressure constraints. 
\textit{Bottom right}: $\alpha=6$, with the pressure constraints imposed. A simple observation shows that increasing $\alpha$ simply expands the grey region away from the magenta region and  shrinks the green and red regions.  Eventually for sufficiently large $\alpha$ there can be only one physical critical point.}
\label{cl_qhsearch1}
\end{figure}

Figure~\ref{cl_qhsearch1} shows the possible number of critical points for four representative $(q,h)$ values in parameter space as $\alpha$ varies. We see that the main difference lies in the grey region, which is the region in parameter space where the critical points become unphysical due to maximum/minimum pressure constraints. A careful analysis reveals that the grey region only applies to the Gauss-Bonnet and Einstein branches, since the Lovelock branch still possesses AdS asymptotics beyond the maximal pressure.  
To study the Lovelock branch, $f_L$, we can simply ignore the grey region, as per the centre diagrams in Figure~\ref{cl_qhsearch1}. 
 We shall not attempt to fully classify the possible thermodynamic phenomena as functions of $\alpha,q,h$ here, but instead focus on the results which differ from the no-hair case. We refer to summary Section~\ref{thermosummary3} for further, more detailed examples which are not explicated upon below. We will focus specifically on three results that do not occur in the absence of scalar hair: (1) reentrant phase transitions, (2) triple points and (3) virtual triple points.

\begin{figure}[htp]
\includegraphics[scale=0.38]{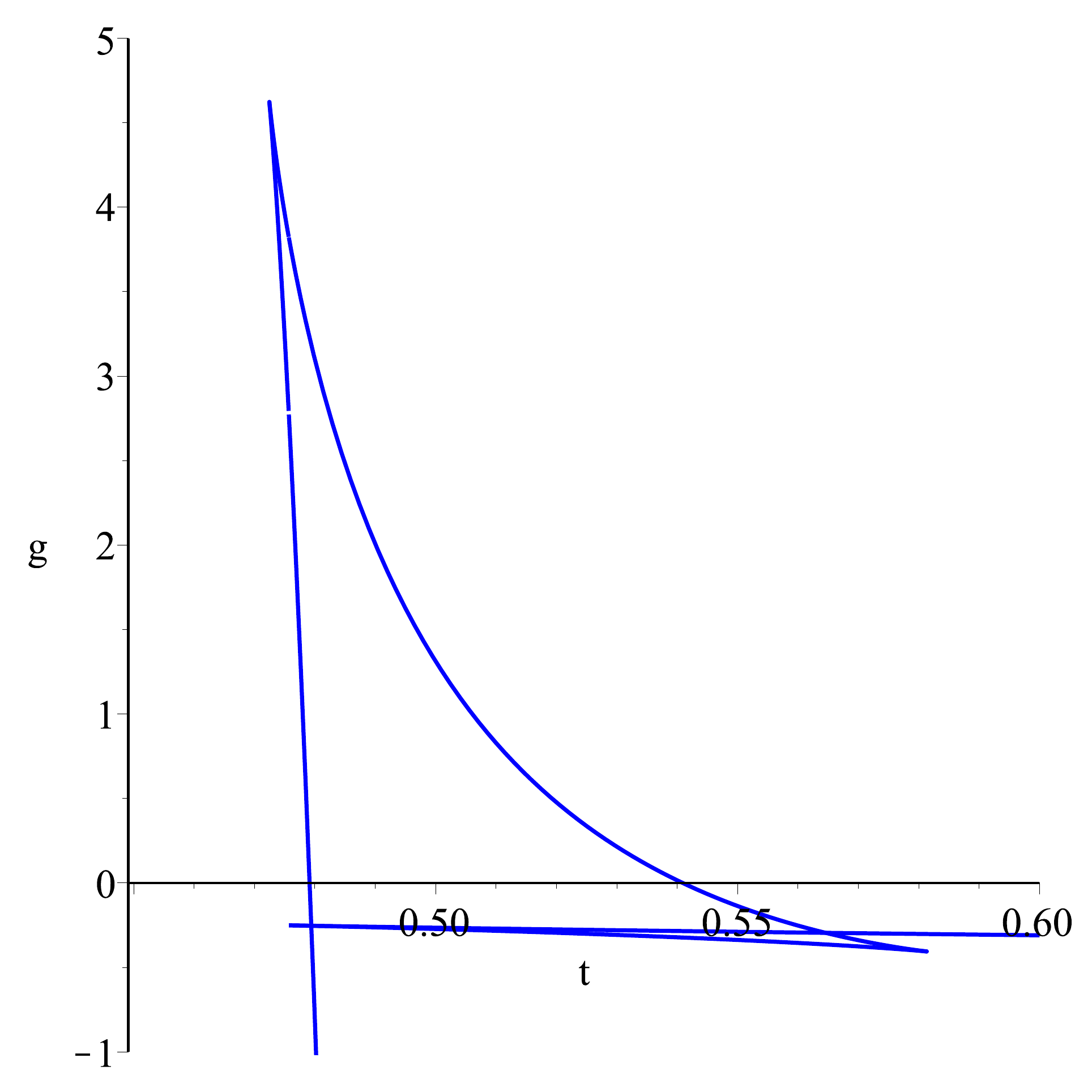}
\includegraphics[scale=0.38]{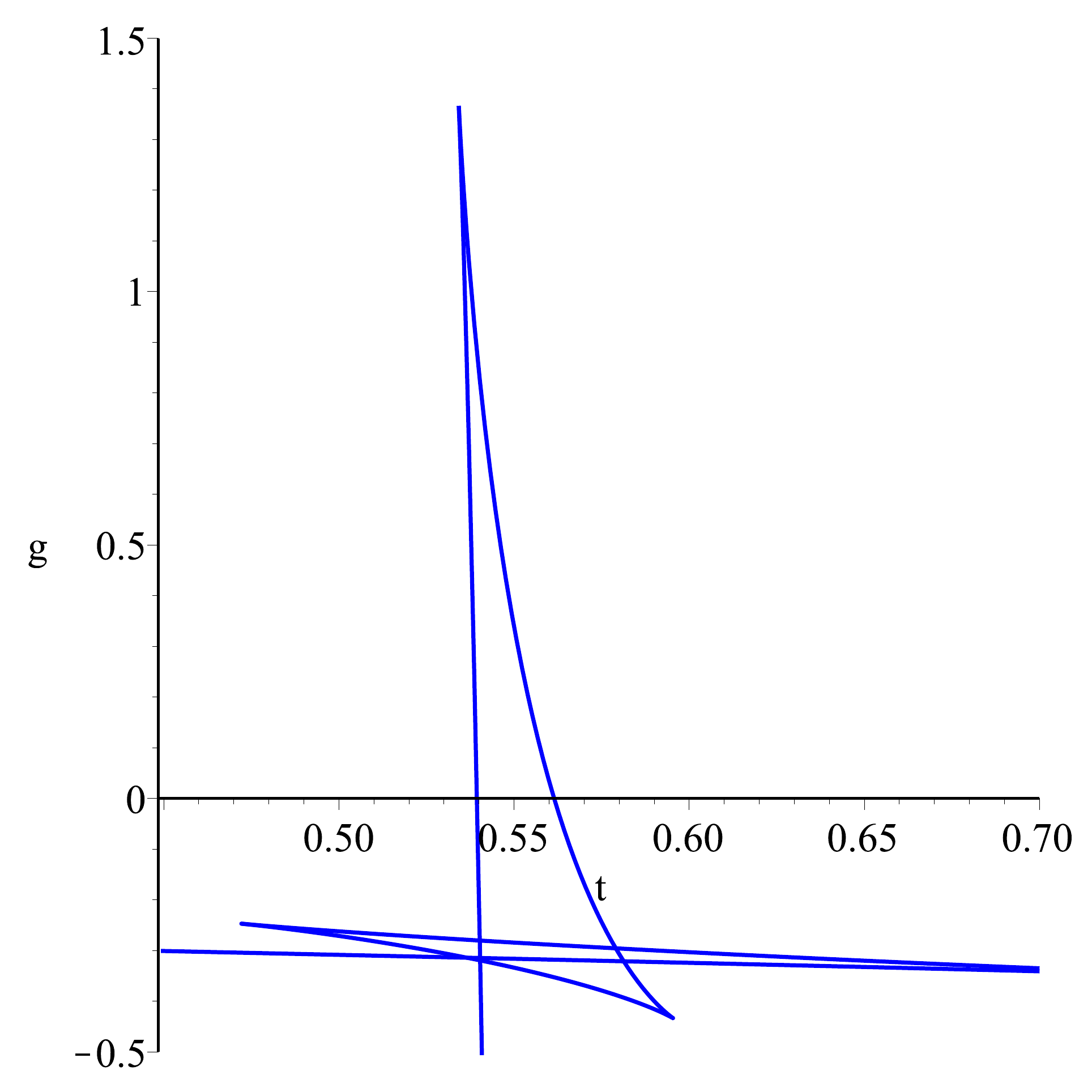}
\caption{\textbf{Reentrant phase transitions and triple points for $d=7$:} \textit{Left}: RPT phenomenon for $\alpha=1,h=0.5,q=0,p=0.0385$. \textit{Right}: triple point phenomenon for $\alpha=1,h=0.5,q=0.1,p=0.051$. }
\label{cl_gibbs1}
\end{figure}

In the absence of scalar hair, the lowest dimension for which RPT can occur for uncharged spherical black holes is $d=8$. Figure~\ref{cl_gibbs1} (left) shows the Gibbs free energy for $h=0.5, q=0, \alpha=1$ demonstrating that $h\neq 0$ allows RPT to occur in uncharged seven-dimensional spherical black holes.  The coexistence plot for this transition is qualitatively no different than that shown in Figure~\ref{gb_cusp}. Furthermore, in the no-hair case, the lowest dimension for a triple point is also $d=8$, and as usual small charge is needed. In contrast to this we find  that hair allows a triple point to be realized in $d=7$, e.g for $h=0.5,q=0.1,\alpha=1$ (c.f. Figure~\ref{cl_gibbs1}).

\begin{figure}[htp]
\includegraphics[scale=0.38]{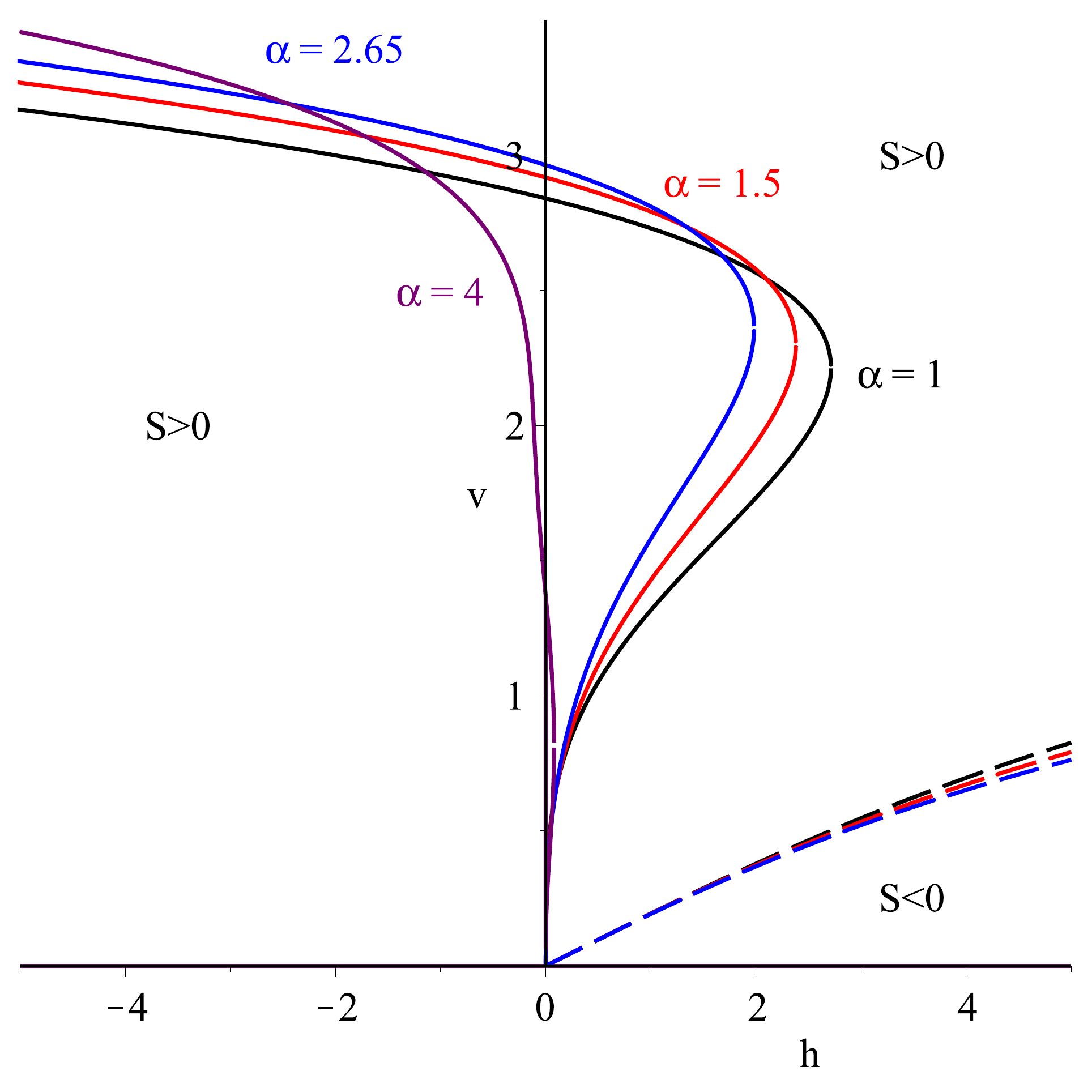}
\includegraphics[scale=0.38]{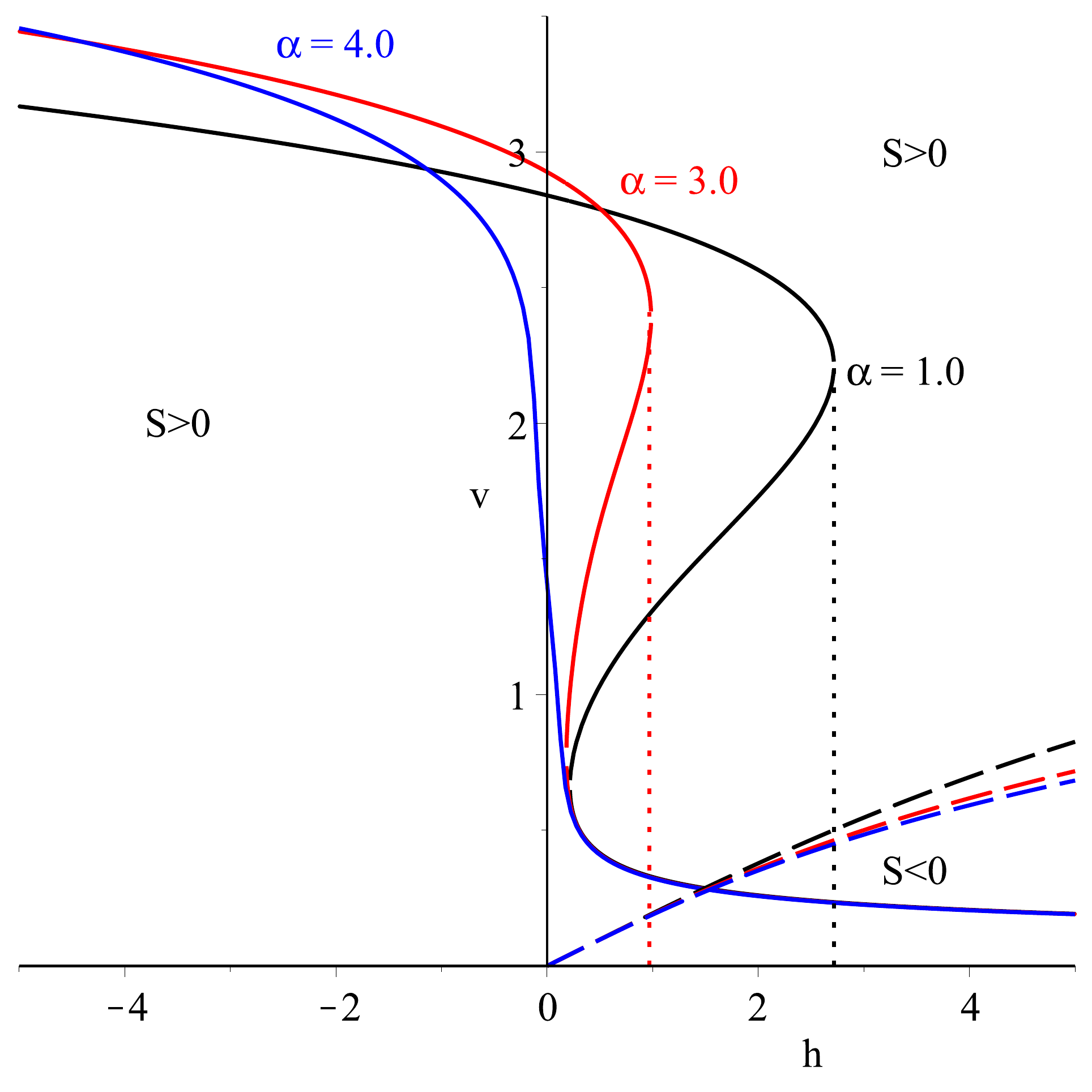}
\caption{ {\bf Critical volume and entropy for $d=7$ spherical black holes ($\sigma=1$)}: the plot of critical volumes as a function of $h$. The solid curves represent the locus of critical volumes and there can be several possible critical points for certain range of $h$ (not necessarily all physical). The dashed curves represent the zero-entropy curve as a function of $\alpha$ and $h$. \textit{Left}: $q = 0$. Increasing $\alpha$ reduces the range of $h$ where more than one critical point is possible. There can be up to two critical points. \textit{Right}: $q = 0.1$. Similar to Gauss-Bonnet case, the presence of sufficiently small charge enables up to three critical points to exist provided $\alpha$ is also sufficiently small. There is a maximum $h=h_{*}$ given by the intersection between the solid curves and the dashed curves, such that no criticality occurs for $h>h_{*}$ due to negative entropy. Note that $h_{*}$ appears to be independent of $\alpha$.}
\label{cl_crit1}
\end{figure}

The virtual triple point phenomenon observed  in Gauss-Bonnet hairy black holes in the previous section (c.f. Figure~\ref{gb_vtp} and accompanying discussion) indicated   that $h$ acts like a `tuning parameter' that relocates various critical points of the system. This is essentially the reason why we can obtain VTP behaviour.   Once a triple point occurs, one can use $h$ to adjust the location of critical points within the triple point phase diagram until a point where a critical point becomes a VTP, i.e. touches one of the first order coexistence lines as in Figure~\ref{gb_vtp}. Therefore, it is not difficult to see that if we can find a triple point for a fixed triple $(q,h,\alpha)$, then $h$ would generate one-parameter family of phase diagrams containing a VTP. For example, a triple point occurs for $\alpha=1, \, h=0.5, \, q=0.1$ and $p\approx 0.51$ and we can obtain the following exact sequence of phase diagrams as $h$ is adjusted:
\begin{equation*}
\text{VdW}\to \text{VdW*}\to \text{VTP}\to \text{\textbf{TP}}\to \text{VTP}\to \text{VdW*}
\end{equation*}
employing the notation of Section~\ref{tab:gb_5d_spherical},
where the arrow denotes an increase in $h$ and each occurs for some pressure $p$.  This example is valid for the Lovelock branch only, since $\alpha=1$, and care must be taken  to ensure the pressure constraints are satisfied for the Gauss-Bonnet and Einstein branches. 
A more general example,  applicable to all three branches, has a triple point and a longer sequence of critical behaviours:
\begin{equation*}
\text{VdW}\to \text{VdW*}\to \text{VTP}\to \text{\textbf{TP}}\to \text{VTP}\to \text{VdW*}\to\text{VdW}  
\end{equation*}
where (in this example)  $\alpha=3, \,  h=0.5, \, q=0.3$ and $ p\approx 0.044$. 
This is because, for $\alpha\in(2.5,3.8)$, there is a small additional range of $h$ with VdW-type behaviour, which we can understand by plotting figures similar to Figure~\ref{cl_crit1}.

This discussion shows that the $d=7$ cubic Lovelock spherical black holes reproduce all of the thermodynamic phenomena found earlier for Gauss-Bonnet spherical hairy black holes as all phenomena previously obtained~\cite{Frassino:2014pha} for $d=8$ hairless cubic Lovelock spherical black holes.  We provide a more detailed representative analysis in Section~\ref{thermosummary3}.

\subsubsection{Hyperbolic case}
\begin{figure}[htp]
\begin{center}
\includegraphics[scale=0.25]{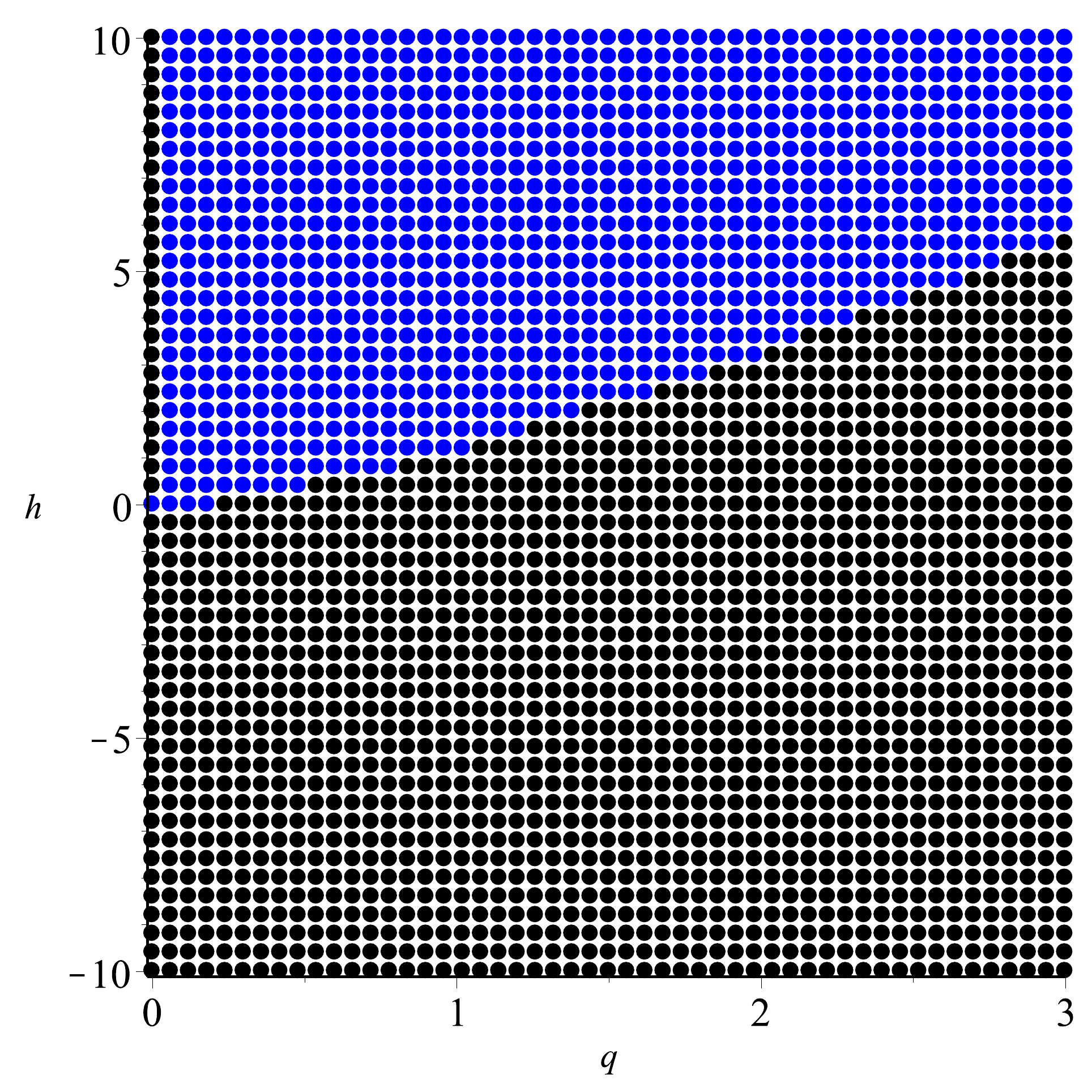}
\includegraphics[scale=0.25]{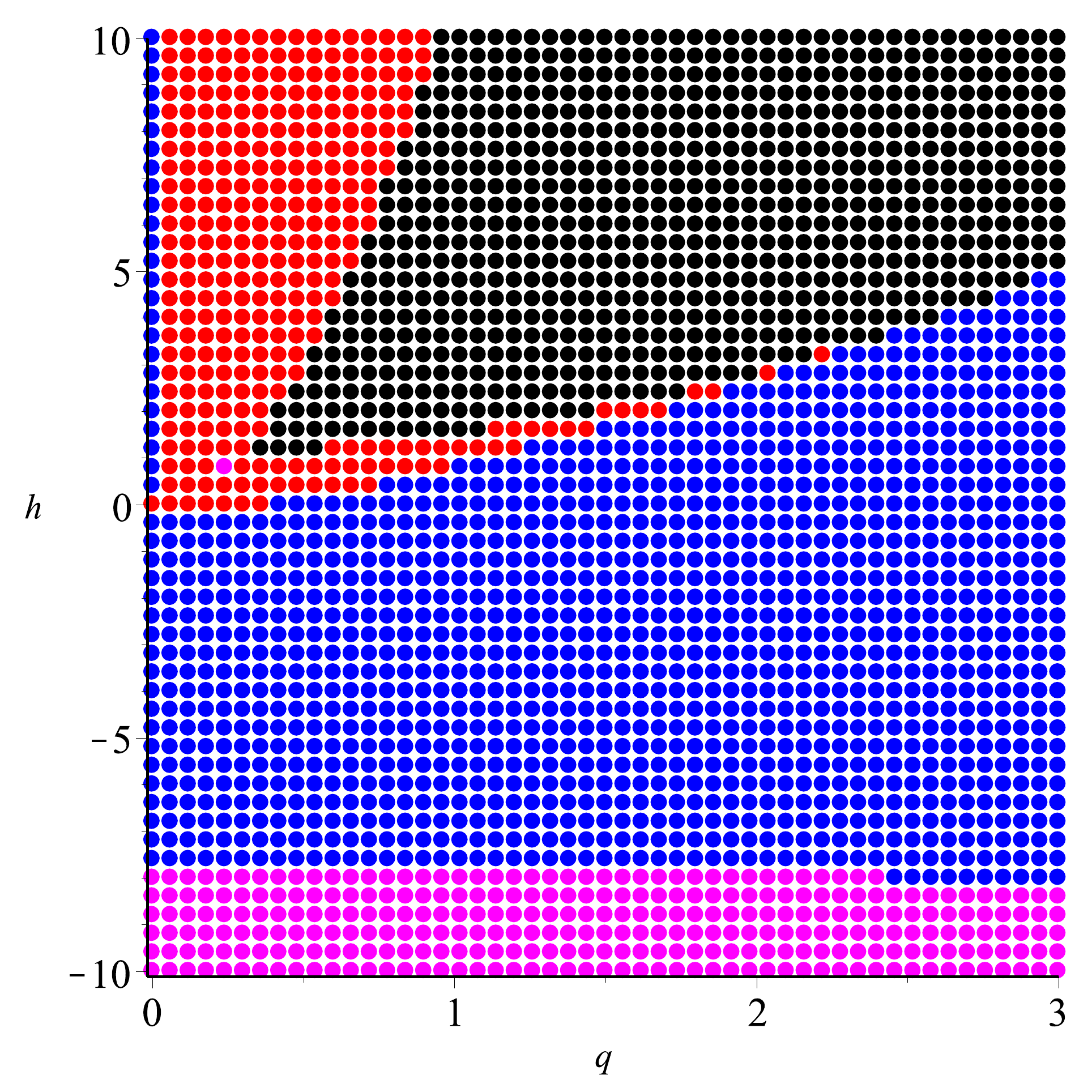}\\
\includegraphics[scale=0.25]{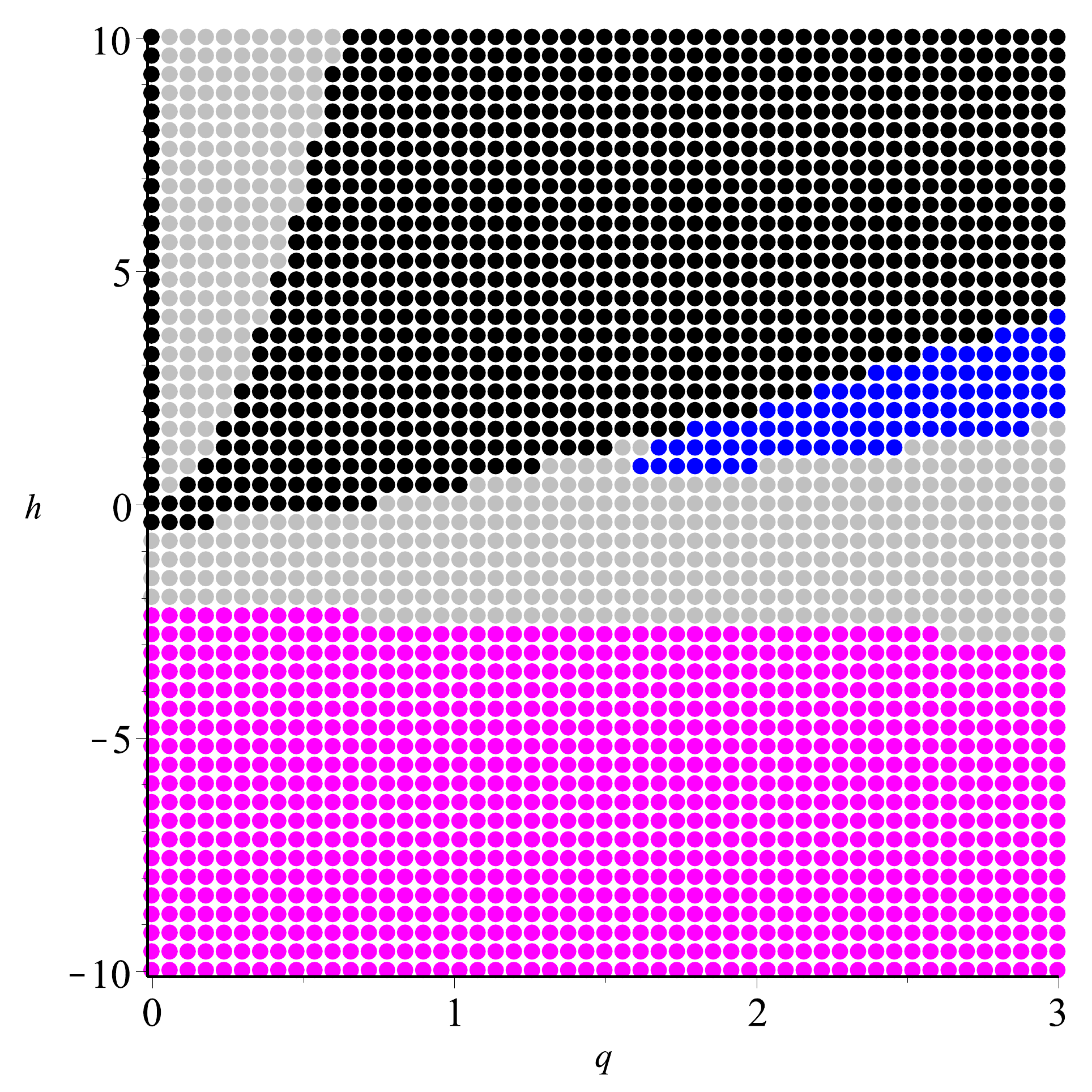}
\includegraphics[scale=0.25]{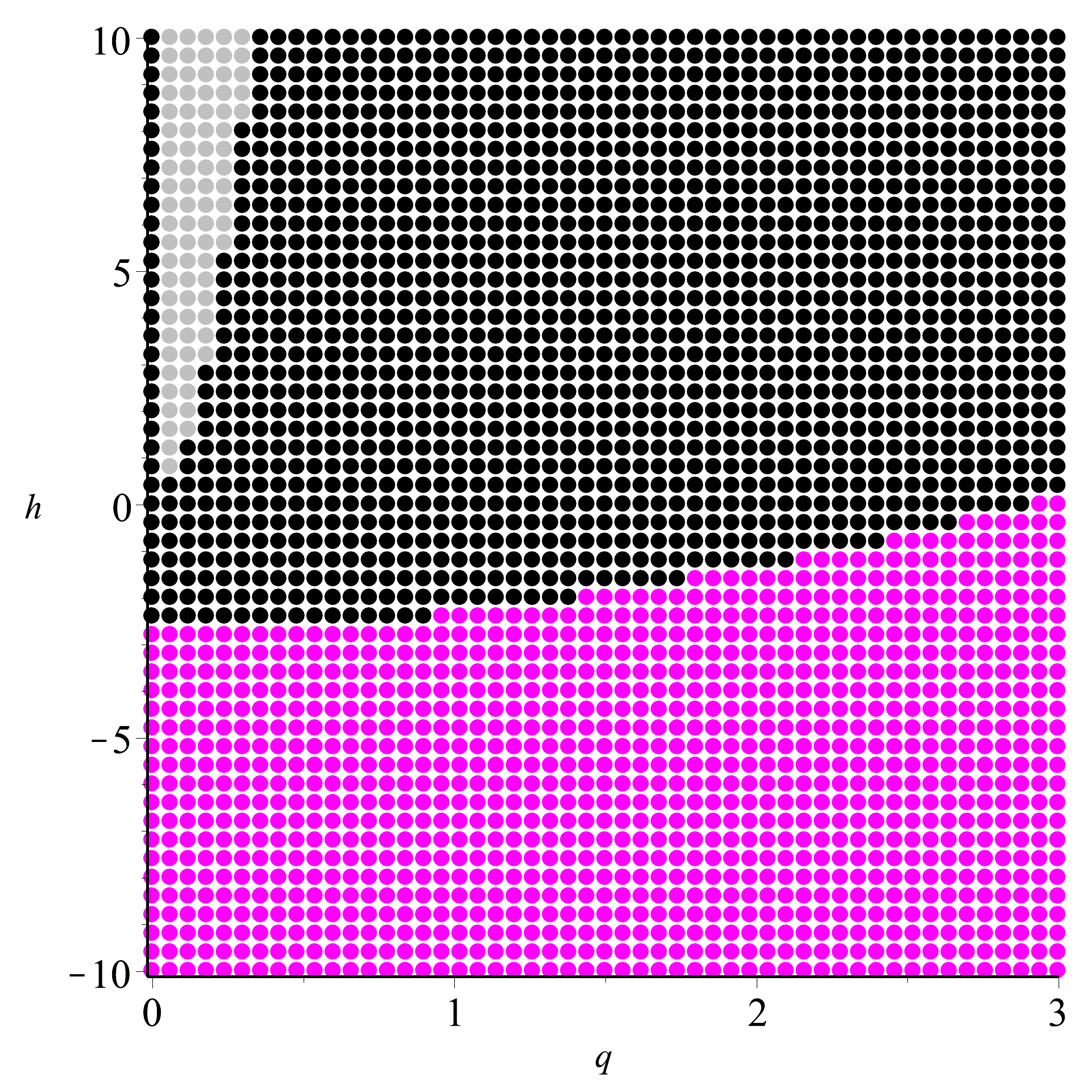}
\end{center}
\caption{{\bf The $(q,h)$ parameter space for $\sigma=-1$}. \textit{Top left}: $\alpha=1<\sqrt{5/3}$. \textit{Top right}: $\alpha=1.5\in(\sqrt{5/3},\sqrt{3})$. Note that there is a small subset of $S<0$ space (magenta) for $h>0$. \textit{Bottom left}: $\alpha=2.0\in(\sqrt{3},3\sqrt{5/3})$.  \textit{Bottom right: $\alpha=4>3\sqrt{5/3}$.} The black, blue, red regions represent zero, one and two critical points respectively, while grey represents the region where the critical points are unphysical for the Gauss-Bonnet and Einstein branches.}
\label{cl_qhsearch2}
\end{figure}

Following \cite{Frassino:2014pha}, we split the range of $\alpha$ into four distinct regions bounded by the following values of $\alpha$: $0, \sqrt{5/3}, \sqrt{3}$ and $3 \sqrt{3/5}$.  Within each region we see distinctive thermodynamic behaviour.  In the no-hair case \cite{Frassino:2014pha}, special attention was given to the point $\alpha = \sqrt{3}$ since at this point an isolated critical point was found at a thermodynamic singularity. However, we will show in this section that ICPs are not a unique property of $\alpha=\sqrt{3}$, but can occur for a much larger parameter space when $h \neq 0$.  Figure~\ref{cl_qhsearch2} provides the $(q,h)$ parameter space in terms of the number of physical critical points at various representative $\alpha$ within the four regions.  These plots show the four qualitatively distinct behaviours in $(q,h)$ space that are characteristic of these partitions of $\alpha$. We do not exhaust all possibilities here, but list some further examples in Tables~\ref{tab:cl_7d_spherical} and  \ref{tab:cl_7d_hyperbolic}. For more details the reader is directed to Section~\ref{thermosummary3}.

For $\alpha<\sqrt{5/3}$ (e.g. $\alpha=1$), we see that the $(q,h)$ parameter space is partitioned into two regions, namely the one with one physical critical point (blue) and no critical point (black). If $q=0$, then for $h=0$ one there is one critical point with standard VdW behaviour and for $h\neq 0$ the behaviour is of ideal gas type and there is no criticality. For $q\neq 0$, the thermodynamic behaviour is VdW for $h>h_0$ where the value of $h_0$ depends on the charge ($h_0$ corresponds to the boundary  between the blue and black regions of Figure~\ref{cl_qhsearch2}) while the system displays ideal gas behaviour if $h<h_0$. Thus for this range of $\alpha$, the thermodynamics is simple: we observe only VdW and ideal gas behaviour. 

\begin{figure}[tp]
\includegraphics[scale=0.38]{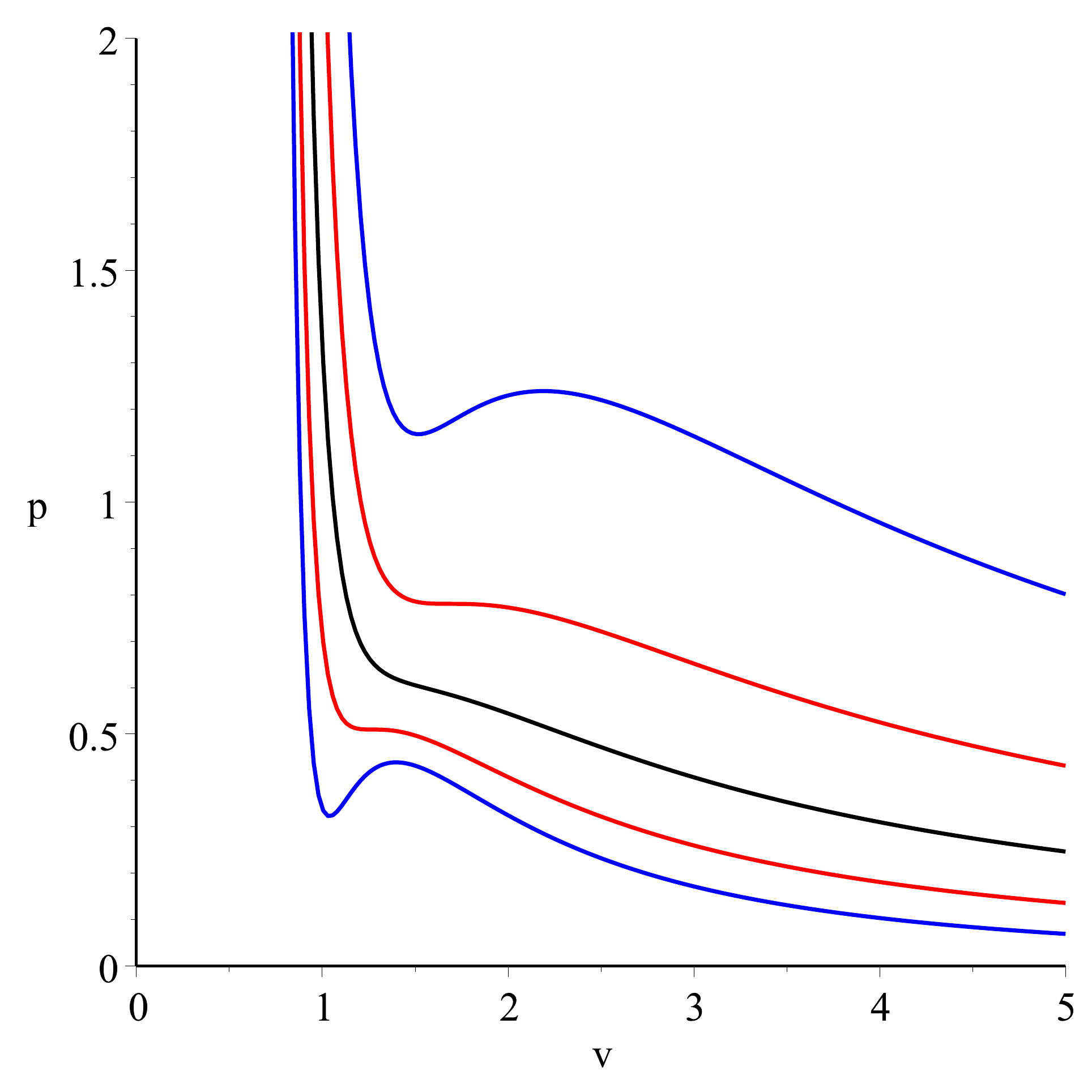}
\includegraphics[scale=0.38]{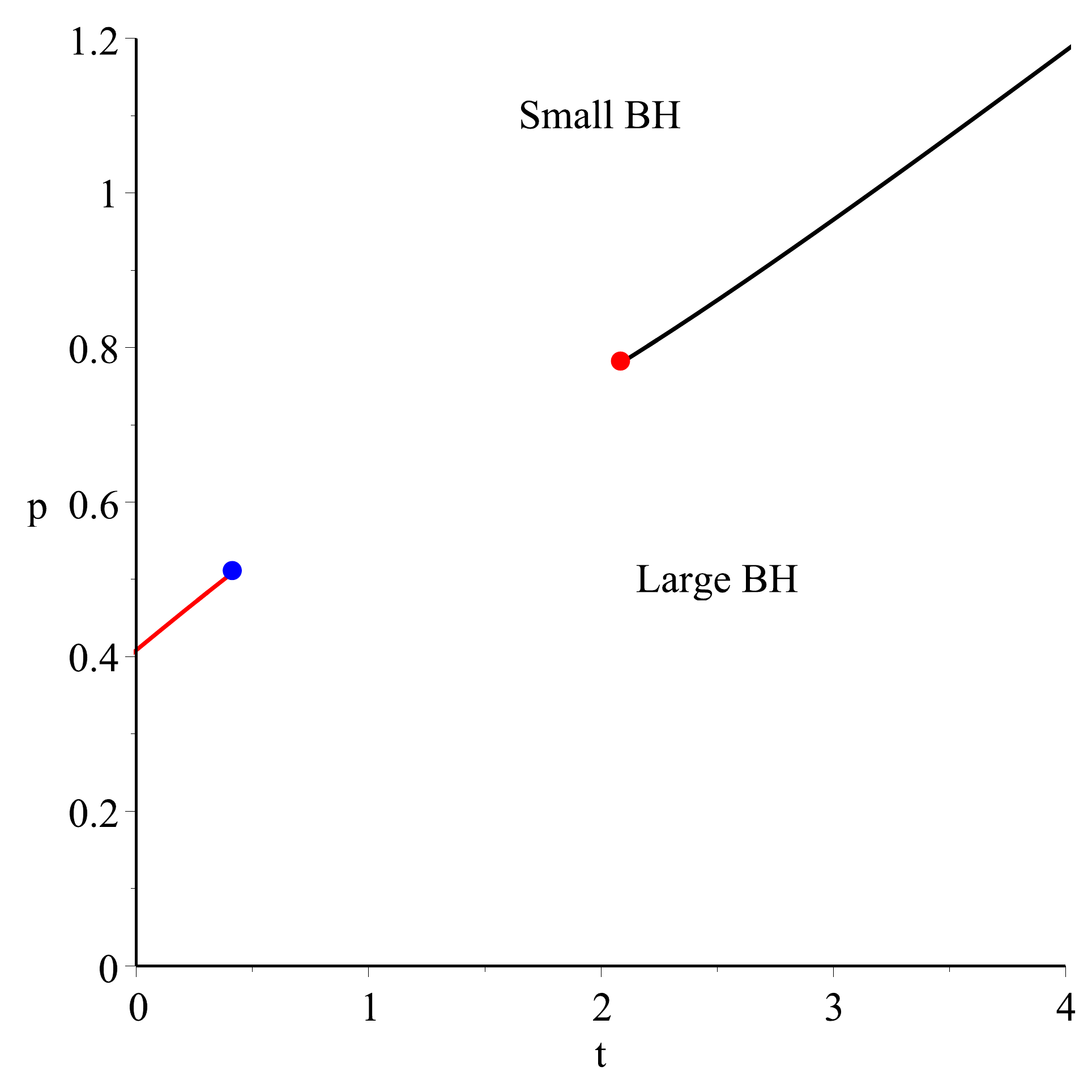}
\caption{{\bf Reverse VdW and VdW behaviour}: \textit{Left}: $p-v$ diagram for $h=1.1,q=1.0,\alpha=1.5$. Note that there are two critical isotherms (red), one corresponding to VdW type and the other corresponds to RVdW type. RVdW is characterized by VdW-type `oscillation' but at temperatures higher than critical temperature while VdW-type oscillation occurs at temperatures below critical temperature (blue). \textit{Right}: the corresponding $p-t$ diagram. The RVdW behaviour manifests itself in an `inverted' coexistence curve (shown here with a black line).} 
\label{cl_VdW_RVdW}
\end{figure}

\begin{figure}[htp]
\includegraphics[scale=0.38]{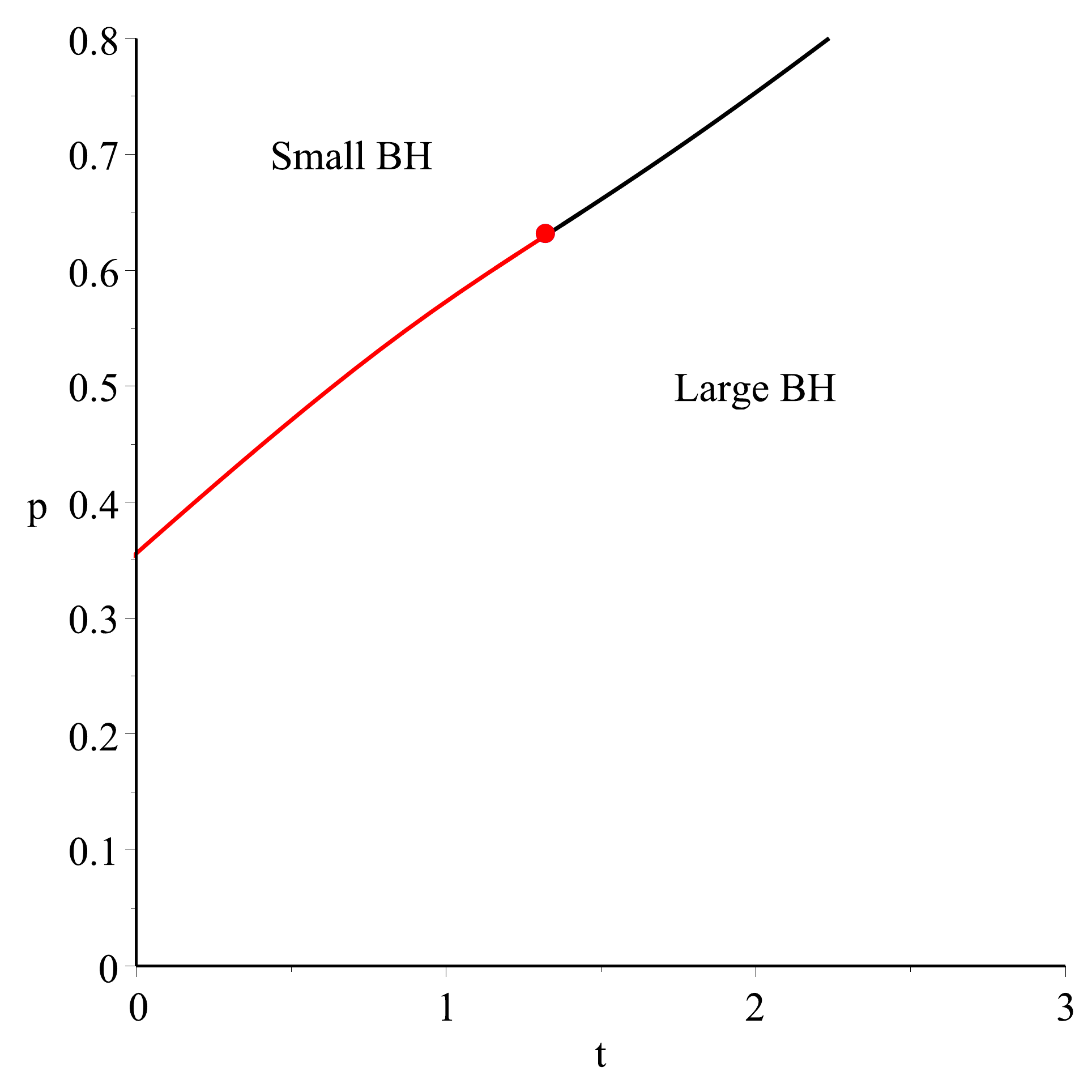}
\includegraphics[scale=0.38]{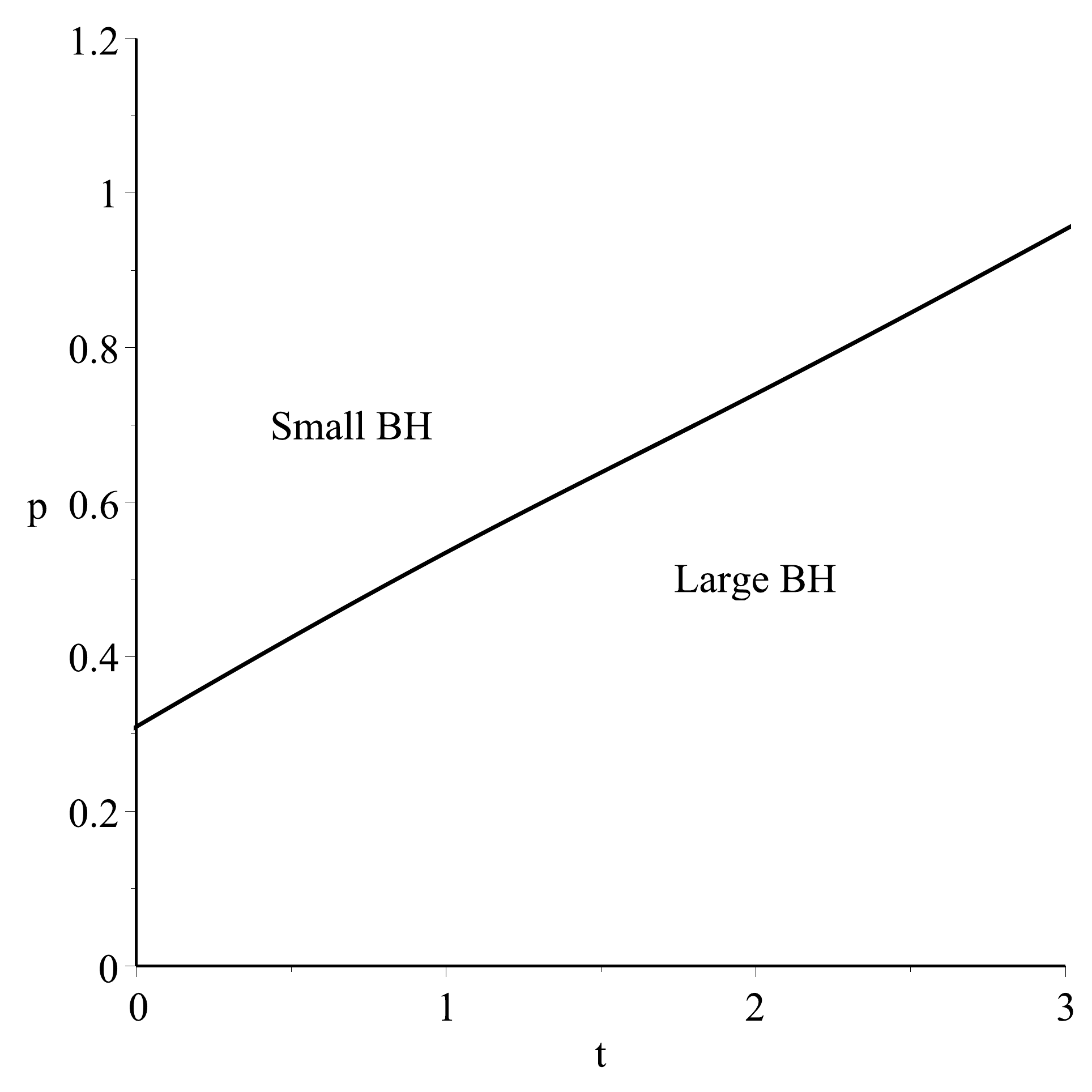}
\caption{{\bf Critical point coalescence and infinite coexistence line}: \textit{Left}: $p-t$ diagram for $h\approx 1.493070374,q=1.0,\alpha=1.5$. Both RVdW and VdW critical points coalesce to form a critical point \textit{within} the coexistence line, which is an instance of \textit{isolated critical point}. \textit{Right}: $p-t$ diagram for $h= 2.0,q=1.0,\alpha=1.5$. There is an infinite coexistence line without any critical point, which we denote as 1PT phenomenon.} 
\label{cl_coalesce}
\end{figure}

For $\sqrt{5/3}<\alpha<\sqrt{3}$ (e.g. $\alpha=1.5$), the situation is more complicated because now it is possible to have more than one critical point (the red regions in Figure~\ref{cl_qhsearch2}). From Figure~\ref{cl_qhsearch2} we see that for all $q$ there is a minimum $h$ below which all critical points have negative entropy. It has been shown in \cite{Frassino:2014pha} that for $q=h=0$ there are two critical points corresponding to VdW and RVdW behaviour with an unphysical thermodynamic singularity.  As $q$ increases (keeping $h=0$), one critical point becomes unphysical, the VdW behaviour ceases while the RVdW remains. In Figure~\ref{cl_qhsearch2} we see that this behaviour persists to some extent for nonzero $h$, though now there is a region with no critical points (black). We will use three exact sequences to represent the thermodynamic behaviour in the three different regions. We have
\begin{equation*}
\begin{aligned}
q&=0.0:\ \ \   \text{RPT}\to\text{RVdW}\to\text{[VdW \& RVdW]}\to\text{RVdW}\\
q&=0.1:\ \ \ \text{RPT}\to\text{RVdW}\to\text{[VdW \& RVdW]}\\
q&=1.0:\ \ \   \text{RPT}\to\text{RVdW}\to\text{[VdW \& RVdW]}\to\text{ICP}\to\text{1PT}
\end{aligned}
\end{equation*}
where the RPT is qualitatively similar to Figure~\ref{gb_RPT} and the square bracket implies both occur in a system with fixed $\alpha,q,h$. Figure~\ref{cl_VdW_RVdW} shows the situation where both VdW and RVdW occur in the same physical region. The 1PT phenomenon corresponds to an infinite coexistence curve where a swallowtail structure persists at all pressures without critical point. Also note that since the exact sequence for $q=1.0$ shows that VdW and RVdW are lost as $h$ increases, we expect that there is a unique $h_*$ where the transition occurs. In particular, at $h=h_*\approx 1.493070374$ the critical points \textit{coalesce} for this value of $q = 1.0$, as show in Figure~\ref{cl_coalesce}. It turns out (see Section~\ref{sec:cl_ICPs} for the explicit calculation) that this is an instance of an \textit{isolated critical point} (ICP) phenomenon which has non-standard critical exponents. 
For $h=0$ this point occurs exactly when $\alpha=\sqrt{3}$ \cite{Frassino:2014pha} and therefore coincides with the thermodynamic singularity. We see from this example that the conformal hair allows ICP to occur for a range $\alpha$; the no-hair $\alpha=\sqrt{3}$ condition is no longer required.  We will 
further elucidate the features of  this isolated critical point  later in this paper. In particular we will show that certain properties of the isolated critical points described in e.g. \cite{Dolan:2014vba, Frassino:2014pha, Hennigar:2015cja} are in fact independent of the critical points themselves.

For $\sqrt{3}<\alpha<3\sqrt{3/5}$ (e.g. $\alpha=2$), the situation is surprisingly simple. Although we can choose four different charges $q=0,0.1,1,3$ corresponding to four regions in $(q,h)$ space with different qualitative features (c.f. Figure~\ref{cl_qhsearch2}), there are only two possible exact sequences of critical behaviour:
\begin{equation*}
\begin{aligned}
q&=0:\ \ \   \text{none}\to\text{1PT}\to\text{none (cusp)}\\
q&\neq 0:\ \ \ \text{none}\to\text{1PT}\\
\end{aligned}
\end{equation*}
where the arrows once again correspond to increasing $h$. Here 1PT here refers to a first-order phase transition with an infinite coexistence line. It is remarkable that $h\neq 0$ does not add any additional features to the system, since 1PT is already a feature of $h=0$ hyperbolic black holes for $\alpha\in(\sqrt{3},3\sqrt{3/5})$.

\begin{figure}[htp]
\includegraphics[scale=0.38]{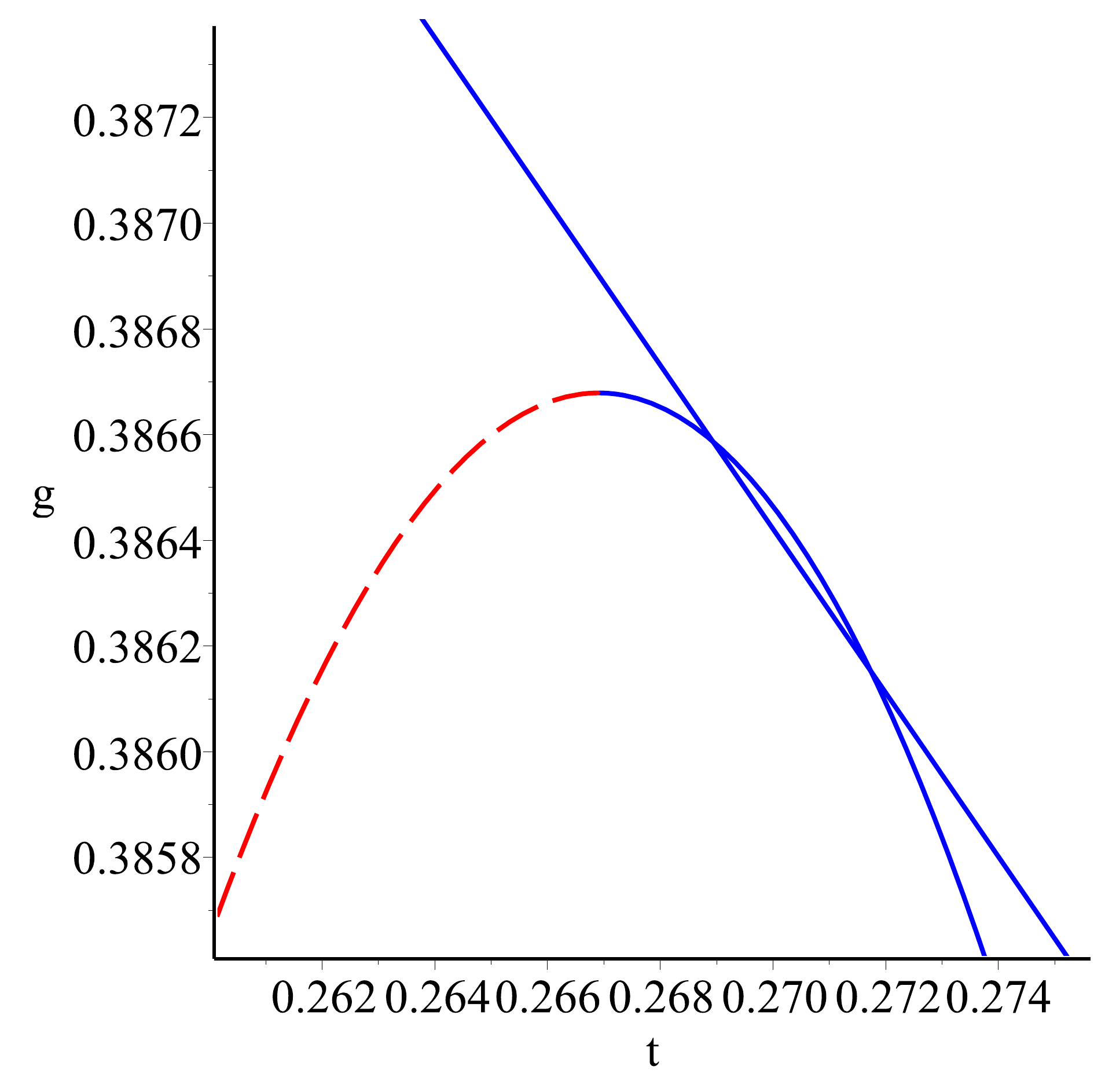}
\includegraphics[scale=0.38]{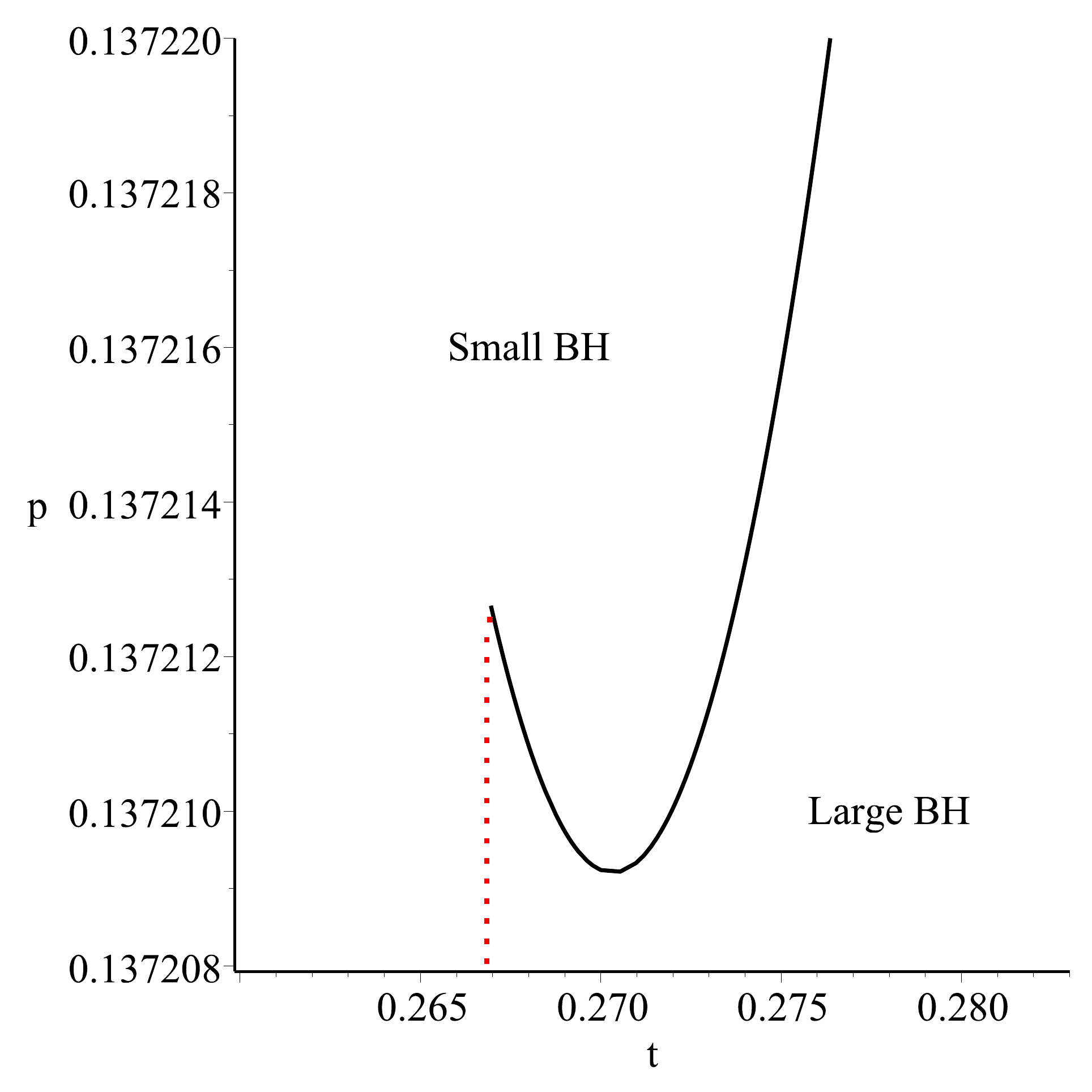}
\caption{{\bf Double reentrant phase transition for hyperbolic black holes}: \textit{Left}: $g-t$ diagram for $h=-1, q=0,\alpha=4, p\approx 0.1372098$. The presence of conformal hair does not destroy multiple RPT phenomenon found in the no-hair case. \textit{Right}: the corresponding $p-t$ diagram, showing the double reentrant phase transition from small/large/small/large black hole phase transition. In the right plot, the dotted line in red is the zeroth-order coexistence curve. } 
\label{cl_RPT}
\end{figure}

Lastly, for $\alpha>3\sqrt{3/5}$ (e.g. $\alpha=4$), we can find instances of RPT2 along with other simpler phenomena such as 1PT (with infinite coexistence line) and also purely 0PT depending on pressure, as shown in Figure~\ref{cl_RPT}. Refering to the leftmost plot: when the pressure is small, the two branches are separated and the `parabolic' part is below the straighter curve, leading to purely 0PT phenomenon. As pressure increases, the curves will eventually intersect and an RPT2 is observed. As the pressure is further increased, a point is reached where there is one intersection point, i.e. purely 1PT corresponding to an infinite coexistence line in the phase diagram. As $h$ is varied, the following sequences of critical behaviour are observed:
\begin{equation*}
\begin{aligned}
q&=0:\ \ \   \text{none}\to\text{RPT2}\to\text{1PT}\\
q&\neq 0:\ \ \ \text{none}\to\text{RPT2}
\end{aligned}
\end{equation*}
where RPT2 refers to double reentrant phase transition. Note that if at some pressure $p_0$ we find an RPT2, lowering the pressure sufficiently will lead to a 0PT,  whereas raising the pressure sufficiently will lead to a 1PT, as can be observed from the rightmost plot of Figure~\ref{cl_RPT}. 

\subsection{$P-v$ criticality in $d=8$}

In eight dimensions, the equation of state is given by
\be
p=\frac{t}{v}-\frac{15\sigma}{2\pi v^2}+\frac{2\alpha\sigma t}{v^3}-\frac{9\alpha}{2\pi v^4}+\frac{3t}{v^5}-\frac{3\sigma}{2\pi v^6}+\frac{q^2}{v^{12}}-\frac{h}{v^8}
\ee
The critical points correspondingly satisfy
\be
t_c=\frac{3}{\pi v_c(v_c^4+6\alpha\sigma v_c^2+15)}\left[3\sigma +6\alpha v_c^2+5\sigma v_c^4-\frac{4\pi q^2}{v_c^6}+\frac{8\pi h}{3v_c^2}\right] 
\ee
and
\be
\begin{aligned}
5v_c^{14}-&12\alpha\sigma v_c^{12}+6(6\alpha^2-35)v_c^{10}-36\alpha\sigma v_c^8+45 v_c^6-\\
&4\pi\sigma q^2(11v_c^4+54\alpha\sigma v_c^2+105)+\frac{8}{3}\sigma \pi v_c^4 h\left(7v_c^4+30\alpha\sigma\ v_c^2 +45\right)=0 
\end{aligned}
\ee
and we note that the presence of hair now introduces additional even powers of $v$.
In the following paragraphs we briefly comment on the thermodynamic properties of spherical and hyperbolic black holes in $d=8$.  As it turns out, there is a close connection between the results in $d=7$ and $d=8$.  We therefore simply provide a summary of the critical behaviour.

\begin{figure}[tp]
\begin{center}
\includegraphics[width=0.25\textwidth]{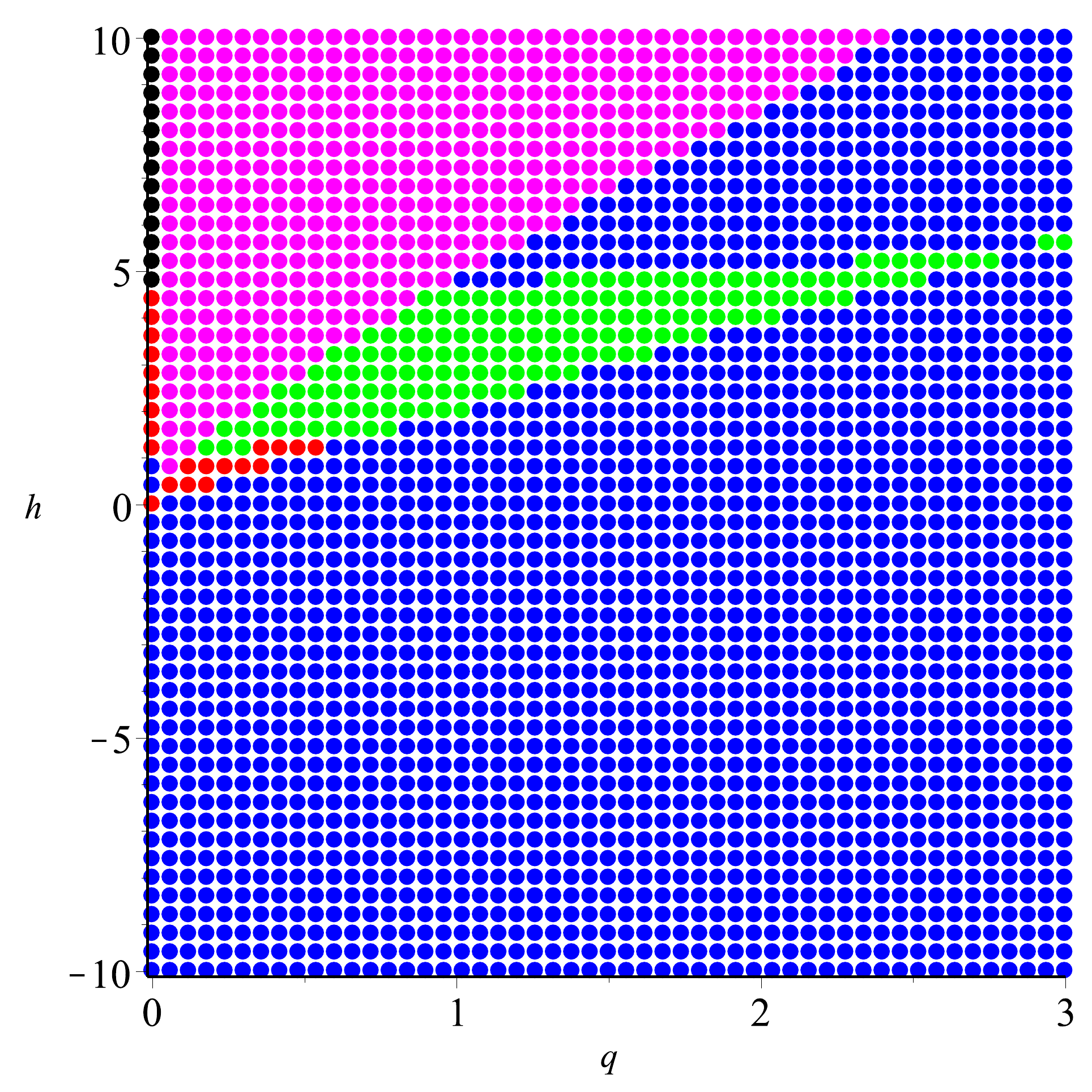}
\includegraphics[width=0.25\textwidth]{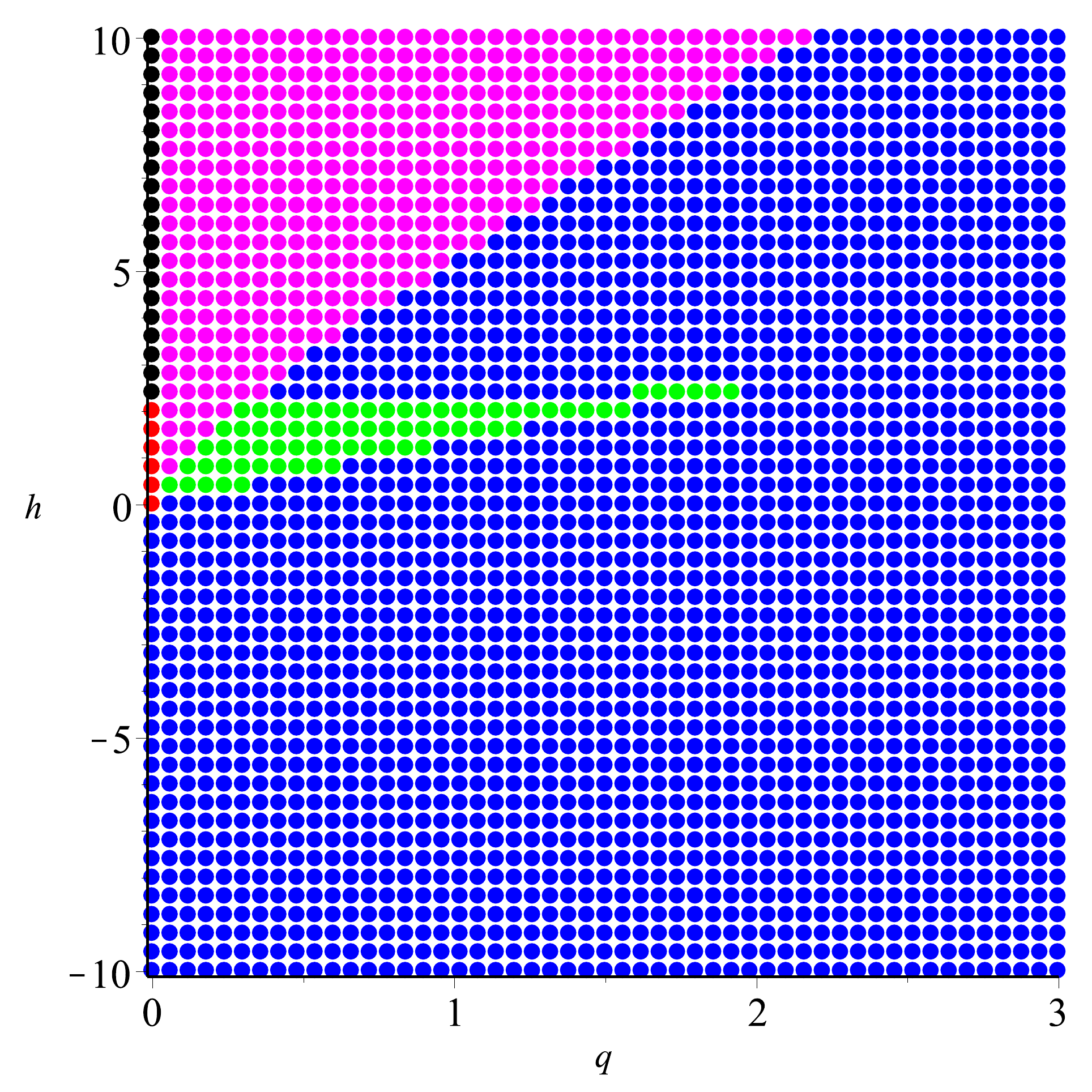}
\includegraphics[width=0.25\textwidth]{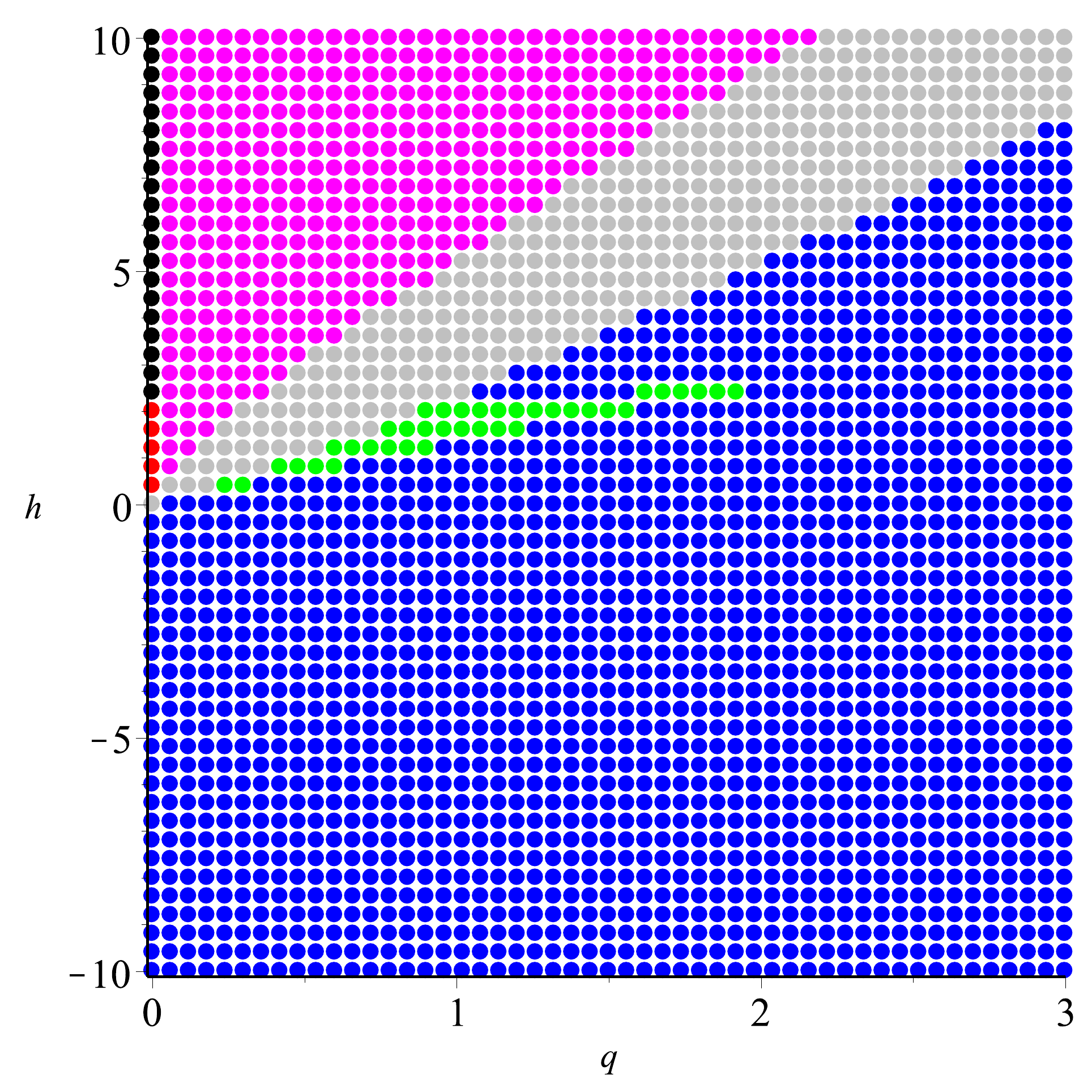}
\\
\includegraphics[width=0.25\textwidth]{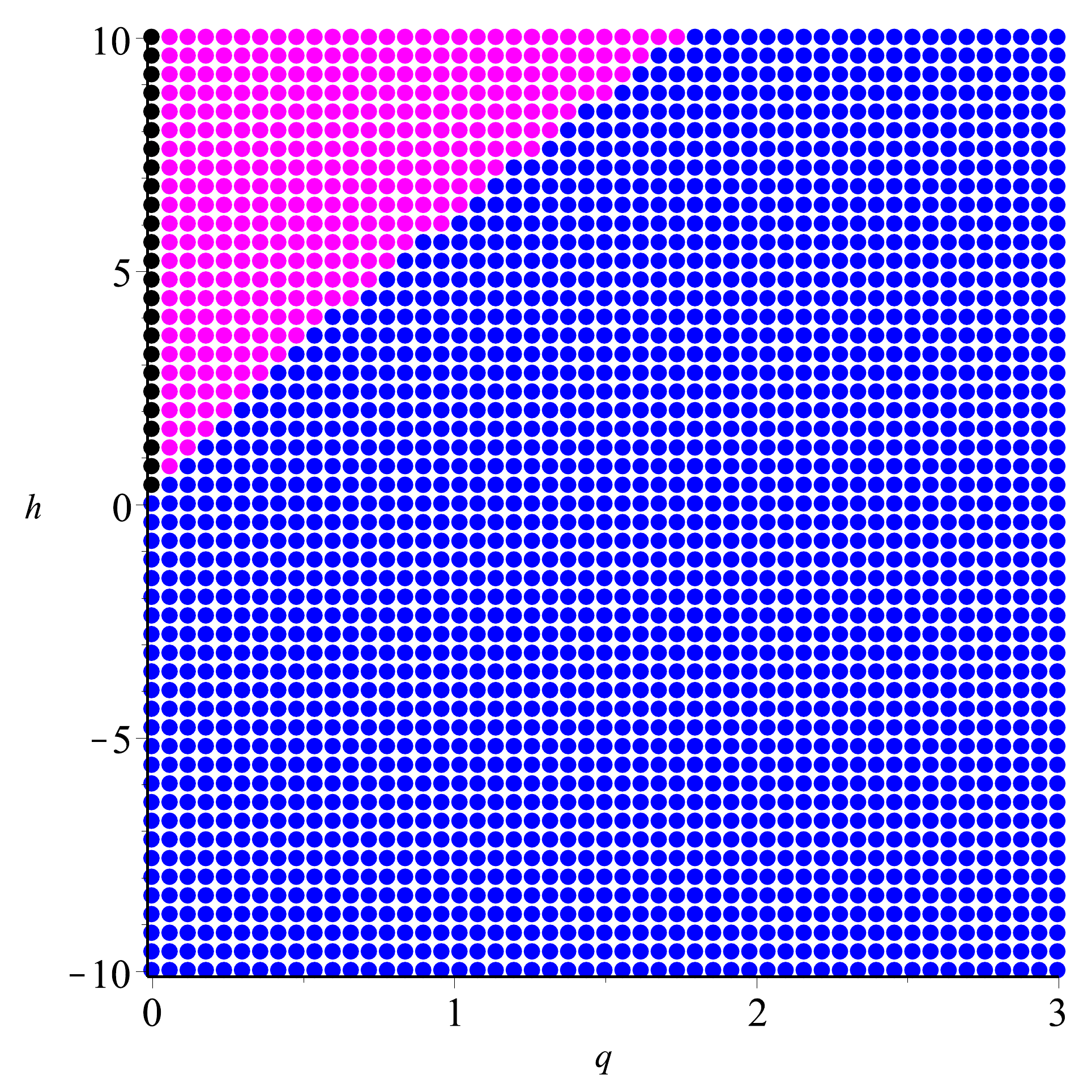}
\includegraphics[width=0.25\textwidth]{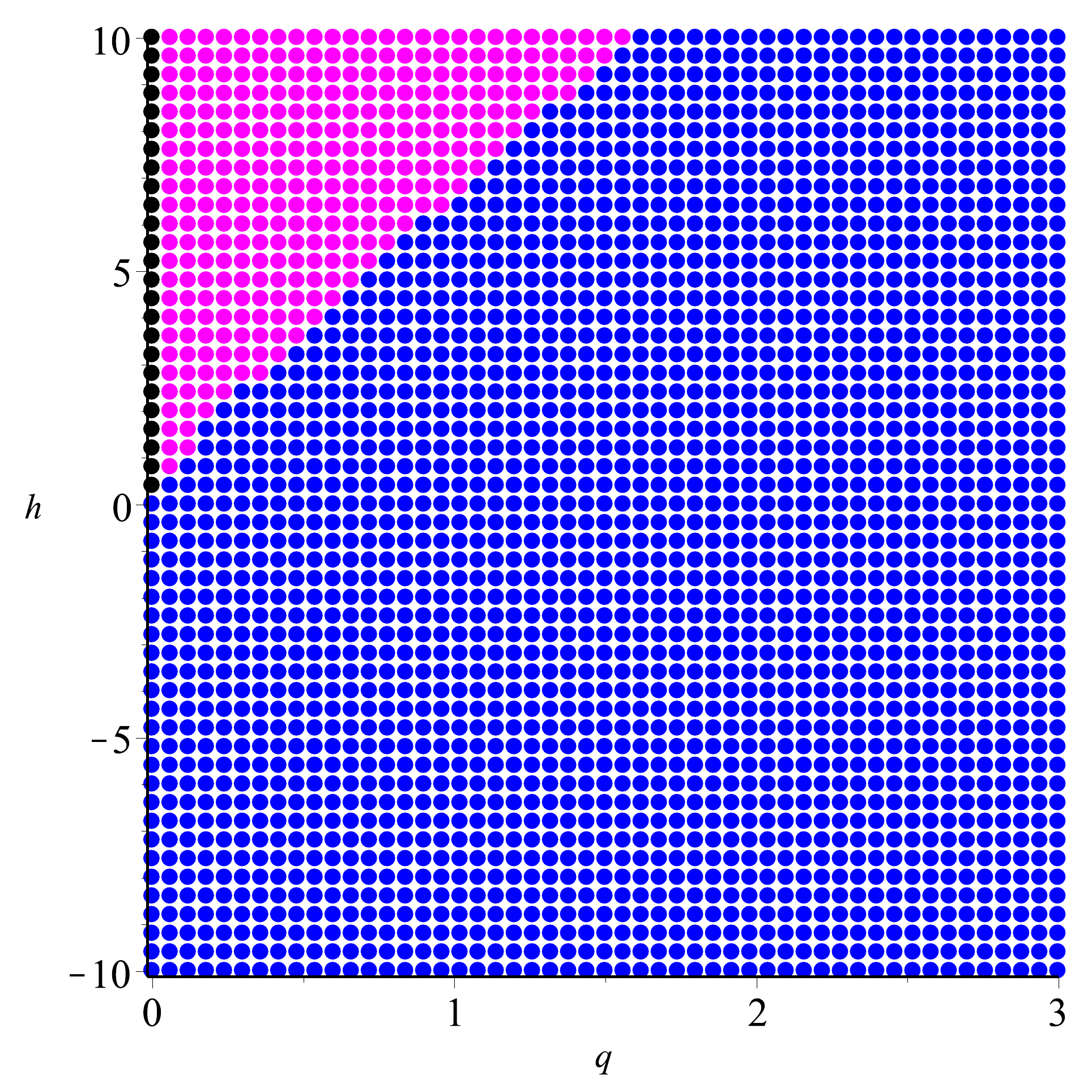}
\includegraphics[width=0.25\textwidth]{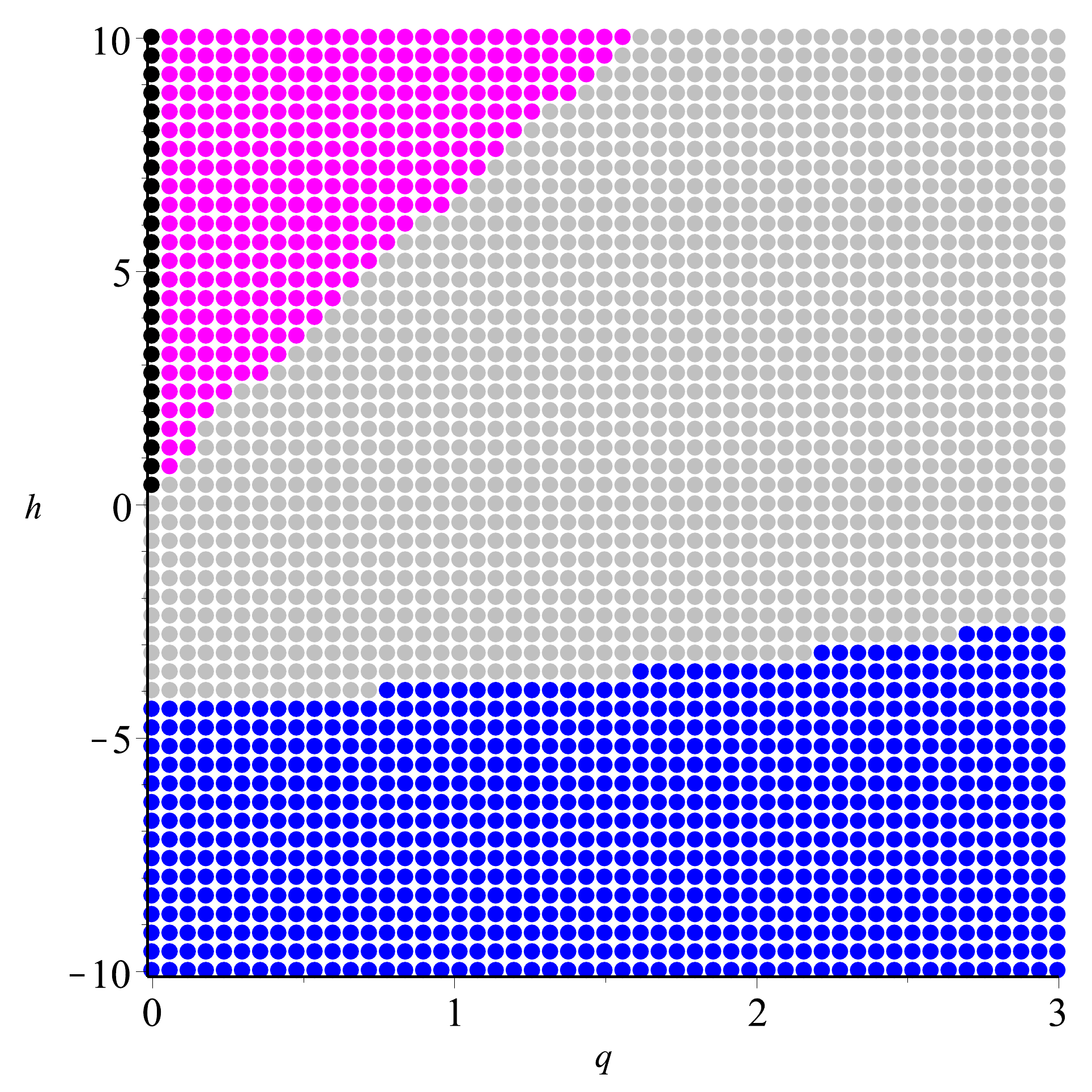}
\end{center}
\caption{{\bf The $(q,h)$ parameter space for $d=8$ and $\sigma = +1$}: The $(q,h)$ parameter space is qualitatively similar to the $d=7$ case in Figure~\ref{cl_qhsearch1}. The black, blue, red, green regions represent zero, one, two and three physical critical points respectively. The grey region represents the part of parameter space where the critical points are unphysical for the Gauss-Bonnet and Einstein branches. For Lovelock branch we simply ignore the grey region, as shown in the centre plots above.
 \textit{Top left}: $\alpha=1$. 
\textit{Top center}: $\alpha=2$, without pressure constraints valid for Lovelock branch.
\textit{Top right}: $\alpha=2$, with pressure constraints imposed. The grey region is the region where critical pressures exceed maximum pressure allowed, applicable for Gauss-Bonnet and Einstein branches.
\textit{Bottom left}: $\alpha=4.55$. 
\textit{Bottom centre}: $\alpha=6$, without pressure constraints.
\textit{Bottom right}: $\alpha=6$, with pressure constraints imposed.}
\label{cl_qhsearch3}
\end{figure}

Considering spherical black holes, in Figure~\ref{cl_qhsearch3} we plot the number of critical points in $(q, h)$ parameter space for various $\alpha$. Inspection of Figure~\ref{cl_qhsearch3} reveals obvious qualitative similarities between $d=7$ and $d=8$, and we find this to be so. That is, we find nothing qualitatively new in $d=8$, with the sequences of critical behaviour reducing to the seven dimensional case identically. For example, Table~\ref{tab:spherical_8d} provides representative parameters values for a reentrant phase transition and triple point occurs.  The associated sequence of critical behaviour which results from varying $h$ is identical to the seven dimensional case.

\begin{table}
\begin{center}
\begin{tabular}{|c|c|}
 \hline
 \multicolumn{2}{|c|}{$d=8, \sigma=1$} \\
 \hline
 Behaviour    & $(\alpha,\, q,\, h,\, p)$  \\
 \hline
 RPT          & $(1,\, 0.2,\, 3.0, \,0.129)$  \\
 \hline
 Triple point & $(1,\, 1.0,\, 3.0, \,0.134)$  \\
 \hline
\end{tabular}
\end{center}
\label{tab:spherical_8d}
\caption{{\bf Thermodynamics in $d=8$ for $\sigma = +1$}: Representative parameter values corresponding to a reentrant phase transition and a triple point.  The associated sequence of critical behaviour is no different from that reported for $d=7$.}
\end{table}

\begin{figure}[tp]
\begin{center}
\includegraphics[scale=0.25]{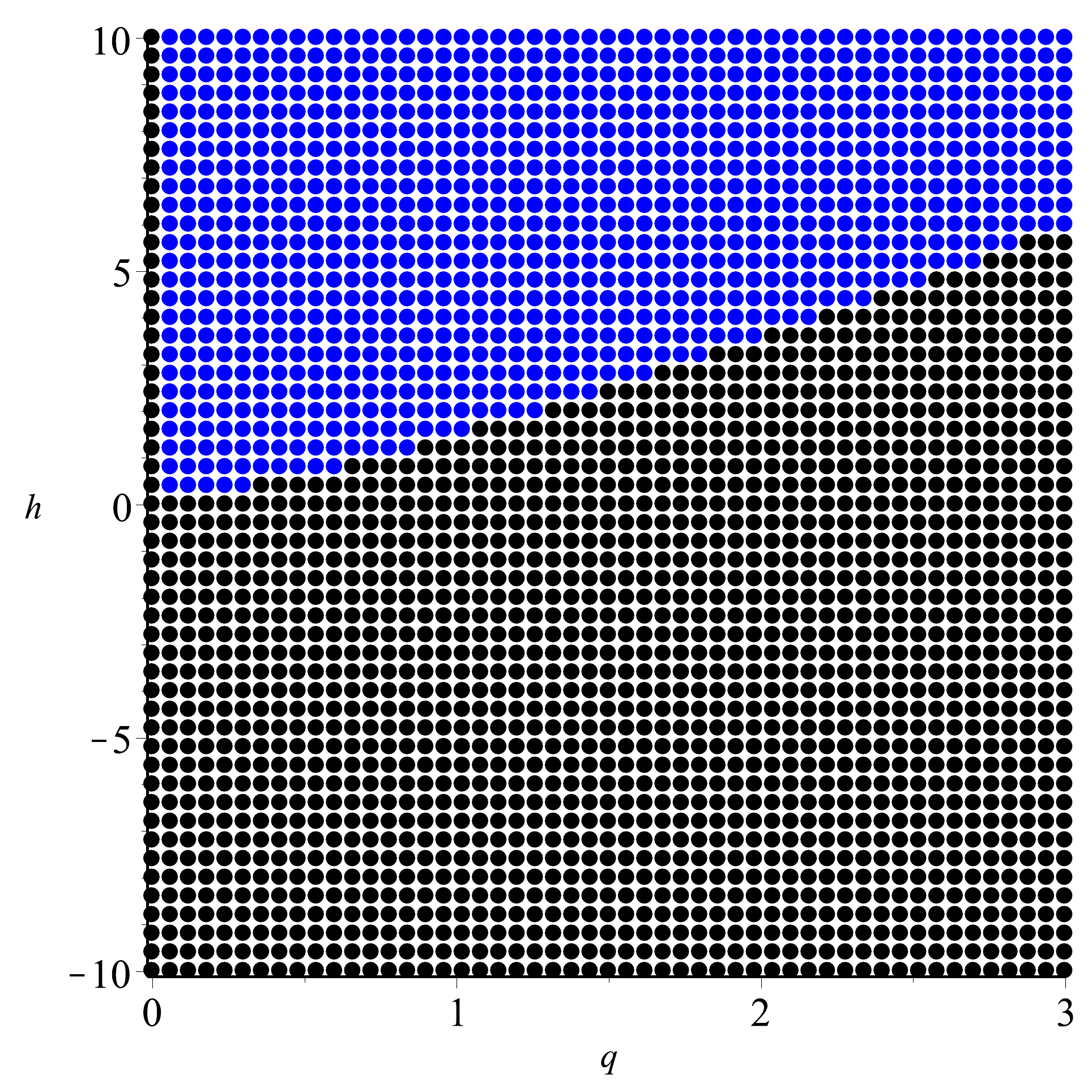}
\includegraphics[scale=0.25]{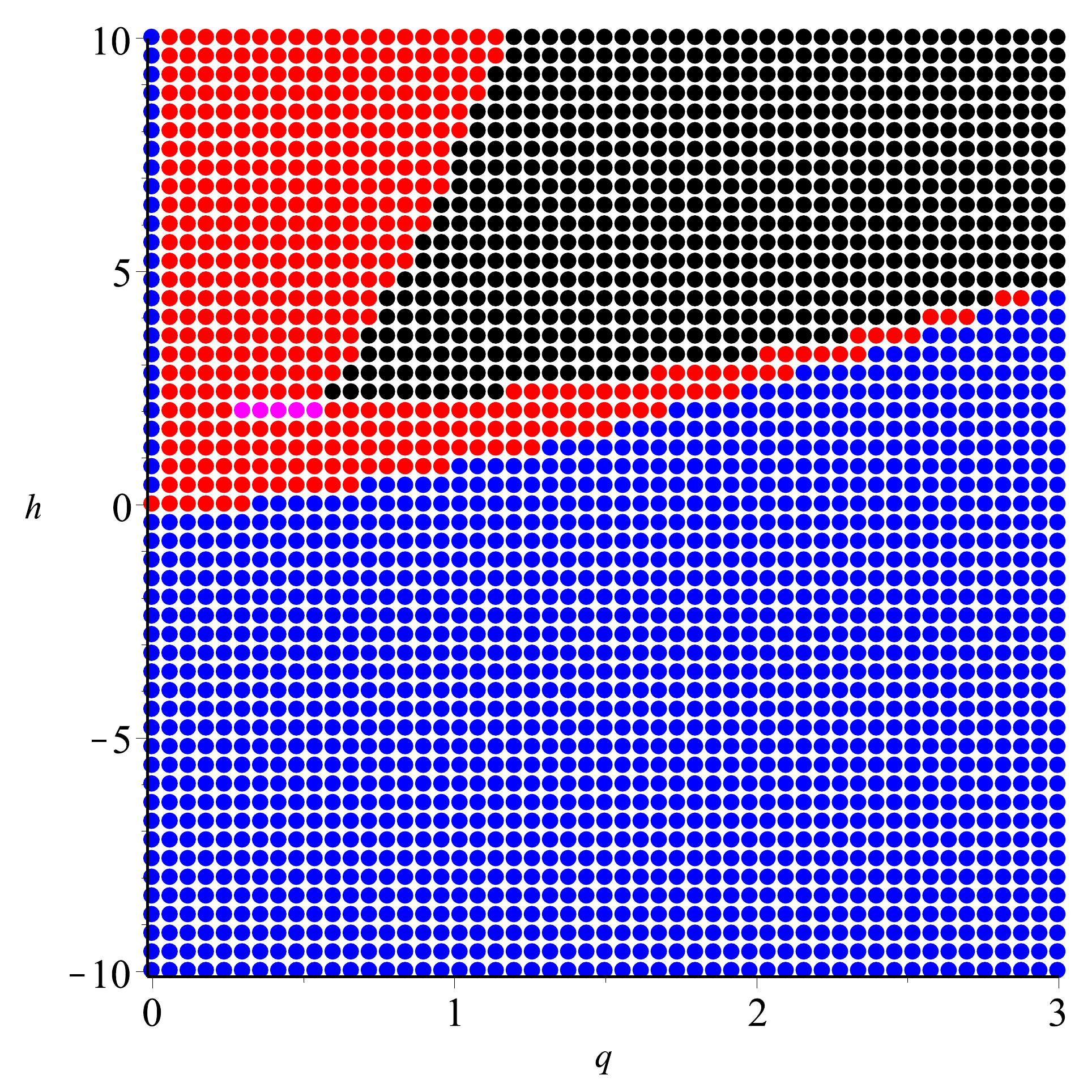}\\
\includegraphics[scale=0.25]{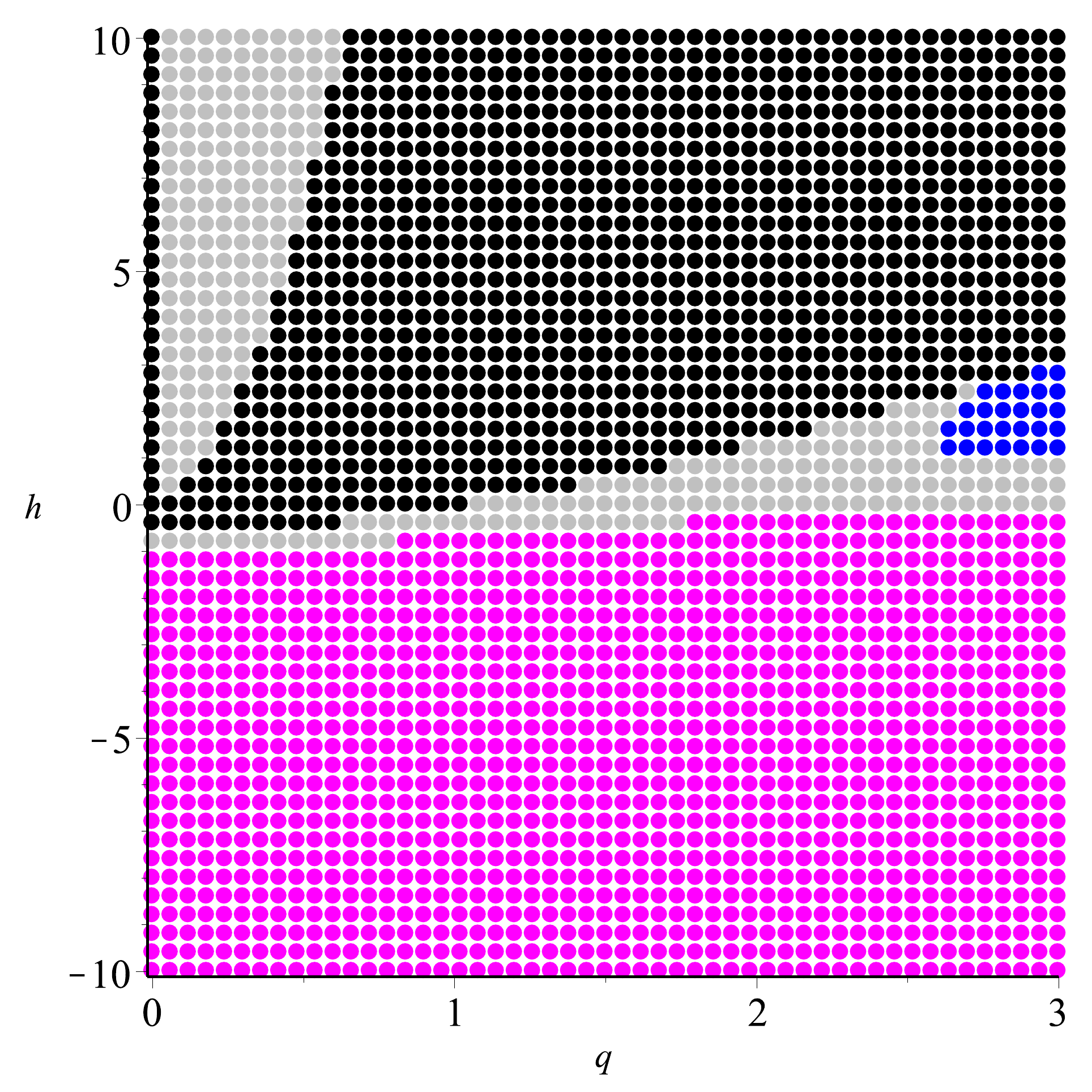}
\includegraphics[scale=0.25]{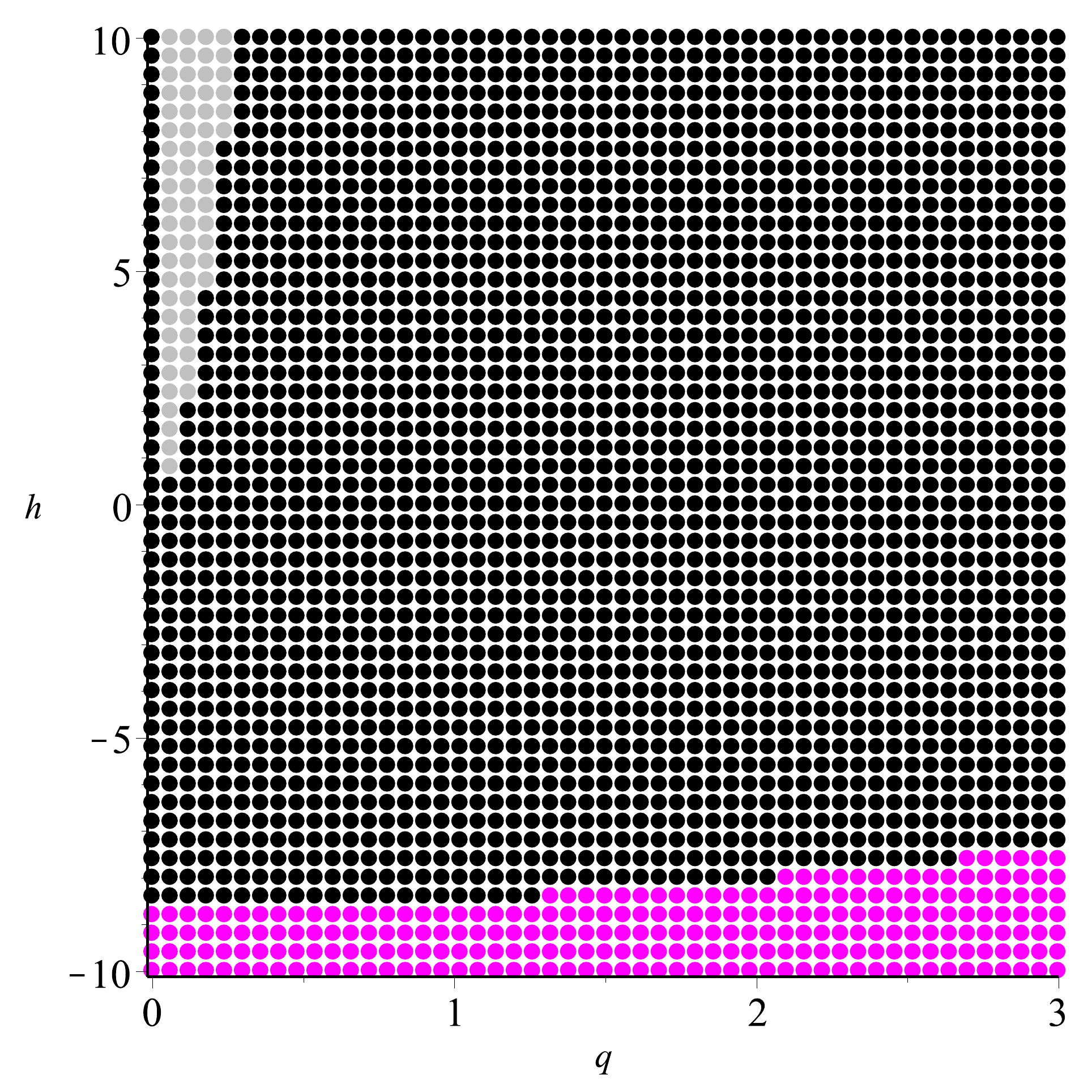}
\end{center}
\caption{{\bf The $(q,h)$ parameter space for $d=8$ and $\sigma = - 1$}: \textit{Top left}: $\alpha=1$. \textit{Top right}: $\alpha=1.5$. \textit{Bottom left}: $\alpha=2$. \textit{Bottom right}: $\alpha=4$. The black, blue, red regions represent zero, one and two physical critical points, respectively, while the grey region represents the region where critical points become unphysical for the Gauss-Bonnet and Einstein branches due to maximum/minimum pressure constraints. The grey region can be ignored for the Lovelock branch.}
\label{cl_crit4}
\end{figure}

Moving on to consider the hyperbolic case, we can see from Figure~\ref{cl_crit4} that the possible critical points and entropy conditions very closely resemble the $d=7$ case once again. Furthermore, our analysis has not revealed any additional critical behaviour different from that reported in the $d=7$ case. Therefore, we do not pursue the criticality analysis any further here.

Once again in the $d=8$ case we find instances of ICPs occurring away from thermodynamic singularities provided $\alpha \in (\sqrt{5/3},\sqrt{3})$ (ensuring the existence of two critical points).
A particular example is given by $\alpha=1.5, \, q=1, \, h=1$, which gives both VdW and RVdW behaviour similar to Figure~\ref{cl_VdW_RVdW}. As $h$ is increased, eventually the two critical points coalesce at $h\approx 2.2884146804856005889$. Performing similar computations as in $d=7$, we find that the critical exponents are given by Eq.~\eqref{cl_critexponent1}. Thus in $d=8$ these black holes belong to the same universality class near the isolated critical point.

\subsection{Summary: thermodynamics of cubic Lovelock hairy black holes}\label{thermosummary3}

Here we summarize the possible thermodynamic behaviours of cubic Lovelock hairy black holes for $d=7$. We find that $d=8$ provides no new critical phenomena, so we do not reported a detailed analysis here. For concreteness we use specific examples of $q,\alpha$ wherever appropriate to highlight representative thermodynamic behaviour.  We mark the boundary between different thermodynamic behaviour by various thresholds $h_a,\, h_b, \, h_c, \, h_d$, which are of course dependent on the values of $q$ and/or $\alpha$.  While the threshold values change with $q$ and $\alpha$, the qualitative thermodynamic behaviour is robust. In many cases we do not provide explicit threshold values as their actual values do not illuminate the physics and they depend on $q,\alpha$ directly.  Instead we focus on describing the underlying sequences of critical behaviour, and provide sample values of $h$ where a given behaviour can be observed for certain values of $q$ and $\alpha$.

We employ the same notation as in the Gauss-Bonnet case and additionally denote the isolated critical point phenomenon as ICP. We ignore the maximum pressure condition as there is always one branch that has AdS asymptotics. Lastly, for the hyperbolic case recall that entropy condition  forces black hole volumes to be bounded from above in general, and there is some maximum $h$ beyond which black holes have negative entropy for any volume. For simplicity we do not list these regions in the table below but remark that this should be taken into consideration.

%%%%%%%%%%%%%%%%%%%%%%%%%%%%%%%%%%%%%%%%% Uncharged Spherical %%%%%%%%%%%%%%%%%%%%%%%%%%%%%%%%%%%%%%%%%%%

\begin{table}[htp]
\centering
\begin{tabular}{c c c c c}
 \toprule
\cellcolor{white} Charge $(q$) \& $\alpha$ & \cellcolor{white} Hair ($h$) range & \cellcolor{white} \# Physical CP & \cellcolor{white} Example $h$ & Behaviour \\
 \midrule
 \cellcolor{white} & $h\leq 0$         & 1 & $-0.5$ & VdW \\

 \cellcolor{white} & $0<h<h_a$         & 1 & $2$    & RPT  \\
 
 \multirow{-3}{*}{\cellcolor{white} $q=0$, $\alpha=1$} & $h>h_a\sim 2.8$   & 0 & $4$    & none (cusp)\\ \midrule

\cellcolor{white} & $h\leq h_a$ & 1       &  $-0.5$ & VdW  \\

\cellcolor{white} & $h_a<h<h_b$ & up to 3 &  $1.0$  & VdW* \\

\cellcolor{white} & $h_b$       & up to 3 &  $\sim 1.0-1.5$ & VTP  \\

\cellcolor{white} & $h_b<h<h_c$ & up to 3 &  $1.5$  & TP   \\

\cellcolor{white} & $h_c$       & up to 3 &  $\sim 1.5-2.0$ & VTP  \\

\cellcolor{white} & $h_c<h<h_d$ & up to 3 &  $2.0$  & VdW* \\

\multirow{-7}{*}{\cellcolor{white} $q=0.4, \alpha=1\lesssim 2.5$} & $h>h_d$     & 0       &  $3.0$  & none \\\midrule

\cellcolor{white} & $h<0$             & 1 & $-0.5$ & VdW \\

\cellcolor{white} & $0\leq h<h_a$     & 1 & $2$    & VdW \\

 \multirow{-3}{*}{\cellcolor{white}$q=2, \alpha=2\lesssim  2.5$} & $h>h_a\sim 8.4$   & 0 & $10$   & none \\\midrule

\cellcolor{white} & $h\leq 0$            & 1 & $-0.5$ & VdW \\
\cellcolor{white} & $0<h<h_a\sim 1.5$    & 1 & $0.5$  & VdW \\

\multirow{-3}{*}{\cellcolor{white}$q=2.0, \alpha=4\gtrsim 3.8$} &  $h>h_a$              & 0 & $5$    & 1PT \\
\bottomrule
\end{tabular}
\caption{{\bf Critical behaviour for spherical black holes in $d=7$}: This table summarizes the various phase transitions that can take place for $d=7$ charged, hairy black holes with $\sigma = +1$.  Note that, due to the difficulties in characterizing a three dimensional parameter space, this table is not complete, but provides a comprehensive survey of the thermodynamic behaviour we found for these black holes.  The column ``Example $h$" provides sample values of $h$ for which this behaviour is observed for the given values of $q$ and $\alpha$. }
\label{tab:cl_7d_spherical}
\end{table}
%%%%%%%%%%%%%%%%%%%%%%%%%%%%%%%%%%%% HYPERBOLIC HORIZON %%%%%%%%%%%%%%%%%%%%%%%%%%%%%%%%%%%%%%%%%%%%%%%%%%%

\begin{landscape}
%\begin{table}
%\centering
\begin{longtable}{c c c c c}
\toprule
\cellcolor{white}Charge $(q$) \& $\alpha$ &\cellcolor{white} Hair ($h$) range & \cellcolor{white}\# Physical CP &\cellcolor{white} Example $h$ & \cellcolor{white}Behaviour \\
\midrule
\endhead 
\cellcolor{white} & $h<0$  & 0 & $-3$ & none \\

\cellcolor{white} & $h=0$  & 1 & $0$  & VdW \\

\multirow{-3}{*}{\cellcolor{white} $q=0$ $\alpha=1<\sqrt{5/3}$} & $h>0$  & 0 & $3$  & none (cusp) \\
\midrule 

\cellcolor{white} & $h<h_a\sim 3.4$ & 0 & $1$ & none \\

\multirow{-2}{*}{\cellcolor{white} $q=2$, $\alpha=1<\sqrt{5/3}$} &  $h>h_a$         & 1 & $2$ & VdW \\
\midrule

\cellcolor{white} & $h<0$         & 1 & $-5$, $-2$ & RPT, RVdW \\ 

\cellcolor{white} & $h=0$         & 2 & $0$  & VdW, RVdW \\

\multirow{-3}{*}{\cellcolor{white} $q=0, \, \alpha=1.5<\sqrt{3}$} & $h>0$         & 1 & $2$  & RVdW \\
\midrule 

\cellcolor{white} & $h<h_a\sim -7.7$         & 0 & $-10$    & none \\

\cellcolor{white} & $h_a<h<h_b\sim -0.2$     & 1 & $-5 , -2$     & RPT, RVdW \\

\multirow{-3}{*}{\cellcolor{white} $q=0.1, \, \sqrt{5/3}<\alpha=1.5<\sqrt{3}$} & $h>h_b$                  & 2 & $5.0$    & VdW, RVdW \\
\midrule

\cellcolor{white} &$h<h_a\sim -7.7$         & 0 & $-10$    & none\\

\cellcolor{white} & $h_a<h<h_b\sim 0.8$      & 1 & $-5 ,-2$     & RPT, RVdW\\

\cellcolor{white} &   $ h = 1.49307$ &      1 & $1.49307$& ICP\\

\cellcolor{white} & $h_b<h<h_c\sim 1.8$      & 2 & $1.1$    & VdW, RVdW \\

\multirow{-5}{*}{\cellcolor{white} $q=1, \sqrt{5/3}<\alpha=1.5<\sqrt{3}$} &  $h>h_c$                  & 0 & $2.0$    & 1PT\\
\midrule 

\cellcolor{white} &  $h<h_a\sim -2.3$        & 0 & $-5$ & none \\

\cellcolor{white} &  $h_a<h<h_b\sim -0.6$    & 1 & $-1$ & 1PT \\

\multirow{-3}{*}{\cellcolor{white} $q=0, \, \sqrt{3}<\alpha=2<3\sqrt{3/5}$} & $h>h_b$   & 0 & $1$  & none (cusp)\\
\midrule 

\cellcolor{white} &  $h<h_a\sim -2.3$         & 0 & $-5$   & none \\

\cellcolor{white} &  $h_a<h<h_b\sim -0.6$     & 1 & $-2$   & 1PT \\

\cellcolor{white} &  $h_b<h<h_c\sim 0.5$      & 0 & $-0.1$ & 1PT \\

\multirow{-4}{*}{\cellcolor{white} $q=0.1, \sqrt{3}<\alpha=2<3\sqrt{3/5}$} &  $h>h_c$  & 1 & $1$    & 1PT \\
\midrule 

\cellcolor{white} & $h<h_a\sim -2.5$       & 0 & $-5$  & none \\

\cellcolor{white} & $h_a<h<h_b\sim 0.1$    & 1 & $-1$  & 1PT  \\

\multirow{-3}{*}{\cellcolor{white} $q=1,\sqrt{3}<\alpha=2<3\sqrt{3/5}$} & $h>h_b$                & 0 & $1$   & 1PT \\
\midrule \pagebreak

\cellcolor{white} & $h<h_a\sim -2.8$         & 0 & $-5$ & none \\

\cellcolor{white} & $h_a<h<h_b\sim 1.9$      & 1 & $-1$ & 1PT\\

\cellcolor{white} & $h_b<h<h_c\sim 4.1$      & 1 & $2$  & 1PT \\

\multirow{-4}{*}{\cellcolor{white} $q=3,\sqrt{3}<\alpha=2<3\sqrt{3/5}$} &  $h>h_c$                  & 0 & $5$  & 1PT \\
\midrule 

\cellcolor{white} & $h<h_a\sim -2.6$         & 0 & $-5$ & none \\

\multirow{-2}{*}{\cellcolor{white} $q=0,\alpha=4>3\sqrt{3/5}$} & $h>h_a$                  & 1 & $-1$ & 0PT, 1PT, RPT2\\
\midrule 

\cellcolor{white} &  $h<h_a\sim -2.6$         & 0 & $-5$ & none \\

\cellcolor{white} &$h_a<h<h_b\sim 1.3$      & 0 & $-1$ & 0PT, 1PT, RPT2\\

\multirow{-3}{*}{\cellcolor{white} $q=0.1,\alpha=4>3\sqrt{3/5}$} &  $h>h_b$                  & 1 & $2$  & 0PT, 1PT, RPT2 \\
\midrule 

\cellcolor{white} &  $h<h_a\sim-2.5$         & 0 & $-5$ & none \\

\multirow{-2}{*}{\cellcolor{white} $q=1,\alpha=4>3\sqrt{3/5}$} & $h>h_a$                 & 0 & $-1,2$ & 0PT, 1PT, RPT2\\
\bottomrule 
\multicolumn{5}{l}{\cellcolor{white}} \\
\caption{ \cellcolor{white} {\bf Critical behaviour for hyperbolic black holes in $d=7$}: This table summarizes the various phase transitions that can take place for $d=7$ charged, hairy black holes with $\sigma = +1$.  Note that, due to the difficulties in characterizing a three dimensional parameter space, this table is not complete, but provides a comprehensive survey of the thermodynamic behaviour we found for these black holes.  The column ``Example $h$" provides sample values of $h$ for which this behaviour is observed for the given values of $q$ and $\alpha$. If more than one sample value of $h$ is provided, the sample values correspond to the different critical behaviours in the same order. }
\label{tab:cl_7d_hyperbolic}
%\end{table}
\end{longtable}
\end{landscape}
%%%%%%%%%%%%%%%%%%%%%%%%%%%%%%%%%%%%%%%%%%%%%%%%%%%%%%%%%%%%%%%%%%%%%%%%%%%%%%%%%%%%%%%%%%%%%%%%%%%%%%%%
%%%%%%%%%%%%%%%%%%%%%%%%%%%%%%%%%%%%%%%%%%%%%%%%%%%%%%%%%%%%%%%%%%%%%%%%%%%%%%%%%%%%%%%%%%%%%%%%%%%%%%%%

\section{Isolated critical points and superfluid transition for cubic Lovelock black holes} \label{sec:cl_ICPs}

Thus far we have explored various critical phenomena resulting from  conformal hair and we have seen that richer thermodynamic behaviour is possible. In this section we focus on two important novel phenomena, namely (1) \textit{thermodynamically non-singular} isolated critical points, and (2) a `superfluid transition' in hyperbolic hairy black hole solutions. Isolated critical points have been previously found in ~\cite{Dolan:2014vba, Frassino:2014pha, Hennigar:2015cja} but we shall clarify their properties in this section. The term `superfluid transition' in our black hole spacetimes will also be motivated here. 

\subsection{Isolated critical points}
In previous work focusing on Lovelock and quasi-topological gravity (without hair) examples of isolated critical points characterized by non-standard critical exponents have been found~\cite{Dolan:2014vba, Frassino:2014pha, Hennigar:2015cja}.  In these studies, the isolated critical point is an extremely special phenomenon, occuring in all cases when the critical point coincides with the thermodynamic singularity and for massless, hyperbolic black holes.  Furthermore, the coupling constants have to be finely tuned to allow the existence of these critical points, e.g. $\alpha = \sqrt{3}$ for the cubic theories.  Here we demonstrate that the thermodynamic singularity in fact has nothing to do with the existence of isolated critical points.   For simplicity, we restrict our discussion to $d=7$ cubic Lovelock gravity.

Recall from \eqref{tempcritcub} that the critical temperature is given in terms of the critical volume by the expression
\be 
t_c=\frac{10\left(\sigma v_c^8 + \alpha v_c^6 -\pi q^2\right) +7\pi h v_c^3}{\pi v_c^5(v_c^4 + 6\sigma \alpha v_c^2 +15)} \, .
\ee
and so for $\sigma=-1$  $t_c$ is ill-defined when $v_c^4 - 6 \alpha v_c^2 +15 = 0$.  For $h\neq 0$,  it is possible to tune $h$ such that the numerator and denominator both approach zero at the same rate as $v_c^4-6\alpha v_c^2 +15 \to 0$, yielding a finite value for $t_c$.  We find that for two roots of $v_c^4 - 6\alpha v_c^2 +15 = 0$, namely $v_{c,\epsilon} = \sqrt{3\alpha + \epsilon \sqrt{9 \alpha^2 - 15}}$ (with $\epsilon = \pm 1)$, yielding  a critical point with positive $v_c, t_c$ and $p_c$ for a wide range of $\alpha$ and $q$.  The precise values of $h$ required are
%which yield a finite value for $t_c$ and $v_c$ in this limit is,
\be\label{h_for_ICP} 
h_\epsilon = \frac{10}{7 \pi } \left[ \frac{225 - 945 \alpha^2 + 540 \alpha^4 - \epsilon (165 \alpha - 180 \alpha^3) \sqrt{9 \alpha^2 - 15} + \pi q^2}{\left(3 \alpha + \epsilon \sqrt{9 \alpha^2 - 15} \right)^{3/2}} \right]
\ee  
with $q$ remaining a free parameter.  It is remarkable that under these conditions, the discriminant of the polynomial determining the critical volume,
\begin{align}
v_c^{12}-3\alpha\sigma v_c^{10}+3(2\alpha^2-15)v_c^8-15\alpha\sigma v_c^6&-3\pi\sigma q^2(3v_c^4+14\sigma\alpha v_c^2+25)\nn\\
&+\left(\frac{21}{5}\pi\sigma v_c^7+\frac{84}{5}\alpha\pi v_c^5+21\pi\sigma v_c^3\right)h=0 \, ,
\end{align}
 vanishes, meaning there are in fact two critical points with $v_{c,\epsilon} = \sqrt{3\alpha + \epsilon \sqrt{9 \alpha^2 - 15}}$ for each $\epsilon$.  In other words, the situation we have outlined corresponds to the coalescence of two critical points.  Furthermore, this critical volume does not in general coincide with the thermodynamic singularity except for in the case where
\be 
\frac{\partial p}{\partial t} \Big |_{v_{c,\epsilon}=v_s} = 0 \Rightarrow v_{c,\epsilon}^4 - 2 \alpha v_{c,\epsilon}^2 + 3 = 0
\ee
which occurs only for  $\alpha \approx 1.847358636$ or $ 4.181314512$.  As a result, neither the Gibbs free energy nor the temperature is problematic near the critical point and there is no pathological behaviour (e.g. curvature singularities) except   at these two values of $\alpha$. For these two values of $\alpha$ the methods used in \cite{Frassino:2014pha} (cf. Sections. 3.3.2 and 4.5) can be employed to make sense of the thermodynamics.

\begin{figure}[htp]
\includegraphics[scale=0.38]{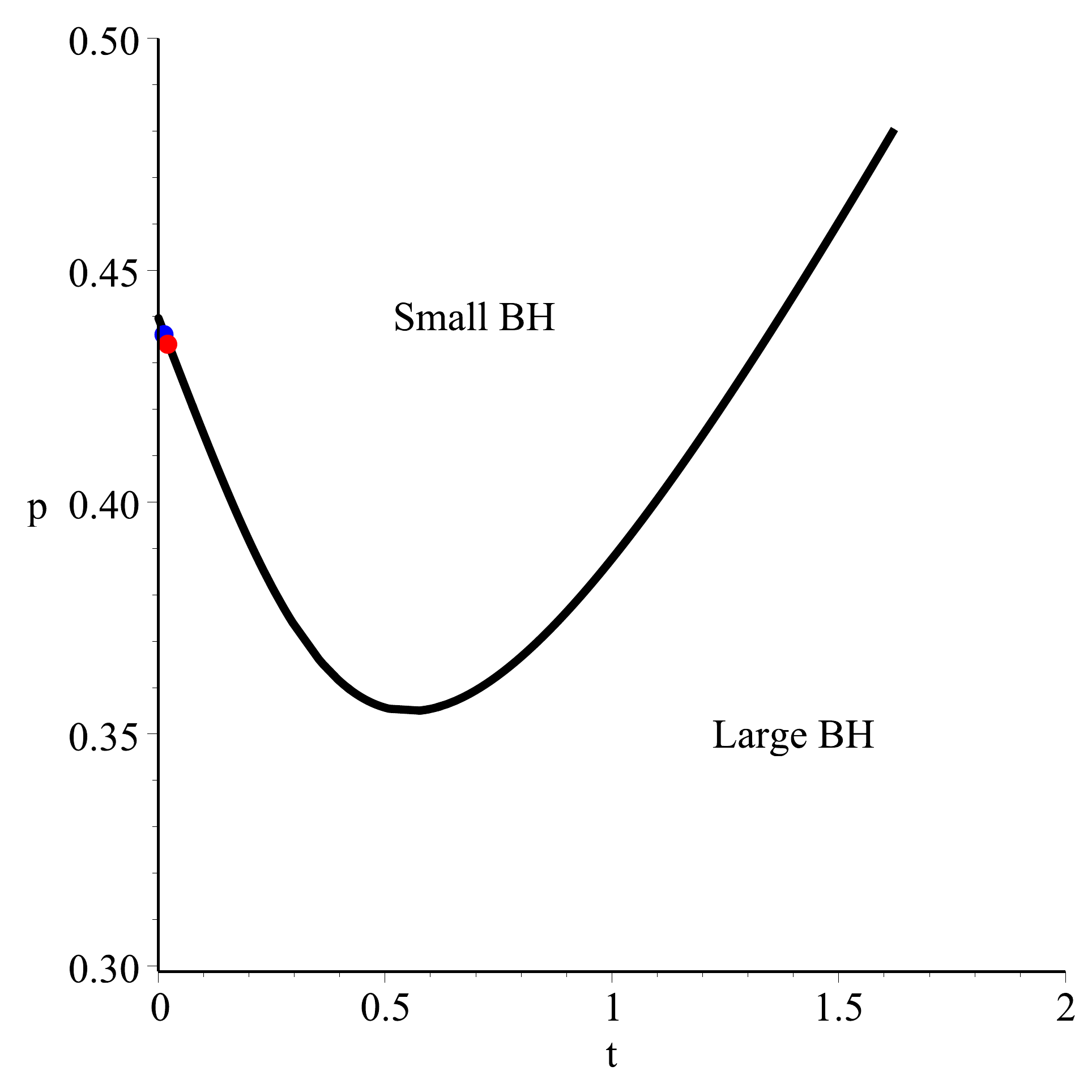}
\includegraphics[scale=0.38]{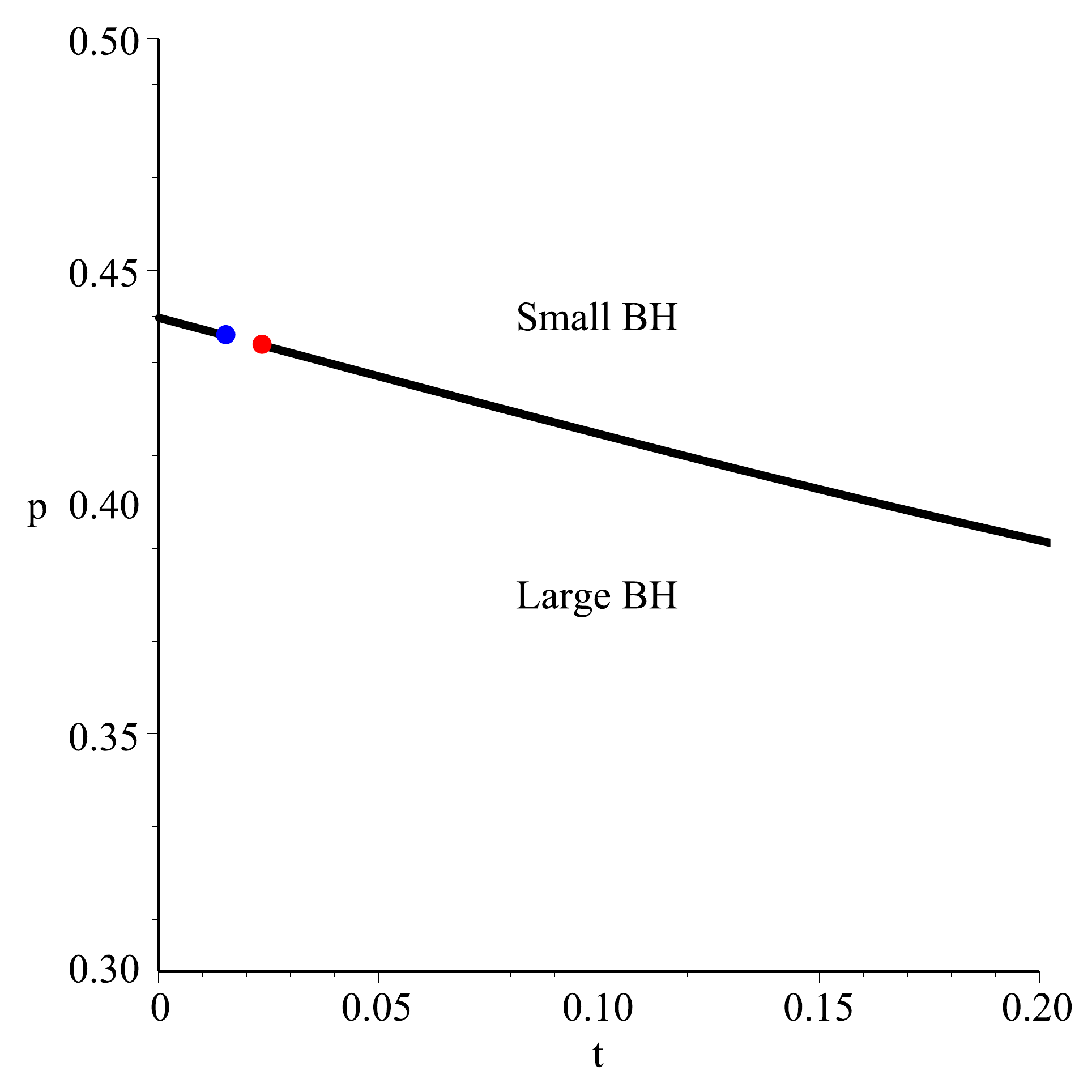}
\includegraphics[scale=0.38]{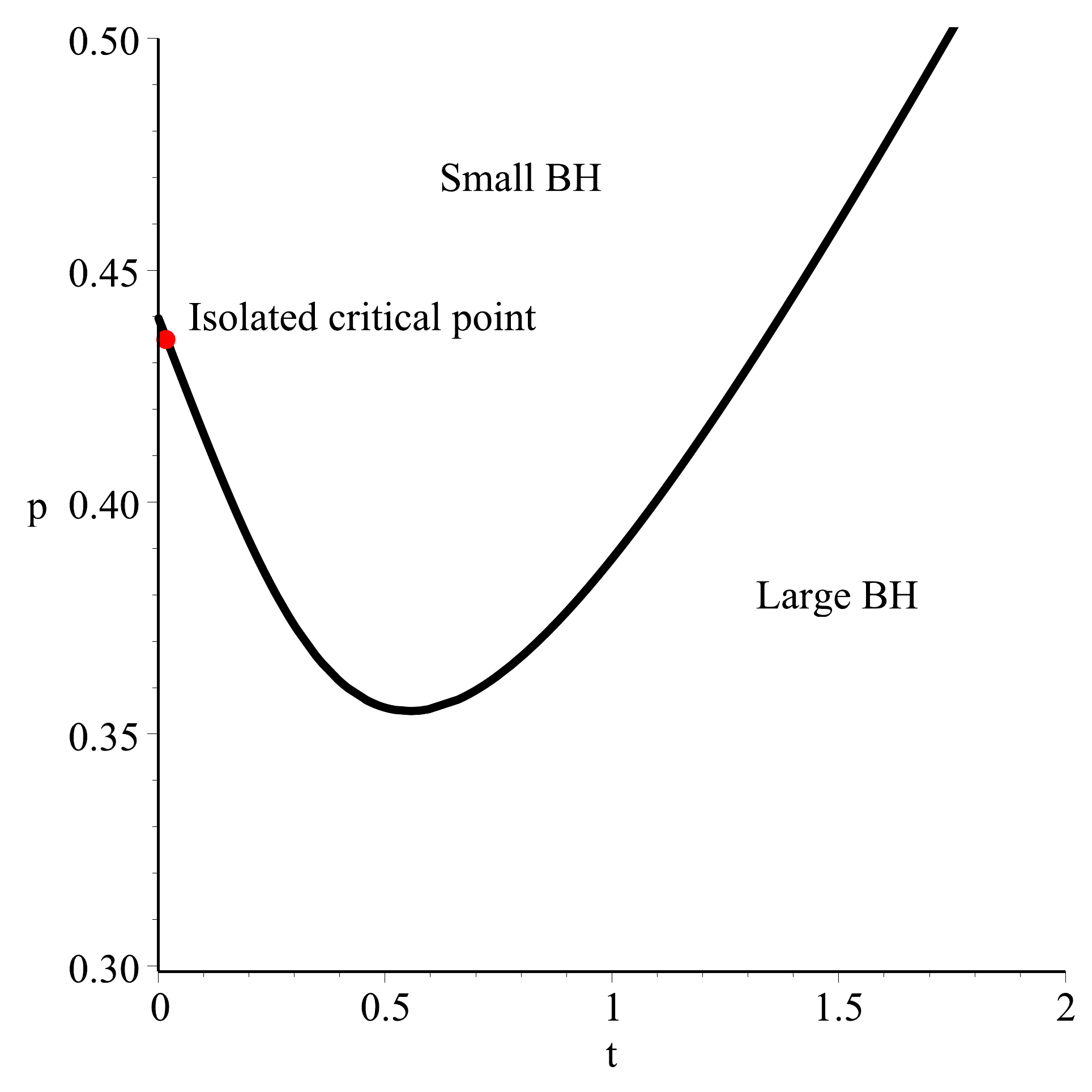}
\includegraphics[scale=0.38]{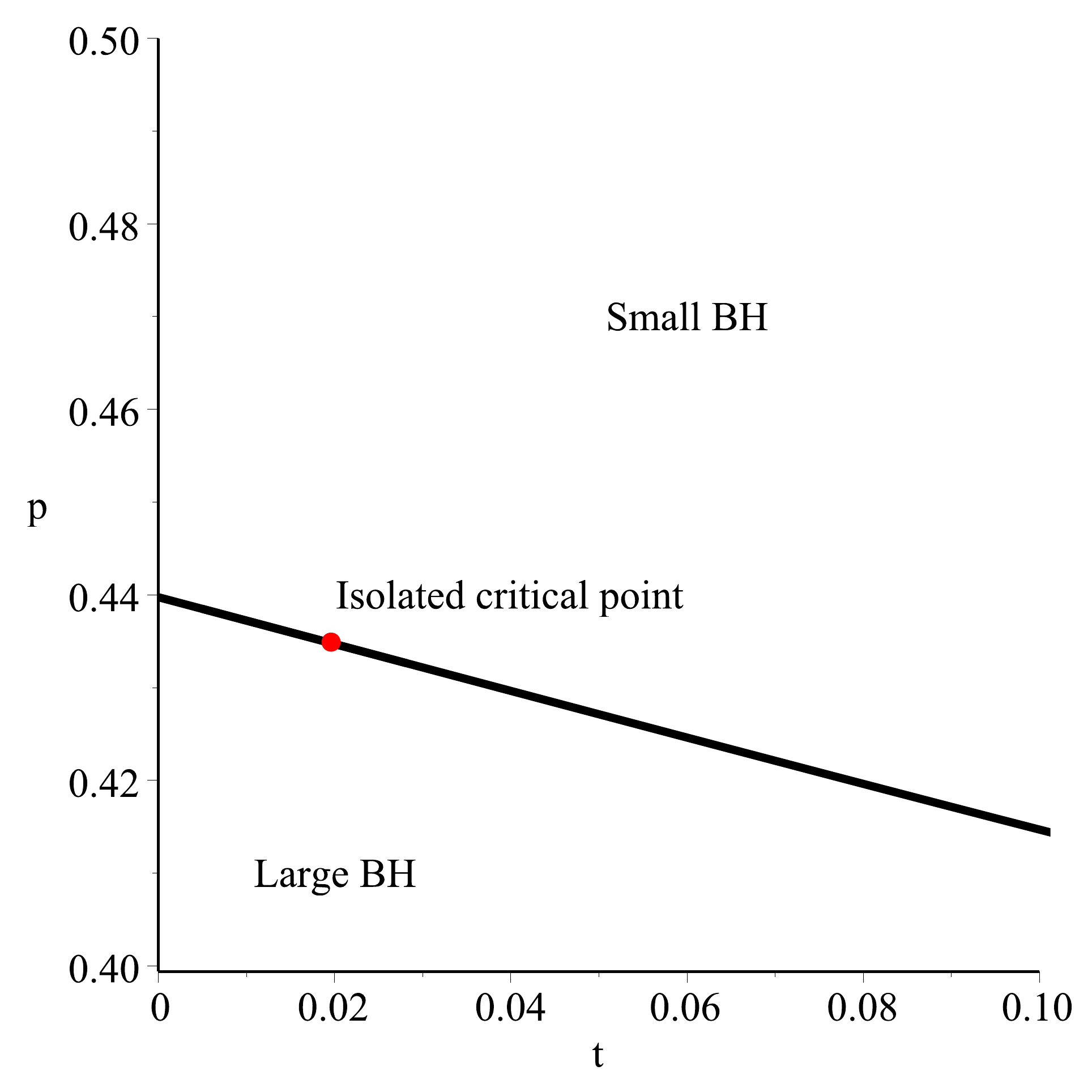}
\caption{ {\bf Isolated critical point for hyperbolic black hole in $d=7$}: the plots correspond to $\alpha=1.98, q=0.59$. \textit{Top left}: $p-t$ diagram when $h=-0.1367<h_0$, where $h_0$ is the value of $h$ when coalescence occurs. Observe that there is RPT phenomenon for this choice of parameter. \textit{Top right}: the zoomed-in version of the top-left diagram, showing the separation between the two critical points which will coalesce at larger $h$. \textit{Bottom left}: the case for $h\approx -0.13668533976582815376$. For this value, we observe coalescence of the two critical points and form an isolated critical point. Note that there is still RPT phenomenon in this phase diagram. \textit{Bottom right}: the zoomed-in version, showing the coalescence. For this choice of parameters the hyperbolic black hole at isolated critical point has a positive (dimensionless) mass $m\approx 0.713$.}
\label{cl_RPT_ICP}
\end{figure}

It is straightforward to expand the equation of state near this critical point.  Although the resulting expression for general $\alpha$ is too cumbersome to present here,  the expression takes the general form
\be\label{p-isocrit}
\frac{p}{p_c} = 1 + A \tau + B \omega^2 \tau + C \omega^3 + \cdots \, ,
\ee
where $A,B,C$ are complicated $\alpha$-dependent coefficients that can be computed exactly.  The most significant feature of this expression is that the $\omega \tau$ term vanishes identically.  It follows from this expansion that the critical exponents for this critical point are
\be\label{cl_critexponent1}
\alpha=0,\, \, \beta=1,\, \, \gamma=2, \, \, \delta=3
\ee
which are the same as those found for the isolated critical point in~\cite{Frassino:2014pha}.  Here the critical point does not occur at the thermodynamic singularity except for the two specific values of $\alpha$ given above, and in these two cases the critical exponents are the same as in \eqref{cl_critexponent1}.

The behaviour in the $p-t$ plane is shown in Figure~\ref{cl_RPT_ICP} for a representative ICP.  We initially see two distinct critical points which then merge as $h$ approaches the special value given by Eq.~\eqref{h_for_ICP}.  From these plots it is abundantly clear that these ICPs are not occurring at the thermodynamic singularity, since it is obviously the case that $\partial p /\partial t \big|_{v=v_c} \neq 0$.  This suggests that all that is required to have an isolated critical point is to have two (or more) critical points coincide.  Here we find that this can occur when both VdW and RVdW behaviour are manifest physically (i.e. respect all of the relevant physicality constraints)---the two critical points can in general be made to coincide through a tuning of $h$.

\begin{figure}[htp]
\centering 
\includegraphics[width=0.4\textwidth]{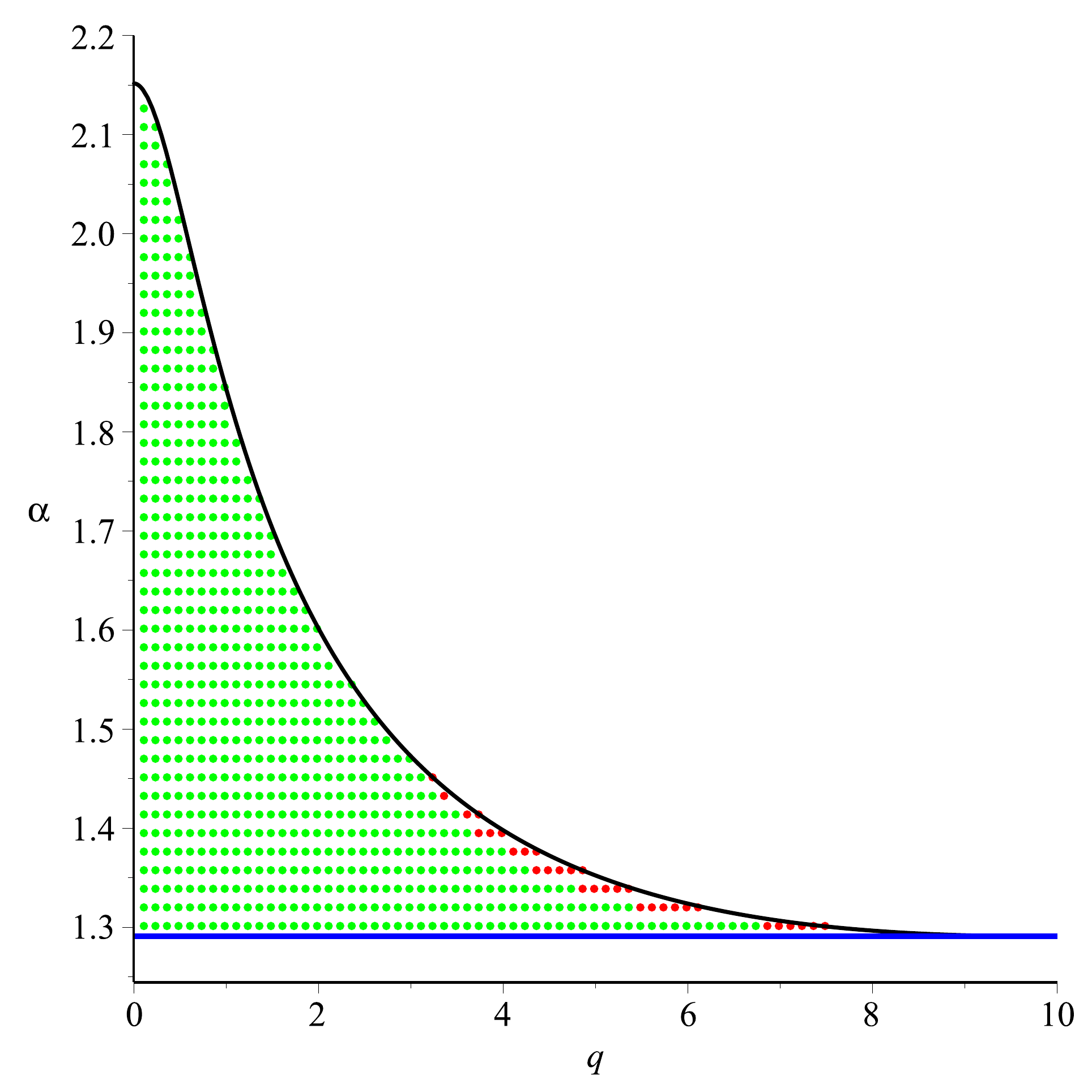}
\includegraphics[width=0.4\textwidth]{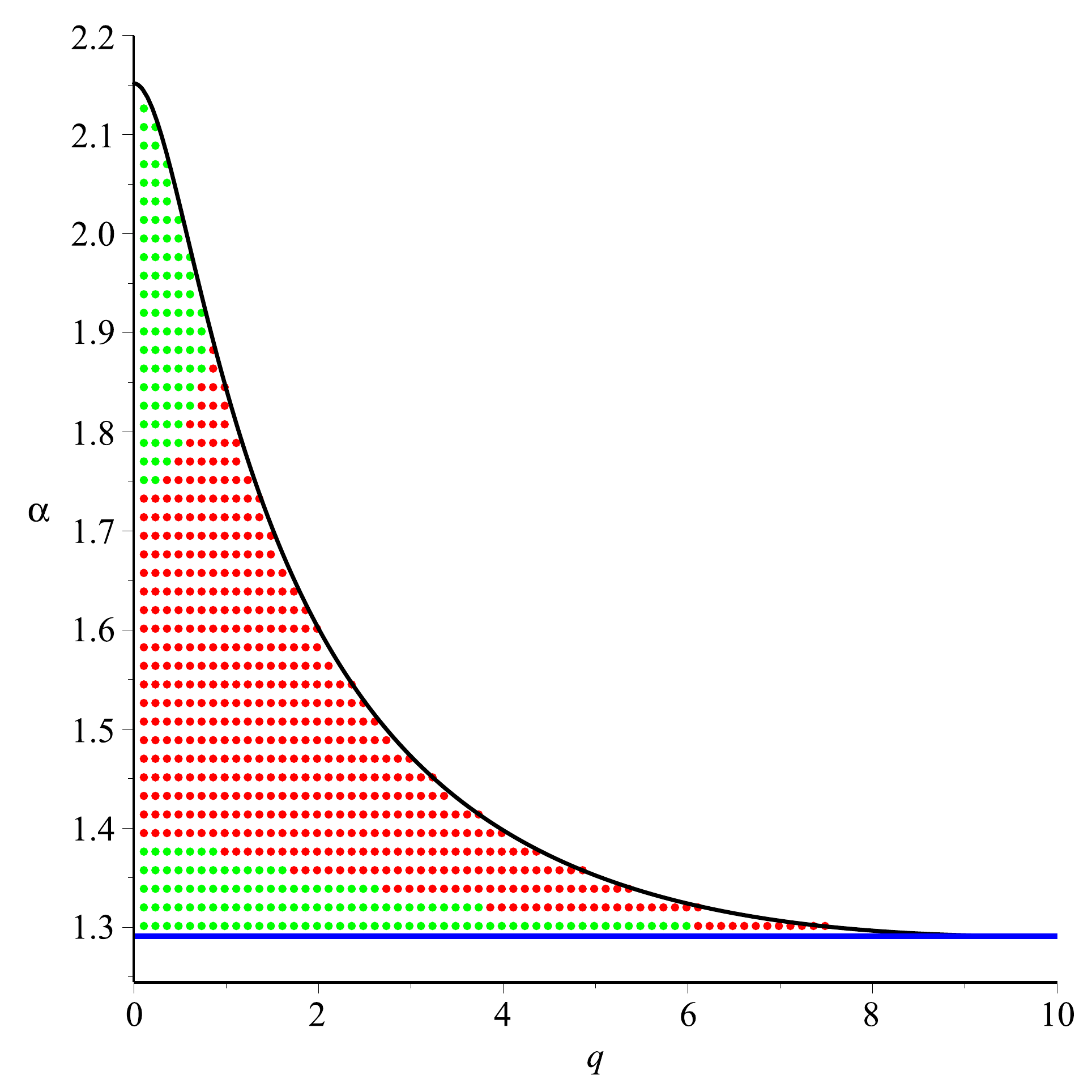}
\caption{{\bf Sign of mass at isolated critical points}: These plots indicate whether the mass is positive (green) or negative (red) at the isolated critical point for $\epsilon=+1$ (left) and $\epsilon = -1$ (right).  The solid blue line corresponds to $\alpha = \sqrt{5/3}$, while the solid black line corresponds to $\alpha$ as given in Eq.~\eqref{alpha_in_terms_of_q}.  This plot was produced for $d=7$ hyperbolic black holes.}
\label{icp_sign_of_mass}
\end{figure}

 Note that in the previous work on isolated critical points, the black holes possessing these non-standard critical exponents are all massless---here this is not the case. The dimensionless mass $m$ for $d=7$ can be easily computed to be
\be 
m = \frac{1}{120v_{c,\epsilon}}\left[\frac{24\pi q^2}{v_{c,\epsilon}^3}-120v_{c,\epsilon}+120\alpha v_{c,\epsilon}^3-120v_{c,\epsilon}^5-16\pi(6h-pv_{c,\epsilon}^7)\right]
\ee
and can be both positive and negative for the critical points we have discussed, as shown in Figure~\ref{icp_sign_of_mass}.  The negative mass solutions are not an issue here since in the case $\sigma = -1$ sensible negative mass black holes exist, provided appropriate identifications are made~\cite{Mann:1997jb}.   This demonstrates that vanishing mass is not a generic feature of black hole systems with isolated critical points. 
 
\subsection{Superfluid-like lambda transition} 

One might be interested in the expansion Eq.~\eqref{p-isocrit} for the cases $B=0$ or $C=0$, since then the critical exponents $\gamma$ and $\delta$ would differ from those presented above in Eq.~\eqref{cl_critexponent1}. While we have not been able to determine any interesting behaviour for $C=0$, we find that  $B=0$ occurs for $\alpha = \sqrt{5/3}$ or
\be\label{alpha_in_terms_of_q} 
\alpha = \frac{\sqrt{3}}{36 \sqrt{\pi} q} \sqrt{ 370 \pi q^2 - 3375 + \sqrt{(24 \pi q^2 + 25)(2 \pi q^2 + 675)^2}} \, .
\ee
The critical temperature vanishes if $\alpha$ takes on the value given in \eqref{alpha_in_terms_of_q}, so we do not consider this case any further.  

\begin{figure}[htp]
\includegraphics[scale=0.38]{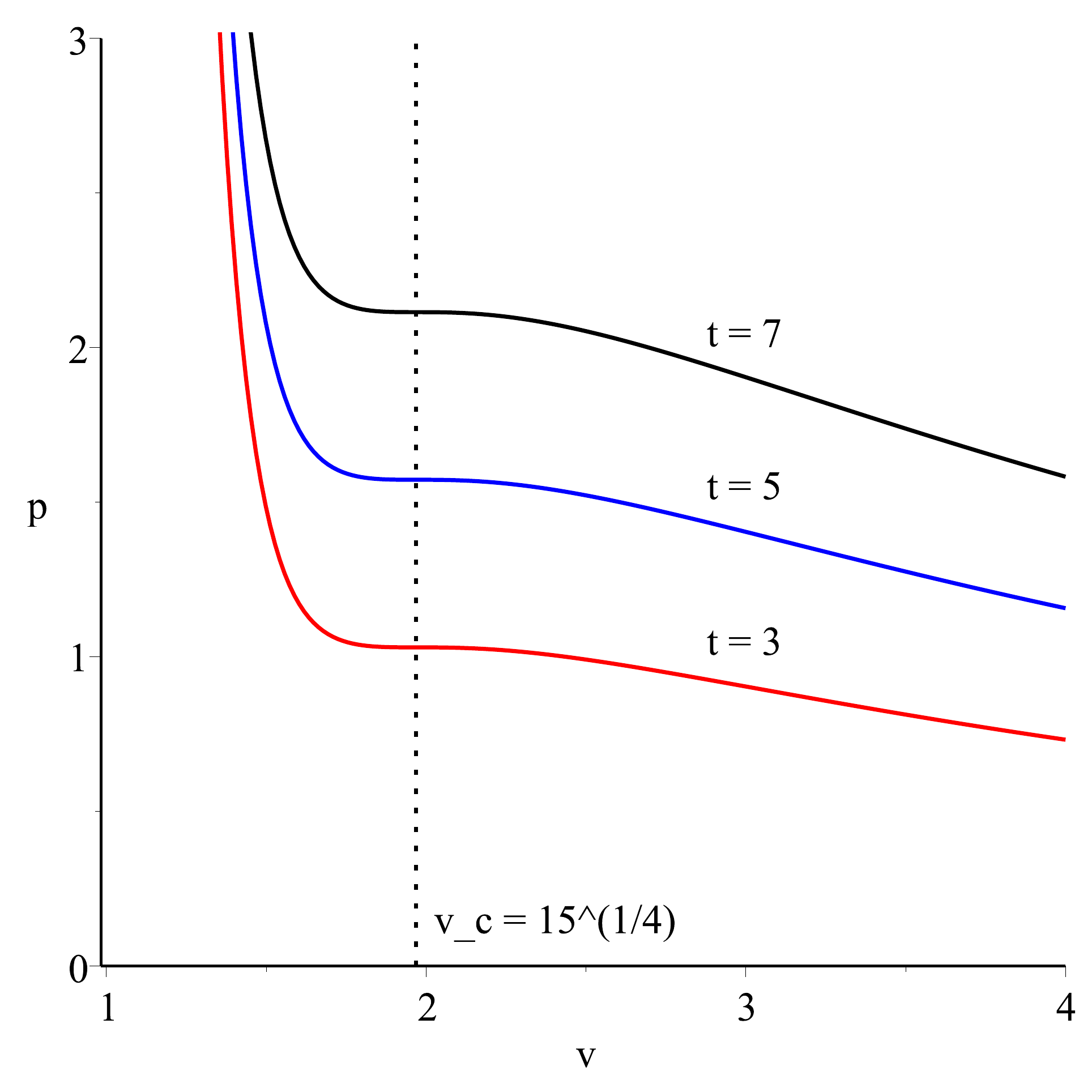}
\includegraphics[scale=0.38]{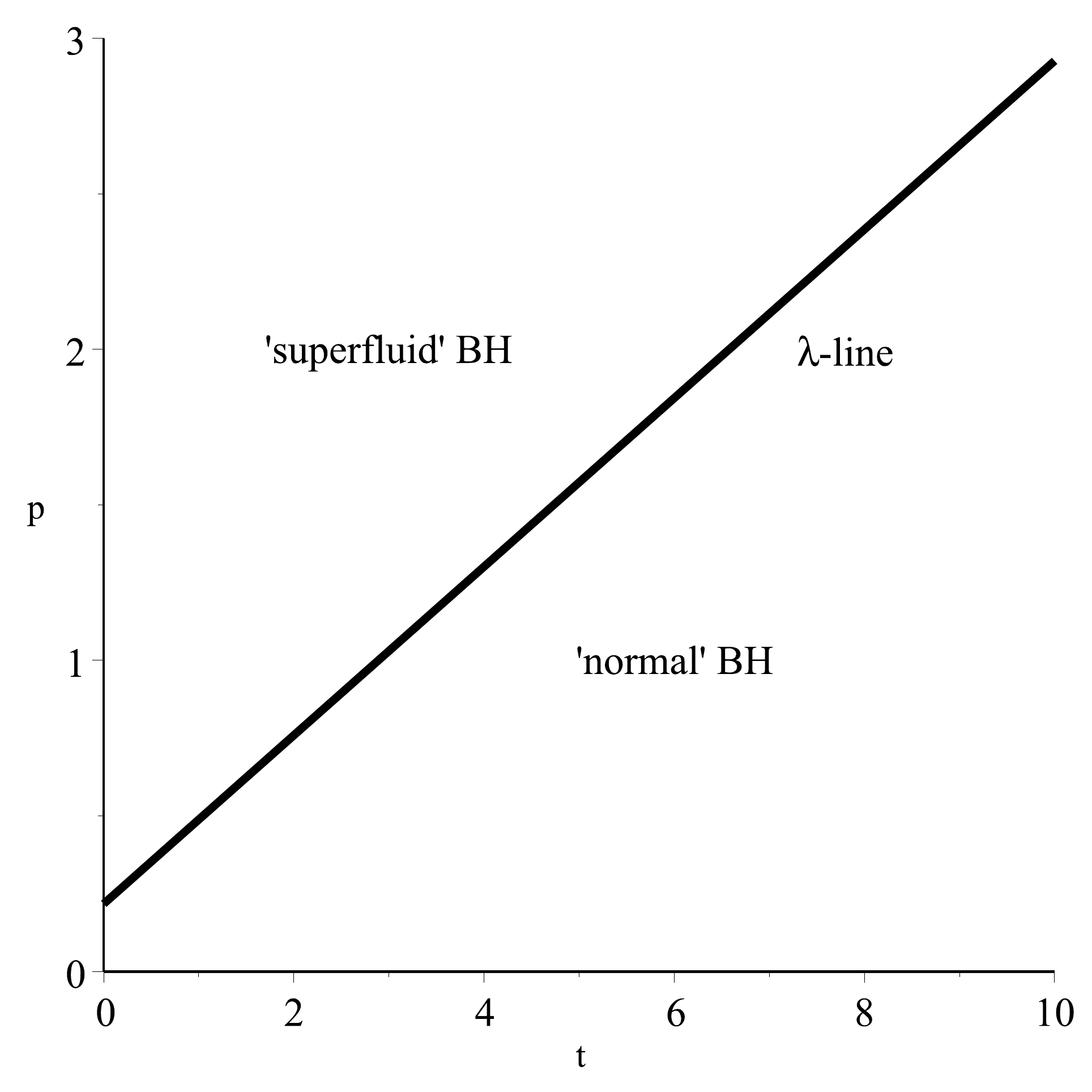}
\caption{\textbf{Infinite number of critical points for $d=7$ hyperbolic black holes}: \textit{Left}: the $p-v$ diagram for various temperatures. Note that all the isotherms here are critical isotherms with an inflection point. \textit{Right}: the $p-t$ coexistence plot. The black line shows the locus of critical points, i.e. a line of second-order phase transitions   known as a `lambda' line in the context of superfluidity. We use the term `superfluid' BH to follow the analogy with superfluid $^4$He, where a lambda line separates the normal fluid/superfluid phases.}
\label{cl_allsecondorder}
\end{figure}

\begin{figure}[htp]
\includegraphics[scale=0.25]{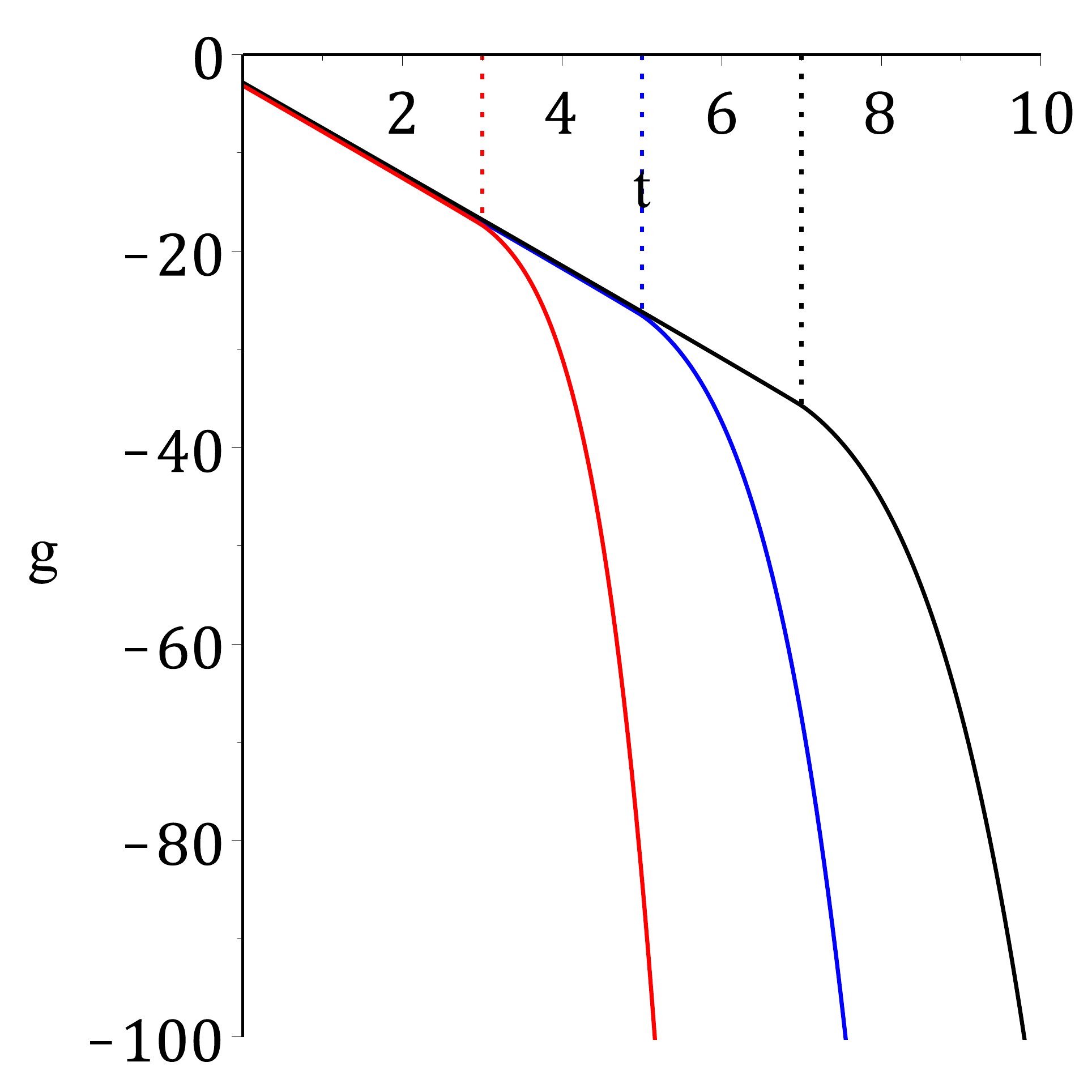}
\includegraphics[scale=0.25]{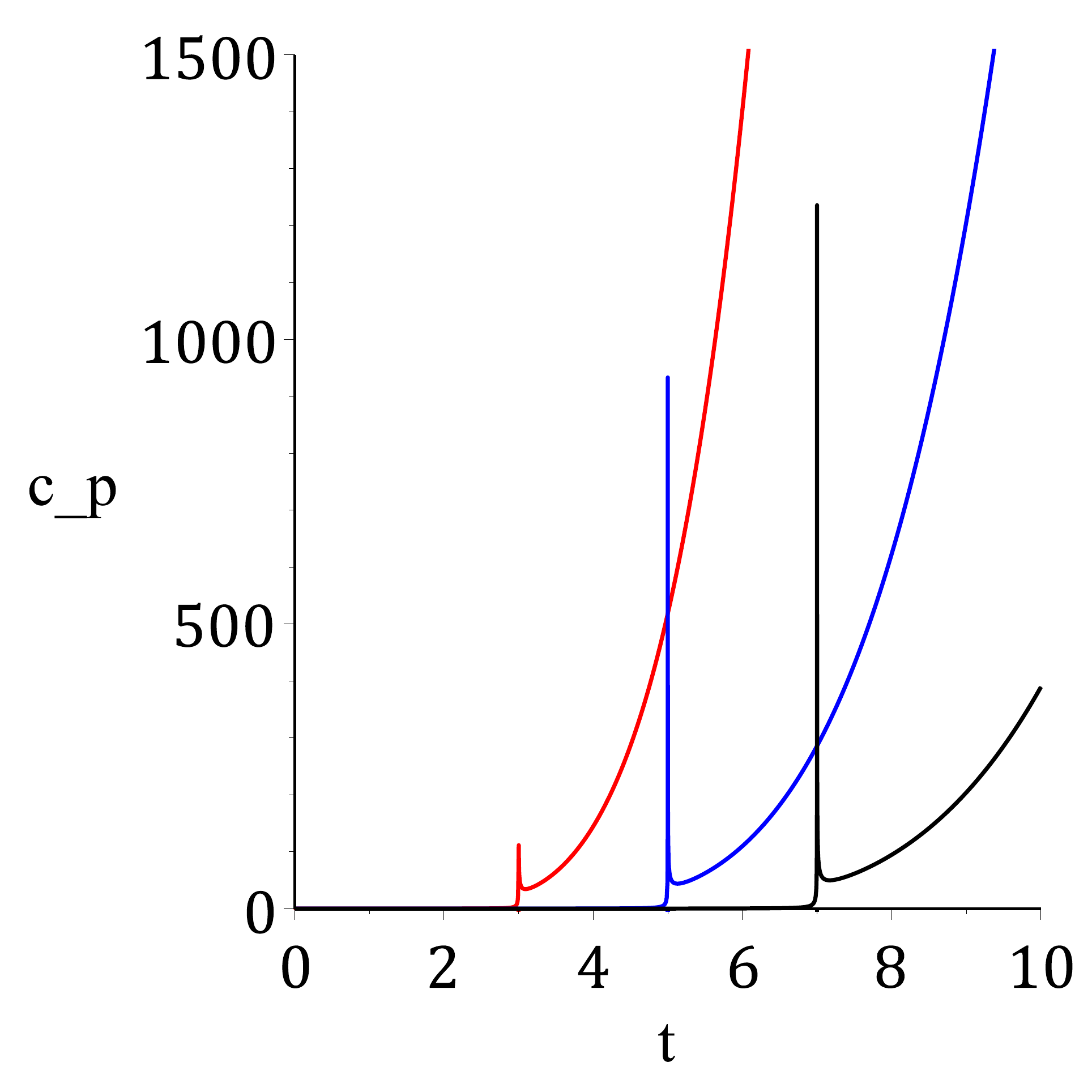}
\includegraphics[scale=0.25]{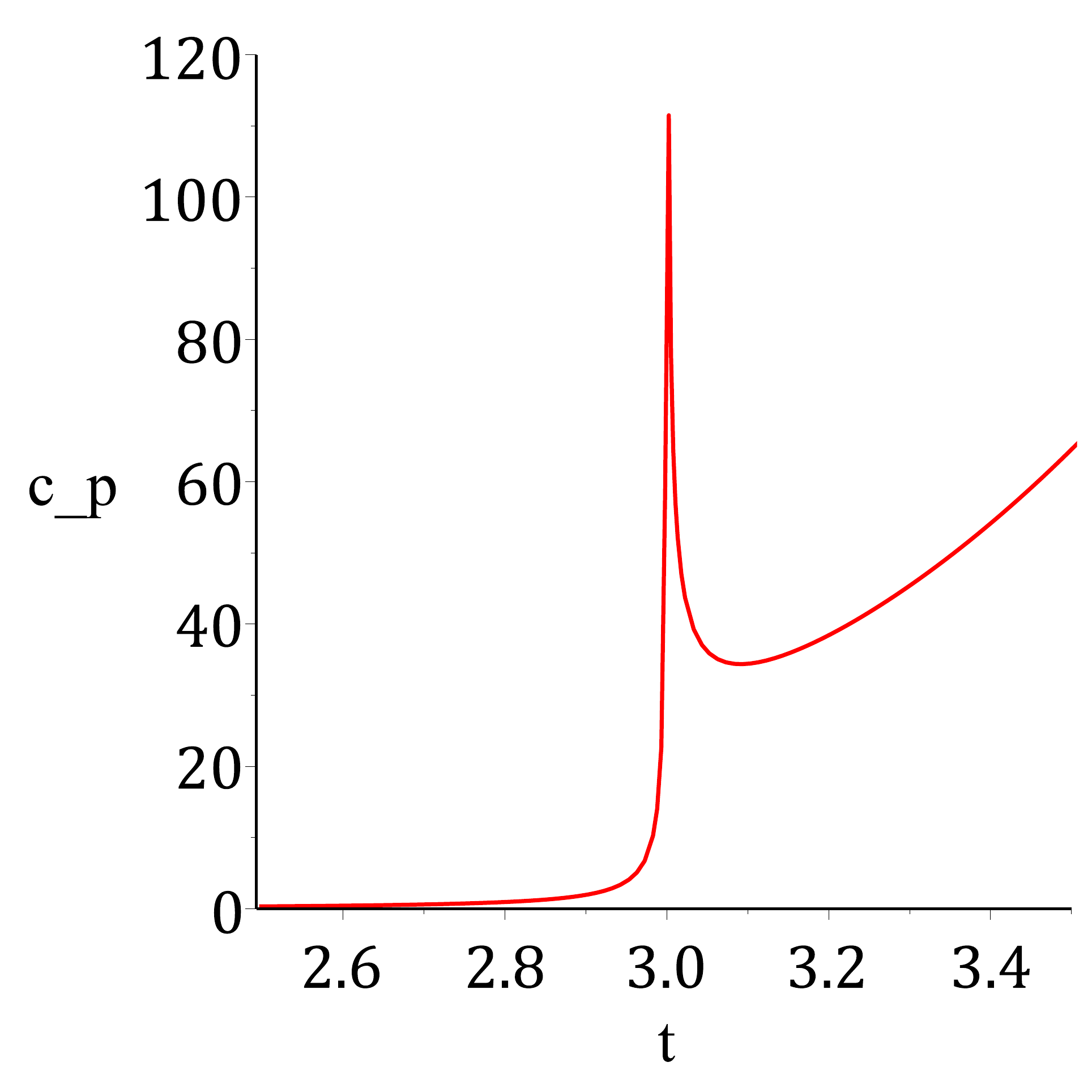}
\caption{\textbf{The Gibbs free energy and specific heat for $d=7$ hyperbolic black holes with infinitely many critical points}: \textit{Left}: the Gibbs free energy diagram for various pressures, chosen such that the critical points occur at $t_c=3,5,7$ which corresponds to red, blue and black curves respectively. \textit{Center}: a plot of specific heat capacity $c_p=-t\frac{\partial^2 g}{\partial t^2}$. Note the divergence in specific heat capacity at $t_c=3,5,7$, indicating a second order phase transitions at these temperatures. \textit{Right}: the zoomed-in version for the case where divergence occurs at $t=3$. This plot bears a strong resemblance to the `lambda phase transition' between fluid/superfluid for $^4$He. }
\label{cl_diverginggibbs}
\end{figure}

However something very interesting happens for $\alpha=\sqrt{5/3}$. Recall that any critical point must satisfy
\be 
\frac{\partial p}{\partial v}=\frac{\partial^2 p}{\partial v^2}=0
\ee 
If we make the choice of parameters
\be 
\alpha=\sqrt{5/3}, \, \, h=\frac{300}{7\pi}(15)^{\frac{1}{4}}, \, \, q=10\sqrt{\frac{3}{\pi}}
\ee
then we find that the solution is given by $v_c=15^{1/4}$ \textit{at arbitrary temperature!} Putting all of these values together into the equation of state, we find that the critical pressure is given by,
\be\label{weirdpressure}
p_c\Bigr|_{v=v_c}=\left[\frac{8}{225}(15)^{\frac{1}{4}}\right]t_c+\frac{37\sqrt{15}}{210\pi}
\ee
where $t_c$ is the critical temperature and is a free parameter.  Therefore, we now have: (1) \textit{infinitely many critical points given by $v_c=15^{1/4}$ at arbitrary temperature}, and (2) the critical pressure scales linearly with critical temperature. The $p-v$ diagram (shown in Figure~\ref{cl_allsecondorder}) shows that \textit{every isotherm is a critical isotherm}, while the phase diagram shows that there is a linear locus of critical points, thus we have a situation where there is no first-order phase transition but a \textit{continuous line of second-order phase transitions}. In Figure~\ref{cl_allsecondorder} we show the isotherms and phase diagram and in Figure~\ref{cl_diverginggibbs} we display the Gibbs free energy  and the specific heat i.e. $c_p = -t \, \partial^2 g/\partial t^2$.  In these latter plots, we see that for a given pressure, at the temperature which satisfies Eq.~\eqref{weirdpressure}, the specific heat diverges, signaling a second order (continuous) phase transition.

Lines of second-order phase transitions occur in condensed matter systems and correspond to, for example, fluid/superfluid transitions~\cite{RevModPhys.73.1}, superconductivity~\cite{superconductor}, and paramagentism/ferromagnetism transtions~\cite{Pathria2011401}. 	Building on the black hole/van der Waals fluid analogy~\cite{Kubiznak:2012wp}, the natural interpretation here is that this second order phase transition between large/small black holes corresponds to a fluid/superfluid type transition.   The resemblance to the fluid/superfluid $\lambda$-line transition of $^4$He (cf. Figure~(2) of \cite{Hennigar:2016xwd}) is striking.  In each case, a line of critical points separates the two phases of fluid where specific heat takes on the same qualitative ``$\lambda"$ structure.  Of course, the phase diagram for helium is more complicated, including solid and gaseous states as well.  This is to be expected since helium is a complicated many body system, while our black hole solutions are comparatively simple being characterized by only four numbers: $v, h, q$ and $\alpha$. However, it is remarkable that with so few parameters we can capture the essence of the $\lambda$-line. Unfortunately, most of the interesting properties of a superfluid are either dynamical or require a full quantum description to understand (see, e.g. \cite{basicSuperfluids, RevModPhys.71.S318} for an introduction and review).  Here we do not have access to a model of the underlying quantum degrees of freedom, and so cannot explore these properties at a deeper level.

In fact we can generalize this result to arbitrary dimension in $d\geq 7$. Starting from the equation of state in arbitrary dimensions given in Eq.~\eqref{lovelockeos}, and considering the values of external parameters
\be 
\begin{aligned}
\alpha=\sqrt{5/3},\,\,\, h=\frac{4(2d-5)(d-2)v_c^{d-6}}{\pi d(d-4)},\,\,\, q^2=\frac{2(d-1)(d-2)v_c^{2d-10}}{\pi(d-10)},
\end{aligned}
\ee
and fixing $\sigma=-1$ i.e. hyperbolic horizon topology, we still obtain a line of second-order phase transitions because there is a continuum of critical points given by
\be 
v_c=15^{1/4},\,\,\, p_c=\left[\frac{8}{225}(15)^{3/4}\right]t_c+\frac{\sqrt{15}(11d-40)(d-1)(d-2)}{900\pi d},\,\,\, t_c\in \mathbb{R}
\ee
It is natural to wonder if there is any pathological behaviour hiding behind the scenes here.  To explore this we consider the Kretschmann scalar evaluated on the horizon,
\be 
K = R_{abcd}R^{abcd} = \left[\left(\frac{d^2 f}{dr^2} \right)^2 + \frac{2(d-2)}{r^2} \left(\frac{df}{dr} \right)^2 + \frac{2(d-2)(d-3)}{r^4} \right]_{r=r_+} \, .
\ee 
The first derivative of $f$ is clearly finite for any finite temperature, so we need only consider $f''$.  For simplicity we consider the case  $d=7$ where  the superfluid solution was first observed~\cite{Hennigar:2016xwd}. Expanding near $v_c, p_c$ we see that,
\be 
f''(v_c + dv, p_c + dp) =  200 \pi^2 t_c^2 + \frac{172 \pi (15)^{1/4}}{75} t_c +  \left[750(15)^{1/4} \pi^2 t_c + \frac{129 \pi}{2}  \right] dp - \frac{172 \sqrt{15} \pi}{75} t_c dv \, .
\ee
This is completely finite both at the critical point and near it; there are no curvature singularities associated with this thermodynamic behaviour.  For thoroughness, we have also examined the explicit solution to the field equations in detail.  Outside the horizon the metric function is well-behaved and the Kretschmann scalar is everywhere finite.  Within the horizon, the metric function extends all the way to $r=0$, but develops an infinite first derivative at finite $r$ for large enough $p$.  The latter is not new behaviour nor is it in any way fatal---similar behaviour occurs for charged Gauss-Bonnet and cubic Lovelock black holes, but there the metric function actually terminates at this point (cf. top right plot of Figure~\ref{gaussmetric}).  Thus, there is nothing pathological about the curvature invariants of these black holes. Furthermore, both the Gibbs free energy and the temperature have smooth expansions near each critical point.  These results are not particularly surprising --- the values of $v_c$ and $\alpha$ used here \textit{do not} correspond to the thermodynamic singularity.  The specific heat is positive (indicating thermodynamic stability) and the positivity of entropy, ghost, and tachyon conditions are all satisfied here.  We note that due to the value of $\alpha = \sqrt{5/3} < \sqrt{3}$, the result is only valid for black holes of the Lovelock branch since the Gauss-Bonnet and Einstein branches do not exist for this value of $\alpha$.

The $\lambda$-line is a line of critical points, and it would be desirable to determine the critical exponents along this line.  The standard method of computing the critical exponents ultimately fails in this case, as we will now explicitly highlight.
 To see this, suppose we proceed to calculate the critical exponents in the naive way (we do this for $d=7$ for concreteness). We can expand the equation of state near any one of the infinitely many critical points to obtain,
\be\label{p-isocrit2} 
\frac{p}{p_c} = 1 + \frac{112 (15)^{1/4} \pi t_c}{112 (15)^{1/4} \pi t_c +  555} \tau - \frac{280 (15)^{1/4} \pi t_c}{112 (15)^{1/4} \pi t_c +  555 } \tau \omega^3 - \frac{140 (85 + 2 \pi (15)^{1/4} t_c)}{112 (15)^{1/4} \pi t_c +  555}  \omega^3 \, .
\ee
We find that $\alpha=0$ governs the behaviour of the specific heat at constant volume near the critical point and $\delta = 3$ governs the behaviour of $|p-p_c| \propto |v-v_c|^\delta$ along any critical isotherm. For $\beta$, we evaluate the following expressions:
\begin{align} 
\frac{p}{p_c}\Big|_{\omega=\omega_1} &= \frac{p}{p_c}\Big|_{\omega=\omega_2} \, , \nn\\
0 &= \int_{\omega_1}^{\omega_2} \omega \frac{d (p/p_c)}{d\omega}
\end{align} 
These equations can only be solved by the trivial solution, $\omega_1=\omega_2$,  or if $\tau$ is constrained to when
\be 
\tau = -\frac{15^{1/4}}{30 \pi t_c} \left[85 \sqrt{15} + 2 \pi (15)^{3/4} t_c\right]
\ee
with $\omega_1$ and $\omega_2$ free.  This latter case is not a sensible solution since in the limit of $\tau \to 0$, we are left with a negative $t_c$, while for the trivial solution we have that the order parameter, $\eta = v_c( \omega_2 - \omega_1)$ vanishes, suggesting that $\beta = 0$.  This result is unchanged by the inclusion of higher order terms in the expansion of the equation of state.    The exponent $\gamma$ governs the behaviour of the isothermal compressibility near criticality, 
\be 
\kappa_T = - \frac{1}{v} \frac{\partial v}{\partial P} \Big|_T \propto |\tau|^{-\gamma} \, .
\ee
Computing this for the above expansion we find,
\be 
\kappa_T = \frac{1}{420 \omega^2} \left[\frac{112 \pi (15)^{1/4} t_c + 555}{2 \pi (15)^{1/4} t_c \tau + 2 \pi  (15)^{1/4} t_c + 85 } \right]
\ee
which in the limit of the critical point is independent of $\tau$, suggesting that $\gamma = 0$.  Therefore, by this argument, it seems that each critical point on this line of criticality is characterized by the critical exponents
\be 
\alpha = 0\, , \quad \beta = 0 \, , \quad \gamma = 0 \, , \quad \delta = 3 \, .
\ee
These critical exponents trivially satisfy the Widom relation,
\be 
\gamma = \beta(\delta-1)
\ee
but violate the Rushbrooke inequality since
\be 
\alpha + 2 \beta + \gamma \not\ge 2 \, .
\ee

While these are manifestly `non-standard' critical exponents, these bizarre results signal that something has gone wrong in the approach.  The problem lies in the assumption that the pressure is still the ordering field.  Here, this is not the case---changing the pressure merely changes the temperature at which the second order phase transition occurs.  Thus, pressure is no longer the appropriate ordering field, a situation similar to that in liquid $^4$He at the $\lambda$ line \cite{RevModPhys.73.1}.  To calculate valid critical exponents, the correct ordering field, $\Theta$,  must be identified and in our case there are three options for $\Theta$: $q$, $h$ or $\alpha$.  It turns out that the obtained critical exponents are the same regardless of which choice is made, but the electric charge $q$ is in some sense the most natural choice since it is easy to imagine adjusting $q$ by throwing charged material into the black hole. To calculate the critical exponents, we proceed as usual, expanding the ordering field near any of the critical points in terms of $\tau$ and $\omega$.  We find,
\be 
\frac{\Theta}{\Theta_c} = 1 - A \tau + B \tau \omega   - C \omega^3  + {\cal O}(\tau \omega^2, \omega^4) \, ,
\ee 
where the values of $A, B, C$ depend on both the pressure, $p$, and will be different (but non-zero) depending on which choice is made for the ordering field.  This expansion yields the following critical exponents,
\be 
\tilde \alpha = 0\, , \quad \tilde \beta = \frac{1}{2} \, , \quad \tilde \gamma = 1 \, , \quad \tilde \delta = 3 \, ,
\ee
which govern the behaviour of the specific heat at constant volume, $C_V \propto |\tau|^{-\tilde\alpha}$, the order parameter $\omega \propto |\tau|^{\tilde\beta}$, the susceptibility/compressibility $(\partial \omega /\partial \Theta)|_{\tau} \propto |\tau|^{-\tilde \gamma}$ and the ordering field $|\Theta-\Theta_c| \propto |\omega|^{\delta}$ near a critical point.  These results coincide with the mean field theory values, which agree with those for a superfluid in a $d > 5$ (cf. Table I of \cite{RevModPhys.73.1}).

\begin{figure}[htp]
\centering
\includegraphics[width=0.5\textwidth]{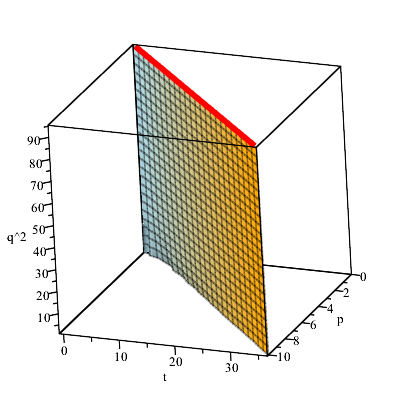}
\caption{{\bf Line of criticality in $(p,t,q)$ space:} If the space of thermodynamic parameters is enlarged, the line of critical points in the $p-t$ plane (the bold red line here) can be thought of as the critical line at which a surface of first order phase transitions terminates. For each constant $p$ slice, this is a first order small/large black hole phase transition as temperature increases.}
\label{qtp_space_3d}
\end{figure}

One way to visualize this result is that the line of critical points in the $p-t$ plane represents a line where a surface of first order phase transitions terminates in some larger space $(p,t, \Theta)$.  Our calculation of critical exponents then represents the behaviour of the system as the line of criticality is approached, not in $(p,t)$ space, but rather in this larger space.  We highlight this in Figure~\ref{qtp_space_3d} for the case $(p,t,q^2)$.  

We can go a bit further and analyse under what situations these type of $\lambda$-lines can be expected for black holes.  The necessary feature of our result is that the conditions for a critical point are satisfied irrespective of the temperature.  Thus, consider a general black hole equation of state of the form,
\be 
P = a_1(r_+, \varphi_i) T + a_2(r_+, \varphi_i)
\ee
where $\varphi_i$ represent additional constants in the equation of state (here they would correspond to $\alpha, q$ and $h$).  Our condition will be met provided the following holds:
\be 
\frac{\partial a_1}{\partial r_+} = 0 \, , \quad  \frac{\partial^2 a_1}{\partial r_+^2} = 0 \, , \quad \frac{\partial a_2}{\partial r_+} = 0 \, , \quad \frac{\partial^2 a_2}{\partial r_+^2} = 0 \,,   
\ee
with both $a_1$ and $a_2$ non-trivial.  From this perspective, it makes sense that we found the behaviour that we did: here we have four equations, and in the case considered in this paper there are a total of four variables.  It is natural then to wonder if other systems exhibit this behaviour.  We have checked this for the rotating black hole of $5d$ minimal gauged super-gravity~\cite{Chong:2005hr} which has four parameters, but have found that no solution to the above equations exist.  Furthermore, in the case of higher order Lovelock gravity with electric charge (but not hair) solving the four equations forces $a_1 = 0$. Hence such a line of critical points does not occur (or, if something similar does, it happens under a different configuration).  It is possible that the superfluid transition could take place in five-dimensions when coupling the scalar field to the quasi-topological density, along the lines considered in~\cite{Chernicoff:2016qrc}.  Thus it remains an interesting line of future work to determine for what other black hole solutions these superfluid-like transitions can occur.

\section{Conclusion}

In this paper we have studied hairy black holes within Lovelock gravity, focusing explicitly on Gauss-Bonnet and cubic Lovelock gravity.  We began this study by examining the AdS vacua of the theory and constraining the coupling constants such that the graviton is not a ghost/tachyon in these backgrounds.  It was found that hyperbolic black holes always satisfy these conditions, but for a non-zero, constant scalar field the spherical black holes require coupling of the scalar field to cubic (or higher) terms to meet this condition.  As a result, for the coupling constants required for spherical black holes in $D< 7$, the theory suffers from the ghost instabilities unless the scalar field vanishes for a constant curvature vacuum.  This could be remedied in $D=5$ (by coupling to the quasi-topological term, for example), but for $D=6$ there is no obvious cure to the problem for non-vanishing scalar field.

We then considered the thermodynamics of these hairy black holes in Gauss-Bonnet gravity. We recover a number of previously found results, such as van der Waals behaviour, reentrant phase transitions, and triple points. In addition to established thermodynamic behaviour for Gauss-Bonnet gravity, we find new behaviour such as virtual triple points.  These correspond to a limiting case of an ordinary triple point with a critical point occurring directly on a first order coexistence line.

In considering the thermodynamics of hairy black holes in cubic Lovelock gravity, we have found two particularly interesting results.  First, we have clarified the connection between isolated critical points and thermodynamic singularities.  In previous work, isolated critical points have been encountered only as very special results, always occurring for massless black holes at the thermodynamic singularity.  The addition of scalar hair permits a family of isolated critical points which occur for black holes away from the thermodynamic singularity.  The key requirement for the existence of such a critical point appears to be the merging of van der Waals and reverse van der Waals behaviour, and therefore has nothing to do with the thermodynamic singularity at all.

Most interestingly, we have found for these hairy black hole solutions the first example of a black hole $\lambda$-line: a line of second order (continuous) phase transitions.  This phase transition bears an interesting resemblance to the superfluid phase transitions which occur, for example, in liquid $^4$He, leading to the name `superfluid black holes' for those which exhibit this behaviour.  We have found that a necessary condition for a black hole $\lambda$-line is three external parameters in the black hole equation of state.  Hence, we have found them to occur for black holes in third (or higher) order Lovelock gravity with scalar hair and electric charge.  However, this is by no means a sufficient condition and determining further examples of black holes that exhibit this behaviour remains an interesting problem for future work.  

\section*{Acknowledgments}
We extend thanks to Julio Oliva for helpful correspondence. E. Tjoa thanks Nanyang Technological University's CN Yang Scholars Programme for financial support.  This work was supported in part by the Natural Sciences and Engineering Research Council of Canada.  
  
\bibliography{LBIB2}
\bibliographystyle{JHEP}

\end{document}